\documentclass[10pt,openany,a4paper]{article}
\usepackage{geometry,color,framed}
\usepackage{amsmath,amssymb,mathtools,slashed,xcolor,mdframed}
\usepackage[vcentermath,enableskew]{youngtab}
\usepackage{cancel, pifont,multirow}
\usepackage{tikz}
\usetikzlibrary{positioning}
\usetikzlibrary{decorations.text}
\usepackage{amstext}
\usepackage{graphicx}
\usepackage[section]{placeins}
\usepackage{verbatim}
\usepackage{bm}
\usepackage{caption, subcaption, float}
\usepackage[colorlinks=true, 
            linkcolor=blue, citecolor=blue,
            menucolor = red,
            urlcolor=blue,linktoc=all]{hyperref}
\makeatletter

\newcommand{\beq}{\begin{eqnarray}}
\newcommand{\eeq}{\end{eqnarray}}
\newcommand{\non}{\nonumber\\}

\newcommand{\ben}{\begin{enumerate}}
\newcommand{\bei}{\begin{itemize}}
\newcommand{\eni}{\end{itemize}}
\newcommand{\enn}{\end{enumerate}}
\newcommand{\ra}{\rightarrow}
\newcommand{\rr}{\right}
\newcommand{\lf}{\left}

\newcommand{\xdownarrow}[1]{%
  {\left\downarrow\vbox to #1{}\right.\kern-\nulldelimiterspace}
}

\def\a{{\alpha}}
\def\b{{\beta}}
\def\g{{\gamma}}
\def\G{{\Gamma}}

\def\O{{\Omega}}
\def\L{{\Lambda}}

\def\d{{\delta}}

\def\l{{\lambda}}
\def\m{{\mu}}

\def\s{{\sigma}}
\def\t{{\theta}}

\usepackage{tikz}
\usetikzlibrary{positioning}
\makeatother
\title{Exploring new variational quantum circuit ansatzes \\for solving $SU(2)$ matrix models}
\author{H. L. Dao\footnote{espoirdujour1162@gmail.com}}
\date{\today}
\date{\today}

\begin{document} 
\maketitle
\begin{abstract}
In this work, we explored and experimented with new forms of parameterized quantum circuits to be used as variational ansatzes for solving the bosonic and supersymmetric $SU(2)$ matrix models at different couplings using the Variational Quantum Eigensolver (VQE) algorithm.  Working with IBM \texttt{Qiskit} quantum computing platform, we show that two types of quantum circuits named \texttt{TwoLocal} and \texttt{EvolvedOperatorAnsatz} can outperform the popular  \texttt{EfficientSU2} circuits which have been routinely used in the recent quantum physics literature to run VQE. With their more customizable constructions that allow for more flexibility beyond choosing the types of parameterized rotation gates, both types of new circuit ansatzes used in this work have led to performances that are either better than or at least comparable to \texttt{EfficientSU2} in the setting of $SU(2)$ matrix models. In particular, in the strong coupling regime of the bosonic model, both \texttt{TwoLocal} and \texttt{EvolvedOperatorAnsatz} circuits provided a better approximation to the exact ground state, while in the supersymmetric model,  shallow \texttt{EvolvedOperatorAnsatz} circuits, with a small number of parameters, attained a comparable or even better performance compared to the much deeper \texttt{EfficientSU2} circuits with around 8 to 9 times more parameters. The results of this work demonstrate conclusively the potential of \texttt{TwoLocal} and \texttt{EvolvedOperatorAnsatz} quantum circuits as efficient new types of variational ansatzes that should be considered more frequently in future VQE studies of quantum physics systems.
\end{abstract}

\tableofcontents
\section{Introduction}
In recent years, there has been a steadily growing interest in the problem of quantum simulation of different systems in high energy physics, especially with the increasingly more accessible quantum computing resources (either in the form of actual quantum computers or quantum simulators) offered by various industrial quantum computing platforms such as IBM-\texttt{Qiskit} \cite{ibm-qiskit} and Google-Cirq \cite{google-cirq}, among others. An important class of examples include the simulation of $\phi^4$ scalar quantum field theory following the seminal works  \cite{JPL-1}, \cite{JPL-2}, \cite{JPL-3}, which opened up new directions that have been investigated in more recent works such as \cite{hardy-2407}, \cite{Zemlevskiy-2411}. The 2021 Snowmass review \cite{snowmass-21} provides an extensive overview on the topics of quantum simulation for quantum field theories. 
Some other interesting examples discuss the quantum simulation of dark energy and dark matter \cite{de-dm-qc}, the simulation of different types of black holes  \cite{qc-bh-2202}, \cite{qc-bh-2207}, the simulation of matrix models \cite{{string-qc-2011}} and quantum field theories with holographic duals  \cite{syk-qc}, \cite{syk-qc2}, the simulation of superconformal quantum mechanics \cite{sqm-qc-2201} on quantum computers. 
\\\\
Among the growing literature of quantum simulation of high energy physics, the work \cite{prx-22} that explored three different approaches involving quantum computing, deep learning and Lattice Monte Carlo to solve the bosonic and minimally supersymmetric $SU(2)$ matrix models is of particular interest to us. In the quantum computing approach, the authors of \cite{prx-22} reported promising results obtained by using a type of IBM \texttt{Qiskit} quantum circuits called \texttt{EfficientSU2} \cite{rinaldi-github}, involving parameterized rotation $R_Y$ and $R_YR_Z$ gates, as variational ansatzes to run the Variational Quantum Eigensolver (VQE) algorithm \cite{vqe-15} \cite{vqe-18}, \cite{vqe-20}, \cite{vqe-21} in order to estimate the ground state energies of the truncated $SU(2)$ matrix models at certain Fock space cutoff $\Lambda$ at four different coupling values. While the  energies calculated by VQE showed a good agreement with the exact ground state energies, the authors of \cite{prx-22} stated that the variational forms of their quantum circuit ansatzes were not specifically optimized for the problem of matrix models, which subsequently might  have led to the larger observed deviation from the exact energies at the strong coupling compared to the weak coupling regime. 
\\\\
Inspired by \cite{prx-22} and the need to identify some better forms of variational quantum circuits that perform well in the strong as well as the weak coupling regimes within the setting of matrix models,  we aim to explore and experiment with additional types of IBM \texttt{Qiskit} quantum circuits in this work. Compared to \cite{prx-22} in which the ansatzes were fixed to be of only two possible forms (\texttt{EfficientSU2} \cite{effsu2} with either $R_Y$ or $R_YR_Z$ parameterized gates), here, we adopt a more ansatz-centric standpoint. In particular, we constructed and experimented with multiple variants of new types of quantum circuit ansatzes called \texttt{TwoLocal} \cite{twolocal} and \texttt{EvolvedOperatorAnsatz} \cite{evop} from  IBM \texttt{Qiskit}, in addition to using multiple variants of \texttt{EfficientSU2} beyond those already introduced in \cite{prx-22}.  For the bosonic $SU(2)$ matrix model at Fock space cutoffs $\L=2$ and $\L=4$ at four coupling values $\l=0.2, 0.5, 1.0, 2.0$, we consistently obtained better performances from \texttt{TwoLocal} and \texttt{EvolvedOperatorAnsatz}  compared to \texttt{EfficientSU2}. When using the results reported in \cite{prx-22}, which use deeper versions of \texttt{EfficientSU2} circuits (than the ones in ours), as benchmarks,  our best results always turned out to be closer to the exact energy values than those reported in \cite{prx-22}. In the supersymmetric case, working only with shallow \texttt{EvolvedOperatorAnsatz} circuits at a small number of parameters,  we obtained results that outperformed those reported in \cite{prx-22} for $\l=0.5$ and $\l=2.0$, while for $\l=0.2$ and $\l=1.0$, our results were quite close to but not as good as those of \cite{prx-22}, which were obtained using much deeper \texttt{EfficientSU2} circuits with 8-9 times larger in terms of the numbers of parameters. With these results, we highlight the promising potential of \texttt{TwoLocal} and \texttt{EvolvedOperatorAnsatz} quantum circuits as new types of variational ansatzes that should be considered more often in future quantum simulation research.  
\\\\
This rest of this paper is organized as follows.
\bei
\item  Section \ref{mm} collects some brief and relevant facts about the $SU(2)$ bosonic (Section \ref{sec-mm-bos}) and supersymmetric matrix models (Section \ref{sec-mm-sup}). 
\item
Section \ref{vqe-s} summarizes the basics of VQE and describes in detail the three components that are essential to VQE. In particular, Section \ref{vqe-estimator} describes the estimator used to simulate the quantum measurements of the Hamiltonian expectation values. Section \ref{qc-ans} discusses in detail the three types of variational quantum circuit ansatzes used to run VQE algorithm for all the experiments in this work. These include \texttt{EfficientSU2} in  \ref{qc-es2}, \texttt{TwoLocal} in \ref{qc-tl} and \texttt{EvolvedOperatorAnsatz} in \ref{qc-evop}. Section \ref{vqe-opt} describes the basics of various types of classical optimizers (\ref{vqe-opt-basics}) and the VQE experiments used to select the optimizers that would be used throughout this work (\ref{vqe-opt-select}). An overview of the whole section is presented in \ref{sec-vqe-summ}.
\item 
 Section \ref{sec_qc_su2_l2} presents the main results of applying the quantum circuit ansatzes introduced in section \ref{qc-ans} to the $SU(2)$ bosonic matrix model at Fock space cutoff $\L=2$.  Within this section, we first present the results obtained by using \texttt{TwoLocal} and \texttt{EfficientSU2} in \ref{sec-l2-es2-tl}, followed by the results obtained by using \texttt{EvolvedOperatorAnsatz} in \ref{sec-l2-evop}, followed by a comparison of the results in this work with those reported in \cite{prx-22} in  \ref{sec_su2_l2_comparison}.
\item Section \ref{sec_qc_su2_l4} presents the results for the case of $SU(2)$ bosonic matrix model at Fock cutoff $\L=4$. This follows the same structure as Section \ref{sec_qc_su2_l2} in which the VQE results obtained by \texttt{EfficientSU2} and \texttt{TwoLocal} ansatzes are first presented in \ref{sec-l4-es2-tl}, followed by the results obtained by using \texttt{EvolvedOperatorAnsatz}  in \ref{sec-l4-evop}, followed by a comparison of all types of ansatzes including the results of \cite{prx-22} in  \ref{sec_su2_l4_comparison}.
\item Section \ref{sec-bmn} presents the results for the case of supersymmetric $SU(2)$ matrix model at Fock cutoff $\L=2$. The VQE results obtained by using \texttt{EvolvedOperatorAnsatz} are presented in \ref{sec-sl2-evop} followed by a comparison of these results against those from \cite{prx-22} in \ref{sec_su2_sl2_comparison}
\item
Section \ref{concl} closes the paper with a summary and some concluding remarks. 
\item
The appendices \ref{sec-L2-full-res}, \ref{sec-L4-full-res}, \ref{sec-L2-bmn-full-res} contain the supplementary material consisting of the convergence curves and the full results from all VQE experiments corresponding to the three truncated $SU(2)$ matrix models (bosonic $\L=2$ in \ref{sec-L2-full-res-1}, \ref{sec-L2-full-res-2}, $\L=4$ in \ref{sec-L4-full-res-1}, \ref{sec-L4-full-res-2} and supersymmetric $\L=2$ in Section \ref{sec-L2-bmn-full-res}).
The appendix \ref{sec-tl-vs-es2} includes the convergence curve plots showing the direct comparisons between the performances of \texttt{TwoLocal} and \texttt{EfficientSU2} circuits, variant by variant, for the cases of bosonic $SU(2)$ models at Fock space cutoffs $\L=2$ (\ref{sec-L2-tl-vs-effsu2}) and $\L=4$ (\ref{sec-L4-tl-vs-effsu2}).
\eni
The Python codes used to construct the quantum circuits and carry out the VQE experiments for this work can be found at the GitHub repository: \href{https://github.com/lorrespz/matrix_model_quantum_computing_vqe}{https://github.com/lorrespz/matrix\_model\_quantum\_computing\_vqe}.\\
We make use of standard Python libraries like \texttt{numpy}, \texttt{pandas}, \texttt{matplotlib} in addition to the specialized \texttt{Qiskit} libraries \texttt{qiskit}, \texttt{qiskit\_aer}, \texttt{qiskit\_algorithms}. 
\section{Matrix models}\label{mm}
Matrix models occupy an important place in string theory, since they are the results of the dimensional reduction of super Yang-Mills  (SYM) theory from higher spacetime dimensions down to just the time dimension \cite{0607-kim-park}. Given the essential role of strongly-coupled $SU(N)$ SYM theory (in the large $N$ limit) as the dual of a weakly coupled supergravity theory in the celebrated AdS/CFT correspondance, various tests of AdS/CFT have been carried out using different versions of SYM, including versions in which the SYM theories are dimensionally reduced to some supersymmetric matrix models. Some notable examples of these tests includes the Monte-Carlo simulation of quantum black holes using matrix model as done in the works \cite{mm-bh-16}, \cite{mm-black-hole} (see also the work \cite{mm-bh-thermo} in which the authors study the thermodynamics of BMN supersymmetric matrix model at strong t'Hooft coupling using the gravity dual).
\\\\
In this section, we only briefly summarize some pertinent facts about matrix models with the practical aim being the derivation of the Hamiltonian to be used in the VQE algorithm. A longer and more detailed discussion of $SU(N)$ matrix models can be found in  \cite{0607-kim-park} and \cite{prx-22}.
\subsection{Bosonic matrix models}\label{sec-mm-bos}
The Hamiltonian of a bosonic $SU(N)$ matrix model in the operator formalism is given by
\beq
\hat H = \text{Tr}\lf[\frac{1}{2}\hat P_I^2 -\frac{m^2}{2} \hat X_I^2 + \frac{g^2}{4}[\hat X_I, \hat X_J]^2\rr]\,,\label{H_su}
\eeq
where $I=1,\ldots, D$ labels the number of matrices. The momentum $\hat P_I$ and position $\hat X_I$ operators can be written in terms of the $(N^2-1)$ $SU(N)$ generators $\tau_\a$ (with $\a = 1, \ldots, N^2-1$ labeling the adjoint representation of $SU(N)$) as
\beq
\hat P_I = \sum_{\a = 1}^{N^2-1} P^\a_I \tau_\a, \hspace{10mm} \hat X_I =\sum_{\a =1}^{N^2-1} \hat X^\a_I \tau_\a. \label{P_X_decomp}
\eeq
The $SU(N)$ generators $\tau_\a$,  normalized as $\text{Tr}(\tau_\a \tau_b) = \delta_{\a\b}$, obey the commutation relations
\beq
[\tau_\a, \tau_b] = f_{\a\b\g}\tau_\g\,,
\eeq
where $f_{\a\b\g}$ are the structure constants of $SU(N)$ group. 
The canonical commutation relation of the Hamiltonian (\ref{H_su}) is
\beq
\lf[ \hat X_{I\a}, \hat P_{J\b}\rr] = i\d_{IJ} \d_{\a\b}\,. \label{can_com}
\eeq
Note that the Hamiltonian (\ref{H_su}) and the canonical commutation relation (\ref{can_com}) are invariant under the $SU(N)$ transformations
\beq
\hat X_I \ra \O \hat X_I \O^{-1},  \hspace{10mm} \hat P_I \ra \O \hat P_I \O^{-1}\,.
\eeq
Using (\ref{P_X_decomp}), the Hamiltonian (\ref{H_su}) can be written as
\beq
\hat H = \sum_{\a, I}\lf(\frac{1}{2}\hat P^2_{I\a} + \frac{m^2}{2} \hat X_{I\a}^2\rr)
+ \frac{g^2}{4}\sum_{\gamma, I, J}\lf( \sum_{\a,\b} f_{\a\b\g} \hat X^\a_I \hat X^\b_J\rr)^2 \,.\label{H_su_ia}
\eeq
in which the total number of (bosonic) degrees of freedom is $D\times (N^2-1)$. 
To use quantum computing, the Hamitonian of the system of interest must be a finite-dimensional matrix of even dimensions. For this purpose, one often uses the discrete Fock space representation involving the creation and annihilation operators in terms of which the Hamiltonian is written.
So, by using the definition of the creation and annihilation operators in terms of the position and momemtum operators
\beq
\hat a^\dagger_{I\a} = \sqrt{\frac{m}{2}}\hat X_{I\a} -\frac{i\hat P_{I\a}}{\sqrt{2m}},
\hspace{10mm}
\hat a_{I\a} = \sqrt{\frac{m}{2}}\hat X_{I\a} + \frac{i\hat P_{I\a}}{\sqrt{2m}}, \label{an_cr_op}
\eeq
and the number operator $\hat n_{I\a} = \hat a^\dagger_{I\a} \hat a_{I\a}$, the Hamiltonian (\ref{H_su_ia}) can be written as
\beq
\hat H = m \sum_{\a,I}\lf(\hat n_{I\a} +\frac{1}{2} \rr) + \frac{g^2}{16m^2} \sum_{\gamma, I, J}\lf(\sum_{\a, \b} f_{\a\b\g}(\hat a_{I\a} + \hat a^\dagger_{I\a})(\hat a_{J\b} + \hat a^\dagger_{J\beta}) \rr)^2\,. \label{H_main}
\eeq
For each $(I,\a)$ mode, the Fock vacuum $|0\rangle_{I\a}$ satisfies
\beq
\hat a_{I\a} |0\rangle_{I\a} = 0
\eeq
and the excited states $|n\rangle_{I\a}$ are created by applying the creation operator $\hat a^\dagger_{I\a}$ on the vacuum state $|0\rangle_{I\a}$
\beq
|n\rangle_{I\a} = \frac{1}{\sqrt{n!}}(\hat a^\dagger_{I\a})^n |0\rangle_{I\a}
\eeq
The Fock vacuum  of the matrix model is the tensor product of each individual $I\a$ mode $|0\rangle = \otimes_{I\a} |0\rangle_{I\a}$. 
Next, we must truncate the system to retain excitations only up to a certain cutoff $\L$ so that the system can be simulated on a quantum computer. This leads to the following definition of the truncated creation, annihilation  and number operators
\beq
\hat a^\dagger_\text{truncated} = \sum_{n=0}^{\L-2} \sqrt{n+1} |n+1\rangle\langle n|, \qquad
\hat a_\text{truncated} = \sum_{n=0}^{\L-2} \sqrt{n+1} |n\rangle\langle n+1|, \qquad
\hat n_\text{truncated} = \sum_{n=0}^{\L-1} n|n\rangle \langle n|
\eeq
In this work, as in the quantum computation part of \cite{prx-22}, the choice of $N=2$ and $D=2$ is made, which leads to the group being $SU(2)$ with $D\times (N^2-1) = 6$ bosonic degrees of freedom.  The  Fock space cutoff $\L$ is taken to be $\L=2$ and $\L=4$. The first case leads to a $2^6=64$-dimensional Hilbert space while the second case leads to a  $4^6=2^{12}=4096$-dimensional Hilbert space. The matrix representation for the $\hat a_i$ annilation operator for the case of $\L=2$ and $\L=4$ is:
\beq
\L=2: &&
\hat a_i = \underbrace{\mathbf{1}_2 \otimes  \ldots\otimes  \mathbf{1}_2}_{\text{$(i-1)$ times}}\otimes \begin{pmatrix} 0 & 1 \\ 0 & 0 \end{pmatrix}\otimes \underbrace{\mathbf{1}_2 \otimes \ldots\otimes  \mathbf{1}_2}_{\text{$(6-i)$ times}}  \label{bos-op-2}\\
\L=4: &&
\hat a_i = \underbrace{\mathbf{1}_4  \otimes  \ldots\otimes  \mathbf{1}_4}_{\text{$(i-1)$ times}}  \otimes \begin{pmatrix} 0 & 1 & 0 & 0 \\ 0 & 0 & \sqrt{2} & 0\\ 0 & 0 & 0 & \sqrt{3}\\ 0 & 0 & 0 & 0 \end{pmatrix}\otimes \underbrace{\mathbf{1}_4 \otimes \ldots \otimes \mathbf{1}_4}_{\text{$(6-i)$ times}} 
\eeq
where $\mathbf{1}_{2}$ and $\mathbf{1}_4$ are the  $2\times 2$ and $4\times 4$ identity matrix, respectively.
\subsection{Supersymmetric matrix models} \label{sec-mm-sup}
The supersymmetric matrix model of interest to us is the mass-deformed version of the one originating from the dimensional reduction of the minimal 3D $SU(N)$ SYM theory  \cite{0607-kim-park}, \cite{prx-22}, with the following
Hamiltonian
\beq
H = \text{Tr}\lf(\frac{1}{2}\hat P_I^2 -\frac{g^2}{4} [\hat X_I, \hat X_J]^2 + \frac{g}{2}\bar \psi \G^I[\hat X_I, \psi] -\frac{3i}{4}\mu \bar\psi \psi + \frac{\mu^2}{2}\hat X_I^2\rr) - (N^2-1)\mu\,,\label{H_su-2}
\eeq
where, as in the bosonic case,  $I= 1,\ldots,D$ labels the number of matrices. $\G^I$ is the $D$-dimensional gamma matrices and  $\psi$ is a two-component Majorana fermion, which can be written as
\beq
\psi = \begin{pmatrix}\zeta \\ i\zeta^\dagger \end{pmatrix}\,.
\eeq
In Eq.(\ref{H_su-2}),  $\mu$ is the mass term that is added to the massless theory resulting from the dimensional reduction of the 3D minimal SYM, and the presence of the term $-(N^2-1)\mu$ forces the ground state energy to be exactly zero.
\\\\
When $N=2$, the minimal $SU(2)$ BMN supersymmetric matrix model has 6 bosonic degrees of freedom ($\hat X_{I\alpha}$ where $I=1,2$ and $\alpha = 1,2,3$) and 3 fermionic degrees of freedom $\zeta_\alpha$, which obey the anticommutation relation $\lf\{\zeta^\dagger_\a, \zeta_\b\rr\} = \d_{\a\b}$. The Hamiltonian for this case is \cite{prx-22}
\beq
H &=& \sum_\a \frac{1}{2}\lf(\hat P^2_{1\alpha} +\hat P^2_{2\alpha} + \mu^2\,\hat X^2_{1\alpha}+ \mu^2\hat X^2_{2\alpha} + 3\mu\,\hat\zeta^\dagger_\a \hat\zeta_\a\rr\}\,,
\non
&& + g^2\sum_{\a\neq\b}\hat X_{1\a}^2 \hat X_{2\b}^2 - 2g^2\sum_{\a <\b}\hat X_{1\a} \hat X_{1\b} \hat X_{2\a} \hat X_{2\b}\,,
\non
&& + \frac{ig}{\sqrt{2}}\sum_{\a,\b,\g}\epsilon_{\a\b\g}\lf[ -(\hat X_{1\a} + i\hat X_{2\a})\hat\zeta^\dagger_\b\hat \zeta^\dagger_\g + \lf(-\hat X_{1\a} + i\hat X_{2\a}\rr)\hat \zeta_\b\hat \zeta_\g\rr] - 3\mu\,.
\eeq
The fermion operators $\zeta_\a$ obey the anticommutation relation $\lf\{\zeta_\a, \zeta_\b\rr\} = \d_{\a\b}$. With the Fock space cutoff chosen to be $\L=2$, the fermion operators are constructed using the Jordan-Wigner transformation involving Pauli spin matrices as follows
\beq
\zeta_\a = \underbrace{\s_z\otimes \ldots \otimes\s_z}_{\text{$\a-1$ times}}\otimes \begin{pmatrix} 0 & 0 \\ 1 & 0 \end{pmatrix} \otimes \mathbf{1}_2 \otimes \ldots \otimes \mathbf{1}_2.
\eeq
In this case, the fermionic Hilbert space sector has dimension $2^3$. Together with the bosonic sector, which has dimension $2^6 = 64$, the total Hilbert space has dimension $2^{9} = 512$. The 3 fermionic operators have to be tensored with $\mathbf{1}_{64}$ - the $64\times 64$ identity matrix. The bosonic operators are the same as defined in Eq.(\ref{bos-op-2}), except that they are now tensored with $\mathbf{1}_8$ ($8\times 8$ identity matrix). Explicitly, for the fermionic part, the three annihilation operators are defined as follows
\beq
c_1 &=& \mathbf{1}_{64}\otimes \begin{pmatrix}  0 & 1\\0 & 0\end{pmatrix} \otimes \mathbf{1}_2 \otimes \mathbf{1}_2\,,
\\
c_2 &=& \mathbf{1}_{64}\otimes \s_z \otimes \begin{pmatrix}  0 & 1\\0 & 0\end{pmatrix}\otimes \mathbf{1}_2\,,
\\
c_3 &=& \mathbf{1}_{64}\otimes \s_z \otimes \s_z\otimes \begin{pmatrix}  0 & 1\\0 & 0\end{pmatrix} \,.
\eeq
Note that for the truncated supersymmetric $SU(2)$ model considered here, the ground state energy is close to, but no longer exactly zero. 
\section{Variational Quantum Eigensolver (VQE)} \label{vqe-s}
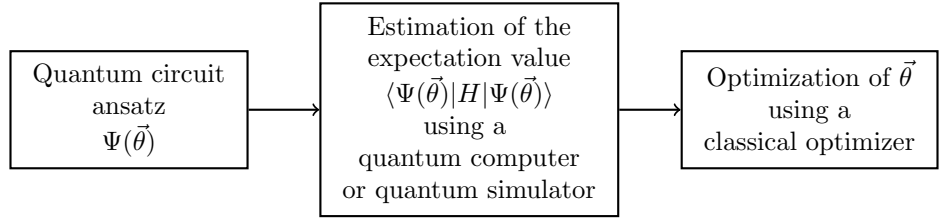
\begin{figure}[H]
\centering
\begin{tikzpicture}[node distance = 4.5cm, thick]%
        \node[draw] (0) {$\begin{array}{c} \text{Estimation of the}\\ \text{expectation value} \\ \langle\Psi(\vec\theta)| H|\Psi(\vec\theta)\rangle\\ \text{using a }\\\text{quantum computer}\\\text{or quantum simulator}\end{array}$};
        \node[draw] (0a) [left of =0]{$\begin{array}{c} \text{Quantum circuit}\\\text{ansatz} \\ \Psi(\vec\theta) \end{array}$};
        \node[draw] (0b) [right of =0]{$\begin{array}{c} \text{Optimization of $\vec\theta$}\\\text{using a}\\\text{classical optimizer} \end{array}$};

        \draw[->] (0a) -- node [right]{} (0);
        \draw[->] (0) -- node [right]{} (0b);
   
    \end{tikzpicture}
    \caption{The main components of VQE:  a quantum circuit ansatz, a quantum device (or a simulator) to estimate the expectation value of the Hamiltonian, and a classical optimizer} \label{vqe-schema}
    \end{figure}

Variational Quantum Eigensolver (VQE) is a popular classical-quantum hybrid algorithm used to estimate the ground state energy of a Hamiltonian system using some form of parameterized quantum circuits as a variational ansatz. Many examples of VQE have been discussed in great detail in the literature \cite{vqe-18}, \cite{vqe-20}, \cite{vqe-21}. Here, for the sake of self-containedness, we will briefly recap some details.
\\\\
Denoting the parameterized quantum circuit  by a unitary operator $\hat U(\vec\theta)$ acting on a collection of qubits initialized to zero\footnote{Some other initializations are possible other than zero.}
\beq
|\mathbf{0}\rangle= \underbrace{|0\rangle \otimes \ldots \otimes |0\rangle}_\text{$n_Q$ times}
\eeq 
where $n_Q$ is the number of qubits, the expectation value of an observable, such as a Hamiltonian $\hat H$, can be measured in terms of a trial wavefunction $\Psi(\vec\theta)$ given by $\Psi(\vec\theta) = \hat U(\vec\theta) \lf|\mathbf{0}\rr\rangle$ as
\beq
\lf\langle \Psi(\vec\theta\lf|\,\hat H\,\rr|\Psi(\vec\theta)\right\rangle \label{qc_M}
\eeq
Using VQE, the ground state energy of the Hamiltonian $\hat H$ is estimated using (\ref{qc_M}) by means of a quantum computer (or a suitable quantum simulator) and is optimized with a classical optimizer. A schematic of the different components of VQE is shown in Fig.\ref{vqe-schema}.
In Eq. (\ref{qc_M}), $\vec\theta$ are the parameters to be varied so that the algorithm returns an upper bound estimate $E_\text{VQE}$ on the exact ground state energy $E_0$:
\beq
E_0 \leq E_\text{VQE} = \frac{\lf\langle \Psi(\vec\theta)| \hat H|\Psi(\vec\theta)\rr\rangle}{\lf\langle \Psi(\vec\theta)|\Psi(\vec\theta)\rr\rangle}\,.
\eeq
 In  order to carry out the estimation involving quantum circuits of the form (\ref{qc_M}), the Hamiltonian has to be written as a sum of the tensor products of Pauli operators, (also called Pauli strings) of the form $Q_1 \otimes Q_2 \otimes \ldots \otimes Q_{n_Q}$ where $Q_i \in \lf(I_2, X, Y, Z\rr)$. If fermionic operators are present, these can be converted to Pauli string operators using the Jordan-Wigner transformation \cite{jordan-wigner}. 
\\\\
 Throughout this work, we perform all VQE experiments exclusively using \texttt{Qiskit}, an IBM quantum computing platform with an extensive suite of libraries for quantum circuits, algorithms, simulators and even access to real quantum hardware (quantum computers with hundreds of qubits) hosted on their cloud servers \cite{ibm-qiskit}. We recall that our settings of interest for all VQE experiments are the following Hamiltonians representing the truncated $SU(2)$ matrix models.
 \bei
\item Bosonic SU(2) model truncated at Fock cutoff $\L=2$: This leads to a model with 6 bosonic modes with $2^6=64$ states and a $64\times 64$-dimensional Hamiltonian - corresponding to 6-qubit circuits used in VQE. 
\item Bosonic SU(2) model truncated at Fock cutoff $\L=4$: This leads to a model with 6 bosonic modes with $4^6=2^{12}=4096$ states and a $4096\times 4096$-dimensional Hamiltonian - corresponding to 12-qubit circuits used in VQE. 
\item Supersymmetric SU(2) model truncated at Fock cutoff $\L=2$: This leads to a model with 6 bosonic modes + 3 fermionic modes with $2^9=512$ states and a $512\times 512$-dimensional Hamiltonian - corresponding to 9-qubit circuits used in VQE. 
\eni
For each of the three cases above, we will look at four different couplings $\lambda = 0.2, 0.5, 1.0, 2.0$ where $\lambda = g^2N$, with $g$ is the actual coupling appearing the in Eqs.(\ref{H_su}), (\ref{H_su-2}), and $N=2$ corresponding to $SU(2)$ group of the matrix model. This leads to four different Hamiltonians per case. In total, there are twelve Hamiltonians $H^\L_\l$ including 8 bosonic Hamiltonians $H^{\L=2,4}_{\l=0.2, 0.5, 1.0, 2.0}$  and 4 supersymmetric Hamiltonians $H^{(S)\,\L=2}_{\l=0.2, 0.5, 1.0, 2.0}$.
 \\\\
While it is desirable to study more complex matrix models such as $SU(2)$ at higher Fock space cutoff $\L$, or matrix models with a larger group $SU(N)$ where $N>2$, we note that such models are drastically more computationally demanding. In general, the number of states for a bosonic $SU(N)$ matrix model (with $N^2-1$ generators) with $d$ bosonic matrices truncated at Fock cutoff $\Lambda$ is $\L^{d(N^2-1)}$. Concretely speaking, when $N=3$, for the $SU(3)$ matrix model with 8 $SU(3)$ generators $\tau_\a$ ($\a=1,\ldots, 8$), the smallest number of bosonic matrices is $d=2$, corresponding to 16 modes $X_{I\a}$ (with $I=1,2$). At the lowest Fock cutoff of $\L=2$, the total number of modes is $2^{16}=65536$ states.  At this level, without access to an actual quantum hardware hosted on a large server, a modern laptop\footnote{such as one with 16Gb - 64Gb RAM (at the time of writing this article)} operating a \texttt{Qiskit} simulator cannot handle this, simply because it will run out of memory before long.
The situation only gets worse: At cutoff $\L=4$, the number of states in the $SU(3)$ matrix model is $4^{16} = 2^{32} = 4.3\times 10^9$. When $N=4$, for $SU(4)$ matrix models, with 15 $SU(4)$ generators $\tau_\a$ with $\a=1,\ldots, 15$, the smallest number of bosonic matrices is $d=2$, corresponding to 30 modes $X_{I\a}$ ($I=1,2$), and at Fock cutoff $\L=2$, the number of states is $2^{30} = 1.07\times 10^9$. Even for the $SU(2)$ models with only 3 generators and 6 bosonic modes at the very least, at Fock space cutoff $\L=8$, the number of states is still $8^6 = 2^{18} = 262144$, which cannot be handled by a modern laptop. The complications arising from the resource intensive nature of the computation with larger and more complex matrix models were also noted in \cite{prx-22} in which the authors chose alternative approaches (rather than quantum computing), such as deep learning involving a classical neural network and lattice Monte Carlo simulation, to deal with $SU(3)$ matrix models.
 \\\\
 In the subsequent sections, we will describe in detail the various components that are integral to the practical implementation of the VQE algorithm in Qiskit.
\subsection{Quantum Circuit Ansatzes} \label{qc-ans}
The first crucial component of VQE that we will focus on is the quantum circuit ansatzes composed of parameterized gates whose parameters can be varied to obtain certain optimized expectation value of the specific Hamiltonian of interest. The main challenge involving quantum circuit ansatzes in VQE is the limited overlap of these ansatzes with the actual quantum states in the corresponding Hilbert space under study, which makes the optimization process rather difficult. A good choice of quantum circuit ansatzes is hence of paramount importance to the overall success of VQE experiments. Two main different approaches exist with respect to the selection of quantum circuit ansatzes, one involving the use of generic, untailored ansatzes chosen for their hardware efficiency for all problem settings, and another involving the use of tailored ansatzes constructed specifically for the particular problem setting of interest. In this work, we will explore both approaches in the context of $SU(2)$ matrix model.
\\\\
One of the building blocks of a quantum circuit ansatz is parameterized rotation gates like $R_X$, $R_Y$, and $R_Z$ given by
\beq
R_X(\theta) = \exp\lf(-i\dfrac{\t}{2}X\rr) &=& \begin{pmatrix} \cos\frac{\t}{2} & -i\sin\frac{\t}{2} \\ \\-i\sin\frac{\t}{2}& \cos\frac{\t}{2}\end{pmatrix}\,,
\non
R_Y(\theta) = \exp\lf(-i\dfrac{\t}{2}Y\rr) &=& \begin{pmatrix} \cos\frac{\t}{2} & -\sin\frac{\t}{2} \\ \\\sin\frac{\t}{2} & \cos\frac{\t}{2}\end{pmatrix}\,,
\non
R_Z(\theta) = \exp\lf(-i\dfrac{\t}{2}Z\rr) &=& \begin{pmatrix} \exp\lf(-i\frac{\t}{2}\rr) & 0\\ \\0& \exp\lf(i\frac{\t}{2}\rr)\end{pmatrix}\,,
\eeq
where $X, Y, Z$ are the Pauli matrices. 
\\\\
Another essential building block comprises the so-called entanglement gates that act on multiple qubits and are used to entangle qubits in the quantum circuits. The most common of such gates are the controlled type of gates, for example the 2-qubit Controlled-$X$ ($CX$) gate (also known as CNOT gate)
\beq
CX(q_0, q_1) =  |0\rangle \langle 0|  \otimes \mathbf{1}_2+ |1\rangle \langle 1|\otimes X \,
= \begin{pmatrix} 1 & 0 & 0 & 0 \\0&1 &0 & 0 \\0 &0&0 &1 \\ 0&0 &1 & 0\end{pmatrix}\,,
\eeq
where $\mathbf{1}_2$ is the 2D identity matrix, and $|0\rangle = (1 , 0)$, $|1\rangle = (0 ,1)$ denote the single qubit state. The parameterized version of the $CX$ gate, the $CRX(\theta)$ gate given by
\beq
CRX(\theta, q_0, q_1) = |0\rangle \langle 0|\otimes \mathbf{1}_2  + |1\rangle \langle 1|\otimes  RX(\theta) = 
\begin{pmatrix} 1&0 &0&0\\0&1&0 &0\\ 0 & 0 & \cos\frac{\theta}{2} & -i\sin\frac{\theta}{2} \\ 0 & 0 & -i\sin\frac{\theta}{2} & \cos\frac{\theta}{2}\end{pmatrix}\,,
\eeq
is also another popular choice. Other choices include $RXX$ (a parameterized 2-qubit $X\otimes X$ rotation gate) given by
\beq
RXX(\theta) = \exp\lf(-i\frac{\theta}{2} \,X\otimes X\rr) \,,
\eeq
$RCCX$ gate (a parameterized simplified Toffoli gate), and $RC3X$ gate (a parameterized simplified 3-controlled Toffoli gate). 
Using some combinations of these gates, we would describe in detail, in subsequent sections, the types of quantum circuits that will be used as variational ansatzes for running all the experiments in this work. All these circuits are implemented in the \texttt{Qiskit} quantum circuit library \texttt{qiskit.circuit.library}\footnote{\href{https://docs.quantum.ibm.com/api/qiskit/circuit_library}{https://docs.quantum.ibm.com/api/qiskit/circuit\_library}}. 
\subsubsection{\texttt{EfficientSU2} circuits} \label{qc-es2}
\texttt{Qiskit} \texttt{EfficientSU2} circuits \cite{effsu2} are hardware efficient quantum circuits that consist of a rotation building block with the default choice being a combination of $R_Y$ and $R_Z$ gates, and an entanglement block with the default choice being $C_X$ gates. This  is the predominant type of circuits used for many recent works in the literature dealing with VQE using \texttt{Qiskit} platform \cite{prx-22}, \cite{de-dm-qc}, \cite{qc-bh-2202}, \cite{qc-bh-2207}. In our experiments, we will vary the rotation block and the scheme of the entanglement block (which can be either `circular' or `full' among other choices). This leads to the eight different \texttt{EfficientSU2} ansatzes listed in Table \ref{effsu2_ans}, which are categorized into eight variants with four different types of rotation blocks and 2 different schemes of entanglement arrangement. The four types of gates in the rotation blocks are $R_Y$, $R_Z$, $R_YR_Z$, $R_YY$, while the two entanglement schemes are either `circular', in which the any qubit in the circuit is entangled with its next nearest neighbor and the last qubit is entangled with the first one, or `full' in which every qubit in the circuit is entangled with every other qubits in the circuit (using only $C_X$ gates). Note that in the work \cite{prx-22} the authors employed 2 variants of \texttt{EfficientSU2}, one consisting of solely $R_Y$ gates in the rotation block and the other consisting of $R_YR_Z$ gates in the rotation block, with the full entanglement scheme for both variants. 
\begin{table}[!ht]
\centering
\begin{tabular}{|l|c|l|cc|}
\hline
\texttt{EfficientSU2} circuits & Parameters & $\begin{matrix} \text{Rotation block} \\ \text{(parameterized)}\end{matrix}$ & $\begin{matrix}\text{Entanglement pattern} \\ \text{(Unparameterized)} \end{matrix}$ &    \\\hline 
\texttt{effsu2\_Ry\_c}\,\, (Fig.\ref{qc_l2_a0a_su2}) & $(d+1)\times n_Q$ &  $R_Y$ & circular &\\
\texttt{effsu2\_Rz\_c}\,\, (Fig.\ref{qc_l2_a0b_su2})  & $(d+1)\times n_Q$ & $R_Z$ & circular &\\
\texttt{effsu2\_RyRz\_c} \,\,(Fig.\ref{qc_l2_a0c_su2}) & $2(d+1)\times n_Q$  & $R_Y$ and $R_Z$ & circular &\\
 \texttt{effsu2\_RyY\_c} \,\,(Fig.\ref{qc_l2_a0d_su2}) & $(d+1)\times n_Q$  &$R_Y$ and $Y$ &circular &\\
 \texttt{effsu2\_Ry\_f}\,\, (Fig.\ref{qc_l2_a1a_su2}) &$(d+1)\times n_Q$  &$R_Y$ & full & \\
  \texttt{effsu2\_Rz\_f}\,\, (Fig.\ref{qc_l2_a1b_su2}) & $(d+1)\times n_Q$  &$R_Z$ & full & \\
   \texttt{effsu2\_RyRz\_f}\,\, (Fig.\ref{qc_l2_a1c_su2})& $2(d+1)\times n_Q$ &$R_Y$ and $R_Z$ & full & \\
    \texttt{effsu2\_RyY\_f} \,\,(Fig.\ref{qc_l2_a1d_su2})& $(d+1)\times n_Q$ &$R_Y$ and $Y$ & full & \\
\hline
\end{tabular}
\caption{Details of the eight variants of \texttt{EfficientSU2} ansatzes used throughout this work. $d$ is the depth of the circuit, and $n_Q$ is the number of qubits in the circuit. }\label{effsu2_ans}
\end{table}
The number of parameters of the \texttt{EfficientSU2} quantum circuits are the same as the number of parameterized gates in the rotation blocks. This amounts to $(d+1)\times n_Q$ parameters where $d$ is the depth of the circuit (the number of repetitions that the basic building block of the circuit is repeated) and $n_Q$ is the number of qubits, for variational forms involving a single type of rotation gates (either $R_Y$ or $R_Z$ or $R_YY$ since $Y$ gate is not parameterized) or $2(d+1)\times n_Q$ for variational forms involving two types of rotation gates ($R_YR_Z$). Given these number of parameters, the scaling property of \texttt{EfficientSU2} for all the 8 variants considered in this work is linear in both the circuit depth $d$ and number of qubits $n_Q$. The variant with the most number of parameters is \texttt{effsu2\_RyRz\_c} and \texttt{effsu2\_RyRz\_f} with $R_YR_Z$ gates in the roation part. 
  \begin{figure*}[h!]
    \centering
    \begin{subfigure}[t]{0.4\textwidth}
        \centering
        \includegraphics[height=1.in]{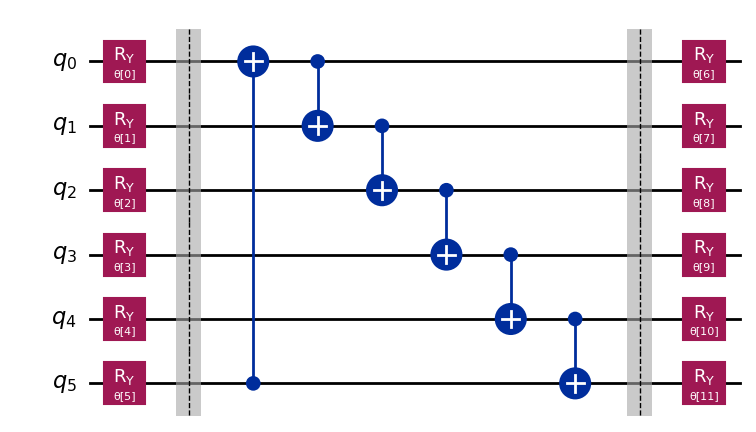}
        \caption{\texttt{effsu2\_Ry\_c}}
        \label{qc_l2_a0a_su2}
    \end{subfigure}%
    \begin{subfigure}[t]{0.4\textwidth}
        \centering
        \includegraphics[height=1.in]{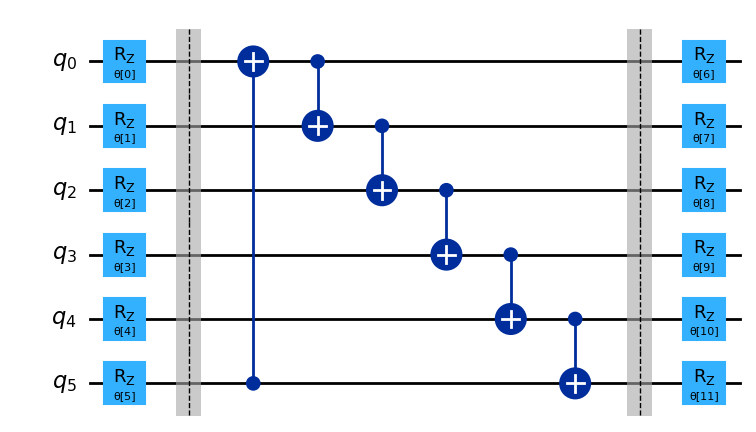}
       \caption{\texttt{effsu2\_Rz\_c}}
\label{qc_l2_a0b_su2}
    \end{subfigure}
\begin{subfigure}[t]{0.4\textwidth}
        \centering
        \includegraphics[height=1.in]{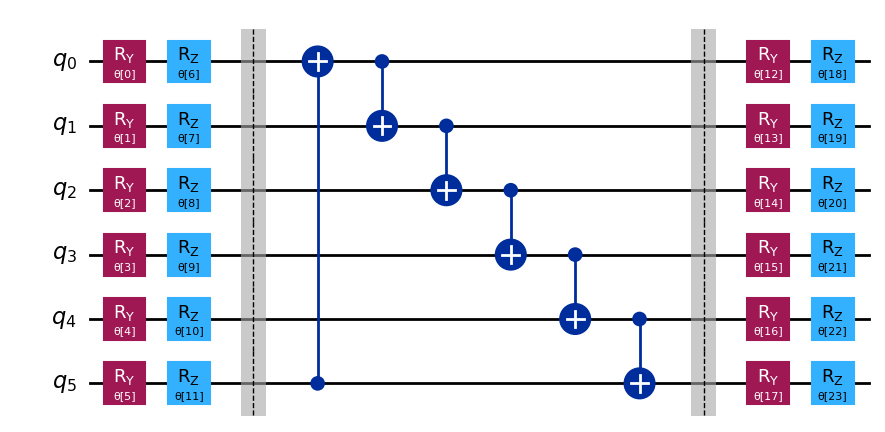}
       \caption{\texttt{effsu2\_RyRz\_c}}
\label{qc_l2_a0c_su2}
    \end{subfigure}
\begin{subfigure}[t]{0.4\textwidth}
\centering
\includegraphics[height=1.in]{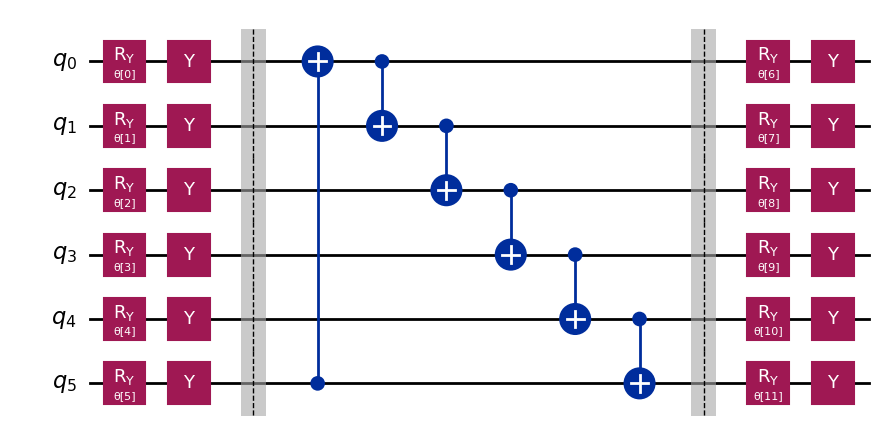}
\caption{\texttt{effsu2\_RyY\_c}}
\label{qc_l2_a0d_su2}
\end{subfigure}
\begin{subfigure}[t]{0.5\textwidth}
\includegraphics[height=1.in]{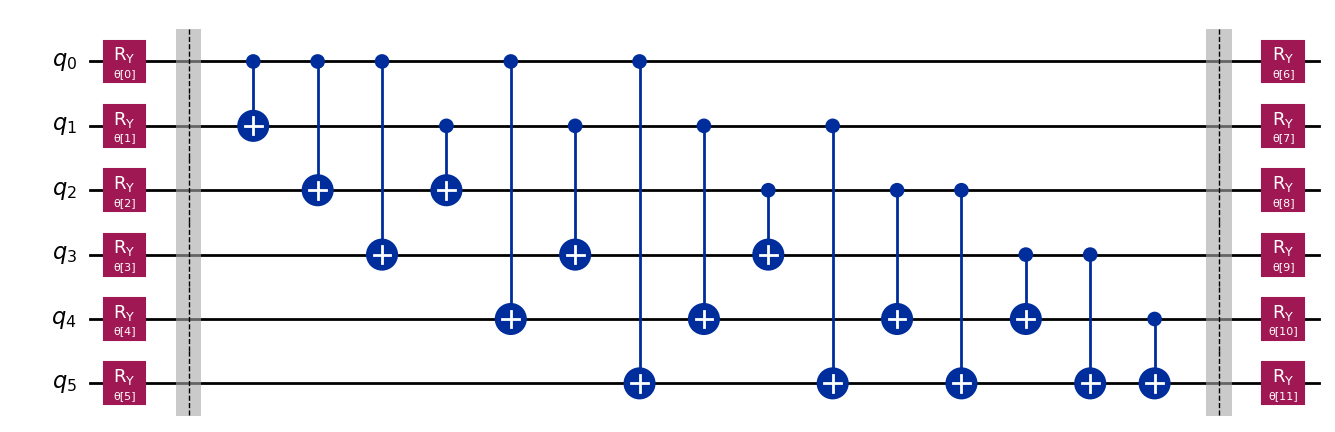}
\caption{\texttt{effsu2\_Ry\_f}}
\label{qc_l2_a1a_su2}
\end{subfigure}
\begin{subfigure}[t]{0.5\textwidth}
\includegraphics[height=1.in]{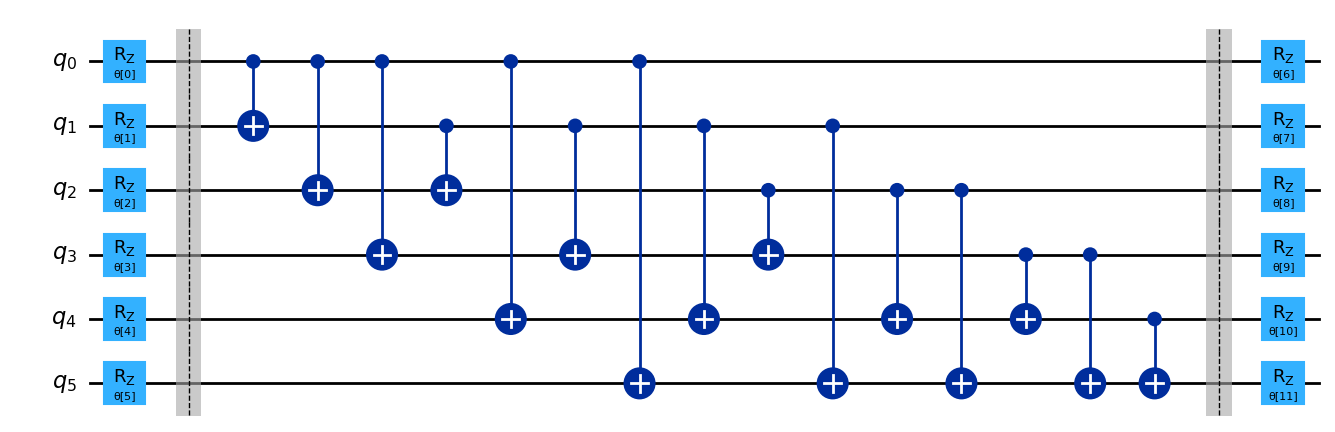}
\caption{\texttt{effsu2\_Rz\_f}}
\label{qc_l2_a1b_su2}
\end{subfigure}
\begin{subfigure}[t]{0.5\textwidth}
\includegraphics[height=1.in]{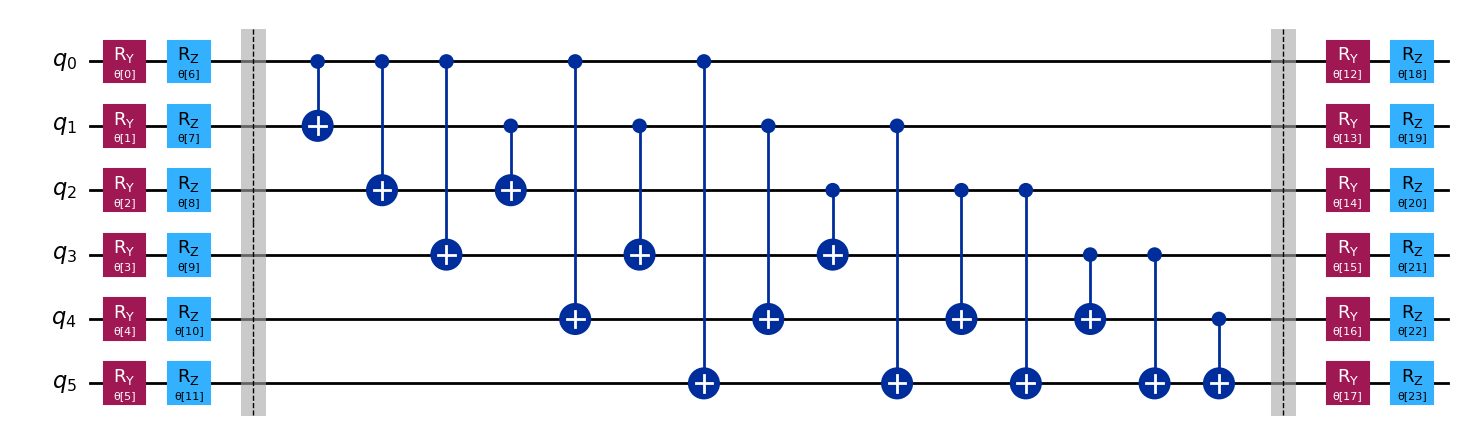}
\caption{\texttt{effsu2\_RyRz\_f}}
\label{qc_l2_a1c_su2}
\end{subfigure}
\begin{subfigure}[t]{0.5\textwidth}
\includegraphics[height=1.in]{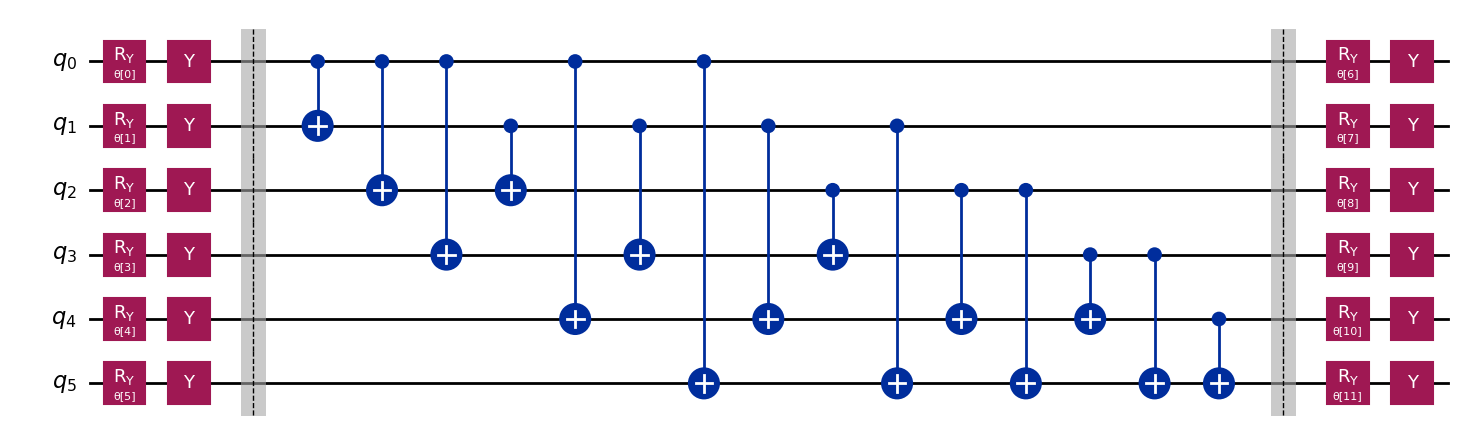}
\caption{\texttt{effsu2\_RyY\_f}}
\label{qc_l2_a1d_su2}
\end{subfigure}
\caption{The eight variants of \texttt{EfficientSU2} ansatzes used throughout this work. Note that for demonstration purpose, we choose the number of qubits to be 6 in the figures above but the actual numbers of qubits are either 6 or 12 depending on the Hamiltonian under study. (a) \texttt{effsu2\_Ry\_c}: $R_Y$ gates and circular entanglement pattern. (b) \texttt{effsu2\_Rz\_c}: $R_Z$ gates and circular entanglement pattern.   (c) \texttt{effsu2\_RyRz\_c}: $R_YR_Z$ gates and circular entanglement pattern. (d) \texttt{effsu2\_RyY\_c}: $R_YY$ gates and circular entanglement pattern. (e) \texttt{effsu2\_Ry\_f}: $R_Y$ gates and full entanglement pattern. (f) \texttt{effsu2\_Rz\_f}: $R_Y$ gates and full entanglement pattern. (f) \texttt{effsu2\_RyRz\_f}: $R_YR_Z$ gates and full entanglement pattern. (g) \texttt{effsu2\_RyY\_f}: $R_YY$ gates and full entanglement pattern}
\label{qc_l2_effsu2}
\end{figure*}
\FloatBarrier
\clearpage
\subsubsection{\texttt{TwoLocal} circuits} \label{qc-tl}
\begin{figure}[!ht]
\centering
\includegraphics[width = .7\textwidth]{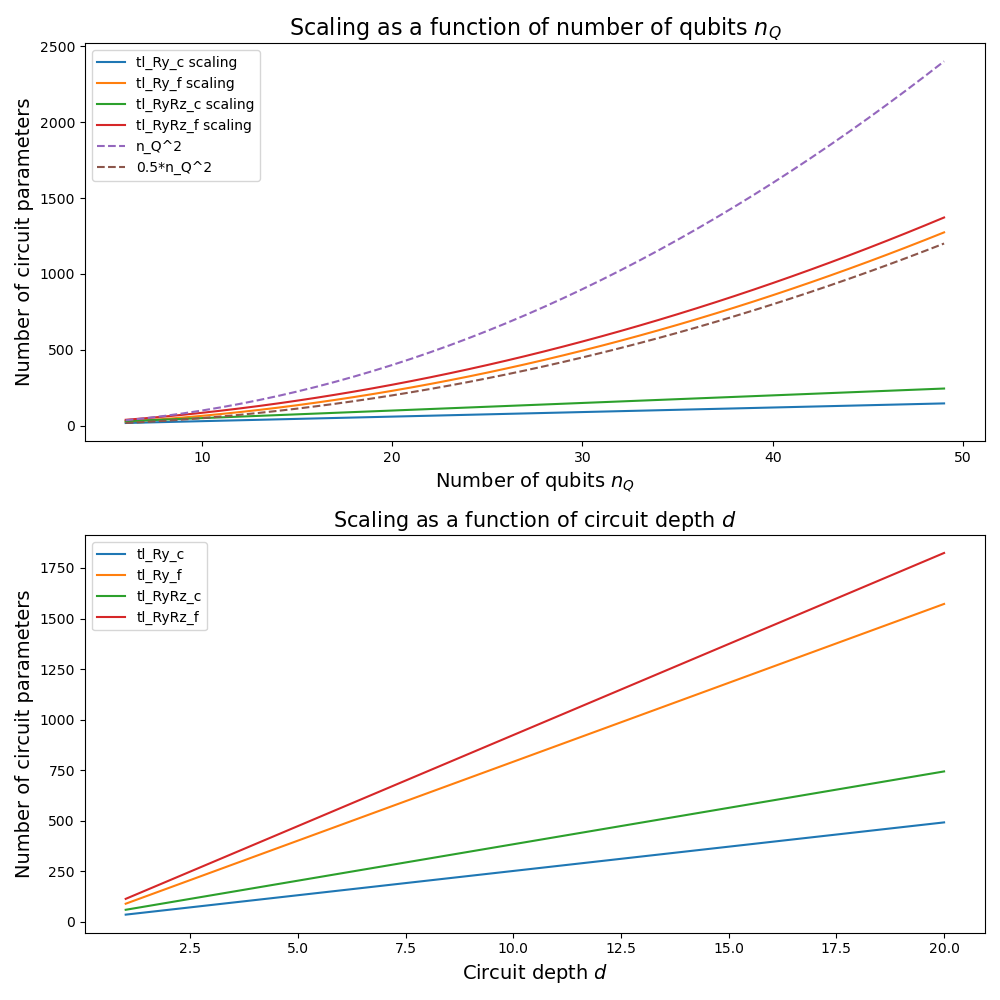}
\caption{Scaling properties of \texttt{TwoLocal} quantum circuits \texttt{tl\_Ry\_c}, \texttt{tl\_Ry\_f}, \texttt{tl\_RyRz\_c}, and \texttt{tl\_RyRz\_f} as a function of the number of qubits $n_Q$ in the range of [6,50] (top figure) at fixed circuit depth $d=1$, and as function of the circuit depth $d$ in the range of [1,20] when the number of qubits is fixed at $n_Q=12$ (bottom figure). As functions of $n_Q$, \texttt{TwoLocal} variants with the full entanglement pattern have a quadratic scaling while \texttt{TwoLocal} variants with the circular entanglement pattern have a linear scaling. As function of the circuit depth $d$, all \texttt{TwoLocal} variants have a linear scaling property.
In the top figure, the two quadratic curves $n^2_Q$ and $0.5 n^2_Q$ are included to compare with the quadratic scaling properties of \texttt{tl\_Ry\_f} and \texttt{tl\_RyRz\_f}, which are shown to be faster than $0.5n^2_Q$ but much slower than $n^2_Q$. 
Furthremore, mote that the scaling properties of \texttt{tl\_RyY\_c} and \texttt{tl\_Rz\_c} are the same as \texttt{tl\_Ry\_c}, while the scaling properties of \texttt{tl\_RyY\_f} and \texttt{tl\_Rz\_f} are the same as \texttt{tl\_Ry\_f}. }\label{tl-scaling}
\end{figure}

\texttt{TwoLocal} circuits \cite{twolocal} have similar structure to, but more general than, \texttt{EfficientSU2} in the sense that they still consist of a rotation block followed by an entanglement block, but there is more freedom in choosing the type of gates in the entanglement block. In particular, we are not limited to the unparameterized $C_X$ but have access to more general gates such as the parameterized $C_{RX}$, $R_{XX}$, $RC2X$ and $RC3X$ gates for entangling the qubits. Analogous to the \texttt{EfficientSU2} case above, we use the eight variants of \texttt{TwoLocal} quantum circuits with the same four types of rotation blocks (consisting of either $R_Y, R_Z, R_YR_Z$ or $R_YY$) and two entanglement schemes (`circular' or `full'), but these \texttt{TwoLocal} circuits employ the parameterized  $C_{RX}$ gates in the entanglement block (see Table \ref{tl_t1_ans}). This increases the number of parameters in \texttt{TwoLocal} quantum circuits but also enhance their expressivity commpared to their \texttt{EfficientSU2} counterparts.
\begin{table}[!ht]
\centering
\begin{tabular}{|l|c|c|l c|}
\hline
\texttt{TwoLocal} circuits & Parameters & $\begin{matrix} \text{Rotation block} \\ \text{(parameterized)}\end{matrix}$ & $\begin{matrix}\text{Entanglement block} \\ \text{(parameterized)} \end{matrix}$ &    \\\hline
\texttt{tl\_Ry\_c} (Fig.\ref{qc_l2_a0a}) & $(2d+1)n_Q $ & $R_Y$ & $CRX$, circular &\\
\texttt{tl\_Rz\_c} (Fig.\ref{qc_l2_a0b}) & $(2d+1)n_Q $ &$R_Z$ & $CRX$, circular &\\
\texttt{tl\_RyRz\_c} (Fig.\ref{qc_l2_a0c})& $(3d+2)n_Q $  &$R_Y$ and $R_Z$ & $CRX$, circular &\\
 \texttt{tl\_RyY\_c} (Fig.\ref{qc_l2_a0d}) & $(2d+1)n_Q $& $R_Y$ and $Y$  &$CRX$, circular &\\
 \hline
 \texttt{tl\_Ry\_f} (Fig.\ref{qc_l2_a1a}) & $\dfrac{1}{2} d\,n^2_Q + \lf(\dfrac{1}{2}d +1\rr)n_Q$ &$R_Y$ & $CRX$, full & \\
  \texttt{tl\_Rz\_f} (Fig.\ref{qc_l2_a1b}) & $\dfrac{1}{2} d\,n^2_Q + \lf(\dfrac{1}{2}d +1\rr)n_Q$ &$R_Z$ & $CRX$, full & \\
   \texttt{tl\_RyRz\_f} (Fig.\ref{qc_l2_a1c})& $\dfrac{1}{2} d\,n^2_Q + \lf(\dfrac{3}{2}d +2\rr)n_Q$  &$R_Y$ and $R_Z$ & $CRX$, full & \\
     \texttt{tl\_RyY\_f} (Fig.\ref{qc_l2_a1d})& $\dfrac{1}{2} d\,n^2_Q + \lf(\dfrac{1}{2}d +1\rr)n_Q$  &$R_Y$ and $Y$ & $CRX$, full & \\
  \hline
 \end{tabular}
\caption{Details of the eight variants of \texttt{TwoLocal} ansatzes used throughout this work. $d$ is the circuit depth, $n_Q$ is the number of qubit in the circuit.}\label{tl_t1_ans}
\end{table}
\begin{figure*}[h!]
\centering
\begin{subfigure}[t]{0.4\textwidth}
        \centering
        \includegraphics[height=1.in]{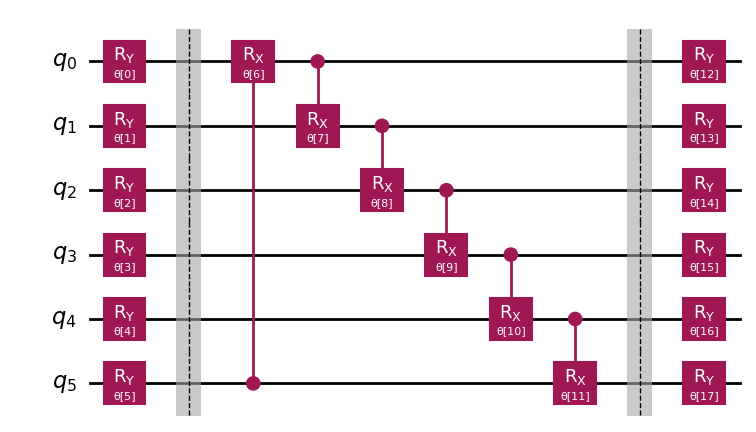}
        \caption{\texttt{tl\_Ry\_c}}
        \label{qc_l2_a0a}
    \end{subfigure}%
    \begin{subfigure}[t]{0.4\textwidth}
        \centering
        \includegraphics[height=1.in]{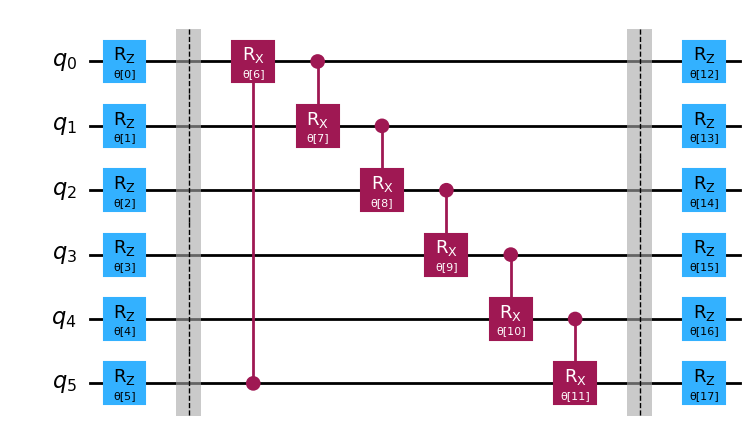}
       \caption{\texttt{tl\_Rz\_c}}
\label{qc_l2_a0b}
    \end{subfigure}
\begin{subfigure}[t]{0.4\textwidth}
        \centering
        \includegraphics[height=1.in]{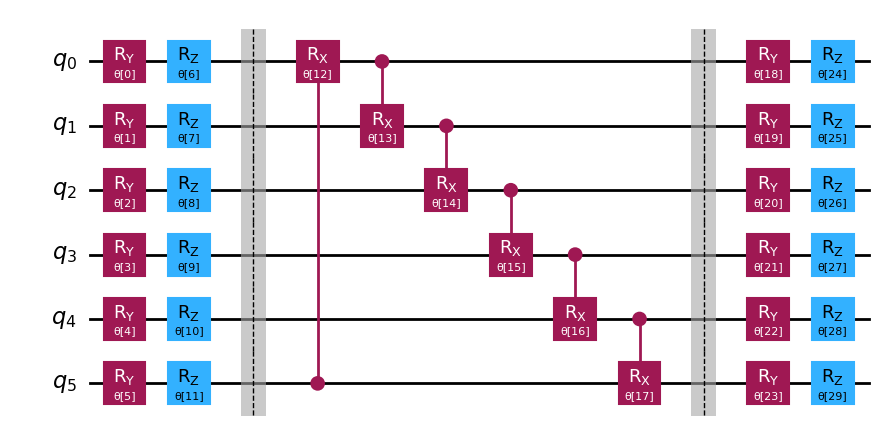}
       \caption{\texttt{tl\_RyRz\_c}}
\label{qc_l2_a0c}
    \end{subfigure}
\begin{subfigure}[t]{0.4\textwidth}
\centering
\includegraphics[height=1.in]{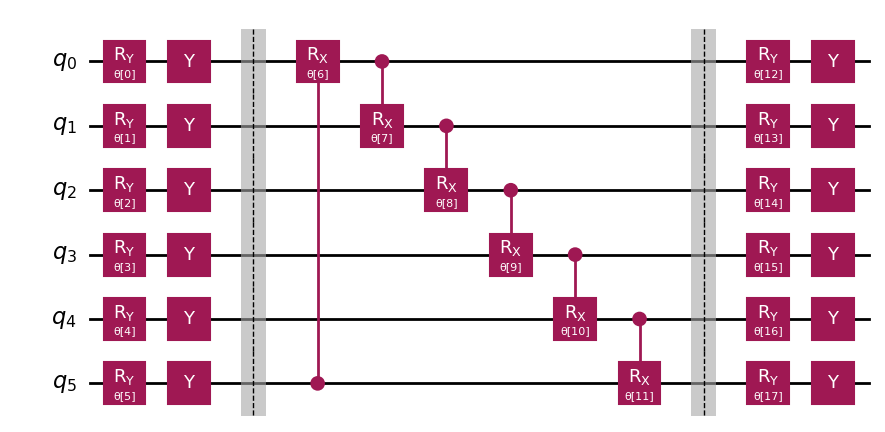}
\caption{\texttt{tl\_RyY\_c}}
\label{qc_l2_a0d}
\end{subfigure}
\begin{subfigure}[t]{0.5\textwidth}
\includegraphics[height=1.in]{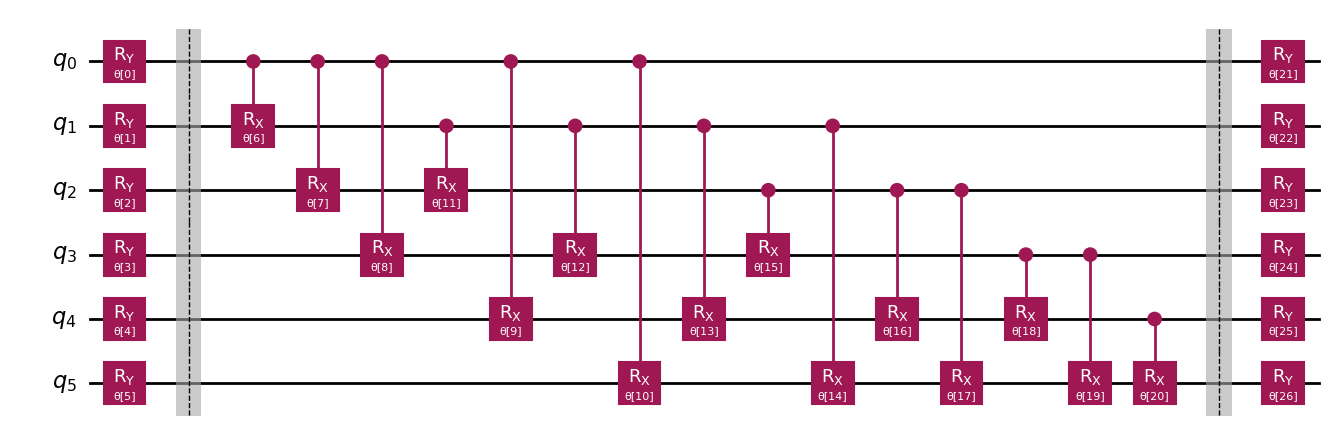}
\caption{\texttt{tl\_Ry\_f}}
\label{qc_l2_a1a}
\end{subfigure}
\begin{subfigure}[t]{0.5\textwidth}
\includegraphics[height=1.in]{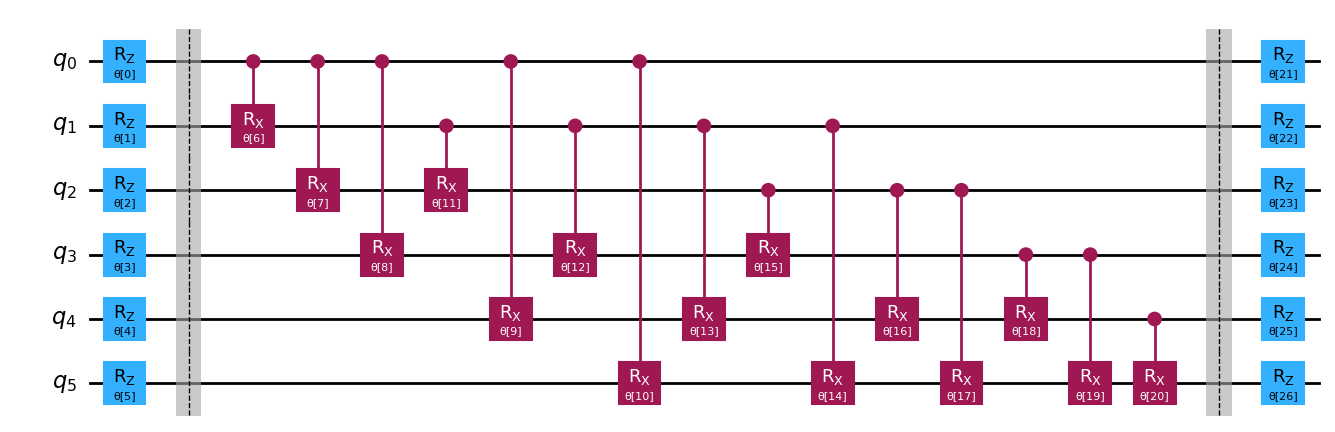}
\caption{\texttt{tl\_Rz\_f}}
\label{qc_l2_a1b}
\end{subfigure}
\begin{subfigure}[t]{0.5\textwidth}
\includegraphics[height=1.in]{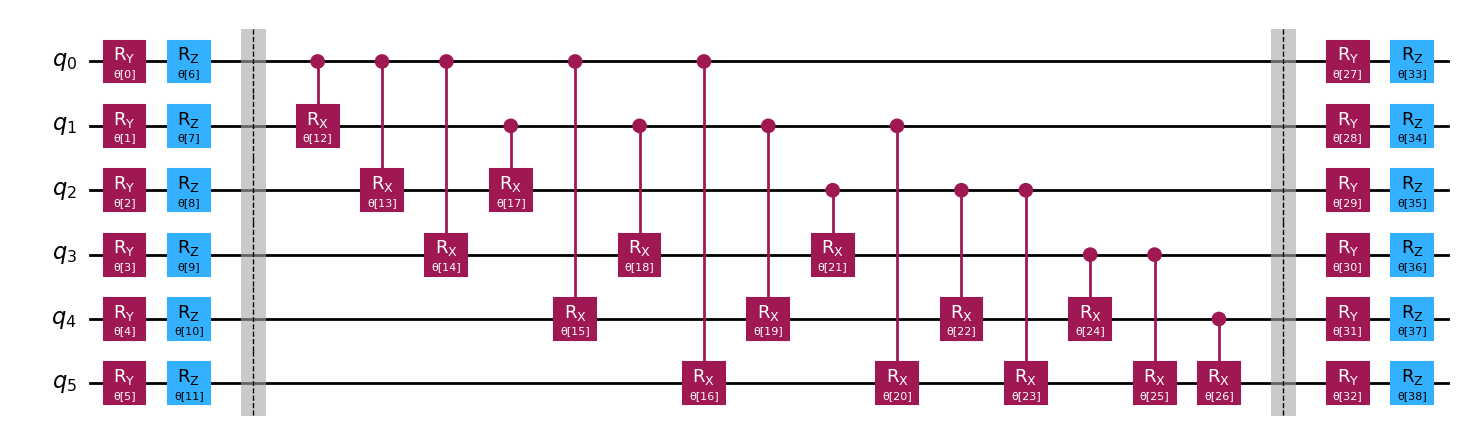}
\caption{\texttt{tl\_RyRz\_f}}
\label{qc_l2_a1c}
\end{subfigure}
\begin{subfigure}[t]{0.5\textwidth}
\includegraphics[height=1.in]{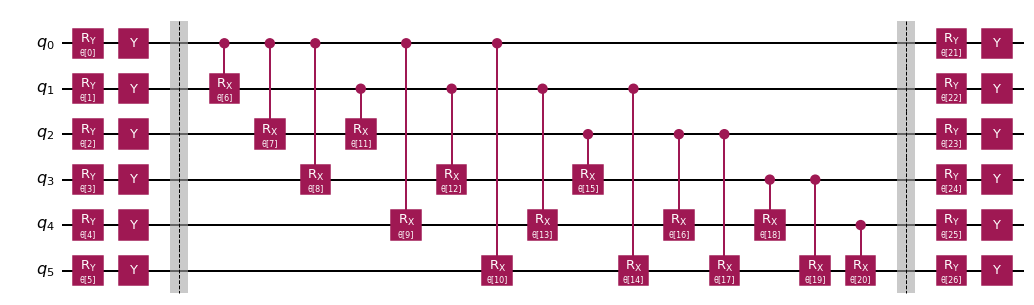}
\caption{\texttt{tl\_RyY\_f}}
\label{qc_l2_a1d}
\end{subfigure}
\caption{The eight variants of \texttt{TwoLocal} ansatzes (with $CRX$ gate in the entanglement block) used throughout this work. Note that for demonstration purpose, we choose the number of qubits to be 6 in the figures above but the actual numbers of qubits are either 6 or 12 depending on the Hamiltonian under study.
 (a) \texttt{tl\_Ry\_c}: $R_Y$ gates and circular entanglement pattern. (b) \texttt{tl\_Rz\_c}: $R_Z$ gates and circular entanglement pattern.   (c) \texttt{tl\_RyRz\_c}: $R_YR_Z$ gates and circular entanglement pattern. (d) \texttt{tl\_RyY\_c}: $R_YY$ gates and circular entanglement pattern. (e) \texttt{tl\_Ry\_f}: $R_Y$ gates and full entanglement pattern. (f) \texttt{tl\_Rz\_f}: $R_Y$ gates and full entanglement pattern. (f) \texttt{tl\_RyRz\_f}: $R_YR_Z$ gates and full entanglement pattern. (g) \texttt{tl\_RyY\_f}: $R_YY$ gates and full entanglement pattern.}
\label{qc_l2_twolocal}
\end{figure*}
In particular, the number of parameters in this type of quantum circuits are the sum of the number of parameterized rotation gates and entanglement gates. A circular entanglement pattern leads to an additional $n_Q$ number of parameters per circuit depth $d$, while a full entanglement pattern leads to an additional $\lf(\sum\limits_{k=1}^{n_Q-1} k\rr) = \frac{1}{2}n_Q(n_Q-1)$ parameters\footnote{Let's consider the term $\sum\limits_{k=1}^{n_Q} k$:
\beq
\lf(\sum\limits_{k=1}^{n_Q} k \rr)&=& 1 + 2 + \ldots + n_Q = \lf[n_Q - (n_Q-1)\rr] +  \lf[n_Q - (n_Q-2)\rr] + \ldots + \lf[n_Q - (n_Q-0)\rr] \non
&=& \underbrace{(n_Q + \ldots + n_Q)}_\text{$n_Q$ times} - \lf[(n_Q-1) + (n_Q-2) + \ldots +1 + 0\rr]
=n_Q^2 - \lf(\sum\limits_{k=1}^{n_Q-1}k\rr)
\eeq
which means
\beq
n^2_Q =\lf(\sum\limits_{k=1}^{n_Q} k\rr) +\lf(\sum\limits_{k=1}^{n_Q-1}k \rr) = 2\lf(\sum\limits_{k=1}^{n_Q-1}k\rr) + n_Q
\,\,\,\,
\Rightarrow \,\, \lf(\sum\limits_{k=1}^{n_Q-1}k\rr)  = \frac{1}{2}n_Q(n_Q-1)\,.
\eeq} per circuit depth $d$.
 Together with the parameters from the rotation gates, which can be either $(d+1)n_Q$ for a single type of rotation gates or $2(d+1)n_Q$ for a double type of rotation gates, the total number of parameters can be moderately large. For example, the variants \texttt{tl\_Ry\_c} and \texttt{tl\_Ry\_f} at depth $d$ have the following total number of parameters
\beq
\texttt{tl\_Ry\_c}: && (d+1)n_Q + d\,n_Q = (2d+1)\,n_Q\,,\non
\texttt{tl\_Ry\_f}: && (d+1)n_Q + d\,\lf(\sum\limits_{k=1}^{n_Q-1}k\rr)= (d+1)n_Q + \frac{1}{2} d\,n_Q(n_Q-1)\non
&& = \,\frac{1}{2} d\,n^2_Q + \lf(\frac{1}{2}d +1\rr)n_Q\,.
\eeq
The exact number of parameters for each of the eight variants of \texttt{TwoLocal} quantum circuits are listed in Table \ref{tl_t1_ans}. Given these number of parameters, the scaling property of \texttt{TwoLocal} quantum circuits for the four variants with circular entanglement pattern is linear in both the circuit depth $d$ and number of qubits $n_Q$, while the scaling property of \texttt{TwoLocal} circuits with full entanglement pattern is quadratic in the number of qubits $n_Q$ but is still linear in $d$, the circuit depth. Compared with \texttt{EfficientSU2} circuits with the same $d$ and $n_Q$, \texttt{TwoLocal} circuits with the circular entanglement pattern can be approximately 1.25-1.5 times larger (when the rotation block is $R_YR_Z$) or 1.5-2 times larger (when the rotation block is $R_Y, R_Z, R_YY$), while \texttt{TwoLocal} circuits with the full entanglement pattern can be several times larger in terms of the number of parameters, depending on the number of qubits $n_Q$ present. Among the \texttt{TwoLocal} variants, those with the full entanglement pattern scale much faster than those with the circular pattern. The variant with the most number of parameters is \texttt{tl\_RyRz\_f} with $R_YR_Z$ rotation gates and full entanglement. For illustration, we plot the number of parameters for \texttt{tl\_Ry\_c}, \texttt{tl\_Ry\_f}, \texttt{tl\_RyRz\_c}, and \texttt{tl\_RyRz\_f} as functions of the circuit depth $d$ and number of qubits $n_Q$ in Fig.\ref{tl-scaling}. In terms of $n_Q$, both \texttt{tl\_RyRz\_f} and \texttt{tl\_Ry\_f} scale quadratically, while \texttt{tl\_Ry\_c} and \texttt{tl\_RyRz\_c} scale linearly. In terms of $d$, all circuits scale linearly. As functions of either $n_Q$ or $d$, \texttt{tl\_RyRz\_f} scales the fastest, followed by \texttt{tl\_Ry\_f}, \texttt{tl\_RyRz\_c} and \texttt{tl\_Ry\_c}.

\FloatBarrier

\subsubsection{\texttt{EvolvedOperatorAnsatz} circuits} \label{qc-evop}
Unlike \texttt{EfficientSU2} and \texttt{TwoLocal} quantum circuits described in the previous sections that are built from parameterized rotation and parameterized/unparameterized controlled type of gates and can serve as generic ansatzes for any VQE problem, \texttt{EvolvedOperatorAnsatz} \cite{evop}, as constructed and used in this work, are quantum circuits tailored to the specific task at hand. 
In general, \texttt{EvolvedOperatorAnsatz} quantum circuits can be written as
\beq
\prod_{r=1}^d \left(\prod^{N_O}_{i=1} \exp\lf\{-i\theta_{i,r}O_i \rr\}\right) \label{eq-ev-op}
\eeq
where $O_i$ is a set of $N_O$ operators $[O_1, \ldots, O_{N_O}]$, $d$ is the depth of the circuit. The $\exp$ term in Eq.\ref{eq-ev-op} is handled using first order Trotterization\footnote{Trotterization is the process in which a term like $U(t)\exp(-iHt)$ representing the evolution operator of a Hamiltonian $H$ can be approximated to order $n$ as \[U(t) \approx \prod_{j=1}^n \prod_{i=1}^k\exp\lf(-iH_i t/n\rr)\] if $H$ can be written as the sum \[H = \sum_{i=1}^k H_i\] where the $H_i$'s do not necessarily commute. Naturally, the larger $n$ is, the better approximation that one gets for $U(t).$}.
\\\\
As written in Eq.(\ref{eq-ev-op}), the number of parameters of the \texttt{EvolvedOperatorAnsatz} circuit is not dependent on the number of qubits $n_Q$, and thus this type of circuit only scales linearly with increasing $d$ but does not scale with increasing $n_Q$, which can be a plus point when $n_Q$ is large.
 In principle, the operators $O_i$ can be chosen randomly, in which case \texttt{EvolvedOperatorAnsatz} can serve as variational ansatzes for any general problem. However, the `tailoredness' of this type of circuits (as used this work) has to do with the fact that the set of operators $O_i$ chosen for each of the cases under study is unique and pertinent only to that case. 
As such, the exact form of this type of quantum circuit ansatzes will be described in detail in  Sections \ref{sec-l2-evop}, \ref{sec-l4-evop}, and  \ref{sec-sl2-evop} for  the $\Lambda=2$, $\Lambda=4$ $SU(2)$ bosonic model and $\L=2$ supersymmetric $SU(2)$ model, respectively. 
\\\\
\textbf{Comparison with ADAPT-VQE algorithm}: While the tailoredness of the  \texttt{EvolvedOperatorAnsatz} circuits  might be reminiscent of the ansatz used in ADAPT-VQE method, conceptually these are two completely different things. It is therefore useful to clarify this point in detail for the benefit of the reader.
ADAPT-VQE, as introduced in the work \cite{adapt-vqe} in the context of quantum chemistry and customized in other works such as \cite{schwinger-model-vqe} in the context of a (1+1)-dimensional gauge theory, constructs a tailored trial wavefunction by relying on a predefined operator pool from which to iteratively adjust the trial wavefunction until convergence is reached. The operator pool contains a selection of various operators that are related to the Hamiltonian $H$ under study. In the case of \cite{adapt-vqe}, this pool includes single and double excitation operators.
With ADAPT-VQE algorithm, the number of operators that are used to act on the wavefunction is continuously and incrementally adjusted during the algorithm run time - by selecting one at a time\footnote{See also \cite{adapt-tetris} for a variant of ADAPT-VQE that allows for multiple operators to be added iteratively instead of a single operator at each iteration.} an operator $O_k$ whose expectation value $\langle [H, O_k]\rangle$ is the largest - to obtain an increasingly improved circuit ansatz.  On the other hand, in this work, \texttt{EvolvedOperatorAnsatz} is a choice of circuit ansatz whose construction starts by defining a set of operators that are completely fixed during the entire run of the VQE algorithm. There is no adjustment of the operators in the ansatz during the VQE run, in direct contrast to the case of ADAPT-VQE.

\subsection{The Estimator module} \label{vqe-estimator}
Once a quantum circuit ansatz $\Psi(\vec\theta) = U(\vec\theta)|\mathbf{0}\rangle$ has been chosen, the second crucial task of VQE is the computation of the expactation value of the Hamiltonian $\langle \Psi(\vec\theta) H |\Psi(\vec\theta)\rangle$. This task can be carried out on actual quantum computer hardware, or on a quantum simulator. 
While the eventual goal is to run the VQE algorithms on actual quantum computers, here - as in the case of \cite{prx-22},  we work with a \texttt{Qiskit} simulator due to the time constraint imposed on the free access to the quantum hardware. The simulator that we use to compute the expectation value of the Hamiltonian is the \texttt{Estimator} module \cite{qiskit-estimator} provided by \texttt{qiskit\_aer} \cite{qiskit-aer}\footnote{note that the versions of \texttt{Qiskit} and \texttt{Qiskit} Aer used in this work are 1.3.0 and 0.15, respectively, while the \texttt{Qiskit} and Aer versions used in \cite{prx-22} were 0.26.2 and 0.8.2, respectively \cite{rinaldi-github}. There are significant differences in the module organizations between these \texttt{Qiskit} versions. One of these  differences is the deprecation of the module \texttt{opflow} that was heavily used in the VQE codes of \cite{prx-22} which allows the Hamiltonian to be declared as a \texttt{MatrixOp} without the need to convert it explicitly to Pauli string opertor form \cite{rinaldi-github}. In later versions of \texttt{Qiskit} (starting from 0.28) and in our codes, the starting point of all VQE runs are the Hamiltonians written in the forms of Pauli string operators. A simple example of how to run VQE (using the latest \texttt{Qiskit} version at the time of this writing)  can be found in the tutorial \cite{qiskit-vqe-tutorial}.}. \texttt{Estimator} can be coded in either a noiseless or noisy setting. 
\bei
\item In the noiseless setting, depending on the \texttt{`approximation'} parameter (either \texttt{True} or \texttt{False}) and the number of \texttt{shots} (either \texttt{None} or \texttt{int}), it is a state vector simulator which either returns the exact expectation value (\texttt{approximation} = \texttt{True, shots} = \texttt{None}) or the expectation value with sampling noise (\texttt{approximation} = \texttt{False, shots} = \texttt{int}) \cite{qiskit-estimator}. In this work, we choose the latter setting with \texttt{shots=1024}. The larger the the number of \texttt{shots}, the smaller the difference will be between the actual value and that of a VQE simulation.
\item In the noisy setting, various available and realistic noise models can be added to the \texttt{Estimator}. These noise models can either be custom built from scratch by adding to the quantum circuit ansatzes several types of preexisting quantum errors defined in \texttt{Qiskit} which include depolarizing quantum error channel, amplitude/phase damping errors, coherent/mixed unitary error, readout errors and thermal relaxation errors among others \cite{qiskit-noise}, or built based on the properties and the noise profiles of real quantum IBM devices such as the 127-qubit \texttt{ibm\_brussels}, \texttt{ibm\_brisbane} or \texttt{ibm\_strasbourg} among others\footnote{The list of active IBM quantum computers can be found at \href{https://quantum.ibm.com/services/resources}{https://quantum.ibm.com/services/resources}, while the list of retired IBM quantum devices can be found at \href{https://docs.quantum.ibm.com/guides/retired-qpus}{https://docs.quantum.ibm.com/guides/retired-qpus}.}. In the presence of noise, there exist several error mitigation techniques \cite{qiskit-emt}, such as TREX (Twirled readout extinction) \cite{emt-trex}, ZNE (Zero-noise extrapolation), PEA (Probabilistic error amplification) \cite{emt-pea} and PEC (Probabilistic error cancellation), that can be coded into the \texttt{Estimator} module \cite{qiskit-emt-2}. It is expected that in the noisy setting, even with the use of the few error mitigation techniques mentioned above, the results of the VQE experiments will be worse compared to those run in the noiseless setting. In particular, the run time of each experiment will be significantly longer before convergence is reached and the values obtained at convergence will not be as close to the actual values as in the noiseless case. In worst case scenarios, convergence might not be reached at all. 

\eni
Due to the nature of this work being an exploration of many different types of quantum circuit ansatzes, thus a proof-of-concept of sort, which requires relatively fast running times (in the order of several hours at most per experiment) in order to quickly establish the most efficient ansatzes, we defer the use of noisy setting for the \texttt{Estimator} (together with some error mitigation techniques listed above) to future works in which a single type, rather than multiple types, of quantum circuit ansatzes are studied.
\subsection{Optimizers} \label{vqe-opt}
\subsubsection{Optimizer basics} \label{vqe-opt-basics}
Optimizers are a crucial component in the VQE algorithm since they perform the essential task of updating the parameters $\vec\theta$ of the quantum circuit ansatzes subjected to a loss or objective function $L(\vec\theta)$\footnote{for VQE,  $L(\vec\theta)$ is the ground state expectation value of the Hamiltonian: $L(\vec\theta) = \lf\langle\Psi(\vec\theta)|H|\Psi(\vec\theta)\rr\rangle$} to be minimized. 
The process of parameter update can be either gradient-based, in which the first derivative or the gradient of $L(\vec\theta)$, $\nabla L(\vec\theta)$ is utilized and needs to be known in exact form, or gradient-free, in which  $\nabla L(\vec\theta)$ is either unnecessary or only needs to be known in approximated forms. The most basic type of gradient-based parameter update is a process known as gradient descent given by the following equation
\beq
\vec\theta_{k+1} = \vec\theta_k - \alpha \nabla L(\vec\theta_k) \label{eq-gd}
\eeq
in which $\vec\theta_{k}$ are the parameters at the $k^{th}$ iteration, and $\alpha$, the learning rate,  is a hyperparameter chosen by the user. Too small $\a$ leads to a slow convergence while too large $\a$ leads to oscillations and overshooting which might prevent convergence \cite{opt-avoiding-minima}. 
 While gradient-based optimizers are the cornerstone of modern classical machine  and deep learning involving artificial neural networks, gradient-free optimizers are popular choices for quantum computing problem settings \cite{opt-spsa-based}, \cite{opt-classical-comparison}. Gradient-free optimizers are more flexible than gradient-based in the sense that they can perform well in situations where $L(\vec\theta)$ is complicated, non-smooth or non-differentiable.
Although the overall efficiency of the VQE algorithm depends greatly on the choices of both quantum circuit ansatzes and optimizers, the convergence quality is almost entirely controlled by the type of optimizers used, given that using the same quantum circuit ansatz with different optimizers lead to different converged results. In this section, we considered the following six optimizers\footnote{The full list of optimizers can be found at \cite{qiskit-algo-opt} and a useful tutorial exploring optimization loops from IBM is at \cite{opt-ibm-course} .}.
\ben
\item SPSA (Stochastic Perturbation Simultaneous Approximation) \cite{spsa} is a  gradient-free stochastic optimization algorithm that performs parameter updates by approximing the gradient of the loss function $\nabla L(\vec\theta_k)$ at iteration $k$ by a function $g(\vec\theta_k)$ obtained by \cite{opt-spsa-based}
\beq
f(\vec\theta_k) = \frac{L(\vec\theta_k + c_k \vec\Delta_k) - L(\vec\theta_k - c_k \vec\Delta_k)}{2 c_k} \vec\Delta_k \,,\label{spsa-g}
\eeq
where $c_k$ is a small positive scaling factor, $\vec\Delta_k$ is a random perturbation vector (at iteration $k$) whose entries are drawn independently from the set $\{-1, 1\} $.
The update rule for SPSA is the same as Eq.(\ref{eq-gd}), but with the derivative $\nabla L(\vec\theta_k)$ replaced by $f(\vec\theta_k$) given in Eq.(\ref{spsa-g})
\beq
\vec\theta_{k+1} = \vec\theta_k - \alpha f(\vec\theta_k)\,.
\eeq
\item COBYLA (Constrained Optimization BY Linear Approximation) \cite{opt-cobyla} is a gradient-free optimization algorithm that performs parameter updates by utilizing a linear approximation of the loss function $L(\vec\theta$) as well as all constraints in the neighborhood of the current point, within a specified trust region, to determine the next point. At each iteration, the algorithm solves a linear programming problem inside a trust region whose radius decreases as certain convergence criterion is reached. An important point to note is that COBYLA treats simple bounds as constraints, which might lead to bound violations.
\item NELDER-MEAD \cite{opt-nm} is a gradient-free optimization algorithm and simplex-based direct search method. Given the loss function $L(\vec\theta)$, this algorithm starts with a set of, say ($n+1$) points $\vec\theta_\text{initial}=(\theta_0, \ldots, \theta_n)\in \mathbb R^n$ that are supposedly the vertices of a simplex $S$ in $\mathbb R^n$, and calculates the loss function value $L(\vec\theta_\text{initial})$ at these vertices. Next, a sequence of transformations is applied to $S$ with the aim of decreasing $L(\vec\theta$) until the simplex $S$ is sufficiently small or a convergence criterion is reached.
\item  L\_BFGS\_B is the limited-memory (subject to bounds) version of BFGS (Broyden–Fletcher–Goldfarb–Shanno), a gradient-based optimization algorithm, which uses the Hessian matrix $\mathcal H$ of the loss function to compute the direction $\vec n_k$
\beq
\vec n_k = \mathcal H^{-1}_k \nabla L(\vec\theta_k)
\eeq
to perform a line search on $\{\vec\theta_k + \eta_k \vec n_k|\eta_k \in \mathbb R\}$ to find an optimal update $\eta_k$. Once this is found, the new parameter $\vec\theta_{k+1}$ is updated to 
\beq
\vec\theta_{k+1} = \vec\theta_{k} + \eta_k \vec n_k\,.
\eeq
 With the updated parameter $\vec\theta_{k+1}$, one can calculate the change in the gradient $D_k = \nabla L(\vec\theta_{k+1}) - \nabla L(\vec\theta_k)$ and use that to update the Hessian:
\beq
\mathcal H_{k+1} = \mathcal H_k + \frac{D_k D_k^T}{\vec n_k D_k^T \vec n_k} - \frac{\mathcal H_k \vec n_k \vec n_k^T \mathcal H_k}{\vec n_k^T \mathcal H_k \vec n_k}\,.
\eeq
\item  SLSQP (Sequential Least Squares Programming) \cite{opt-slsqp} is a gradient-based optimization
based on sequential quadratic programming (SQP) which involves the construction of a Lagrangian $\mathcal L$ from the loss function $L(\vec\theta)$ and the equality and inequality constraints $h_i(\vec\theta)$, $g_i(\vec\theta)$
\beq
\mathcal L(\vec\theta, \vec\lambda, \vec\mu)  = L(\vec \theta) + \sum_i \lambda_i h_i(\vec\theta) + \sum_i\mu_i g_i(\vec\theta)
\eeq
where  $\l_i$ and $\mu_i$ are the Lagrange multipliers associated with $h_i$ and $g_i$. The parameter update at iteration $k^{th}$ process not only involves $\vec\theta$ but also $\vec\l$ and $\vec\mu$.
\beq
\begin{pmatrix} \vec\theta_{k+1}\\ \vec\l_{k+1} \\ \vec\m_{k+1}\end{pmatrix}  = \begin{pmatrix} \vec\theta_{k}\\ \vec\l_{k} \\ \vec\m_{k}\end{pmatrix} - \frac{\nabla \mathcal{L}(\vec\theta_k, \vec\l_k, \vec\m_k)}{\nabla^2 \mathcal{L}(\vec\theta_k, \vec\l_k, \vec\m_k)}
\eeq
where
\[\nabla \mathcal{L} = \left(\nabla_{\vec\theta} \mathcal L, \nabla_{\vec\l} \mathcal L, \nabla_{\vec\mu} \mathcal L  \rr)\,.\]

\item ADAM (Adaptive Moment Estimation) \cite{opt-adam} is a gradient-based optimization algorithm that is very popular in  machine and deep learning involving classical neural networks. To perform a parameter update at step $k$, ADAM uses the running estimates of the  first and the second moment  of the gradient $\nabla L(\vec\theta_k)$  \cite{opt-spsa-based} 
\beq
\vec\mu_{k+1} &=& \beta_1 \vec\mu_k + (1-\beta_1) \nabla L(\vec \theta_k) \non
\vec\s_{k+1} &=& \beta_2 \vec\s_k + (1-\beta_2) \nabla L(\vec \theta_k) \odot \nabla L(\vec \theta_k) \label{eq-adam-1}
\eeq
where $\vec\mu_k$ denotes the mean estimate and $\s_k$ the variance, $\b_1, \b_2 \in [0,1)$ are the decay rates. The final parameter update equation for ADAM is
\beq
\vec\theta_{k+1} = \vec\theta_k -\frac{\a\,\vec{\hat{\mu}}_{k+1}}{\sqrt{\vec{\hat{\s}}_{k+1}} + \epsilon}
\eeq
where $\a$ is a positive scaling factor, and $\vec{\hat{\mu}}$ and $\vec{\hat{\s}}$ are the scaled or corrected versions of $\vec\mu$ and $\vec\s$ (due to the fact that these estimates can be biased when the iteration $k$ is small)
\beq
\vec{\hat{\m}}_k = \frac{\vec\m_k}{1-\b_1^{k+1}}, \qquad \vec{\hat{\s}}_k = \frac{\vec\s_k}{1-\b_2^{k+1}}\,.
\eeq
\enn
While gradient-based and gradient-free optimizers have their own advantages as well as disadvantages, gradient-based optimizers such as ADAM are directly impacted by the barren plateau phenomenon \cite{opt-bp} in which the gradient $\nabla L(\vec\theta)$ of the loss function $L(\vec\theta)$ vanishes exponentially in the number of qubits. To a lesser extent, gradient-free optimizers such as COBYLA have been shown to be affected \cite{opt-bp-gf} by this phenomenon but this is highly dependent on the specific setting under study. A viable method to prevent the occurrence of barren plateaus in VQE experiments utilizing gradient-based optimizers involves a special initialization of the parameters as reported in \cite{opt-bp-init}. The factors inducing the occurrences of barren plateaus, ranging from the effects of the loss function, the form of the ansatzes, to the presence of noise, are an active area of research whose results have been reported in recent works such as  \cite{opt-bp}, \cite{opt-bp-noise}, \cite{opt-bp-loss}, \cite{opt-bp-ansatz}. Furthermore, depending on the complexity of the cost function $L(\vec\theta)$ and the structure of quantum circuit ansatzes, multiple local minima might exist and could cause the optimization to be stuck. This is a prevalent problem that affects all types of optimizers, especially when the number of parameters to be optimized is high.
 While these issues can negatively impact the training process in VQE experiments generally, they are not of too much concern for us, since our VQE experiments involve a noiseless quantum simulator as well as shallow quantum circuits with small number of parameters, similar to the case in \cite{prx-22}.

\subsubsection{Optimizer selection}\label{vqe-opt-select}
\begin{figure}[!ht]
\centering
\includegraphics[width=.6\textwidth]{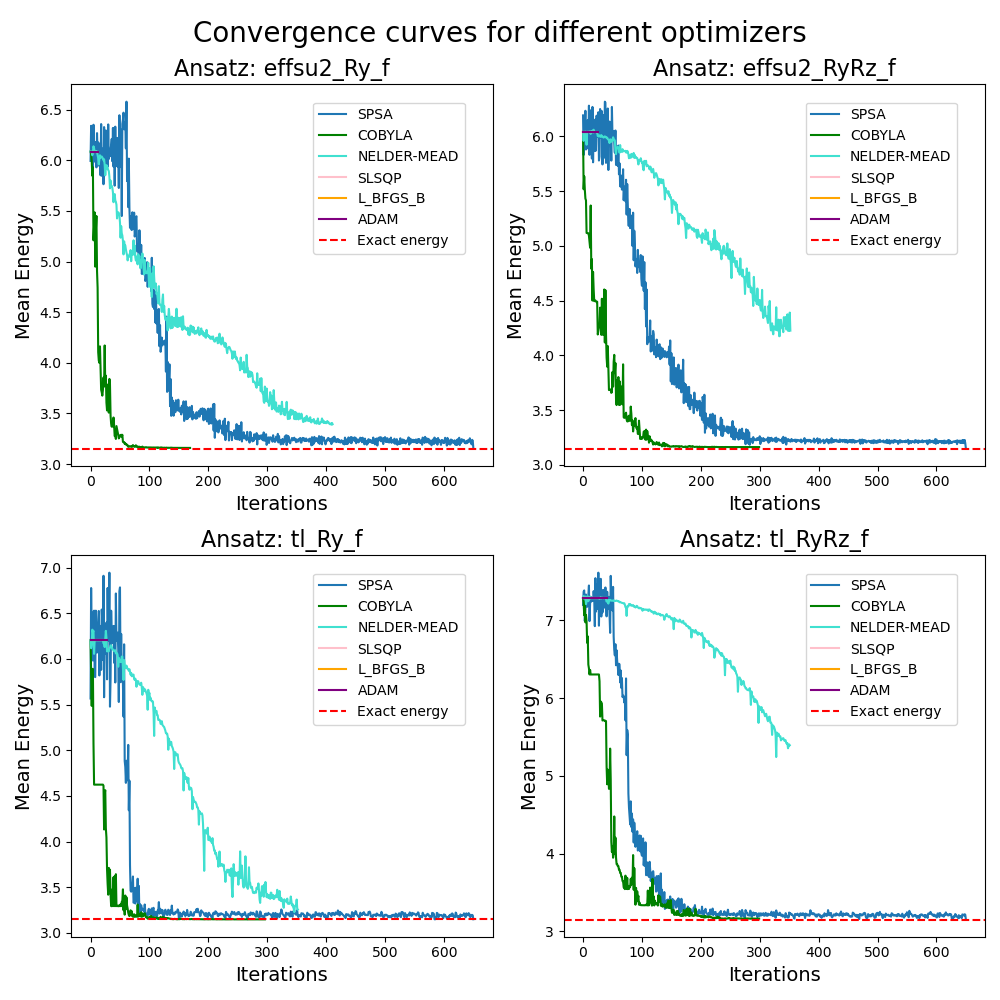}
\caption{Convergence curves from the VQE experiments involving the $SU(2)$ matrix model at Fock cutoff $\L=2$ at coupling $\l=0.2$, using six different optimizers SPSA, COBYLA, NELDER-MEAD, L-BFGS-B, SLSQP, ADAM  for four ansatzes: Clockwise from left: \texttt{effsu2\_Ry\_f}, \texttt{effsu2\_RyRz\_f}, \texttt{tl\_Ry\_f} and \texttt{tl\_RyRz\_f}.}
\label{fig_opt_check}
\end{figure}

To make an informed choice of the eventual optimizers to be used in our VQE experiments, we will perform several experiments to check the performances of all six optimizers using some of the ansatzes introduced in Section \ref{qc-ans}  within the setting of the $SU(2)$ matrix model. For these experiments, we use the $64\times 64$ Hamiltonian at Fock cutoff $\L=2$ at weak coupling $\l=0.2$, $H^{\L=2}_{\l=0.2}$ with the exact ground state energy $E_\text{exact} = 3.14808$,  and four quantum circuit ansatzes, \texttt{effsu2\_Ry\_f}, \texttt{effsu2\_RyRz\_f}, \texttt{tl\_Ry\_f} and \texttt{tl\_RyRz\_f}. For each of the ansatzes, six VQE experiments will be run using the six optimizers SPSA, COBYLA, NELDER-MEAD, L-BFGS-B, SLSQP, ADAM. Most of these optimizers, except SPSA,  automatically end the optimization process when convergence is reached. The results are listed in Table \ref{opt-select} and the convergence curves are shown in Fig.\ref{fig_opt_check}. 
\begin{table}[!ht]
\centering
\begin{tabular}{ccccccc}
\hline\hline
Ansatz & SPSA & COBYLA & NELDER-MEAD & SLSQP & L-BFGS-B & ADAM\\
\hline\hline
\texttt{effsu2\_Ry\_f} & 3.15449& 3.15918& 3.38867&  6.08574 &   6.08574&6.08574\\
\texttt{effsu2\_RyRz\_f} & 3.15020& 3.16211& 4.17441&  6.03906& 6.03906 &6.03906\\
\texttt{tl\_Ry\_f} & 3.14902 & 3.14785 & 3.22715 & 6.20488 & 6.20488 & 6.20488\\
\texttt{tl\_RyRz\_f} &3.15371 & 3.16211 &  5.24121 & 7.28477 & 7.28477 &  7.28477\\
\hline
\end{tabular}
\caption{Results of the 24 VQE experiments involving the $\L=2, \l=0.2$ Hamiltonian $H^{\L=2}_{\l=0.2}$  and six optimizers SPSA, COBYLA, NELDER-MEAD, L-BFGS-B, SLSQP, ADAM with four ansatzes \texttt{effsu2\_Ry\_f}, \texttt{effsu2\_RyRz\_f},
\texttt{tl\_Ry\_f} and \texttt{tl\_RyRz\_f}.} \label{opt-select}
\end{table}
\begin{figure}[!ht]
\centering
\includegraphics[width=.6\textwidth]{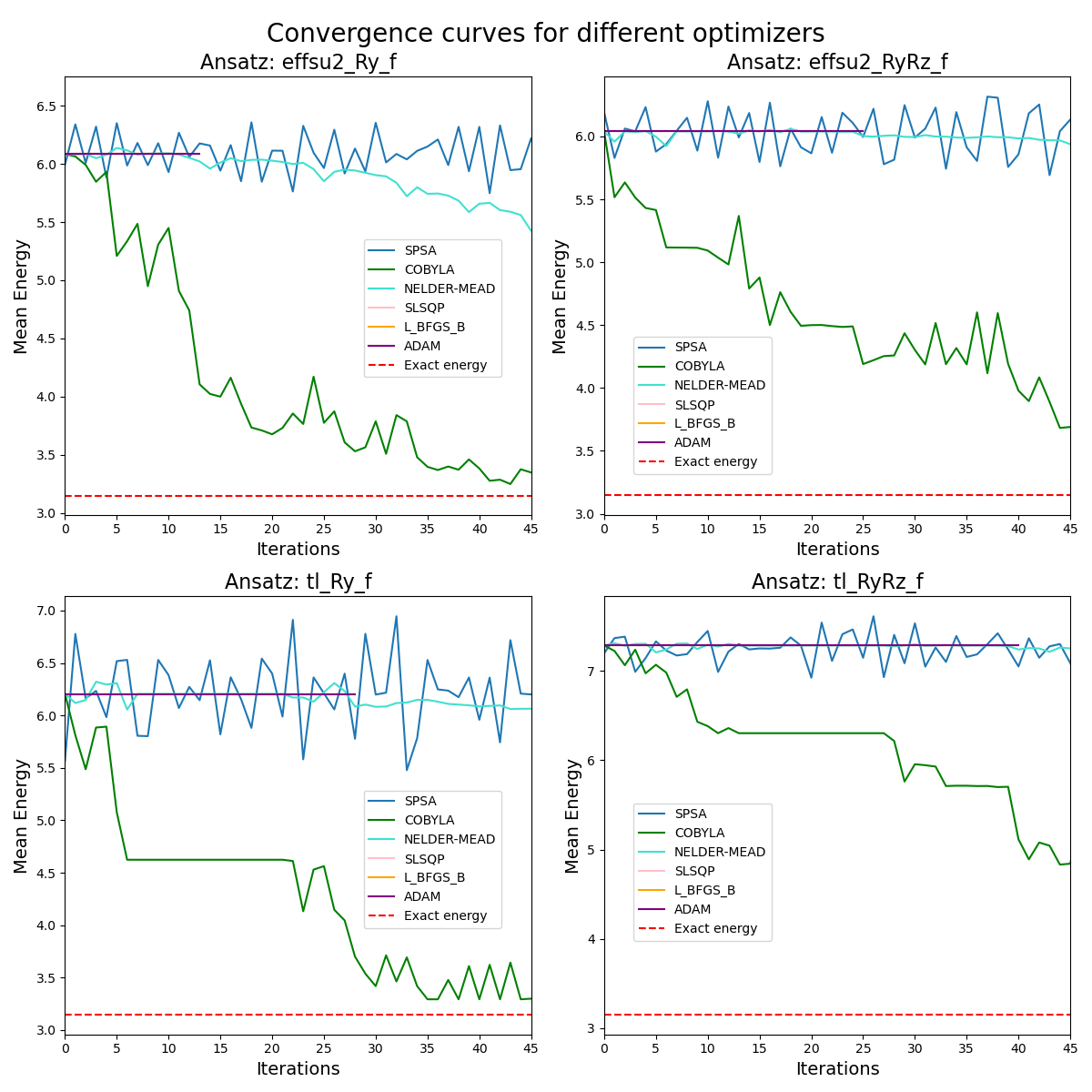}
\caption{A zoomed-in version of Fig.\ref{fig_opt_check} showing only the first 45 iterations of the VQE experiments. The convergence curves of the gradient-based optimizers SLSQP, L-BFGS-B, ADAM are just straight lines exactly coinciding with one other, hence only the purple line is visible in all 4 plots.}
\label{fig_opt_check-2}
\end{figure}

Using Table \ref{opt-select} and Fig.\ref{fig_opt_check}, the clear trends that emerged from the VQE experiments involving the six optimizers are the following.
\bei
\item Three of the gradient-based optimizers (SLSQP,  L-BFGS-B, ADAM) performed extremely poorly in the sense that no parameters update occurred and the optimization process was terminated after fewer than 40 iterations (corresponding to the fact that the convergence curves were just short  straight lines (colored pink, purple, orange) coinciding with one another in Fig.\ref{fig_opt_check-2}, which is a zoomed in version of Fig.\ref{fig_opt_check}) showing only the first 45 iterations.
\item The gradient-free optimizers SPSA, COBYLA, NELDER-MEAD performed much better than the gradient-based optimizers with convergences reached, albeit at different values for different optimizers. Among the three, NELDER-MEAD is the worst performer, with convergence values far above the correct ones.  COBYLA and SPSA consistently yielded results at convergence that are quite close to the exact values.
\eni
Based on the results shown in Table \ref{opt-select} and Figs.\ref{fig_opt_check}, \ref{fig_opt_check-2}, we select SPSA and COBYLA for all the VQE experiments carried out in the main part of this work.
\FloatBarrier
\subsection{Overview of VQE experiments} \label{sec-vqe-summ}
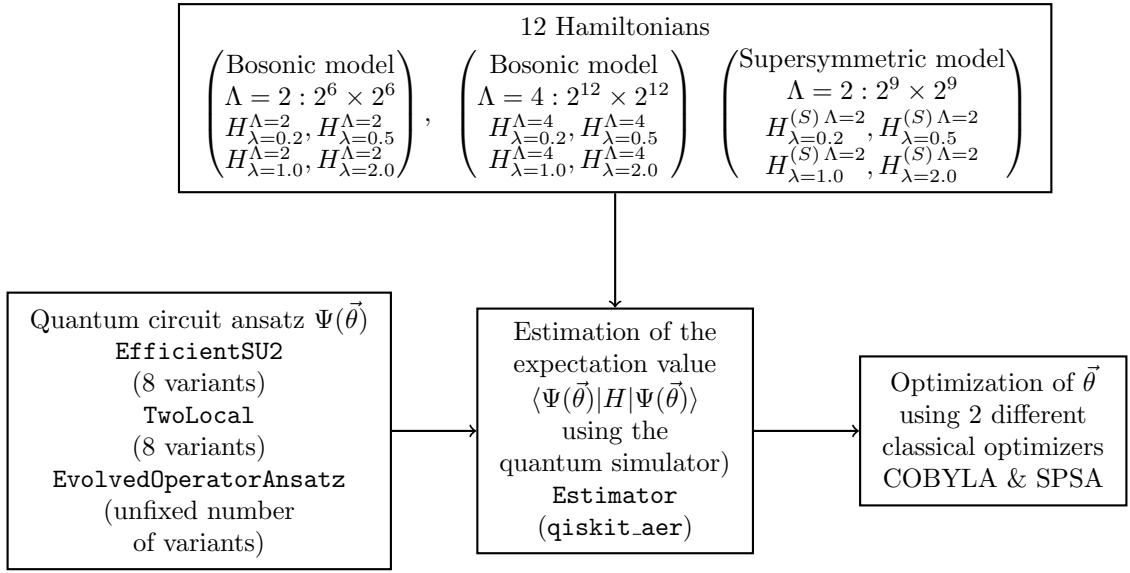
\begin{figure}[H]
\centering
\begin{tikzpicture}[node distance = 5cm, thick]%
        \node[draw] (0c) {$\begin{array}{c}\text{12 Hamiltonians}
       \\\begin{pmatrix}  \text{Bosonic model}\\\L=2: 2^6\times 2^6\\ H^{\L=2}_{\l=0.2},H^{\L=2}_{\l=0.5}\\ H^{\L=2}_{\l=1.0},H^{\L=2}_{\l=2.0}\end{pmatrix},\,\, \,\,
       \begin{pmatrix} \text{Bosonic model}\\\L=4: 2^{12}\times 2^{12}\\H^{\L=4}_{\l=0.2},H^{\L=4}_{\l=0.5}\\H^{\L=4}_{\l=1.0},H^{\L=4}_{\l=2.0} \end{pmatrix}\,\,\,\,
       \begin{pmatrix}\text{Supersymmetric model}\\\L=2: 2^9\times 2^9\\ H^{(S)\,\L=2}_{\l=0.2},H^{(S)\,\L=2}_{\l=0.5}\\ H^{(S)\,\L=2}_{\l=1.0},H^{(S)\,\L=2}_{\l=2.0}\end{pmatrix}\end{array}$};
        \node[draw] (0)[below of = 0c, yshift = 0.6cm] {$\begin{array}{c} \text{Estimation of the}\\ \text{expectation value} \\ \langle\Psi(\vec\theta)| H|\Psi(\vec\theta)\rangle\\ \text{using the}\\\text{quantum simulator)}\\ \text{\texttt{Estimator}}\\\text{(\texttt{qiskit\_aer})}\end{array}$};
        \node[draw] (0a) [left of =0, xshift = -0.5cm]{$\begin{array}{c} \text{Quantum circuit ansatz $\Psi(\vec\theta)$} \\ \texttt{EfficientSU2}\\\text{(8 variants)} \\\texttt{TwoLocal} \\ \text{(8 variants)}\\\texttt{EvolvedOperatorAnsatz} \\ \text{(unfixed number}\\\text{of variants)}\end{array}$};
        \node[draw] (0b) [right of =0]{$\begin{array}{c} \text{Optimization of $\vec\theta$}\\\text{using 2 different}\\\text{classical optimizers}\\ \text{COBYLA \& SPSA} \end{array}$};

        \draw[->] (0a) -- node [right]{} (0);
        \draw[->] (0c) -- node [right]{} (0);
        \draw[->] (0) -- node [right]{} (0b);
   
    \end{tikzpicture}
    \caption{The main components of VQE as selected in this work: three types of quantum circuit ansatzes which include 8 variants of \texttt{EfficientSU2}, 8 variants of \texttt{TwoLocal} for the bosonic $SU(2)$ model experiments and an unfixed number of variants of \texttt{EvolvedOperatorAnsatz} depending on the $SU(2)$ model, a quantum simulator called \texttt{Estimator} (state vector simulator with sampling) to estimate the expectation value of the Hamiltonian, and two different classical optimizers  (COBYLA and SPSA). There are four different Hamiltonians $H^{\L}_\l$ for each of the three $SU(2)$ models, corresponding to the four couplings $\lambda=0.2, 0.5, 1.0, 2.0$, for a total of 12 Hamiltonians.} \label{vqe-schema-2}
    \end{figure}

In this section, we provide an overview of all the VQE experiments that will be carried out in this work. 
With all the components for VQE in place as described in the previous sections, we update Fig.\ref{vqe-schema} to reflect our specific choices of these VQE components in Fig.\ref{vqe-schema-2}. There are 12 Hamiltonians $H^{\L}_{\l}$ in total corresponding to the three truncated $SU(2)$ models (bosonic at Fock cutoff $\L=2,4$ and supersymmetric at Fock cutoff $\L=2$) at four different couplings $\l=0.2, 0.5, 1.0, 2.0$. 
For quantum circuit ansatzes, we work with the 8 variants of \texttt{EfficientSU2} in Table \ref{effsu2_ans}, the 8 variants \texttt{TwoLocal} in Table \ref{tl_t1_ans} and the various variants of \texttt{EvolvedOperatorAnsatz} whose construction are described in detail in later sections. The problem-agnostic, or generic, ansatzes \texttt{EfficientSU2}, \texttt{TwoLocal} are used in the bosonic models only, while the tailored \texttt{EvolvedOperatorAnsatz} circuits are used in both the bosonic and supersymmetric models. Two optimizers are used throughout: COBYLA and SPSA. Given these parameters, each VQE experiment is uniquely specified by the tuple of choices denoted as (Hamiltonian,  ansatz, optimizer)
\footnote{For example,
\bei
\item ($H^{\L=2}_{\l=0.2}$, \texttt{effsu2\_Ry\_c}, COBYLA) corresponds to the VQE experiment involving the bosonic $\L=2$ model at coupling $\l=0.2$  with the \texttt{EfficientSU2} ansatz variant \texttt{effsu2\_Ry\_c} and COBYLA optimizer. ($H^{(S)\L=2}_{\l=0.5}$, \texttt{ev\_op\_Hp15}, SPSA)  refers to the VQE experiment involving the supersymmetric $\L=2$ model at coupling $\l=0.5$  with the \texttt{EvolvedOperatorAnsatz} variant \texttt{ev\_op\_Hp15} with SPSA optimizer.
\item ($H^{(\L=4}_{\l}$, \texttt{EfficientSU2}\&\texttt{TwoLocal}, COBYLA/SPSA) refers to the VQE experiments involving the bosonic $\L=4$ model at all couplings $\l=0.2, 0.5, 1.0, 2.0$  with all the variants of \texttt{EfficientSU2} and all the variants of \texttt{TwoLocal} with both COBYLA and SPSA optimizers.
\item ($H^{(\L=4}_{\l}$, \texttt{EvolvedOperatorAnsatz}, COBYLA/SPSA) refers to the VQE experiments involving the bosonic $\L=4$ model at all couplings $\l=0.2, 0.5, 1.0, 2.0$  with all the variants of \texttt{EvolvedOperatorAnsatz} with both COBYLA and SPSA optimizers.
\eni}
where the Hamiltonian choice includes either $H^{\L}_{\l}$ or $H^{(S)\L}_{\l}$ corresponding to the bosonic or supersymmetric $SU(2)$ model at Fock cutoff $\L$ and coupling $\l$.
The total numbers of VQE experiments for each truncated $SU(2)$ matrix model are listed below.
\ben
\item Bosonic $SU(2)$ at $\L=2$: At each coupling $\l$, there are 16 \texttt{EfficientSU2} and \texttt{TwoLocal} ansatzes, together with 9 \texttt{EvolvedOperatorAnsatz} variants, for a total of 25 ansatz variants. Together with the usage of two optimizers (COBYLA \& SPSA), there are 50 VQE experiments per $\l$. With 4 values of $\l$, we have 200 VQE experiments in total. 
\item Bosonic $SU(2)$ at $\L=4$: At each coupling $\l$, there are 16 \texttt{EfficientSU2} and \texttt{TwoLocal} ansatzes, together with 8 \texttt{EvolvedOperatorAnsatz} variants, for a total of 24 ansatz variants, which lead to 48 VQE experiments using either COBYLA or SPSA optimizer. With 4 values of $\l$, this leads to 192 experiments.
\item
Supersymmetric $SU(2)$ at $\L=2$: At each coupling $\l$, there are 12 \texttt{EvolvedOperatorAnsatz} variants (\texttt{EfficientSU2} and \texttt{TwoLocal} ansatzes are not used) which lead to 24 VQE experiments using either COBYLA or SPSA optimizer. With 4 values of $\l$, this leads to 96 experiments.
\enn

In the following sections, we present and discuss in detail the results obtained from runnning the VQE experiments involving the three $SU(2)$ matrix models (bosonic models at $\L=2$, $\L=4$ and supersymmetric model at $\L=2$) using three different types of ansatzes and two different types of optimizers. For each $SU(2)$ model at four different couplings $\l=0.2, 0.5, 1.0, 2.0$,  we will highlight the best results for each type of ansatzes and compare these best results among one another, as well as among those reported from \cite{prx-22} in order to benchmark the performances of our proposed ansatzes against the \texttt{EfficientSU2} ansatzes used in \cite{prx-22}. The full results for all experiments, together with the corresponding convergence curves for the energy of each of the ansatzes will be included as supplementary materials in the Appendix, since the these are not necessary for the qualitative discussion in the main text. Due to the presence of numerous Tables and Figures, we summarize the structure of our results in Table \ref{overview-table}. 

\begin{table}[!ht]
\centering
\begin{tabular}{|c| c| c |c |}
\hline
 $SU(2)$ Model& Ansatz &  Main Results& Supplementary Results \\
\hline
$\begin{array}{c} \L=2\,\\
 \text{(Bosonic)}\\
\text{Section \ref{sec_qc_su2_l2}}
 \end{array}$ 
&$\begin{array}{c}
\texttt{EfficientSU2}\\ \text{Table \,\ref{effsu2_ans}}\\
\texttt{TwoLocal}\\\text{Table \ref{tl_t1_ans}} 
\\\\
\texttt{EvolvedOperator} \\ \text{Table\,\ref{L2-ev-op-ansatze}}
\end{array}$ &
$\begin{array}{c}
\\\\
\text{Table \ref{L2_best_es2_tl}}\\
\text{Fig.\ref{fig_L2_all_ef_tl}}
\\\\\\
\text{Table \ref{L2_best_ev_op}}
\\
\text{Fig.\ref{fig_L2_all_ev_op}} 
\\\\
\text{Best overall results:}\\
\text{Table \ref{L2-all-res}}
\\\\
\end{array}$
&
$\begin{array}{c}
\begin{array}{c}\text{Table}\, \ref{qve-l2-l02-ef-tl}\\\text{(F-S)} \\ 
\text{Fig.}\ref{fig_L2_l02_ef_tl} \\\text{(CC)}\\\\
 \text{Table} \,\ref{qve-l2-l02-ev-op} \\\text{(F-S)}
\end{array}  \,\,
\begin{array}{c} \text{Table}\, \ref{qve-l2-l05-ef-tl}\\\text{(F-S)}\\
 \text{Fig.}\ref{fig_L2_l05_ef_tl} \\\text{(CC)}\\\\
 \text{Table}\, \ref{qve-l2-l05-ev-op} \\\text{(F-S)}
 \end{array}   \,\,
 \begin{array}{c} \text{Table}\, \ref{qve-l2-l10-ef-tl}\\\text{(F-S)}\\
 \text{Fig.}\ref{fig_L2_l10_ef_tl}\\\text{(CC)} \\\\
 \text{Table}\, \ref{qve-l2-l10-ev-op} \\\text{(F-S)}
 \end{array}   \,\,
 \begin{array}{c} \text{Table}\, \ref{qve-l2-l20-ef-tl}\\\text{(F-S)}\\
 \text{Fig.}\ref{fig_L2_l20_ef_tl} \\\text{(CC)}\\\\
 \text{Table}\, \ref{qve-l2-l20-ev-op} \\\text{(F-S)}
 \end{array}
 \\  
 \begin{array}{c}
\\ \text{Fig.\ref{fig_L2_ev_op_cobyla}} \,\text{(CC)},\,\,\text{Fig.\ref{fig_L2_ev_op_spsa}} \,\text{(CC)}
 \end{array}
 \end{array}$\\
\hline
$\begin{array}{c}\\ \L=4 
\\\text{(Bosonic)}\\
\text{Section \ref{sec_qc_su2_l4}}
 \end{array}$&
 $\begin{array}{c}
 \texttt{EfficientSU2}\\ \text{Table \,\ref{effsu2_ans}}\\
  \texttt{TwoLocal}\\\text{Table \ref{tl_t1_ans}}\\
  \\\texttt{EvolvedOperator}\\
\text{Table}\, \ref{L4_ev_op_quantum circuits}
  \end{array}$ 
  &
$\begin{array}{c}
\\
\text{Table \ref{L4_best_es2_tl}}\\
\text{Fig.\ref{fig_L4_all_es_tl}}
\\\\\\
\text{Table \ref{L4_best_ev_op}}\\
\text{Fig.\ref{fig_L4_all_ev_op}}
\\\\
\text{Best overall results:}\\ 
\text{Table \ref{L4-all-res}}
\\\\
 \end{array}$
   &
$\begin{array}{c}
\begin{array}{c}
\text{Table}\, \ref{qve-l4-l02-es-tl}\\\text{(F-S)}\\
\text{Fig.}\ref{fig_L4_l02_ef_tl}\\\text{(CC)}\\
\\\text{Table}\, \ref{L4_l02_eo}\\\text{(F-S)}
\end{array}
\,\, \begin{array}{c}
\text{Table}\, \ref{qve-l4-l05-es-tl}\\\text{(F-S)}\\
\text{Fig.}\ref{fig_L4_l05_ef_tl}\\\text{(CC)}\\\\
\text{Table}\, \ref{L4_l02_eo}\\\text{(F-S)}
\end{array}
\,\,\begin{array}{c}
\text{Table} \,\ref{qve-l4-l10-es-tl}\\\text{(F-S)}\\
\text{Fig.}\ref{fig_L4_l10_ef_tl}\\\text{(CC)} \\\\
\text{Table}\, \ref{L4_l10_eo} \\\text{(F-S)}
\end{array}
\,\,\begin{array}{c}
\text{Table}\, \ref{qve-l4-l20-es-tl}\\\text{(F-S)}\\
\text{Fig.}\ref{fig_L4_l20_ef_tl} \\\text{(CC)} \\\\
\text{Table}\, \ref{L4_l20_eo} \\\text{(F-S)}
\end{array}\\
\begin{array}{c}
\text{Fig. \ref{fig_L4_eo_curves}}\,\text{(CC)}\end{array}
\end{array}$  \\
\hline
$\begin{array}{c} \\\L=2 \\
\text{(BMN)}\\
\text{Section \ref{sec-bmn}}\end{array}$& 
$\begin{array}{c}\texttt{EvolvedOperator}\\
\text{Table}\, \ref{bmn-ev-op}
 \end{array}$ 
& $\begin{array}{c} 
\\ \\
 \text{Table \ref{sup_l2_best}}\\
\text{Fig.\ref{fig_L2_BMN_all_ev_op}}
\\\\
\text{Best overall results:}\\
 \text{Table\,\ref{L2-BMN-all-res}}
 \\\\
 \end{array}$
& 
$\begin{array}{c}
 \begin{array}{c}\text{Table}\,\ref{bmn_L2_l02_eo} \\ \text{(F-S)} \end{array}
\,\,\begin{array}{c}\text{Table}\,\ref{bmn_L2_l05_eo} \\ \text{(F-S)}   \end{array}
\,\,\begin{array}{c}\text{Table}\,\ref{bmn_L2_l10_eo}\\  \text{(F-S)}  \end{array}
\,\,\begin{array}{c}\text{Table}\,\ref{bmn_L2_l20_eo} \\ \text{(F-S)} \end{array}
 \\\\
\begin{array}{c}  
\text{Fig.\ref{fig_L2_bmn_eo_curves_cobyla}},\,\,\text{Fig.\ref{fig_L2_bmn_eo_curves_spsa_1f}\,-\,Fig.\ref{fig_L2_bmn_eo_curves_spsa_4f}}
\,\,\text{(CC)} \end{array}
\end{array}$\\
\hline
\end{tabular}
\caption{An overview of the Tables and Figures containing the main results for all VQE experiments run in this work for the three types of quantum circuit ansatzes (\texttt{EfficientSU2}, \texttt{TwoLocal}, \texttt{EvolvedOperatorAnsatz}) at 4 different couplings $\l=0.2, 0.5, 1.0, 2.0$ for the cases of $\L=2$, $\L=4$ bosonic and $\L=2$ supersymmetric $SU(2)$ matrix model. For each of the three $SU(2)$ models, the main results include a summary table listing only the best result from each type of optimizers for each type of ansatzes (\texttt{EfficientSU2/TwoLocal} are counted together), the plots showing the comparisons of all variants within the ansatz type considered, and a table containing best overall results for all ansatzes (including the ones reported in \cite{prx-22}). The supplementary results for each model include the four tables (labeled F-S for Full-Supplementary) listing the full results for all ansatz variants at each coupling $\l$, and the convergence curve plots (labeled CC for Convergence Curves) of all variants. All tables with the F-S label and figures with the CC label are supplementary material included in the Appendix.} \label{overview-table}
\end{table}
\FloatBarrier
\clearpage
\section{$\Lambda = 2$ bosonic model} \label{sec_qc_su2_l2}
The Hamiltonian for the $SU(2)$ bosonic matrix model at $\Lambda=2$ cutoff is a  $2^6\times 2^6$ or $64\times 64$ matrix that can be expressed as a sum of 10 Pauli string operators  whose coefficients change at different couplings $\lambda$. 
The list of the 10 Pauli string operators and their coefficients are shown in Table \ref{L2-H-pauliOps}. 
\begin{table}[!ht]
\centering
\begin{tabular}{c c c c c}
\hline\hline
Operator & $\lambda=0.2$ & $\lambda=0.5$ & $\lambda=1.0$ & $\lambda=2.0$ \\
\hline\hline
IIIIII &  6.15 & 6.375 & 6.75& 7.5\\
\text{IIIIIZ} & $-0.5$ &$-0.5$& $-0.5$& $-0.5$ \\
\text{IIIIZI} & $-0.5$ &$-0.5$& $-0.5$& $-0.5$  \\
\text{IIIZII} & $-0.5$ &$-0.5$& $-0.5$& $-0.5$ \\
\text{IIZIII} & $-0.5$ &$-0.5$& $-0.5$& $-0.5$ \\
\text{ZIIIII} & $-0.5$ &$-0.5$& $-0.5$& $-0.5$ \\
\text{IZIIII} & $-0.5$ &$-0.5$& $-0.5$& $-0.5$ \\
\text{IXXIXX} & $-0.05$ & $-0.125$ &  $-0.25$& $-0.5$\\
\text{XIXXIX} & $-0.05$ & $-0.125$& $-0.25$ & $-0.5$\\
\text{XXIXXI} & $-0.05$& $-0.125$ & $-0.25$  &$-0.5$\\
\hline
\end{tabular}
\caption{The 10 Pauli string operators forming the $SU(2)$ matrix model Hamiltonians at four couplings $\l=0.2, 0.5, 1.0, 2.0$.}
\label{L2-H-pauliOps}
\end{table}
The exact form of the Hamiltonian $H^{\L=2}_\l$ at any of the four couplings $\l$ can be read off from the correct $\l$ column of Table \ref{L2-H-pauliOps}. For example, 
 at $\l=0.2$, the Hamiltonian $H^{\L=2}_{\l=0.2}$ is
 \beq
 H^{\L=2}_{\l=0.2} &=& 6.15\text{IIIIII} - 0.5\lf(\text{IIIIIZ}+ \text{IIIIZI}+\text{IIIZII}+ \text{IIZIII} +\text{ZIIIII}+\text{IZIIII}\rr) 
 \non
 && - \,0.05\lf(\text{IXXIXX}+ \text{XIXXIX}+\text{XXIXXI}\rr)\,.
 \eeq
 Going from the weak coupling regime at $\lambda=0.2, 0.5$ to the strong coupling regime at $\lambda =1.0, 2.0$, the first 7 rows of Table \ref{L2-H-pauliOps} corresponding to the diagonal operators of the $\L=2$ Hamiltonian do not change their coefficients, while the three operators IXXIXX, XIXXIX, XXIXXI accounting for the interaction part (the off-diagonal components) in the $\L=2$ Hamiltonian increase in values from $-0.05$ to $-0.5$. 
 \\\\
The exact energies at the four couplings, obtained by diagonalization, are
\beq
E^{\L=2}_{\l=0.2} = 3.14808, \qquad E^{\L=2}_{\l=0.5} = 3.36254,\qquad E^{\L=2}_{\l=1.0} = 3.69722,\qquad E^{\L=2}_{\l=2.0} = 4.26795.
\eeq
\subsection{\texttt{EfficientSU2} \& \texttt{TwoLocal}} \label{sec-l2-es2-tl}
\begin{figure}[!ht]
\centering
\includegraphics[width=.8\textwidth]{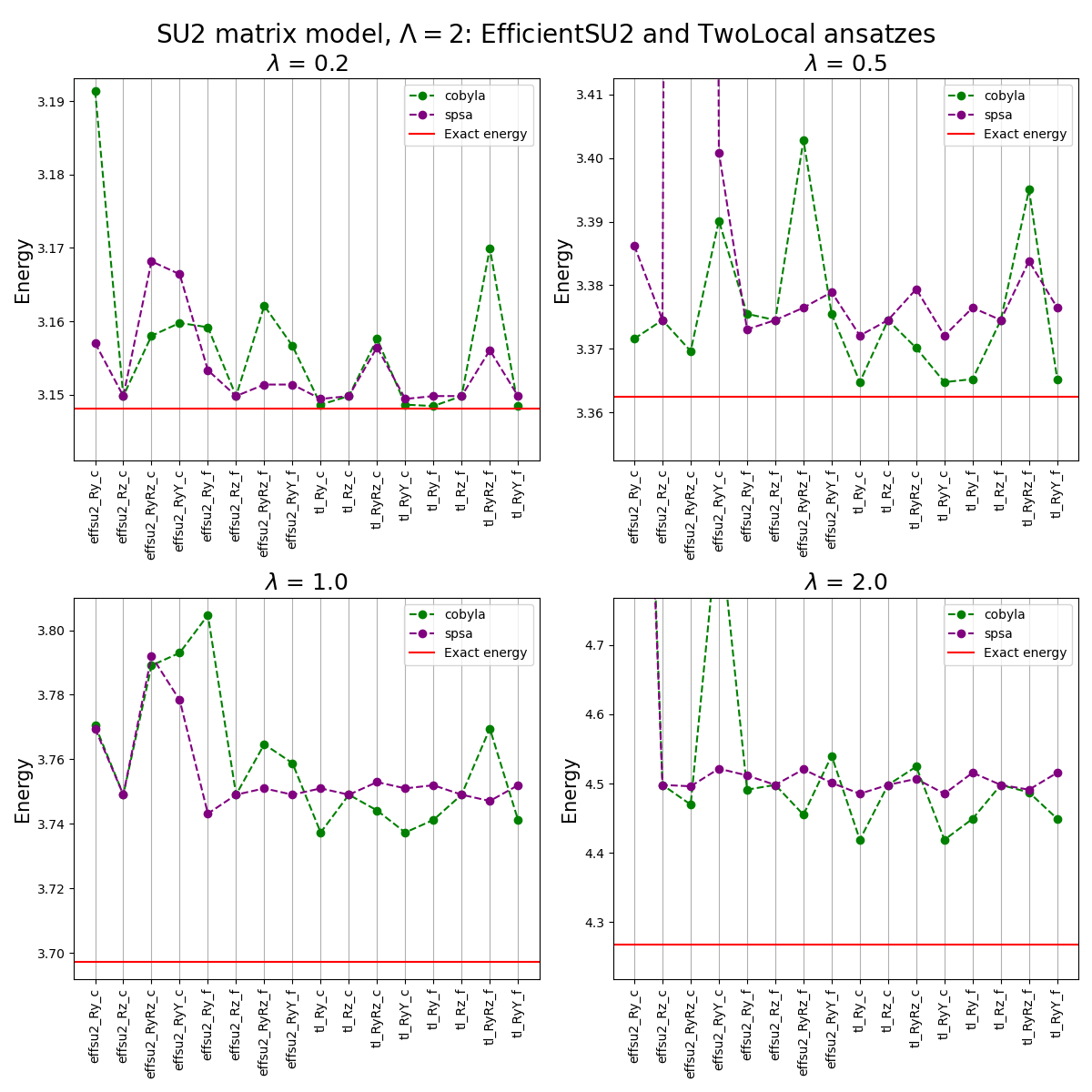}
\caption{ VQE experiments $\lf(H^{\L=2}_{\l}, \text{\texttt{EfficientSU2} \& \texttt{TwoLocal}, COBYLA/SPSA}\rr)$ - clockwise from top left $\l=0.2$, $\l=0.5$, $\l=2.0$, $\l=1.0$:  Comparison of the 8 variants of \texttt{EfficientSU2} and 8 variants \texttt{TwoLocal} ansatzes at each $\l$. The data points in each of the 4 subfigures above are generated from the full results included in Tables \ref{qve-l2-l02-ef-tl} - \ref{qve-l2-l20-ef-tl} found in the appendix (Section \ref{sec-L2-full-res-1}).}
\label{fig_L2_all_ef_tl}
\end{figure}

\begin{figure}[!ht]
\centering
\includegraphics[width=.7\textwidth]{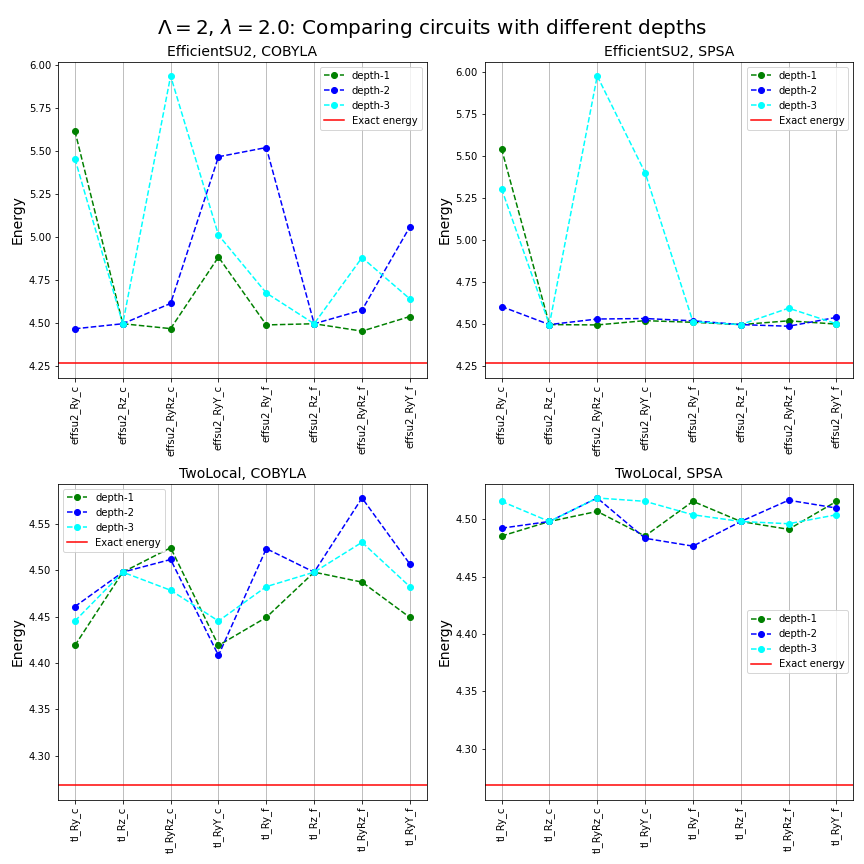}
\caption{Comparison of the energy results for the VQE experiments involving $H^{\L=2}_{\l=2.0}$ using depth-1, depth-2 and depth-3 \texttt{EfficientSU2} and \texttt{TwoLocal} quantum circuits, with COBYLA and SPSA optimizers. Clockwise from top left: 8 variants of depth-1, depth-2, depth-3 \texttt{EfficientSU2} circuits with COBYLA optimizer, same \texttt{EfficientSU2} circuits with SPSA optimizer, 8 variants of depth-1, depth-2, depth-3 \texttt{TwoLocal} circuits with SPSA optimizer, same \texttt{TwoLocal} circuits with COBYLA optimizer. A noticeable trend is that deeper circuits have comparable or even worse performances compared to the depth-1 version.}
\label{fig_L2_l20_ef_tl_mf}
\end{figure}

The 16 variants of the depth-1 ($d=1$) \texttt{EfficientSU2} and \texttt{TwoLocal} quantum circuits used in this problem, implemented with $n_Q = 6$ qubits, have the exact forms as those shown in Fig.\ref{qc_l2_effsu2} and Fig.\ref{qc_l2_twolocal}. Depending on the rotation gates and the entanglement pattern, the numbers of parameters due to these structures are
\beq
\text{Rotation}: &&
\begin{dcases}(d+1)n_Q = 12, &(R_Y, R_Z, R_YY)\\
\\2(d+1)n_Q = 24, &(R_YR_Z) \end{dcases}\non  \non
\text{Entanglement}:&& \begin{dcases} n_Q = 6 & \text{(circular)} \\ \sum\limits_{k=1}^{n_Q-1}k =\frac{1}{2}n_Q(n_Q-1)= 15 & \text{(full)} \end{dcases} \nonumber
\eeq
Table \ref{l2_es2_tl_params} recaps the structure of all 16 ansatz variants and lists their numbers of parameters. Variant-wise, \texttt{EfficientSU2} circuits have the same parameters for both full and circular entanglement patterns, since the entanglement part of these circuits does not include any parameterized gates. On the other hand, variant-wise, \texttt{TwoLocal} circuits whose entanglement part includes the parameterized $CRX$ gate have more parameters for the full entanglement than for the circular entanglement. The circuit with the largest number of parameters is \texttt{tl\_RyRz\_f}.
\begin{table}[!ht]
\centering
\begin{tabular}{ccccc}
\hline\hline
Ansatz & Rotation block & Entanglement pattern & Number of parameters & \\
\hline\hline
$\begin{array}{c}
 \texttt{effsu2\_Ry\_c} \\ \texttt{effsu2\_Rz\_c} \\\texttt{effsu2\_RyY\_c} \\\texttt{effsu2\_RyRz\_c}\end{array}$ &
$\begin{array}{c}R_Y \\ R_Z\\ R_YY \\ R_YR_Z\end{array}$
& circular
& $\begin{array}{c} 12\\ 12 \\ 12 \\ 24\end{array}$ & \\ \hline
$\begin{array}{c}
\texttt{effsu2\_Ry\_f}  \\\texttt{effsu2\_Rz\_f}\\ \texttt{effsu2\_RyY\_f}\\\texttt{effsu2\_RyRz\_f}\end{array}$
& $\begin{array}{c}R_Y\\ R_Z\\ R_YY \\R_YR_Z \end{array}$
& full
& $\begin{array}{c} 12\\ 12 \\ 12 \\ 24\end{array}$ & \\
\hline
$\begin{array}{c} \texttt{tl\_Ry\_c} \\ \texttt{tl\_Rz\_c} \\\texttt{tl\_RyY\_c} \\\texttt{tl\_RyRz\_c}\end{array}$ &
$\begin{array}{c}R_Y \\ R_Z\\ R_YY \\ R_YR_Z\end{array}$
& circular
& $\begin{array}{c} 18\\ 18 \\ 18 \\ 30\end{array}$ & \\\hline
$\begin{array}{c}
\texttt{tl\_Ry\_f}  \\\texttt{tl\_Rz\_f}\\ \texttt{tl\_RyY\_f}\\\texttt{tl\_RyRz\_f}\end{array}$
& $\begin{array}{c}R_Y\\ R_Z\\ R_YY \\R_YR_Z \end{array}$
& full
& $\begin{array}{c} 27\\ 27 \\ 27 \\ 39\end{array}$ & \\
\hline
\end{tabular}
\caption{VQE experiments $\lf(H^{\L=2}_{\l}, \text{\texttt{EfficientSU2} \& \texttt{TwoLocal}, COBYLA/SPSA}\rr)$: The list of the 8 variants of \texttt{EfficientSU2} and 8 variants of \texttt{TwoLocal} detailing their structures and number of parameters.} \label{l2_es2_tl_params}
\end{table}
\FloatBarrier
The best results obtained by running the  VQE experiments $\lf(H^{\L=2}_{\l}, \text{\texttt{EfficientSU2} \& \texttt{TwoLocal}, COBYLA/SPSA}\rr)$ using the eight variants of \texttt{EfficientSU2} and eight variants of \texttt{TwoLocal} quantum circuits with COBYLA and SPSA optimizers are summarized in Table \ref{L2_best_es2_tl}, in which the column `COBYLA' lists the best performing ansatz, with its asociated energy, obtained with COBYLA optimizer, the `SPSA' column lists the best ansatz and its associated energy obtained using SPSA optimizer. The column `Full results' lists the supplementary Tables/Figures (in the appendix) containing the full energy results for all 16 ansatzes together with their convergence curves. The performances of all \texttt{EfficientSU2} and \texttt{TwoLocal} ansatzes for all four coupling values are shown in Fig.\ref{fig_L2_all_ef_tl}.
\begin{table}[!ht]
\centering
\begin{tabular}{cccccc}
\hline\hline
Coupling & Exact & COBYLA & SPSA& Full results & \\
\hline\hline
 $\lambda = 0.2$ & 3.14808 & 
 $\begin{array}{c} \textbf{3.14844}\\ \texttt{tl\_Ry\_f}\end{array}$  & 
$\begin{array}{c}3.14941 \\ \texttt{tl\_Ry\_c} \end{array}$ 
& $\begin{array}{c} \text{Table \ref{qve-l2-l02-ef-tl} (F-S)} \\\text{Fig.\ref{fig_L2_l02_ef_tl} (CC)} \end{array}$ &
\\ \hline
 $\lambda = 0.5$ & 3.36254 &
 $\begin{array}{c} \textbf{3.36475} \\ \texttt{tl\_Ry\_c}\end{array}$  &
$\begin{array}{c} 3.37207\\ \texttt{tl\_Ry\_c} \end{array}$ 
& $\begin{array}{c} \text{Table \ref{qve-l2-l05-ef-tl} (F-S)} \\\text{Fig.\ref{fig_L2_l05_ef_tl} (CC)} \end{array}$&
\\\hline
 $\lambda = 1.0$ & 3.69722 & 
 $\begin{array}{c}\textbf{3.7373} \\\texttt{tl\_Ry\_c}  \end{array}$  & 
$\begin{array}{c} 3.74316\\ \texttt{effsu2\_Ry\_f}\end{array}$ 
& $\begin{array}{c} \text{Table \ref{qve-l2-l10-ef-tl} (F-S)} \\\text{Fig.\ref{fig_L2_l10_ef_tl} (CC)} \end{array}$&
\\ \hline
 $\lambda = 2.0$ & 4.26795 & 
 $\begin{array}{c} \textbf{4.41895}\\ \texttt{tl\_Ry\_c} \end{array}$  & 
$\begin{array}{c} 4.48535\\ \texttt{tl\_Ry\_c} \end{array}$ 
& $\begin{array}{c} \text{Table \ref{qve-l2-l20-ef-tl} (F-S)} \\\text{Fig.\ref{fig_L2_l20_ef_tl} (CC)} \end{array}$&
\\\hline
\end{tabular}
\caption{VQE experiments $\lf(H^{\L=2}_{\l}, \text{\texttt{EfficientSU2} \& \texttt{TwoLocal}, COBYLA/SPSA}\rr)$: Summary of the best results for each of the optimizer at the four couplings $\l$. See main text for the description of the columns. The best results are noted in bold. (F-S) denotes Full-Supplementary, and CC denotes `Convergence Curves'. Tables with the label (F-S) and Figures with the label (CC) can be found in Section \ref{sec-L2-full-res-1} in the appendix.} 
\label{L2_best_es2_tl}
\end{table}
\FloatBarrier
The main observations regarding the results are noted below. Of particular importance for the analyses of the results are the details concerning the best type of ansatzes, the overlap of the ansatzes with the true wavefunction and the performances of \texttt{TwoLocal} versus those of \texttt{EfficientSU2}. 
\ben
\item \textit{Best ansatzes}:  For the set of VQE experiments $\lf(H^{\L=2}_{\l}, \text{\texttt{EfficientSU2} \& \texttt{TwoLocal}, COBYLA/SPSA}\rr)$, the best ansatz type is always \texttt{TwoLocal}, obtained with COBYLA optimizer. At weak coupling ($\lambda=0.2$), \texttt{TwoLocal} ansatzes with full entanglement, \texttt{tl\_Ry\_f}, perform best, while at stronger couplings ($\l= 0.5, 1.0, 2.0)$, \texttt{TwoLocal} ansatzes with circular entanglement, \texttt{tl\_Ry\_c}, perform best (Table \ref{L2_best_es2_tl}). It is interesting to note that \texttt{EfficientSU2/TwoLocal} variants with $R_YR_Z$ rotation block perform poorly compared to those with $R_Y, R_Z, R_YY$ rotation block (see Fig.\ref{fig_L2_all_ef_tl}). 
 Furthermore, different optimizers yield different results for the same quantum circuit ansatz used  as obvious from Fig.\ref{fig_L2_all_ef_tl}, where different data points corresponding to either COBYLA or SPSA are observed for the same ansatz. 
\item \textit{Overlaps with the exact ground state}:
\bei
\item At $\l=0.2$ (see Fig.\ref{fig_L2_all_ef_tl}, first row, left subfigure),  there is a close overlap of the majority of ansatzes with the exact energy such as \texttt{effsu2\_Rz\_f}, \texttt{tl\_Ry\_c}, \texttt{tl\_RyY\_c}, \texttt{tl\_Ry\_f}, \texttt{tl\_Rz\_f}, \texttt{tl\_RyY\_f} (using SPSA), and \texttt{tl\_Ry\_c}, \texttt{tl\_Ry\_f}, \texttt{tl\_RyY\_f} (using COBYLA).
\item
At $\l=0.5$ (see Fig.\ref{fig_L2_all_ef_tl}, first row, right subfigure), there is no overlap between any ansatz using SPSA optimizer, but some close overlap for several ansatzes such as \texttt{tl\_Ry\_c}, \texttt{tl\_RyY\_c}, \texttt{tl\_Ry\_f}, \texttt{tl\_RyY\_f} using COBYLA optimizer.
\item
At $\l=1.0, 2.0$ (Fig.\ref{fig_L2_all_ef_tl}, second row, left and right subfigures), with either optimizer, all ansatzes yielded values far above the correct energy value. This trend of the results obtained at weak couplings being more acurrate than those at strong couplings is indicative of the fact that the problem-agnostic, generic \texttt{EfficientSU2} and \texttt{TwoLocal} quantum circuit ansatzes used have more overlap with the actual wavefunction at weak couplings than those at strong couplings. This observation was already made in the work \cite{prx-22}.
\eni
\item \textit{Effect of circuit depths}: For the $\l=2.0$ case, since there is no overlap of the depth-1 circuits used with the exact ground state wavefunction, we performed some additional VQE experiments involving deeper versions of the 16 ansatzes to determine whether more parameters would lead to better performance. Our results, plotted in Fig.\ref{fig_L2_l20_ef_tl_mf} indicate that increasing the depths of the circuits does not lead to better results, as we obtained mostly similar or worse results with the depth-2 and depth-3 versions of all 16 circuits. Each subfigure of Fig.\ref{fig_L2_l20_ef_tl_mf}  shows the results of the VQE experiments involving a particular combination of ansatzes and optimizer, clockwise from the top left, we have the 8 variants of depth-1, depth-2, depth-3 \texttt{EfficientSU2} used with COBYLA optimizer, followed by the same \texttt{EfficientSU2} circuits used with SPSA optimizer, followed by the 8 variants of depth-1, depth-2, depth-3 \texttt{TwoLocal} used with SPSA optimizer, followed by the same \texttt{TwoLocal} circuits used with COBYLA optimizer. In each of the subfigures, the green line denoting the depth-1 circuits are almost always closer to the exact energy than the blue and cyan lines denoting depth-2 and depth-3 circuits.
\FloatBarrier
\item \textit{\texttt{TwoLocal} versus \texttt{EfficientSU2}}: Out of 64 comparisons made between the 8 variants of \texttt{EfficientSU2} and the corresponding 8 variants of \texttt{TwoLocal} using 2 different optimizers at 4 different couplings, \texttt{TwoLocal} quantum circuit ansatzes almost always outperform or at least are on par with \texttt{EfficientSU2} ansatzes of the same variant, using either COBYLA or SPSA. This is evident from the convergence curve plots of Figs. \ref{fig_L2_l02_tl_vs_es_cobyla}, \ref{fig_L2_l02_tl_vs_es_spsa}, \ref{fig_L2_l05_tl_vs_es_cobyla}, \ref{fig_L2_l05_tl_vs_es_spsa}, \ref{fig_L2_l10_tl_vs_es_cobyla}, \ref{fig_L2_l10_tl_vs_es_spsa}, \ref{fig_L2_l20_tl_vs_es_cobyla}, \ref{fig_L2_l20_tl_vs_es_spsa} in  Section \ref{sec-L2-tl-vs-effsu2} of the appendix, in which the orange curve representing \texttt{TwoLocal} ansatzes always converge at the same values as or at lower values than the blue curve representing \texttt{EfficientSU2} ansatzes.
The faster convergence and better performance of \texttt{TwoLocal} ansatzes compared to their  \texttt{EfficientSU2} counterparts  can  probably be attributed to the fact that the latter contain the parameterized entanglement block that enhances their expressivity while the former do not.
\item A peculiar trend to note is that circuits involving $R_Z$ in the rotation blocks of either type (\texttt{EfficientSU2} or \texttt{TwoLocal}) have flat convergence curve with almost no variations (using COBYLA) or very few oscillations (using SPSA) in values. This form of ansatz is almost impervious to the variational process (especially using COBYLA). Furthermore, at all couplings, the convergence curves are identical for \texttt{EfficientSU2} and \texttt{TwoLocal} circuits involving $R_Z$ rotation block (as evident from the complete overlap of these curves in Figs. \ref{fig_L2_l02_tl_vs_es_cobyla} - \ref{fig_L2_l20_tl_vs_es_spsa} in Section \ref{sec-L2-tl-vs-effsu2}).
\enn
\subsection{\texttt{EvolvedOperatorAnsatz}} \label{sec-l2-evop}
\begin{figure}[!ht]
\centering
\begin{subfigure}[t]{0.4\textwidth}
\includegraphics[height=1.in]{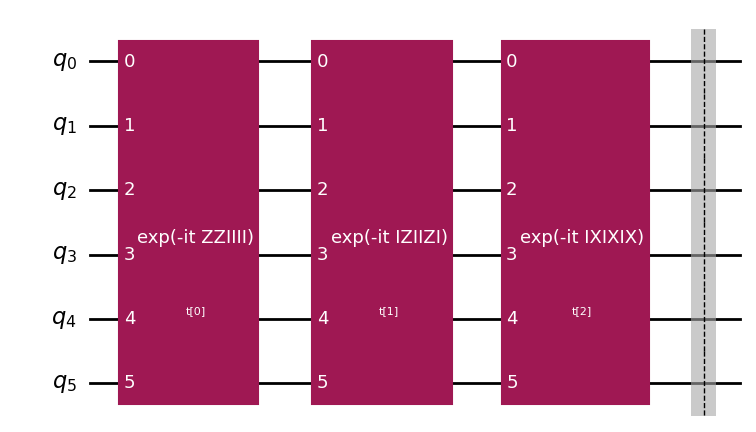}
\caption{\texttt{ev\_op\_r}}
\label{qc-l2-ev-op-r}
\end{subfigure}
\begin{subfigure}[t]{0.4\textwidth}
\includegraphics[height=1.in]{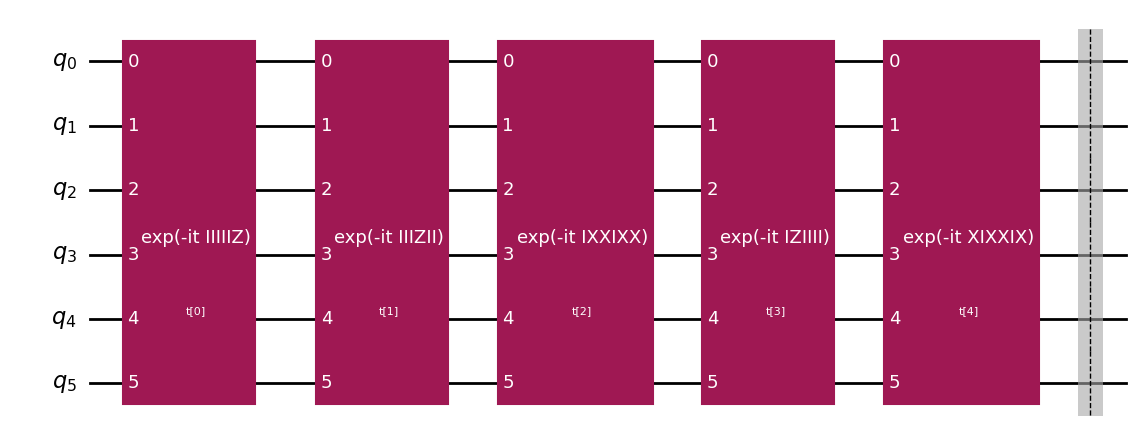}
\caption{\texttt{ev\_op\_Hp}}
\label{qc-l2-ev-op-Hp}
\end{subfigure}
\begin{subfigure}[t]{0.7\textwidth}
\centering
\includegraphics[height=2.2in]{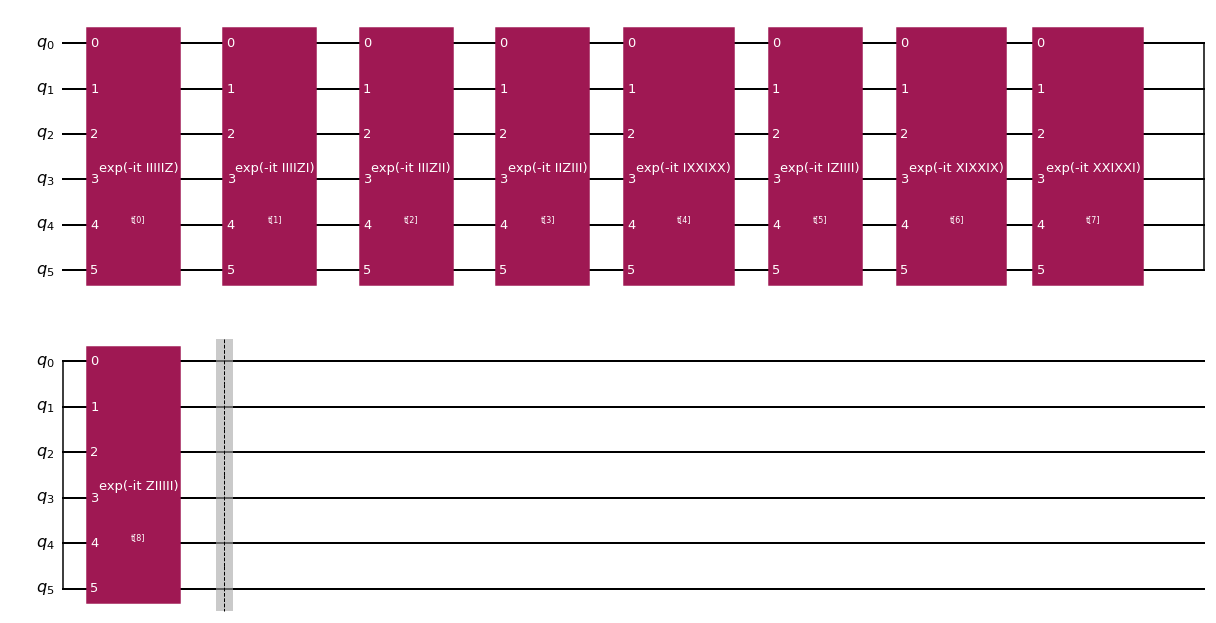}
\caption{\texttt{ev\_op\_Hp}}
\label{qc-l2-ev-op-Hp}
\end{subfigure}
\caption{The three unique depth-1 \texttt{EvolvedOperator} quantum circuits from Table \ref{L2-ev-op-ansatze}.}
\label{L2-eo-fig}
\end{figure}

\begin{figure}[!ht]
\centering
\includegraphics[width=.8\textwidth]{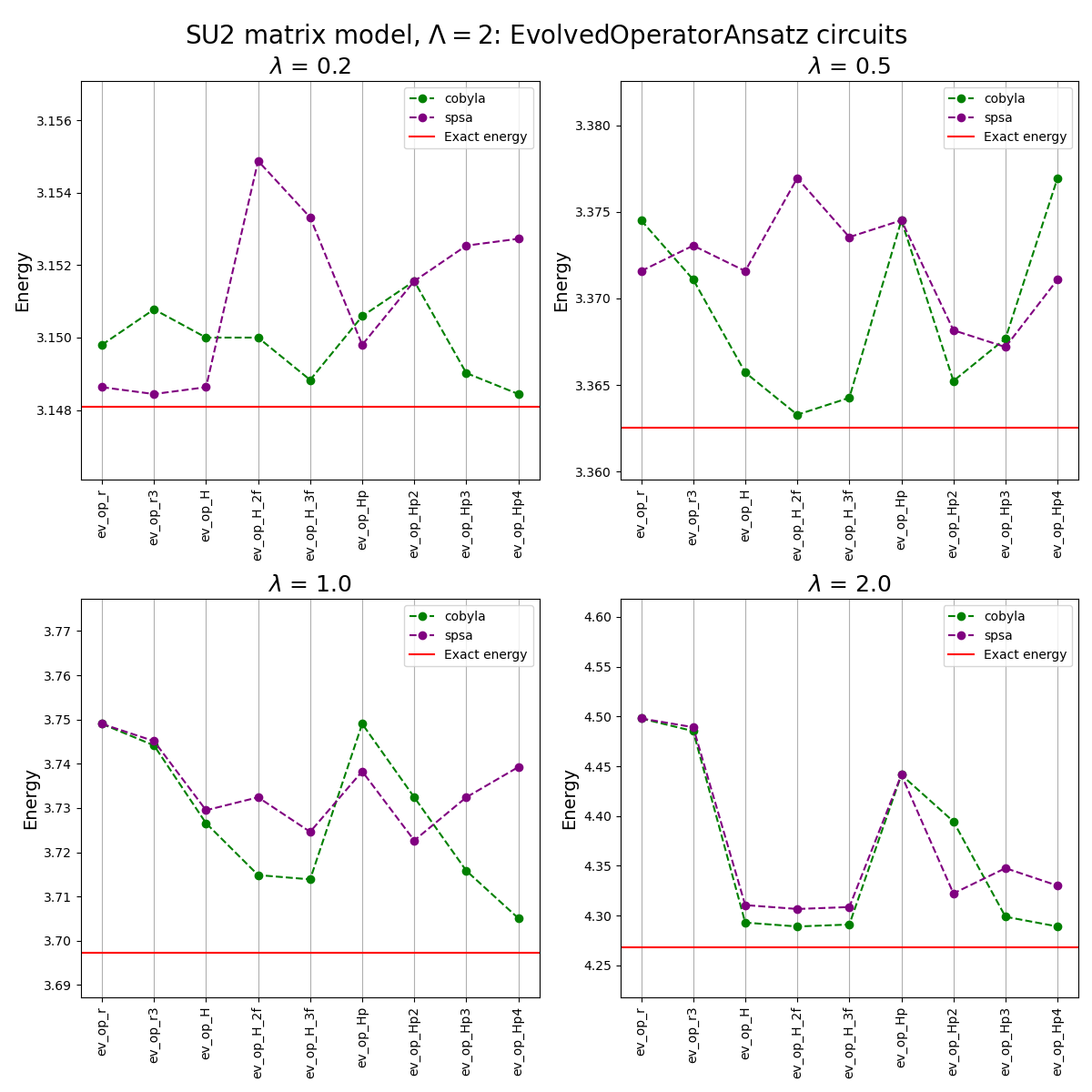}
\caption{Bosonic $SU(2)$ matrix model at Fock cutoff $\L=2$ at different couplings (clockwise from top left $\l=0.2$, $\l=0.5$, $\l=2.0$, $\l=1.0$): Comparison of all \texttt{EvolvedOperatorAnsatz} quantum circuit ansatzes at each $\l$. The data points in the four subfigures above are from the Tables \ref{qve-l2-l02-ev-op} - \ref{qve-l2-l20-ev-op} in Section \ref{sec-L2-full-res-2} in the appendix.}
\label{fig_L2_all_ev_op}
\end{figure}

For the case of $SU(2)$ bosonic matrix model at Fock cutoff $\L=2$, we construct the nine tailored \texttt{EvolvedOperator}\footnote{We use the shortened form \texttt{EvolvedOperator} to refer to \texttt{EvolvedOperatorAnsatz} occasionally in this paper  since the meaning is clear with little chance of confusion.} quantum circuits listed in Table \ref{L2-ev-op-ansatze}. Out of these eight ansatzes, only three quantum circuits (shown in Fig.\ref{L2-eo-fig}) are unique, with the rest being the deeper versions of these. In particular,
\bei
\item \texttt{ev\_op\_r} uses three random Pauli string operators ZZIIII, IZIIZI, IXIXIX as building blocks. These operators are random in the sense that they are not related to the operators listed in Table \ref{L2-H-pauliOps} that make up the $SU(2)$ $\L=2$ Hamiltonian. 
\item \texttt{ev\_op\_H} uses the 9 Pauli string operators listed in Table \ref{L2-H-pauliOps} as building blocks - These 9 operators make up  almost the entirety of the  $SU(2)$ $\L=2$ Hamiltonian matrix. In selecting the operators for this ansatz, we could also include the identity operator IIIIII, but that makes no difference in the VQE algorithm since the identity operator cannot be parameterized. 
\item \texttt{ev\_op\_Hp} uses the 5 Pauli string operators IIIIIZ, IIIZII, IXXIXX, IZIIII, XIXXIX as building blocks - These 5 operators are a subset of the 9 operators used in  \texttt{ev\_op\_H}. 
\eni
All nine ansatzes with their structures and corresponding numbers of paramters are listed in Table \ref{L2-ev-op-ansatze}. 
\begin{table}[!ht]
\centering
\begin{tabular}{ccccc}
\hline\hline
& Ansatz & Parameters & Operators & \\
\hline\hline
&\texttt{ev\_op\_r} & 3 &[\text{ZZIIII}, \text{IZIIZI}, \text{IXIXIX}] &  \\ \\
&\texttt{ev\_op\_H} & 9&$\begin{pmatrix}\text{IIIIIZ}, & \text{IIIIZI}, & \text{IIIZII},\\
        \text{IIZIII}, & \text{IXXIXX}, & \text{IZIIII},\\
        \text{XIXXIX}, & \text{XXIXXI}, & \text{ZIIIII}\end{pmatrix}$ & 
\\\\
&\texttt{ev\_op\_Hp} & 5&[\text{IIIIIZ}, \text{IIIZII},
        \text{IXXIXX}, \text{IZIIII}, 
        \text{XIXXIX}] &
\\\\
&\texttt{ev\_op\_r3} &  9&depth-3 version of \texttt{ev\_op\_r}   & 
\\\\
&\texttt{ev\_op\_H\_2f} &18& depth-2 version of \texttt{ev\_op\_H}& 
\\\\
&\texttt{ev\_op\_H\_3f}&27 & depth-3 version of \texttt{ev\_op\_H}& 
\\\\
&\texttt{ev\_op\_Hp2} &10& depth-2 version of \texttt{ev\_op\_Hp}  &
\\\\
&\texttt{ev\_op\_Hp3} & 15& depth-3 version of \texttt{ev\_op\_Hp}  &
\\\\
&\texttt{ev\_op\_Hp4} & 20&depth-4 version of \texttt{ev\_op\_Hp}  &
\\
\hline
\end{tabular}
\caption{The list of all  \texttt{EvolvedOperatorAnsatz} circuit variants used for the VQE experiments $\lf(H^{\L=2}_{\l}, \texttt{EvolvedOperatorAnsatz}, \text{COBYLA \& SPSA}\rr)$.}\label{L2-ev-op-ansatze}
\end{table}
\newpage
The best results obtained by running the  VQE experiments $\lf(H^{\L=2}_{\l}, \texttt{EvolvedOperatorAnsatz}, \text{COBYLA/SPSA}\rr)$ using the nine variants of \texttt{EvolvedOperatorAnsatz} from Table \ref{L2-ev-op-ansatze}  with COBYLA and SPSA optimizers are summarized in Table \ref{L2_best_ev_op}. Similar in structure to Table \ref{L2_best_es2_tl}, the column `COBYLA' lists  the best performing ansatz obtained when using COBYLA optimizer, the column `SPSA' lists the  best ansatz  obtained when using SPSA optimizer. The last column lists the supplementary Tables containing the full results (all of which can be found in the appendix, Section \ref{sec-L2-full-res-2}). Furthermore, Fig.\ref{fig_L2_all_ev_op} shows the performances of all nine \texttt{EvolvedOperatorAnsatz} circuits at all four couplings.
\begin{table}[!ht]
\centering
\begin{tabular}{cccccc}
\hline\hline
Coupling & Exact & COBYLA & SPSA& Full results & \\
\hline\hline
 $\lambda = 0.2$ & 3.14808 & 
 $\begin{array}{c} 3.14844 \\ \texttt{ev\_op\_Hp4} \end{array}$  & 
$\begin{array}{c} \textbf{3.14844}\\ \texttt{ev\_op\_r3} \end{array}$ &
 $\begin{array}{c} \text{Table \ref{qve-l2-l02-ev-op} (F-S)}  \end{array}$ &
\\ \hline
 $\lambda = 0.5$ & 3.36254 & 
 $\begin{array}{c} \textbf{3.36328}\\\texttt{ev\_op\_H\_2f} \end{array}$  & 
$\begin{array}{c} 3.36719\\ \texttt{ev\_op\_Hp3}\end{array}$ &
 $\begin{array}{c} \text{Table \ref{qve-l2-l05-ev-op} (F-S)} \end{array}$&
\\\hline
 $\lambda = 1.0$ & 3.69722 & 
 $\begin{array}{c}\textbf{3.70508} \\\texttt{ev\_op\_Hp4} \end{array}$  & 
$\begin{array}{c} 3.72266\\\texttt{ev\_op\_Hp2} \end{array}$ &
 $\begin{array}{c} \text{Table \ref{qve-l2-l10-ev-op} (F-S)}  \end{array}$&
\\ \hline
 $\lambda = 2.0$ & 4.26795 & 
 $\begin{array}{c}  \textbf{4.28906}\\\texttt{ev\_op\_H\_2f} \end{array}$  & 
$\begin{array}{c} 4.30664\\\texttt{ev\_op\_H\_2f}  \end{array}$ &
 $\begin{array}{c} \text{Table \ref{qve-l2-l20-ev-op} (F-S)}  \end{array}$&
\\\hline
\end{tabular}
\caption{VQE experiments $\lf(H^{\L=2}_{\l}, \text{\texttt{EvolvedOperatorAnsatz}, COBYLA/ SPSA}\rr)$: Summary of the best results for each type of optimizers at the four couplings $\l$. See the main text for the description of the columns.  The best results are noted in bold. (F-S) denotes Full-Supplementary. Tables with the label (F-S) can be found in the appendix (Section \ref{sec-L2-full-res-2}). The convergence curves for this set of VQE experiments are plotted in Fig.\ref{fig_L2_ev_op_cobyla} for COBYLA optimizer and  Fig.\ref{fig_L2_ev_op_spsa} for SPSA optimizer (Section \ref{sec-L2-full-res-2}).} \label{L2_best_ev_op}
\end{table}
\begin{itemize}
\item \textit{Best ansatzes}: For the VQE experiments $\lf(H^{\L=2}_{\l}, \text{\texttt{EvolvedOperatorAnsatz}, COBYLA \& SPSA}\rr)$ at all four couplings, conforming to expections, the best performing ansatzes are almost always those quantum circuits comprising operators that are part of the $\L=2$ Hamiltonian (either the 9-operator or the 5-operator variants), and not the ones formed by random operators. The only exception is the case of $\l=0.2$, with the best performing ansatz being \texttt{ev\_op\_r3}, the depth-3 version of \texttt{ev\_op\_r} (Fig \ref{L2-eo-fig}(a)) - the variant containing 3 random operators\footnote{At $\l=0.2$, \texttt{ev\_op\_r3} and \texttt{ev\_op\_Hp4} otained the same results of 3.14844, but \texttt{ev\_op\_r3} has only 9 parameters versus the 20 parameters of \texttt{ev\_op\_Hp4}, making it the better variant, since a more performing variant is always one with fewer parameters.}.  At $\l=0.5$ and $\l=2.0$, the best ansatz variant is \texttt{ev\_op\_H\_2f}, the depth-2 version of \texttt{ev\_op\_H} (Fig.\ref{L2-eo-fig}(c)) containing 9 operators making up the $SU(2)$ Hamiltonian. At $\l=1.0$, the best variant is \texttt{ev\_op\_Hp4}, the depth-4 version of \texttt{ev\_op\_Hp} containing 5 operators making up the Hamiltonian (Fig.\ref{L2-eo-fig}(b))
\item \textit{Overlaps with the exact ground state}:
\bei
\item At $\l=0.2$ (Fig.\ref{fig_L2_all_ev_op}, first row, left subfigure), all depth-1 unique circuits used with SPSA, together with \texttt{ev\_op\_Hp4} used with COBYLA, have good overlaps with the exact wavefunction.
\item At $\l=0.5$ (Fig.\ref{fig_L2_all_ev_op}, first row, right subfigure), only \texttt{ev\_op\_H\_2f} and \texttt{ev\_op\_H\_3f} used with COBYLA have some close overlaps with the exact wavefunction.
\item
At $\l=1.0$ (Fig.\ref{fig_L2_all_ev_op}, second row, left subfigure), only \texttt{ev\_op\_Hp4} used with COBYLA has a relatively close overlap with the ground state.
\item  At $\l=2.0$ (Fig.\ref{fig_L2_all_ev_op}, second row, right subfigure), several variants (including \text{ev\_op\_H}, \texttt{ev\_op\_H\_2f} and \texttt{ev\_op\_H\_3f} used with either COBYLA or SPSA) have relatively close overlaps with the exact ground state.
\item Despite the fact that none of the ansatzes have truly good overlaps with the exact ground state at strong couplings $\l=1.0, 2.0$, these overlaps of the \texttt{EvolvedOperatorAnsatz} circuits considered in this section with the ground state are still significantly better than those of \texttt{EfficientSU2} and \texttt{TwoLocal} ansatzes in the previous section, as evident by comparing the top left and top right subfigures in Fig.\ref{fig_L2_all_ev_op} with the corresponding ones in Fig.\ref{fig_L2_all_ef_tl}. This is indicative of a better approximation to the true wavefunction using the tailored \texttt{EvolvedOperatorAnsatz} circuits.
\eni
\end{itemize}
\newpage
\subsection{Comparison of all quantum circuits}\label{sec_su2_l2_comparison}
In this section, we collect the best results obtained by running the VQE experiments with the three different types of ansatzes and two different types of optimizers in Table \ref{L2-all-res}. For reference and later as benchmarks, we also include the best results reported in \cite{prx-22}, obtained by using the depth-3 \texttt{EfficientSU2} quantum circuits with the rotation block being $R_Y$ (for all couplings) with different optimizers. As the circuits of \cite{prx-22} are all depth 3 ($d=3$), the numbers of parameters are $(d+1)n_Q = 24$ for the $R_Y$ variational form and $2(d+1)n_Q = 48$ for the $R_YR_Z$ variational form. Each entry in the first three rows of Table \ref{L2-all-res} is a tuple ($E$, ansatz, number of parameters, optimizer) listing the energy at convergence, the ansatz variant, the number of parameters in the ansatz, and the optimizer used to obtain the result. The entries in the second last row corresponding to the results of \cite{prx-22} have a slightly different format,  $(E$,  number of parameters, optimizer), in which the  ansatz is not listed, since the authors of \cite{prx-22} exclusively used depth-3 \texttt{EfficientSU2} circuits with $R_Y$ rotation block for these experiments. 
\ben
\item Within this work, among the three types of quantum circuits used, the tailored \texttt{EvolvedOperator} circuits emerged as the best performing type of ansatzes for all four couplings $\l=0.2, 0.5, 1.0, 2.0$, followed by the \texttt{TwoLocal} quantum circuits and finally by \texttt{EfficientSU2} quantum circuits. Among \texttt{EfficientSU2} and \texttt{TwoLocal}, an interesting trend to note is the better performance of variants with circular entanglement pattern compared to those with full entanglement pattern, as three out of four best variants of either \texttt{EfficientSU2} or \texttt{TwoLocal} have circular entanglement pattern.
\item 
Including the results of \cite{prx-22} as benchmarks (the second last row of Table \ref{L2-all-res}), \texttt{EvolvedOperatorAnsatz} circuits still emerged as the best performers. In particular, for the cases of $\l=0.2, 0.5$, both \texttt{TwoLocal} and  \texttt{EvolvedOperatorAnsatz} variants  yielded better results than \cite{prx-22}, while for the cases of $\l=1.0, 2.0$, only \texttt{EvolvedOperatorAnsatz} circuits achieved better results than \cite{prx-22}. It is also noteworthy that in all four instances, \texttt{EvolvedOperatorAnsatz}, with at most 20 parameters, outperformed \texttt{EfficientSU2} with 24 parameters. 
\enn
\begin{table}[!ht]
\centering
\begin{tabular}{l l l l l l}
\hline\hline
 Ansatz Type & $\lambda = 0.2$ & $\lambda = 0.5$ & $\lambda = 1.0$ & $\lambda = 2.0$ & \\
\hline\hline
 \texttt{EfficientSU2} & 
 $\begin{array}{l} 3.14980\\ \texttt{effsu2\_Rz\_c}\\\text{(12 params)}\\ \text{COBYLA/SPSA}\end{array}$ &
 $\begin{array}{l} 3.36963\\ \texttt{effsu2\_RyRz\_c}\\\text{(24 params)}\\ \text{COBYLA}\end{array}$  &
 $\begin{array}{l} 3.74902 \\\texttt{effsu2\_Rz\_c}\\\text{(12 params)}\\ \text{COBYLA/SPSA}\end{array}$ &
  $\begin{array}{l}4.45508 \\\texttt{effsu2\_RyRz\_f}\\\text{(24 params)}\\\text{COBYLA}\end{array}$&   \\
\hline
 \texttt{TwoLocal}  &
 $\begin{array}{l}  3.14844\\ \texttt{tl\_Ry\_f}\\\text{(27 params)}\\\text{COBYLA}\end{array}$  & 
 $\begin{array}{l}  3.36475\\ \texttt{tl\_Ry\_c}\\\text{(18 params)}\\\text{COBYLA}\end{array}$&
 $\begin{array}{l}  3.73730 \\ \texttt{tl\_Ry\_c}\\\text{(18 params)}\\\text{COBYLA}\end{array}$ & 
  $\begin{array}{l}  4.41895\\ \texttt{tl\_Ry\_c}\\\text{(18 params)}\\\text{COBYLA}\end{array}$ &  \\\hline
 \texttt{EvolvedOperator}& 
 $\begin{array}{l} \textbf{3.14844} \\ \texttt{ev\_op\_r3}\\\text{(9 params)}\\ \text{COBYLA}\end{array}$& 
  $\begin{array}{l}\textbf{3.36328} \\ \texttt{ev\_op\_H\_2f}\\\text{(18 params)}\\ \text{COBYLA}\end{array}$& 
 $\begin{array}{l}\textbf{3.70508} \\\texttt{ev\_op\_Hp4}\\\text{(20 params)}\\ \text{COBYLA}\end{array}$&
 $\begin{array}{l}\textbf{4.28906} \\ \texttt{ev\_op\_Hp\_2f}\\\text{(18 params)}\\ \text{COBYLA}\end{array}$ & \\\hline
$\begin{array}{l} \text{Results from \cite{prx-22}}\\\texttt{EfficientSU2}\\\text{$R_Y$ (depth 3)}
\end{array}$ & 
$\begin{array}{l} 3.14897\\\text{(24 params)}\\\text{NELDER-MEAD} \end{array}$ &
 $\begin{array}{l} 3.36675\\\text{(24 params)}\\ \text{SLSQP}\end{array}$& 
  $\begin{array}{l} 3.71463 \\\text{(24 params)}\\ \text{COBYLA}\end{array}$ & 
   $\begin{array}{l} 4.33636\\\text{(24 params)}\\ \text{SLSQP}\end{array}$
&\\\hline 
 Exact energy & 3.14808& 3.36254&  3.69722& 4.26795& \\

\hline
\end{tabular}
\caption{VQE experiments involving $\L=2$ bosonic $SU(2)$ matrix model: Summary of the best results from each type of ansatzes (\texttt{EfficientSU2}, \texttt{TwoLocal}, \texttt{EvolvedOperatorAnsatz}) obtained from this work, as well as those reported in \cite{prx-22}, at different couplings $\l$ for $SU(2)$ matrix model at cutoff $\Lambda=2$. The absolute best results obtained from comparing all results in this table are noted in bold. } \label{L2-all-res}
\end{table}
\newpage
\section{$\Lambda = 4$ bosonic model} \label{sec_qc_su2_l4}
\begin{figure}[!ht]
\centering
\includegraphics[width=.6\textwidth]{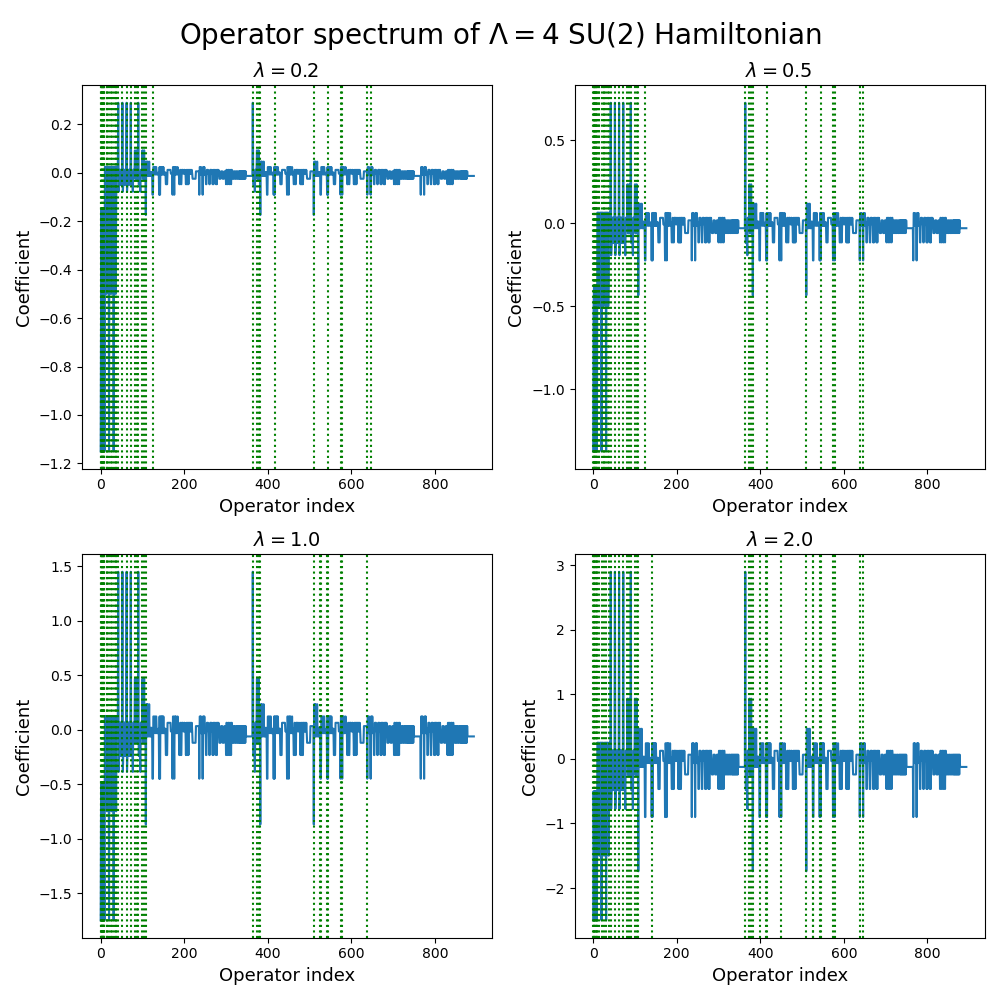}
\caption{Graphical representations of the set of 895 Pauli string operators forming the $\L=4$ Hamiltonian for four couplings $\l=0.2$ (first row, left subfigure), $\l=0.5$ (first row, right subfigure), $\l=1.0$ (second row, left subfigure) and $\l=2.0$ (second row, right subfigure). The $x$-axis labels  the operator index (the order of appeareance of the operator in the sum forming the Hamiltonian whose full expression is available at \cite{lorrespz-l4}) ranging from 1 to 895, the $y$ axis labels the operator coefficient. The green vertical lines in each subfigures correspond to the 40 largest operators (at each of the coupling $\l$) listed in Table \ref{l4_ops_40}.}\label{l4_op_sp}
\end{figure}

For the case of $SU(2)$ bosonic matrix model at Fock cutoff $\Lambda =4$, the Hamiltonian is a $2^{12}\times 2^{12}$, or $4096\times 4096$ matrix. When expressed as a sum of Pauli operator strings, the final expression contains 895 terms and can be downloaded as a text file which is available at this GitHub link \cite{lorrespz-l4}, since it is too long to be included in full here. 
The exact energies by diagonalization for the four Hamiltonians  at different couplings are:
\beq
E_{\l=0.2} = 3.13406,\qquad E_{\l=0.5} = 3.29894, \qquad E_{\l=1.} = 3.52625, \qquad E_{\l=2.} = 3.89548\,.
\label{Ee_L4}
\eeq
In Table \ref{l4_ops_40}, we list the 40 largest operators (by absolute values) and their coefficients for the $\L=4$ Hamiltonian at four different couplings $\l=0.2, 0.5, 1.0, 2.0$. These operators correspond to the vertical green lines in Fig.\ref{l4_op_sp} which shows graphically the magnitudes of the coefficients of all 895 operators for each of the 4 couplings. In Table \ref{l4_ops_40}, Group (E), Group (G) and Group (C) operators, which are Pauli strings made of the tensor products of the various combinations of the identity matrix and Pauli `Z' matrix, account partly for the diagonal components of the $\L=4$ Hamiltonian. The remaining operators, from Group (A) to Group (K), which are Pauli strings made of tensor products of various combinations of the identity matrix with the Pauli `X', `Y'  matrices, account partly for the interaction part, or the off-diagonal components, of the $\L=4$ Hamiltonian. 
\begin{table}[!ht]
\centering
\begin{tabular}{c l c c c c c}
\hline\hline
Group & Operator & $\l=0.2$ & $\l=0.5$ & $\l=1.0$ & $\l=2.0$ & \\
\hline\hline
&  IIIXYYIIIXIX &-0.0901& -0.2253 &-0.4506 &-0.9012&\\
(A) &   IIIXIXIIXXIX    &-0.0901& -0.2253 &-0.4506 &-0.9012&\\
&  IIIXIXIIYYIX &-0.0901& -0.2253 &-0.4506 &-0.9012&\\
&  IXIIIXIXIIXX &-0.0901& -0.2253 &-0.4506 &-0.9012&\\
\hline
&  IIXIIIXIIIII &0.0933& 0.2333  &0.4665 & 0.9330&\\
&   IIIIXIXIIIII    &0.0933& 0.2333  &0.4665 & 0.9330&\\
(B) &  XIIIIIIIXIII &0.0933& 0.2333  &0.4665 & 0.9330&\\
&  IIIIXIIIXIII &0.0933& 0.2333  &0.4665 & 0.9330&\\
& XIIIIIIIIIXI  &0.0933& 0.2333  &0.4665 & 0.9330&\\
&   IIXIIIIIIIXI    &0.0933& 0.2333  &0.4665 & 0.9330&\\
\hline
& ZZIIIIIIIIII  &-0.1500& -0.3750 &-0.7500 &-1.5000&\\
& IIZZIIIIIIII  &-0.1500& -0.3750 &-0.7500 &-1.5000&\\
(C) & IIIIZZIIIIII  &-0.1500& -0.3750 &-0.7500 &-1.5000&\\
& IIIIIIZZIIII  &-0.1500& -0.3750 &-0.7500 &-1.5000&\\
& IIIIIIIIZZII  &-0.1500& -0.3750 &-0.7500 &-1.5000&\\
& IIIIIIIIIIZZ  &-0.1500& -0.3750 &-0.7500 &-1.5000&\\
\hline
& IXIXIIIXIXII  &-0.1741& -0.4353 &-0.8705 &-1.7410&\\
(D)&  IXIIIXIXIIIX &-0.1741& -0.4353 &-0.8705 &-1.7410&\\
&  IIIXIXIIIXIX &-0.1741& -0.4353 &-0.8705 &-1.7410&\\
\hline
&  ZIIIIIIIIIII &-1.1500 &-1.3750 &-1.7500 &-2.5000&\\
&  IIZIIIIIIIII &-1.1500 &-1.3750 &-1.7500 &-2.5000&\\
&  IIIIZIIIIIII &-1.1500 &-1.3750 &-1.7500 &-2.5000&\\
(E)&  IIIIIIZIIIII &-1.1500 &-1.3750 &-1.7500 &-2.5000&\\
&  IIIIIIIIZIII &-1.1500 &-1.3750 &-1.7500 &-2.5000&\\
&  IIIIIIIIIIZI &-1.1500 &-1.3750 &-1.7500 &-2.5000&\\
\hline
&  XIIIIIIIIIII &0.2898& 0.7244 &1.4489  &2.8978&\\
&  IIXIIIIIIIII &0.2898& 0.7244 &1.4489  &2.8978&\\
&   IIIIXIIIIIII    &0.2898& 0.7244 &1.4489  &2.8978&\\
(F)& IIIIIIXIIIII  &0.2898& 0.7244 &1.4489  &2.8978&\\
&  IIIIIIIIXIII &0.2898& 0.7244 &1.4489  &2.8978&\\
&   IIIIIIIIIIXI    &0.2898& 0.7244 &1.4489  &2.8978&\\
\hline
& IZIIIIIIIIII  &-0.5000& -0.5000 &-0.5000 &---&\\
&  IIIZIIIIIIII &-0.5000&-0.5000 &-0.5000  &---&\\
&  IIIIIZIIIIII &-0.5000&-0.5000 &-0.5000  &---&\\
(G)&  IIIIIIIZIIII &-0.5000&-0.5000 &-0.5000  &---&\\
&  IIIIIIIIIZII &-0.5000&-0.5000 &-0.5000  &---&\\
&  IIIIIIIIIIIZ &-0.5000&-0.5000 &-0.5000  &---&\\
\hline
(H)&   XXIXIIIXIXII    &-0.0901& -0.2253 &--- & --- &\\
\hline
 (I)&   IXIIYYIXIIIX    &-0.0901& -0.2253 & --- & -0.9012&\\
&  IXIIIXIXIIYY &-0.0901& -0.2253 & --- & -0.9012&\\
\hline
&   IIXXIXIIIXIX    & ---& ---&-0.4506 & -0.9012&\\
(J) &  IIYYIXIIIXIX & ---& ---&-0.4506 &-0.9012&\\
&   IIIXXXIIIXIX    & ---& ---&-0.4506 &-0.9012&\\
\hline
&   IXXXIIIXIXII    & ---& ---& --- & -0.9012 &\\
(K) &   YYIIIXIXIIIX    & ---& ---& --- &-0.9012&\\
&   IXIIXXIXIIIX    & ---& ---& --- &-0.9012&\\
&   IXIIIXYYIIIX    & ---& ---& --- &-0.9012&\\
\hline
\end{tabular}
\caption{The largest 40 operators by absolute values for the $\L=4$ Hamiltonian for 4 couplings $\l=0.2, 0.5, 1.0, 2.0$. The dashed lines `---' refer to the absence of a particular operator in the set under consideration. For example, the 31 operators in groups from (A) to (F) are common to all couplings, while the operators in group (G) are common only to $\l=0.2, 0.5$ and $\l=1.0$ Hamiltonians (not present in the  $\l=2.0$ case). The operators in group (J) are only present in $\l=1.0$ and $\l=2.0$ cases, etc.  }\label{l4_ops_40}
\end{table}
\FloatBarrier
\clearpage

\subsection{\texttt{EfficientSU2} \& \texttt{TwoLocal}}\label{sec-l4-es2-tl}
\begin{figure}[!ht]
\centering
\includegraphics[width=.8\textwidth]{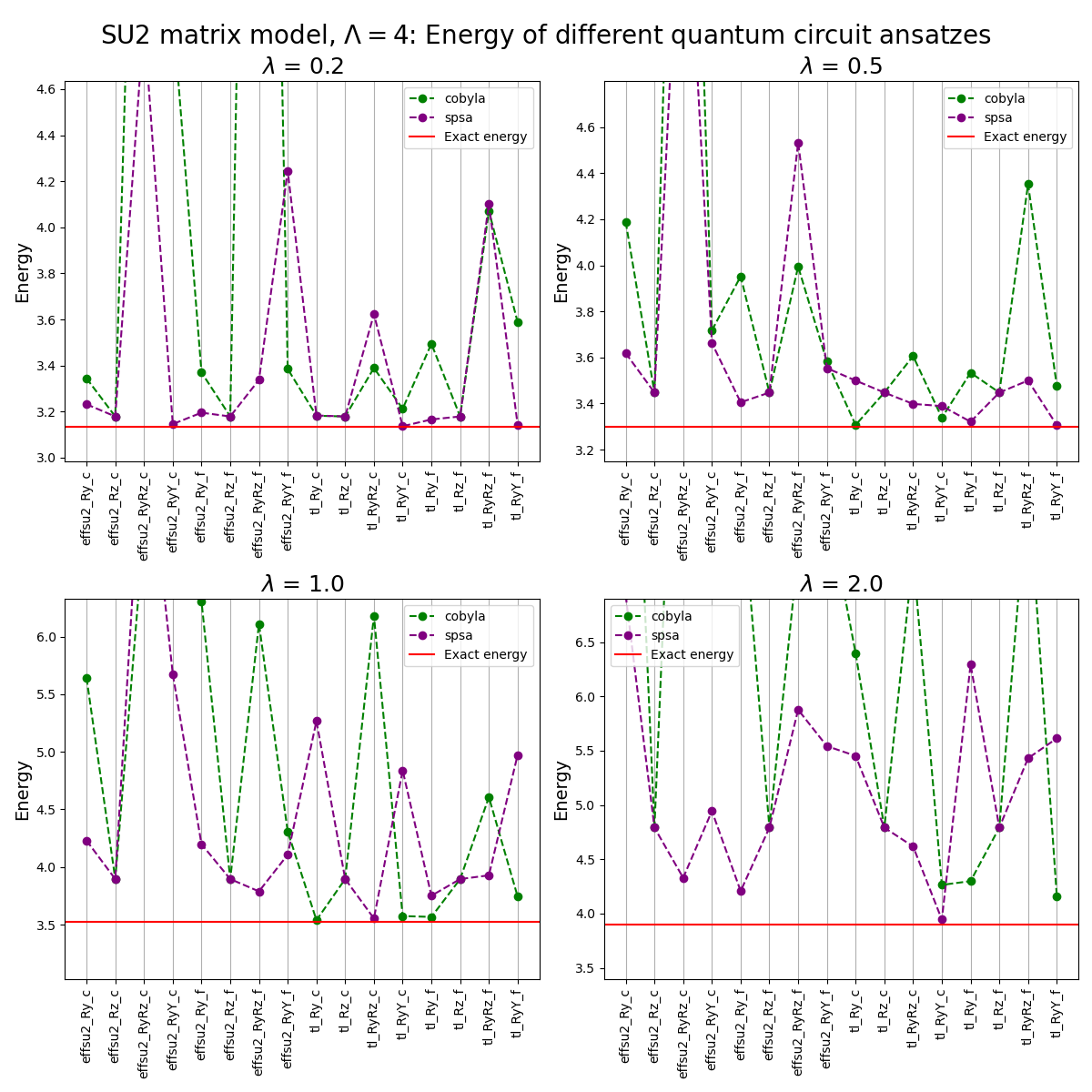}
\caption{Bosonic $SU(2)$ matrix model at Fock cutoff $\L=4$ at different couplings (clockwise from top left $\l=0.2$, $\l=0.5$, $\l=2.0$, $\l=1.0$): Comparison of all \texttt{EfficientSU2}/\texttt{TwoLocal} ansatzes at each $\l$ with the $y$-axis zoomed in to the vicinity of the exact energy value. The data points for each $\l$ from the Figure above are from Table \ref{qve-l4-l02-es-tl}, \ref{qve-l4-l05-es-tl}, \ref{qve-l4-l10-es-tl}  and \ref{qve-l4-l20-es-tl} found in the Appendix.}
\label{fig_L4_all_es_tl}
\end{figure}

We use the same 8 variants of the depth-1 \texttt{EfficientSU2} and 8 variants of the depth-1 \texttt{TwoLocal} quantum circuit ansatzes as in the case of $\Lambda=2$, but now each circuit consists of $n_Q=12$ qubits instead of $n_Q=6$ qubits.  The numbers of parameters due to the rotation and entanglement parts for this case are
\beq
\text{Rotation}: &&
\begin{dcases}(d+1)n_Q = 24, &(R_Y, R_Z, R_YY)\\
\\2(d+1)n_Q = 48, &(R_YR_Z) \end{dcases}\non  \non
\text{Entanglement}:&& \begin{dcases} n_Q = 12 & \text{(circular)} \\ \sum\limits_{k=1}^{n_Q-1}k =\frac{1}{2}n_Q(n_Q-1)= 66 & \text{(full)} \end{dcases} \nonumber
\eeq
Table \ref{l4_es2_tl_params} recaps the structure of the 16 ansatz variants and lists their numbers of parameters. As already noted in the case of $\L=2$, \texttt{EfficientSU2} circuits have the same parameters for both full and circular entanglement patterns (variant-wise), since the entanglement part of these circuits does not include any parameterized gates. On the other hand, variant-wise, \texttt{TwoLocal} circuits whose entanglement part includes the parameterized $CRX$ gate have more parameters for the full entanglement than for the circular entanglement. The circuit with the largest number of parameters is \texttt{tl\_RyRz\_f} with 114 parameters.
\begin{table}[!ht]
\centering
\begin{tabular}{ccccc}
\hline\hline
Ansatz & Rotation block & Entanglement pattern & Number of parameters & \\
\hline\hline
$\begin{array}{c}
 \texttt{effsu2\_Ry\_c} \\ \texttt{effsu2\_Rz\_c} \\\texttt{effsu2\_RyY\_c} \\\texttt{effsu2\_RyRz\_c}\end{array}$ &
$\begin{array}{c}R_Y \\ R_Z\\ R_YY \\ R_YR_Z\end{array}$
& circular
& $\begin{array}{c} 24\\ 24 \\ 24 \\ 48\end{array}$ & \\ \hline
$\begin{array}{c}
\texttt{effsu2\_Ry\_f}  \\\texttt{effsu2\_Rz\_f}\\ \texttt{effsu2\_RyY\_f}\\\texttt{effsu2\_RyRz\_f}\end{array}$
& $\begin{array}{c}R_Y\\ R_Z\\ R_YY \\R_YR_Z \end{array}$
& full
& $\begin{array}{c}  24\\ 24 \\ 24 \\ 48\end{array}$ & \\
\hline
$\begin{array}{c} \texttt{tl\_Ry\_c} \\ \texttt{tl\_Rz\_c} \\\texttt{tl\_RyY\_c} \\\texttt{tl\_RyRz\_c}\end{array}$ &
$\begin{array}{c}R_Y \\ R_Z\\ R_YY \\ R_YR_Z\end{array}$
& circular
& $\begin{array}{c} 36\\ 36 \\ 36 \\ 60\end{array}$ & \\\hline
$\begin{array}{c}
\texttt{tl\_Ry\_f}  \\\texttt{tl\_Rz\_f}\\ \texttt{tl\_RyY\_f}\\\texttt{tl\_RyRz\_f}\end{array}$
& $\begin{array}{c}R_Y\\ R_Z\\ R_YY \\R_YR_Z \end{array}$
& full
& $\begin{array}{c} 90\\ 90 \\ 90 \\ 114\end{array}$ & \\
\hline
\end{tabular}
\caption{VQE experiments $\lf(H^{\L=4}_{\l}, \text{\texttt{EfficientSU2} \& \texttt{TwoLocal}, COBYLA/SPSA}\rr)$: The list of the 8 variants of \texttt{EfficientSU2} and 8 variants of \texttt{TwoLocal} detailing their structures and number of parameters.} \label{l4_es2_tl_params}
\end{table}
\FloatBarrier
The best results obtained by running the VQE exeriments $\lf(H^{\L=4}_\l, \texttt{EfficientSU2}\&\texttt{TwoLocal}, \text{COBYLA/SPSA}\rr)$ using the 8 variants of \texttt{EfficientSU2} and 8 variants of \texttt{TwoLocal} quantum circuits with COBYLA and SPSA optimizers are summarized in Table \ref{L4_best_es2_tl}, in which the column `COBYLA' lists the best performing ansatz (at each coupling) together with the associated energy $E$ obtained by using COBYLA optimizer, the column `SPSA' lists the best ansatz with the associated $E$ obtained by using SPSA optimizer. The column `Full results' lists the supplementary Tables/Figures (in the appendix) containing the full energy results for all 16 ansatzes together with their convergence curves.
 The performances of all \texttt{EfficientSU2} and \texttt{TwoLocal} ansatzes for all four coupling values are shown in Fig.\ref{fig_L4_all_es_tl}.
\begin{table}[!ht]
\centering
\begin{tabular}{ccccccc}
\hline\hline
Coupling &  Exact &  COBYLA & SPSA & Full results & \\
\hline\hline
$\l = 0.2$ & 3.13406 &  
$\begin{array}{c}3.1791\\ \texttt{effsu2\_Rz\_c} \end{array}$ &
$\begin{array}{c} \textbf{3.13679}\\\texttt{tl\_RyY\_c}\\ \end{array}$ &
$\begin{array}{c} \text{Table \ref{qve-l4-l02-es-tl} (F-S)} \\\text{Fig.\ref{fig_L4_l02_ef_tl} (CC)}\end{array}$ & \\
\hline
$\l = 0.5$ & 3.29894 &  
$\begin{array}{c}3.27478\\ \texttt{tl\_RyY\_c} \end{array}$ &
$\begin{array}{c} \textbf{3.30641}\\\texttt{tl\_RyY\_f}\\ \end{array}$ &
 $\begin{array}{c}\text{Table \ref{qve-l4-l05-es-tl} (F-S)} \\\text{Fig.\ref{fig_L4_l05_ef_tl} (CC)} \end{array}$ & \\
\hline
$\l = 1.0$ & 3.52625 &  
$\begin{array}{c}\textbf{3.53869}\\ \texttt{tl\_Ry\_c} \end{array}$ &
$\begin{array}{c} 3.55374\\\texttt{tl\_RyRz\_c}\\ \end{array}$ &
 $\begin{array}{c} \text{Table \ref{qve-l4-l10-es-tl} (F-S)} \\\text{Fig.\ref{fig_L4_l10_ef_tl} (CC)}\end{array}$ & \\
\hline
$\l = 2.0$ & 3.89548&  
$\begin{array}{c}4.16062\\ \texttt{tl\_RyY\_f} \end{array}$ &
$\begin{array}{c} \textbf{3.94466}\\\texttt{tl\_RyY\_c}\\ \end{array}$ &
 $\begin{array}{c} \text{Table \ref{qve-l4-l20-es-tl} (F-S)} \\\text{Fig.\ref{fig_L4_l20_ef_tl} (CC)}\end{array}$ & \\
\hline
\end{tabular}
\caption{VQE experiments $\lf(H^{\L=4}_{\l}, \text{\texttt{EfficientSU2} \& \texttt{TwoLocal}, COBYLA \& SPSA}\rr)$: Summary of the best results for each of the coupling $\l$. See main text for the description of the columns. For each row corresponding to a coupling $\l$, the best result (which is closest to the exact energy) is noted in bold. (F-S) denotes Full-Supplementary, and CC denotes `Convergence Curves'. Tables with the label (F-S) and Figures with the label (CC) can be found in Section \ref{sec-L4-full-res-1} in the appendix.} \label{L4_best_es2_tl}
\end{table}
\FloatBarrier
The main observations regarding the results are noted below. Trends concerning the best ansatzes, the overlap of the ansatzes with the true wavefunction, and the performances of \texttt{TwoLocal} versus those of \texttt{EfficientSU2} are the most important details to note. 
 \bei
\item \textit{Best ansatzes}: 
The best performing ansatz for 2 out of 4 couplings ($\l=0.2, 2.0$) is the \texttt{TwoLocal} variant \texttt{tl\_RyY\_c} involving the $R_YY$ rotation block with circular entanglement . For $\l=1.0$, the best performing ansatz is the \texttt{TwoLocal} variant \texttt{tl\_Ry\_c} with $R_Y$ rotation block with circular entanglement. For $\l=0.5$, the best performing ansatz is \texttt{tl\_RyY\_f} with $R_YY$ rotation block and full entanglement (see Table \ref{L4_best_es2_tl}).
\item \textit{Overlaps with the true wavefunction}:
\bei
\item
 At $\l=0.2$ (Fig.\ref{fig_L4_all_es_tl}, first row, left subfigure), using SPSA, many variants from both \texttt{EfficientSU2} such as \texttt{effsu2\_Ry\_c}, \texttt{effsu2\_Rz\_c}, \texttt{effsu2\_RyY\_c}, \texttt{effsu2\_RyY\_f}, \texttt{effsu2\_Rz\_f},\\ and \texttt{TwoLocal} such as \texttt{tl\_Ry\_c},
 \texttt{tl\_Rz\_c}, \texttt{tl\_RyY\_c}, \texttt{tl\_Ry\_f}, \texttt{tl\_Rz\_f},  \texttt{tl\_RyY\_f} have good overlaps with the exact ground state.
 \item At $\l=0.5$ (Fig.\ref{fig_L4_all_es_tl}, first row, right subfigure), only a few
 \texttt{TwoLocal} variants such as \texttt{tl\_Ry\_c} (with COBYLA), \texttt{tl\_Ry\_f} and \texttt{tl\_RyY\_f} (with SPSA) have good overlaps with the ground state.
 \item At $\l=1.0$ (Fig.\ref{fig_L4_all_es_tl}, second row, left subfigure), more \texttt{TwoLocal} ansatzes have good overlaps with the ground state, including \texttt{tl\_Ry\_c} with COBYLA, \texttt{tl\_RyRz\_c} with SPSA, \texttt{tl\_RyY\_c} and \texttt{tl\_Ry\_f} with COBYLA.
 \item At $\l=2.0$ (Fig.\ref{fig_L4_all_es_tl}, second row, right subfigure), the only variant with good overlap is \texttt{tl\_RyY\_c} with SPSA.
 \eni
\item \textit{Optimizer performances}: For weak couplings $\l=0.2, 0.5$, a wide range of fluctuations in the obtained $E$ values can be observed for the 8 variants of \texttt{EfficientSU2} with both COBYLA and SPSA, while the 8 variants of \texttt{TwoLocal} show a much smaller range of fluctuation with SPSA (and to a smaller extent, COBYLA). For $\l=1.0$, wider range of fluctuations across all variants of \texttt{EfficientSU2} (and to a smaller extent, \texttt{TwoLocal}) are seen with either optimizer. For $\l=2.0$, COBYLA optimizer yields a large range of fluctuation across all 16 ansatzes while SPSA has a relatively better performance (see Fig.\ref{fig_L4_all_es_tl}).
 \item \textit{\texttt{TwoLocal} versus \texttt{EfficientSU2}}: Similar to the $\L=2$ case above, out of the 64 comparisons made between the 8 variants of \texttt{TwoLocal} and the corresponding 8 variants of \texttt{EfficientSU2} at 4 different couplings using 2 different optimizers, \texttt{TwoLocal} circuits consistently outperform \texttt{EfficientSU2} using either optimizer for all couplings.  This is evident from the convergence curve plots in Figs. \ref{fig_L4_l02_tl_vs_es_cobyla}, \ref{fig_L4_l02_tl_vs_es_spsa}, \ref{fig_L4_l05_tl_vs_es_cobyla}, \ref{fig_L4_l05_tl_vs_es_spsa}, \ref{fig_L4_l10_tl_vs_es_cobyla}, \ref{fig_L4_l10_tl_vs_es_spsa}, \ref{fig_L4_l20_tl_vs_es_cobyla}, \ref{fig_L4_l20_tl_vs_es_spsa} (included in Section \ref{sec-L4-tl-vs-effsu2} of the appendix) in which the orange curves representing \texttt{TwoLocal} ansatzes almost always converge faster and at values below  the blue curves representing \texttt{EfficientSU2} ansatzes. The only few exceptions to this are:
 \bei
 \item the $\l=0.2$ case with SPSA involving circuits \texttt{tl\_RyRz\_f} and \texttt{effsu2\_RyRz\_f} with $R_YR_Z$ rotation blocks and full entanglement pattern (see Fig.\ref{fig_L4_l02_tl_vs_es_spsa}).
 \item the $\l=1.0$ case (see Fig.\ref{fig_L4_l10_tl_vs_es_spsa}) with SPSA  involving circuits with $R_Y$ rotation block with circular entanglement (\texttt{tl\_Ry\_c} \& \texttt{effsu2\_Ry\_c}), and $R_YY$ rotation block with full entanglement pattern (\texttt{tl\_RyY\_f} \& \texttt{effsu2\_RyY\_f}). 
 \item  the $\l=2.0$ case with SPSA involving circuits \texttt{tl\_Ry\_f} and \texttt{effsu2\_Ry\_f} with $R_Y$ rotation block with full entanglement (see Fig.\ref{fig_L4_l20_tl_vs_es_spsa}).
 \eni
The fact that all of the exceptions above occur with SPSA optimizer, which has a fixed number of iterations, and not COBYLA optimizer, which has a variable number of iterations might indicate that the observed exceptions above are attributable to the optimizer performance rather than the actual ansatz performance.
  \item The same peculiar trend noted in the $\L=2$ case is observed here:  Circuits involving $R_Z$ in the rotation block of either \texttt{EfficientSU2} or \texttt{TwoLocal} are almost impervious to the variational process (especially using COBYLA), as their convergence curves are practically straight lines which overlap completely for the two types of circuits (as can be seen from the complete overlap of these curves in Figs.\ref{fig_L4_l02_tl_vs_es_cobyla} - \ref{fig_L4_l20_tl_vs_es_spsa}).
\eni
\newpage
\subsection{\texttt{EvolvedOperatorAnsatz}} \label{sec-l4-evop}
\begin{figure}[!ht]
\centering
\includegraphics[width=.8\textwidth]{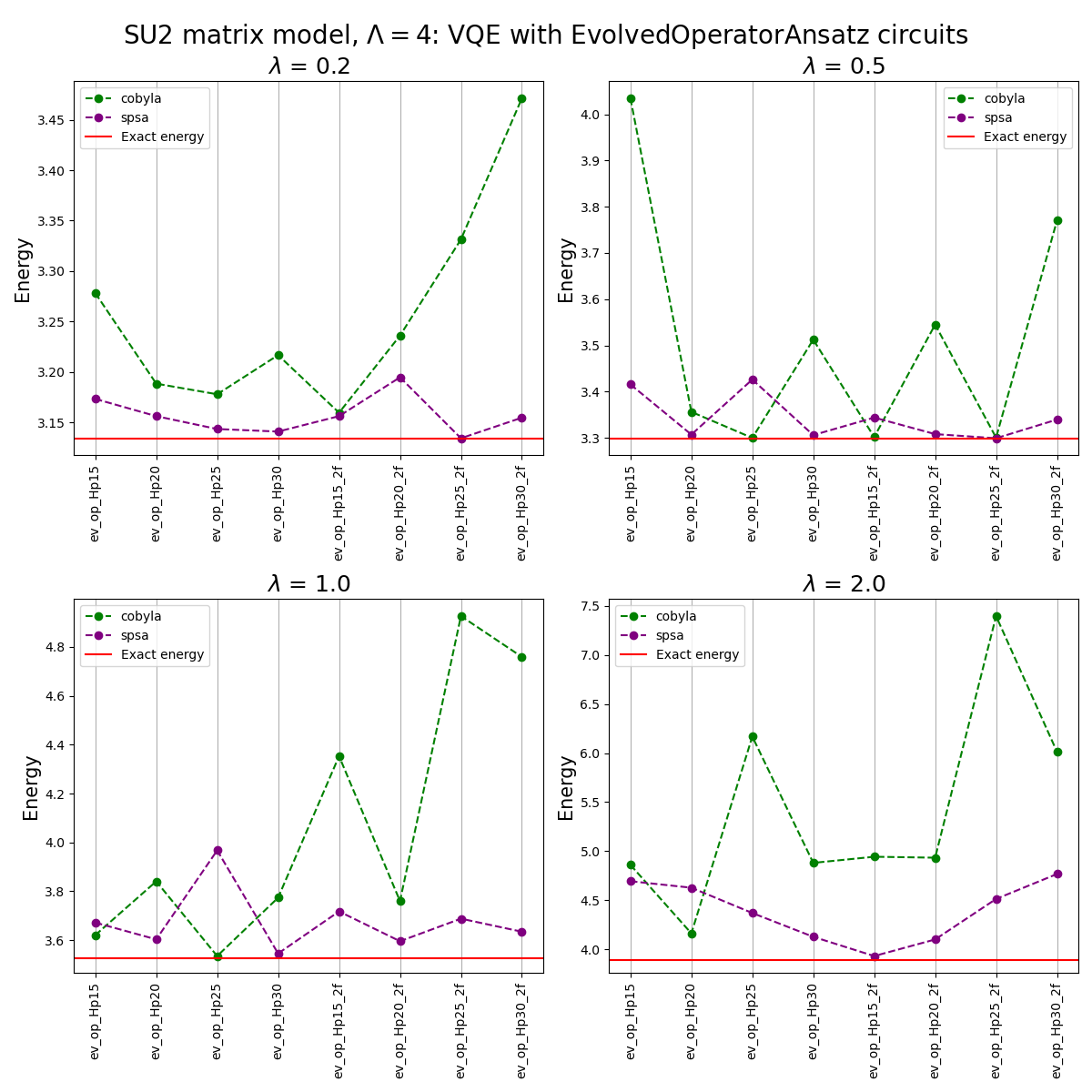}
\caption{Bosonic $SU(2)$ matrix model at Fock cutoff $\L=4$ at different couplings (clockwise from top left $\l=0.2$, $\l=0.5$, $\l=2.0$, $\l=1.0$): Comparison of all \texttt{EvolvedOperatorAnsatz} quantum circuit ansatzes. The data points for the 4 subfigures above are from the Tables \ref{L4_l02_eo} - \ref{L4_l20_eo} in Section \ref{sec-L4-full-res-2} of the appendix.}
\label{fig_L4_all_ev_op}
\end{figure}

To build the tailored \texttt{EvolvedOperatorAnsatz} quantum circuits for the $SU(2)$ bosonic matrix model at Fock cutoff $\L=4$, we use the same approach as the $\L=2$ case in which the quantum circuits are created by selecting a subset of operators that form the Hamiltonian to be the building blocks. However, unlike the $\L=2$ case where the Hamiltonian is only a $64\times 64$ matrix  and can be expressed as the sum of only 10 Pauli string operators in which  9 out of  these 10 operators can be picked to build the tailored \texttt{EvolvedOperatorAnsatz} circuits, the $\L=4$ Hammiltonian is a $4096\times 4096$ matrix and is the sum of 895 Pauli string operators. It is out of the question to use all 895 operators or even a much smaller number of 100 operators to build the tailored circuit, due to the exponentially slow running time of the VQE algorithms when dealing with circuits of that size.
Because of this setback, we will work with various smaller subsets, containing $N=15, 20, 25, 30$ operators chosen by the largest absolute values of their coefficients. 
\\\\
For each set of $N$ operators (where $N=15,20,25,30$), we created two \texttt{EvolvedOperatorAnsatz} quantum circuits, a depth-1 version and a depth-2 version. This led to the eight quantum circuit ansatzes  which are listed in Table \ref{L4_ev_op_quantum circuits}. Note that for each coupling $\l$, the content of the set of $N$ largest operators is different, i.e. the set of  $N=15$ operators at $\lambda=0.2$ (weak coupling), consisting of operators from Group (E)+ Group (G) + half of Group (F) in Table \ref{l4_ops_40}, is not the same as the set of $N=15$ operators at $\l=2.0$ (strong coupling), consisting of operators from Group (F) + Group (E) + Group (D) in Table \ref{l4_ops_40}. This leads to \texttt{EvolvedOperatorAnsatz} quantum circuits having different building blocks at each coupling $\lambda$, although they may have the same name. The specific building blocks for each variant of the \texttt{EvolvedOperatorAnsatz} quantum circuits are listed in full in Table \ref{L4_ev_op_quantum circuits}.
\begin{table}[!ht]
\centering
\begin{tabular}{c c l c}
\hline\hline
Ansatz & Parameters & Description & \\
\hline\hline
\texttt{ev\_op\_Hp15} & 15 & 
$\begin{array}{l} \text{Largest 15 operators} 
\\ \l=0.2: (E)+(G)+\frac{1}{2}(F)
\\ \text{excl. [IIIIIIXIIIII, IIIIIIIIXIII, IIIIIIIIIIXI] in (F)}\\
\\ \l=0.5: (E) + (F)+\frac{1}{2}(G)
\\\text{excl. [IZIIIIIIIIII, IIIIIIIIIZII, IIIIIIIIIIIZ] in (G)}\\
\\\l=1.0:  (E) + (F) + (D)\\
\\\l=2.0: (E) + (F) + (D)
\end{array}$ &\\
\hline
\texttt{ev\_op\_Hp20} & 20 & 
 $\begin{array}{l} \text{Largest 20 operators} 
 \\ \l=0.2: (E) + (F) + (G) + (2/3)(D)
 \\\text{excl. IXIXIIIXIXII in (D)}\\
\\ \l=0.5: (E) + (F) + (G) + (2/3)(D)
\\\text{excl. IXIXIIIXIXII in (D)}\\
\\\l=1.0: (E) + (F) + (D) + (5/6)(C)
\\ \text{excl. IIZZIIIIIIII in (C)}\\
\\\l=2.0: (E) + (F) + (D) + (5/6)(C)\\
\text{excl. ZZIIIIIIIIII in (C)}
 \\ \end{array}$  &\\
 \hline
\texttt{ev\_op\_Hp25} & 25 & 
$\begin{array}{l} \text{Largest 25 operators}
 \\ \l=0.2: (E) + (F) + (G) + (D)+ (2/3)(C)
 \\\text{excl. [IIIIIIZZIIII, IIIIIIIIZZII] in (C)}\\
\\ \l=0.5: (E) + (F) + (G) + (D)+ (2/3)(C)
\\\text{excl. [IIIIIIZZIIII, IIIIIIIIZZII] in (C)}\\
\\\l=1.0: (E) + (F) + (C) + (D) + (2/3)(G)
\\\text{excl. [IZIIIIIIIIII, IIIIIIIIIZII] in (G)}\\
\\\l=2.0: (E) + (F) + (C) + (D) + (2/3)(B)
\\\text{excl. [IIXIIIXIIIII, XIIIIIIIXIII] in (B)}
 \\ \end{array}$   &\\
 \hline
\texttt{ev\_op\_Hp30} & 30 &  
 $\begin{array}{l} \text{Largest 30 operators} 
  \\ \l=0.2: (E) + (F) + (G) + (D)+ (C) + (1/2)(B)
  \\\text{excl. [IIXIIIXIIIII, IIIIXIXIIIII, XIIIIIIIXIII] in (B)}\\
\\ \l=0.5: (E) + (F) + (G) + (D)+ (C) + (1/2)(B)
\\\text{excl. [IIXIIIXIIIII, IIIIXIXIIIII, XIIIIIIIXIII] in (B)}\\
\\\l=1.0: (E) + (F) + (C) + (D) + (G)+ (1/2)(B) 
\\\text{excl. [IIXIIIXIIIII, XIIIIIIIXIII, IIIIXIIIXIII] in (B)}\\
\\\l=2.0: (E) + (F) + (C) + (D) + (B) + (3/4)(A)
\\\text{excl. IIIXYYIIIXIX in (A)} 
 \\ \end{array}$  &\\
 \hline
\texttt{ev\_op\_Hp15\_2f} & 30 & depth-2 version of \texttt{ev\_op\_Hp15}&\\
\texttt{ev\_op\_Hp20\_2f} & 40 & depth-2 version of \texttt{ev\_op\_Hp20}&\\
\texttt{ev\_op\_Hp25\_2f} & 50 & depth-2 version of \texttt{ev\_op\_Hp25}&\\
\texttt{ev\_op\_Hp30\_2f} & 60 & depth-2 version of \texttt{ev\_op\_Hp30}&\\
\hline
\end{tabular}
\caption{The list of eight \texttt{EvolvedOperatorAnsatz} quantum circuits used for running the VQE for $\L=4$ Hamiltonian. In the `Description' column, the building blocks of each  variant is listed in the notation of Table \ref{l4_ops_40}, i.e. (A),\ldots, (G) refer to the Group (A),\ldots, (G) to which certain types of Pauli string operators are labeled. The fractions before some groups mean that only those fractions specfified (and not all operators from the groups) are selected. The full list of operators for each $N$ at each coupling can be found in the \texttt{Jupyter} notebook available at this GitHub link \cite{lorrespz-ops}.}\label{L4_ev_op_quantum circuits}
\end{table}
The best results for each type of optimizers for $\L=4$ $SU(2)$ matrix model at all four couplings are summarized in Table \ref{L4_best_ev_op}, which has the same format as previous sections. The column `COBYLA'/`SPSA' lists the best ansatz and associated energy result obtained with COBYLA/SPSA for each coupling. The performances of all \texttt{EvolvedOperatorAnsatz} variants are visually presented in Fig.\ref{fig_L4_all_ev_op}. 
\begin{table}[!ht]
\centering
\begin{tabular}{ccccccc}
\hline\hline
Coupling &  Exact &  COBYLA & SPSA & Full results & \\
\hline\hline
$\l = 0.2$ & 3.13406 &  
$\begin{array}{c} 3.15952\\ \texttt{ev\_op\_Hp15\_2f} \end{array}$ &
$\begin{array}{c} \textbf{3.13421}\\\texttt{ev\_op\_Hp25\_2f}\\ \end{array}$ &
 \text{Table \ref{L4_l02_eo}  (F-S)} & \\
\hline
$\l = 0.5$ & 3.29894 &  
$\begin{array}{c}3.29968 \\ \texttt{ev\_op\_Hp25} \end{array}$ &
$\begin{array}{c} \textbf{3.29896}\\\texttt{ev\_op\_Hp25\_2f}\\ \end{array}$ &
 \text{Table \ref{L4_l05_eo}  (F-S)} & \\
\hline
$\l = 1.0$ & 3.52625 &  
$\begin{array}{c}\textbf{3.53512}\\ \texttt{ev\_op\_Hp25} \end{array}$ &
$\begin{array}{c} 3.54551\\\texttt{ev\_op\_Hp30}\\ \end{array}$ &
 \text{Table \ref{L4_l10_eo}  (F-S)} & \\
\hline
$\l = 2.0$ & 3.89548 &  
$\begin{array}{c} 4.16425\\ \texttt{ev\_op\_Hp20} \end{array}$ &
$\begin{array}{c} \textbf{3.93348}\\\texttt{ev\_op\_Hp15\_2f}\\ \end{array}$ &
 \text{Table \ref{L4_l20_eo}  (F-S)} & \\
\hline
\end{tabular}
\caption{VQE experiments $\lf(H^{\L=4}_{\l}, \text{\texttt{EvolvedOperatorAnsatz}, COBYLA \& SPSA}\rr)$: Summary of the best results for each of the optimizer at four couplings $\l$. See main text for the description of the columns.  The best results are noted in bold. (F-S) denotes Full-Supplementary. Tables with the label (F-S) can be found in the appendix. The convergence curves of the energy for all \texttt{EvolvedOperatorAnsatz} can be found in Fig. \ref{fig_L4_eo_curves}.} \label{L4_best_ev_op}
\end{table}
Some observations regarding the best ansatzes and the overlaps of the ansatzes with the true ground state are noted below. 
\bei
\item \textit{Best ansatzes}: Rather contrary to the expectation that performance should improve as more operators (forming the Hamiltonian) are added in the circuits, the best performing quantum circuits are not those with the largest number of operators.  As can be seen from Table \ref{L4_best_ev_op}, the best ansatz at each coupling is never \texttt{ev\_op\_Hp30} or its depth-2 version \texttt{ev\_op\_Hp30\_2f}, which have the largest number of operators out of all variants considered. Instead, the best performing variant for $\l=0.2, 0.5, 1.0$ is the one with 25 operators (the second-largest number of operators), \texttt{ev\_op\_Hp25\_2f} and \texttt{ev\_op\_Hp25}. For $\l=2.0$, it is \texttt{ev\_op\_Hp15\_2f} (with 15 operators).
\item \textit{Overlaps of the ansatzes with the exact ground state}:
\bei
\item At $\l=0.2$ (Fig.\ref{fig_L4_all_ev_op}, first row, left subfigure), using SPSA, \texttt{ev\_op\_Hp25}, \texttt{ev\_op\_Hp30} and \texttt{ev\_op\_Hp25\_2f} show good overlaps with the exact ground state.
\item At $\l=0.5$ (Fig.\ref{fig_L4_all_ev_op}, first row, right subfigure), multiple variants show good overlaps with the ground state, including \texttt{ev\_op\_Hp20}, \texttt{ev\_op\_Hp30}, \texttt{ev\_op\_Hp20\_2f}, \texttt{ev\_op\_Hp25\_2f} with SPSA, and \texttt{ev\_op\_Hp25}, \texttt{ev\_op\_Hp15\_2f}, \texttt{ev\_op\_Hp25\_2f}  using COBYLA.
\item At $\l=1.0$ (Fig.\ref{fig_L4_all_ev_op}, second row, left subfigure), only \texttt{ev\_op\_Hp30} with SPSA and \texttt{ev\_op\_Hp25} with COBYLA show good overlaps with the ground state.
\item
At $\l=2.0$ (Fig.\ref{fig_L4_all_ev_op}, second row, right subfigure), the only variant with a good overlap with the ground state is \texttt{ev\_op\_H15\_2f} with SPSA.
\eni
\item
\textit{Optimizer performances}:
For all couplings, it is evident that SPSA optimizer yields a more stable and accurate performance for all quantum circuits compared to COBYLA as can be seen from Fig.\ref{fig_L4_all_ev_op} in which the purple line representing the results obtained with SPSA are almost always closer to the exact energy line than the green line representing the results obtained with COBYLA.
\eni
\clearpage
\subsection{Comparison of all quantum circuits}\label{sec_su2_l4_comparison}
In this section, we collect the best results obtained by running the experiments with the three different types of ansatzes and two different types of optimizers for the case of $\L=4$ in Table \ref{L4-all-res}. 
As was done in the $\L=2$ case, we also included the best results reported from the work \cite{prx-22} in the second last row of Table \ref{L4-all-res}. These results were obtained by using the L-BFGS-B optimizer and the depth-3  \texttt{EfficientSU2} quantum circuits with the rotation block being $R_Y$ (for $\l=0.5, 1,0, 2.0$) and $R_YR_Z$ for $\l=0.2$. This means that the number of parameters in these circuits are $(d+1)n_Q = 48$ for the $R_Y$ variational form and $2(d+1)n_Q =96$ for the $R_YR_Z$ variational form.  Each entry in the first three rows of Table \ref{L4-all-res} is a tuple ($E$, ansatz, number of parameters, optimizer) listing the best energy at convergence, the best ansatz variant, the number of parameters in the ansatz, and the optimizer used to obtain the result. The entries in the second last row corresponding to the results of \cite{prx-22} has a slightly different format,  $(E$, rotation block, number of parameters, optimizer), in which rotation block type used in the \texttt{EfficientSU2} ansatz is listed, since the authors of \cite{prx-22} exclusively used \texttt{EfficientSU2} and no other types of ansatzes.
\bei
\item Within this work, for all four couplings $\l=0.2, 0.5, 1.0, 2.0$, the best quantum circuit ansatz type is \texttt{EvolvedOperator}, followed by \texttt{TwoLocal} and \texttt{EfficientSU2} (see the first three rows of Table \ref{L4-all-res}). Among \texttt{EfficientSU2} and \texttt{TwoLocal}, an interesting trend to note is the better performance of variants with circular entanglement pattern compared to those with full entanglement pattern, as three out of four best variants of either \texttt{EfficientSU2} or \texttt{TwoLocal} have circular entanglement pattern. This trend was already noted in the case of bosonic $SU(2)$ model at $\L=2$ in Section \ref{sec_su2_l2_comparison}.
\item Using as benchmarks the best results reported in the work \cite{prx-22}, our results are really competitive. In particular, for the $\l=0.2, 05$ cases, both \texttt{TwoLocal} (at 3.13679 \& 3.30641 for $\l=0.2$ and $\l=0.5$, respectively) and \texttt{EvolvedOperatorAnsatz} circuits (3.13421 \& 3.29896, respectively) yield better results than \cite{prx-22} (at 3.13705 \& 3.30869, respectively). For the case of $\l=1.0$, only \texttt{EvolvedOperatorAnsatz} circuit (at 3.53512) yields a better result than \cite{prx-22} (at 3.54748). For the case of $\l=2.0$, our best result, which was obtained by an \texttt{EvolvedOperator} ansatz at 3.93348 is the same as that obtained by \cite{prx-22}, but our ansatz has only 30 parameters in contrast to the 48 parameters of the depth-3 \texttt{EfficientSU2} circuit used by \cite{prx-22}.
\eni
\begin{table}[!ht]
\centering
\begin{tabular}{l l l l l l}
\hline\hline
 Ansatz Type & $\lambda = 0.2$ & $\lambda = 0.5$ & $\lambda = 1.0$ & $\lambda = 2.0$ & \\
\hline\hline
 \texttt{EfficientSU2} & 
 $\begin{array}{l} 3.14605 \\ \texttt{effsu2\_RyY\_c}\\\text{(24 params)} \\ \text{SPSA}\end{array}$ &
 $\begin{array}{l} 3.44775 \\ \texttt{effsu2\_Rz\_c} \\\text{(24 params)}\\\text{COBYLA/SPSA}\end{array}$  &
 $\begin{array}{l} 3.89550\\ \texttt{effsu2\_Rz\_c} \\\text{(24 params)} \\\text{COBYLA/SPSA}\end{array}$ &
  $\begin{array}{l} 4.20670\\ \texttt{effsu2\_Ry\_f}\\\text{(24 params)}\\\text{SPSA}\end{array}$&   \\
\hline
 \texttt{TwoLocal}  &
 $\begin{array}{l} 3.13679\\ \texttt{tl\_RyY\_c}\\\text{(36 params)} \\\text{SPSA} \end{array}$  & 
 $\begin{array}{l} 3.30641 \\\texttt{tl\_RyY\_f} \\\text{(90 params)}\\\text{SPSA}\end{array}$&
 $\begin{array}{l} 3.53869 \\ \texttt{tl\_Ry\_c}\\\text{(36 params)} \\\text{COBYLA}\end{array}$ &
  $\begin{array}{l} 3.94466\\\texttt{tl\_RyY\_c} \\\text{(36 params)}\\\text{SPSA}\end{array}$ &  \\\hline
 \texttt{EvolvedOperator}& 
 $\begin{array}{l} \textbf{3.13421}\\\texttt{ev\_op\_Hp25\_2f} \\\text{(50 params)}\\\text{SPSA}\end{array}$& 
  $\begin{array}{l}\textbf{3.29896}\\\texttt{ev\_op\_Hp25\_2f} \\\text{(50 params)} \\ \text{SPSA}\end{array}$& 
 $\begin{array}{l}\textbf{3.53512}\\ \texttt{ev\_op\_Hp25} \\\text{(25 params)}\\\text{COBYLA}\end{array}$&
 $\begin{array}{l} \textbf{3.93348}\\\texttt{ev\_op\_Hp15\_2f} \\\text{(30 params)}\\\text{SPSA}\end{array}$ & \\\hline
  $\begin{array}{l}\text{Results from \cite{prx-22}} \\ \texttt{EfficientSU2} \\\text{(depth-3)} \end{array}$
 &    $\begin{array}{l} 3.13705\\ R_YR_Z \\\text{(96 params)}\\\text{L-BFGS-B}\end{array}$ & 
  $\begin{array}{l} 3.30869 \\ R_Y\,\,\text{(48 params)}\\\text{L-BFGS-B}\end{array}$&  
  $\begin{array}{l}3.54748 \\R_Y\,\,\text{(48 params)}\\\text{L-BFGS-B}\end{array}$ & 
   $\begin{array}{l} 3.93348 \\R_Y\,\,\text{(48 params)}\\\text{L-BFGS-B}\end{array}$ & 
 \\\hline
 Exact energy & 3.13406& 3.29894&  3.52625& 3.89548& \\
\hline
\end{tabular}
\caption{VQE experiments involving $\L=4$ bosonic $SU(2)$ matrix model: Summary of the best results from the three types of quantum circuit ansatzes (\texttt{EfficientSU2}, \texttt{TwoLocal}, \texttt{EvolvedOperatorAnsatz}) from this work, as well as those reported in \cite{prx-22}, at different couplings for $SU(2)$ matrix model at cutoff $\Lambda=4$. The absolute best results obtained by comparing all ansatzes from our work and those from \cite{prx-22} are noted in bold.} \label{L4-all-res}
\end{table}
\newpage
\section{ $\Lambda = 2$ supersymmetric model} \label{sec-bmn}
The Hamiltonian for the $SU(2)$ supersymmetric matrix model at Fock cutoff $\L=2$ is a $2^9\times 2^9$ matrix with the following exact energies obtained by diagonalization
\beq
E_{\l=0.2} = 0.003287, \qquad E_{\l=0.5} = 0.01690, \qquad E_{\l=1.0} = 0.04829, \qquad E_{\l=2.0} = 0.08385\,.
\label{Ee_bmn}
\eeq
At each coupling $\lambda$, the $2^9\times 2^9$ Hamiltonian can be written as the sum of 25 Pauli string operators as shown in Table \ref{bmn-Hpauli}. Operators in Group (A) are all those contributing to the diagonal elements of the Hamiltonian, and are the tensor products of the identity `I' and Pauli `Z' operators. Their values remain unchanged as the coupling constant $\l$ varies. Operators in Group (B) and (C) , which are the tensor products of various combinations of the identity, Pauli `X' and Pauli `Y' operators, are those contributing to the interaction part, or the off-diagonal components, of the Hamiltonian. Their values steadily increase as the coupling constant $\l$ varies from weak ($\l=0.2$) to strong ($\l$=2.0). At $\l=2.0$, the values of these off-diagonal operators reach the maximum values are equal to those in Group (A).
\begin{table}[!ht]
\centering
\begin{tabular}{l l c c c c l}
\hline\hline
Group  & Operator & $\lambda = 0.2$ & $\lambda = 0.5$ & $\lambda = 1.0$ & $\lambda = 2.0$ & \\
\hline\hline
&IIIIIIIII & 5.4 &5.625& 6.0 & 6.75 &\\
\hline
&ZIIIIIIII &-0.5&-0.5& -0.5 & -0.5&\\
&IZIIIIIII &-0.5&-0.5& -0.5 &-0.5 &\\
&IIZIIIIII&-0.5&-0.5& -0.5 &-0.5 &\\
&IIIZIIIII &-0.5&-0.5& -0.5 &-0.5 &\\
(A)&IIIIZIIII &-0.5&-0.5& -0.5 &-0.5 &\\
&IIIIIZIII &-0.5&-0.5& -0.5 &-0.5 &\\
&IIIIIIZII &-0.75&-0.75 & -0.75 &-0.75&\\
&IIIIIIIZI &-0.75&-0.75 & -0.75 &-0.75&\\
&IIIIIIIIZ &-0.75&-0.75 & -0.75 &-0.75&\\
\hline
& XXIXXIIII &0.05&-0.125& -0.25 & -0.5&\\
(B)& XIXXIXIII &0.05&-0.125& -0.25 &-0.5&\\
& IXXIXXIII &0.05&-0.125& -0.25 &-0.5&\\
\hline
& IIXIIIYXI &0.158&0.25& 0.354 & 0.5&\\
& IIXIIIXYI &0.158&0.25&  0.354& 0.5&\\
& IIIIIXXXI &0.158&0.25& 0.354& 0.5&\\
& IIIIXIYZY &0.158&0.25& 0.354& 0.5&\\
& XIIIIIIYX &0.158&0.25& 0.354& 0.5&\\
(C)& XIIIIIIXY &0.158&0.25& 0.354& 0.5&\\
& IIIXIIIXX &0.158&0.25& 0.354& 0.5&\\
& IIIIIXYYI &-0.158&-0.25& -0.354& -0.5&\\
& IXIIIIYZX &-0.158&-0.25& -0.354& -0.5&\\
& IXIIIIXZY &-0.158&-0.25& -0.354& -0.5&\\
& IIIIXIXZX &-0.158&-0.25& -0.354& -0.5&\\
& IIIXIIIYY &-0.158&-0.25& -0.354& -0.5&\\
 \hline
\end{tabular}
\caption{The 25 Pauli string operators, together with their coefficients at each coupling $\lambda$, making up the supersymmetric $SU(2)$ matrix model at $\L=2$. }\label{bmn-Hpauli}
\end{table}
\FloatBarrier
From Table \ref{bmn-Hpauli}, the supersymmetric $\L=2$ Hamiltonian at any of the four couplings $\l$ can be read off using the corresponding column for $\l$. For example, the Hamiltonian at $\l=0.5$ reads
\beq
H^{(S)\L=2}_{\l=0.5} &=& -0.5\lf(\text{IZIIIIIII + IIZIIIIII+ IIIZIIIII+
IIIIZIIII+ IIIIIZIII}\rr)\non
&& - 0.75 \lf(\text{IIIIIIZII + IIIIIIIZI + IIIIIIIIZ}\rr)
\non 
&&- 0.125\lf(\text{XXIXXIIII + XIXXIXIII + IXXIXXIII}\rr) 
\non && + 0.25\lf(\text{IIXIIIYXI + IIXIIIXYI + IIIIIXXXI + IIIIXIYZY}\rr.
\non &&\lf.\hspace{1.2cm}
+\text{XIIIIIIYX+ XIIIIIIXY +IIIXIIIXX}\rr)
\non
&& - 0.25\lf(\text{IIIIIXYYI + IXIIIIYZX+ IXIIIIXZY + IIIIXIXZX + IIIXIIIYY}\rr)
\eeq
\newpage
\subsection{\texttt{EvolvedOperatorAnsatz}}\label{sec-sl2-evop}
\begin{figure}[!ht]
\centering
\includegraphics[width=.8\textwidth]{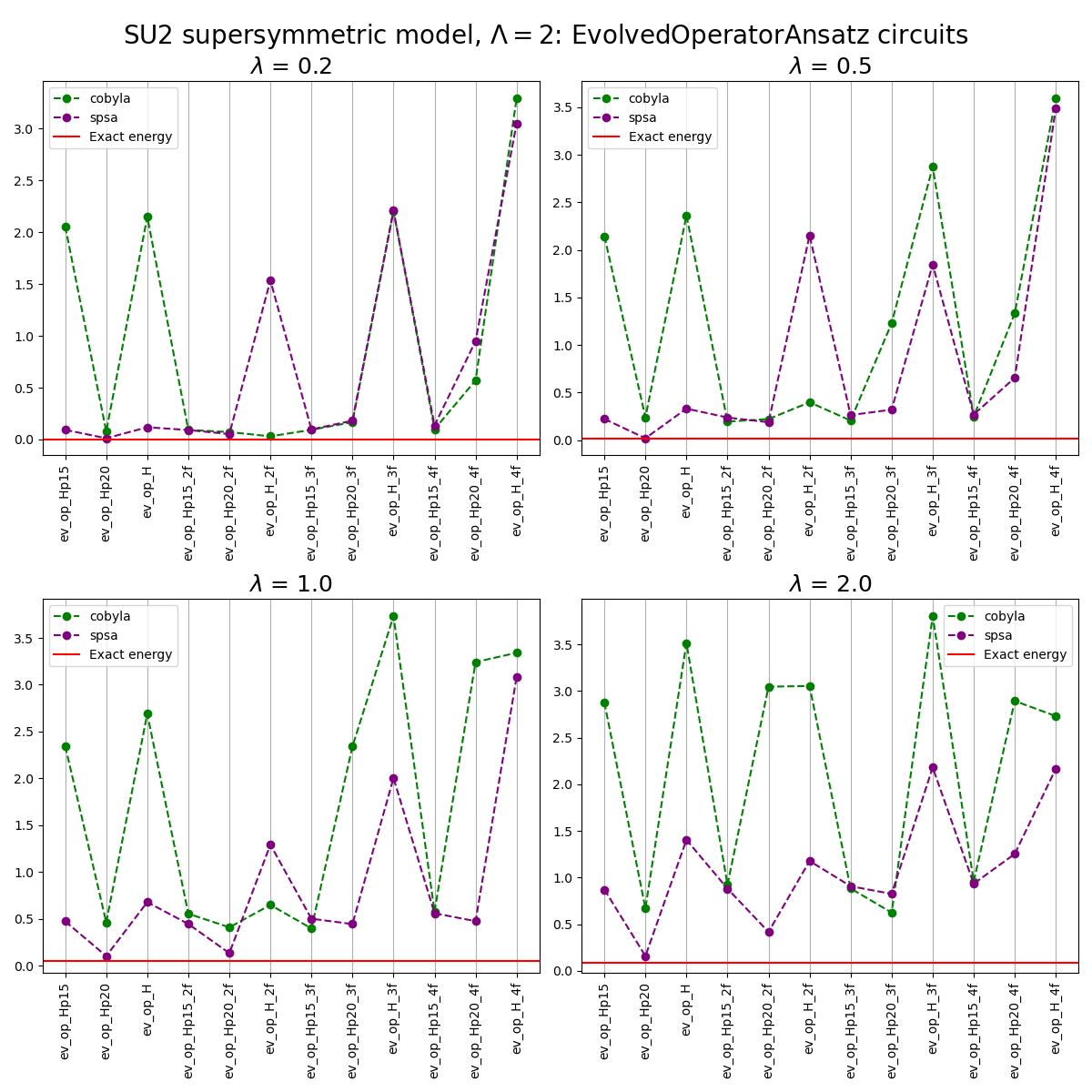}
\caption{Supersymmetric $SU(2)$ model at Fock cutoff $\L=2$ at different couplings (clockwise from top left $\l=0.2$, $\l=0.5$, $\l=2.0$, $\l=1.0$): Comparison of all \texttt{EvolvedOperatorAnsatz} quantum circuits. The data points in the subfigures above are from Tables \ref{bmn_L2_l02_eo} -  \ref{bmn_L2_l20_eo}.}
\label{fig_L2_BMN_all_ev_op}
\end{figure}

In this case, we work only with \texttt{EvolvedOperatorAnsatz}, as our goal is to keep the quantum circuit ansatzes as shallow as possible. This is not achievable with either \texttt{EfficientSU2} or \texttt{TwoLocal}, since the depth-1 versions of these circuits fail to yield results that are close enough to the exact values, which can only be reached with much deeper circuits of around 8-9 layers.
\\\\
For the construction of the \texttt{EvolvedOperatorAnsatz} circuits, we work with the largest 15, 20 and 24 operators chosen from Table \ref{bmn-Hpauli}. This leads to the 12 ansatzes (listed in Table \ref{bmn-ev-op}) which include the depth-1 circuits \texttt{ev\_op\_15}, \texttt{ev\_op\_20}, \texttt{ev\_op\_H} with 15, 20, and 24 building blocks, respectively, together with their depth-2, depth-3, depth-4 versions. Since the supersymmetric $\L=2$ Hamiltonian only contains 25 Pauli string operators, excluding the identity operator (which cannot be parameterized anyway), those circuits whose building blocks use 24 operators (\texttt{ev\_op\_H} and their higher-depth versions) practically contain the whole $\L=2$ Hamiltonian. Although we have 12 ansatzes in total, structurally, there are only three unique variants.
\begin{table}[!ht]
\centering
\begin{tabular}{l l c l l } 
\hline\hline
& Ansatz & Parameters & Operators & \\
\hline\hline
& \texttt{ev\_op\_Hp15} &  15 & Largest 15 operators by absolute values & \\
& &  & $\l=0.2$: (A) + $\left[\begin{array}{l} \text{IIXIIIYXI, IIIXIIIXX, IIIXIIIYY}\\
                                       \text{IIXIIIXYI, IIIIIXXXI, IIIIIXYYI}\end{array}\right]$ & \\
& &  & $\l=0.5$: Same as $\l=0.2$ & \\
& &  & $\l=1.0$: Same as $\l=0.2$ & \\
& &  & $\l=2.0$: Same as $\l=0.2$ & \\
\hline
& \texttt{ev\_op\_Hp20} &  20 &  Largest 20 operators by absolute values  & \\
& &  & $\l=0.2$: (A) + (C) (\text{excl. XIIIIIIXY}) & \\
& &  & $\l=0.5$: Same as $\l=0.2$  & \\
& &  & $\l=1.0$: Same as $\l=0.2$  & \\
& &  & $\l=2.0$: Same as $\l=0.2$ & \\
\hline
& \texttt{ev\_op\_Hp} &  24 & All operators in Table \ref{bmn-Hpauli} except IIIIIIIII & \\
\hline
& \texttt{ev\_op\_Hp15\_2f} &  30 & Depth-2 version of  \texttt{ev\_op\_Hp15}& \\
& \texttt{ev\_op\_Hp20\_2f} &  40 &   Depth-2 version of  \texttt{ev\_op\_Hp20}& \\
& \texttt{ev\_op\_Hp\_2f} &  48 &  Depth-2 version of  \texttt{ev\_op\_Hp} & \\
\hline
& \texttt{ev\_op\_Hp15\_3f} &  45 & Depth-3 version of  \texttt{ev\_op\_Hp15}& \\
& \texttt{ev\_op\_Hp20\_3f} &  60 &   Depth-3 version of  \texttt{ev\_op\_Hp20}& \\
& \texttt{ev\_op\_Hp\_3f} &  72 &  Depth-3 version of  \texttt{ev\_op\_Hp} & \\
\hline
& \texttt{ev\_op\_Hp15\_4f} &  60 & Depth-4 version of  \texttt{ev\_op\_Hp15}& \\
& \texttt{ev\_op\_Hp20\_4f} &  80 &   Depth-4 version of  \texttt{ev\_op\_Hp20}& \\
& \texttt{ev\_op\_Hp\_4f} &  96 &  Depth-4 version of  \texttt{ev\_op\_Hp} & \\
\hline
\end{tabular}
\caption{\texttt{EvolvedOperatorAnsatz} quantum circuits used to run VQE for the case of $SU(2)$ supesrsymmetric model with $\L=2$.}\label{bmn-ev-op}
\end{table}
\FloatBarrier
The best results of the VQE experiments $\lf(H^{(S)\L=2}_{\l}, \text{\texttt{EvolvedOperatorAnsatz}, COBYLA \& SPSA}\rr)$ are summarized in Table \ref{sup_l2_best}. A comparison of the performances of all 12 variants at four couplings can be found in Fig.\ref{fig_L2_BMN_all_ev_op}. For all couplings, the best quantum circuit ansatz is \texttt{ev\_op\_Hp20} with 20 parameters.
\begin{table}[!ht]
\centering
\begin{tabular}{ccccccc}
\hline\hline
Coupling &  Exact &  COBYLA & SPSA & Full results & \\
\hline\hline
$\l = 0.2$ &  0.003287 &
$\begin{array}{c} 0.03099\\ \texttt{ev\_op\_Hp\_2f} \end{array}$ &
$\begin{array}{c} \textbf{0.01228}\\\texttt{ev\_op\_Hp20}\\ \end{array}$ &
 \text{Table \ref{bmn_L2_l02_eo}  (F-S)} & \\
\hline
$\l = 0.5$ & 0.01690 &
$\begin{array}{c}0.19482 \\ \texttt{ev\_op\_Hp15\_2f} \end{array}$ &
$\begin{array}{c} \textbf{0.01953}\\\texttt{ev\_op\_Hp20}\\ \end{array}$ &
 \text{Table \ref{bmn_L2_l05_eo}  (F-S)} & \\
\hline
$\l = 1.0$ & 0.04829 &
$\begin{array}{c}0.39722\\ \texttt{ev\_op\_Hp15\_3f} \end{array}$ &
$\begin{array}{c} \textbf{0.10229}\\\texttt{ev\_op\_Hp20}\\ \end{array}$ &
 \text{Table \ref{bmn_L2_l10_eo}  (F-S)} & \\
\hline
$\l = 2.0$ & 0.08385 &
$\begin{array}{c} 0.6250\\ \texttt{ev\_op\_Hp20\_3f} \end{array}$ &
$\begin{array}{c} \textbf{0.15918}\\\texttt{ev\_op\_Hp20}\\ \end{array}$ &
 \text{Table \ref{bmn_L2_l20_eo}  (F-S)} & \\
\hline
\end{tabular}
\caption{VQE experiments $\lf(H^{(S)\L=2}_{\l}, \text{\texttt{EvolvedOperatorAnsatz}, COBYLA \& SPSA}\rr)$: Summary of the best results for each type of optimizers for each of the four coupling $\l$. The best results are noted in bold.  (F-S) denotes Full-Supplementary. Tables with the label (F-S) can be found in Section \ref{sec-L2-bmn-full-res} in the appendix.} \label{sup_l2_best}
\end{table}

We note the following observations regarding the best ansatzes, the overlaps of the ansatzes with the exact ground state, and the effect of increasing the depths of the ansatzes on the convergence results.
\bei
\item \textit{Best ansatzes}:
At all couplings, the best \texttt{EvolvedOperatorAnsatz} variant is \texttt{ev\_op\_Hp20} with 20 parameters (see Table \ref{sup_l2_best}).
Interestingly, circuits with 24 operators in their building blocks perform quite poorly compared to those containing fewer operators. They are in fact among the worst performers at all four couplings (see also Fig.\ref{fig_L2_bmn_eo_curves_cobyla} and  Figs.\ref{fig_L2_bmn_eo_curves_spsa_1f}-\ref{fig_L2_bmn_eo_curves_spsa_4f}).
\item
 \textit{Overlaps with the exact ground state}:
\bei
\item At $\l =0.2$ (Fig.\ref{fig_L2_BMN_all_ev_op}, first row, left subfigure), 4 variants from \texttt{ev\_op\_Hp15} to \texttt{ev\_op\_Hp20\_2f} as well as \texttt{ev\_op\_Hp15\_3f} , \texttt{ev\_op\_Hp15\_4f} - all with SPSA -  show good overlaps with the exact ground state.. With COBYLA, \texttt{ev\_op\_Hp15\_2f}, \texttt{ev\_op\_H\_2f}, \texttt{ev\_op\_Hp15\_3f}, \texttt{ev\_op\_Hp15\_4f} also show good overlaps with the exact ground state..
\item At $\l =0.5$ (Fig.\ref{fig_L2_BMN_all_ev_op}, first row, right subfigure) \texttt{ev\_op\_Hp20}, \texttt{ev\_op\_Hp15\_2f}  and \texttt{ev\_op\_Hp20\_2f} - all with SPSA - show good overlaps with the exact ground state..
\item At $\l =1.0$ (Fig.\ref{fig_L2_BMN_all_ev_op}, second row, left subfigure), only \texttt{ev\_op\_Hp20}  and \texttt{ev\_op\_Hp20\_2f} - both with SPSA -  show good overlaps with the exact ground state.
\item At $\l =2.0$ (Fig.\ref{fig_L2_BMN_all_ev_op}, second row, right subfigure), only \texttt{ev\_op\_Hp20} with SPSA shows a good overlap with the exact ground state.
\eni
\item
\textit{Effect of circuit depths}:
As supplementary material, the full convergence curves at different couplings obtained by running VQE algorithms using COBYLA for all 12 circuits  are shown in Fig.\ref{fig_L2_bmn_eo_curves_cobyla}, while the convergence curves obtained using SPSA are plotted seperately for circuits of different depths. Convergence curves of depth-1 circuits (comprising \texttt{ev\_op\_15}, \texttt{ev\_op\_20}, \texttt{ev\_op\_H}) using SPSA are plotted in
Fig.\ref{fig_L2_bmn_eo_curves_spsa_1f}. Convergence curves of depth-2 circuits (\texttt{ev\_op\_15\_2f}, \texttt{ev\_op\_20\_2f}, \texttt{ev\_op\_H\_2f}) are shown in Fig.\ref{fig_L2_bmn_eo_curves_spsa_2f}, those of depth-3 circuits  (\texttt{ev\_op\_15\_3f}, \texttt{ev\_op\_20\_3f}, \texttt{ev\_op\_H\_3f}) are shown in Fig.\ref{fig_L2_bmn_eo_curves_spsa_3f}, and those of depth-4 circuits (\texttt{ev\_op\_15\_4f}, \texttt{ev\_op\_20\_4f}, \texttt{ev\_op\_H\_4f}) are shown in Fig.\ref{fig_L2_bmn_eo_curves_spsa_4f}. All the curves are included in Section \ref{sec-L2-bmn-full-res} in the appendix.
Within the depth-1 circuits comprising the three variants \texttt{ev\_op\_Hp15}, \texttt{ev\_op\_Hp20}, \texttt{ev\_op\_H}, for both COBYLA and SPSA optimizers, the order of best to worst performing, for all four couplings, is
$\texttt{ev\_op\_Hp20} \ra \texttt{ev\_op\_Hp15}\ra \texttt{ev\_op\_H}$.
As the depth of the circuits is increased from 1 to 4, a clear trend is the general decrease in performance of all variants compared to their shallower versions, which is evident in the convergence curves that end in higher and higher values for the case of COBYLA optimizer in Fig.\ref{fig_L2_bmn_eo_curves_cobyla}  and become more and more widespread for the case of SPSA optimizer which can be seen in Fig.\ref{fig_L2_bmn_eo_curves_spsa_1f}-Fig.\ref{fig_L2_bmn_eo_curves_spsa_4f}.
\eni

\FloatBarrier

\subsection{Comparison of all quantum circuits}\label{sec_su2_sl2_comparison}
In this section, we compare the best results obtained using \texttt{EvolvedOperatorAnsatz} circuits with those obtained in \cite{prx-22} using  \texttt{EfficientSU2} $R_YR_Z$ circuits either with depth-8 ($2(d+1)n_Q = 18\times 9 = 162$ parameters) or depth-9 ($2(d+1)n_Q = 20\times 9 = 180$ parameters). The results are tabulated in Table \ref{L2-BMN-all-res}, in which the first row contains the best results from using \texttt{EvolvedOperatorAnsatz} while the second row lists the results reported by \cite{prx-22}.  Each entry in the first row is a tuple $(E$, ansatz, depth, number of parameters, optimizer) corresponding to the best ansatz variant and its characteristics. The entries in the second row have a similar format, $(E$, depth, number of parameters, optimizer), in which the ansatz is not listed since it is always the variant of \texttt{EfficientSU2} with $R_YR_Z$ rotation block and full entanglement pattern.
\\\\
For $\l=0.2$ and $\l=1.0$, depth-8 and depth-9 \texttt{EfficientSU2} ansatzes achieved slightly better results than depth-1 \texttt{ev\_op\_Hp20}. For $\l=0.5$ and $\l=2.0$, the same depth-1 \texttt{ev\_op\_Hp20} ansatz outperformed the depth-9 \texttt{EfficientSU2} ansatz. The fact that \texttt{ev\_op\_Hp20} with only 20 parameters can perform on par or better than \texttt{EfficientSU2} with 162 or 180 parameters is a very promising result which shows the clear advantage of tailored ansatzes over generic ones.
\begin{table}[!ht]
\centering
\begin{tabular}{l l l l l l } 
\hline\hline
 Ansatz Type & $\lambda = 0.2$ & $\lambda = 0.5$ & $\lambda = 1.0$ & $\lambda = 2.0$ & \\
\hline\hline
 \texttt{EvolvedOperator}& 
 $\begin{array}{l} 0.012277\\ \texttt{ev\_op\_Hp20} \\ \text{depth-1}\\\text{(20 params)} \\\text{SPSA}\end{array}$
 & $\begin{array}{l} \textbf{0.01953}\\ \texttt{ev\_op\_Hp20} \\ \text{depth-1}\\ \text{(20 params)}\\\text{SPSA}\end{array}$
 & $\begin{array}{l} 0.10229\\ \texttt{ev\_op\_Hp20} \\ \text{depth-1}\\ \text{(20 params)}\\\text{SPSA}\end{array}$
 & $\begin{array}{l} \textbf{0.15918}\\ \texttt{ev\_op\_Hp20} \\ \text{depth-1}\\ \text{(20 params)}\\\text{SPSA}\end{array}$
 &\\
 \hline
  $\begin{array}{l}\text{Results from \cite{prx-22}} \\ \texttt{EfficientSU2} \\R_YR_Z \end{array}$
  &   $\begin{array}{l} \textbf{0.010126}\\ \text{depth-8}\\ \text{(162 params)} \\\text{SLSQP}\end{array}$ 
  &   $\begin{array}{l} 0.02744\\ \text{depth-9} \\ \text{(180 params)} \\\text{SLSQP}\end{array}$
  & $\begin{array}{l} \textbf{0.07900}\\ \text{depth-9} \\ \text{(180 params)} \\\text{SLSQP}\end{array}$ 
   &   $\begin{array}{l} 0.17688\\ \text{depth-9} \\ \text{(180 params)} \\\text{SLSQP}\end{array}$
  &\\
  \hline
 Exact energy & 0.003287 & 0.01690 &  0.04829 & 0.08385 & \\
\hline
\end{tabular}
\caption{Comparison of the \texttt{EvolvedOperatorAnsatz} quantum circuits at different couplings for the supersymmetric $SU(2)$ matrix model at cutoff $\Lambda=2$ with the results reported in \cite{prx-22}. The absolute best results are noted in bold.} \label{L2-BMN-all-res}
\end{table}
\section{Summary and concluding remarks} \label{concl}
In this work, we revisited the problem of solving for the ground state energy of $SU(2)$ matrix models (both bosonic and supersymmetric) with Variational Quantum Eigensolver (VQE) algorithm involving variational quantum circuit ansatzes using the IBM quantum computing platform \texttt{Qiskit} \cite{ibm-qiskit}. With the aim of exploring and identifying new variational ansatzes to extend the well-known  \texttt{EfficientSU2} quantum circuits used in \cite{prx-22}, we first experimented with \texttt{TwoLocal} circuits - a more general form of \texttt{EfficientSU2} with the same underlying architecture, and later with \texttt{EvolvedOperatorAnsatz} - a type of circuits with different architecture that we tailor-made for each specific Hamiltonians of interest, in addition to experimenting with more variants of \texttt{EfficientSU2}  beyond those used in \cite{prx-22}. We referred to both \texttt{EfficientSU2} and \texttt{TwoLocal} as generic ansatzes on account of the fact that their structures, whose building blocks consist of a rotation part and an entanglement part, are essentially the same in all problem settings, while \texttt{EvolvedOperatorAnsatz} had to be constructed by choosing the suitable operators that go into each building block. 
\bei
\item In total, for the cases of $SU(2)$ bosonic matrix model at Fock space cutoffs $\L=2$ and $\L=4$, we explored eight different variants of \texttt{EfficientSU2} that are combinations of four possible choices of rotation block, involving the parameterized $R_Y, R_Z, R_YR_Z, R_YY$ gates,  and two possible choices of  entanglement arrangments (full or circular) involving the unparameterized $C_X$ gate (see Table \ref{effsu2_ans} and Fig.\ref{qc_l2_effsu2}). Corresponding to these eight \texttt{EfficientSU2} variants are the eight variants of \texttt{TwoLocal} ansatzes with the same four combinations of rotation gates and two possible entanglement arrangements involving the parameterized $C_{RX}$ gates (see Table \ref{tl_t1_ans} and Fig.\ref{qc_l2_twolocal}). To keep the number of variational parameters small, all circuits used are depth-1\footnote{except the case of bosonic $SU(2)$ model at $\L=2, \l=2.0$ in which we used deeper \texttt{EfficientSU2} and \texttt{TwoLocal} circuits to evaluate the effect of circuit depth on the results}. Regarding the \texttt{EvolvedOperatorAnsatz}, we created nine variants for the bosonic $SU(2)$ matrix model with Fock cutoff $\L=2$  and eight variants for the $\L=4$ case. For $\L=2$, the nine variants include one with random operators, one with a full set of operators making up the Hamiltonian (with the exception of the identity), and one with a partial set of operators making up the Hamiltonian, together with their higher-depth versions (see Table \ref{L2-ev-op-ansatze} and Fig.\ref{L2-eo-fig}). For $\L=4$, the variants include circuits whose building blocks are made from the 15, 20, 25, 30 operators with largest coefficients by absolute values out of the 895 operators making up the full $\L=4$ Hamiltonian, together with their higher depth versions (see Table \ref{L4_ev_op_quantum circuits}).
\\\\
With these different variants within each type of quantum circuit ansatzes, for the $\L=2$ and $\L=4$ $SU(2)$ cases, we performed 32 VQE runs using \texttt{EfficientSU2} and \texttt{TwoLocal} at each coupling for a total of four different couplings $\l=0.2, 0.5, 1.0, 2.0$ using two different optimizers: COBYLA and SPSA (see Fig.\ref{fig_L2_all_ef_tl} and Fig.\ref{fig_L4_all_es_tl})\footnote{In total, for the $SU(2)$ bosonic matrix model, this resulted in 128 VQE experiments for $\L=2$ and 128 experiments for $\L=4$, using \texttt{EfficientSU2} and \texttt{TwoLocal} ansatze}. With \texttt{EvolvedOperatorAnsatz}, at each coupling, there were 18 VQE runs for the $\L=2$ case (see Fig.\ref{fig_L2_all_ev_op}), and 16 VQE runs for the $\L=4$ case\footnote{All together, for the $SU(2)$ bosonic matrix model, there were 72 VQE runs for $\L=2$ and 64 VQE runs for $\L=4$ using \texttt{EvolvedOperatorAnsatz} circuits.} (see Fig.\ref{fig_L4_all_ev_op}).
The obtained results  show a consistent trend for both $\L$ at all couplings: The best performing quantum circuit ansatz type is always the tailor-made \texttt{EvolvedOperatorAnsatz}, followed by the generic \texttt{TwoLocal} ansatzes, followed by \texttt{EfficientSU2} (as documented in Table \ref{L2-all-res} and Table \ref{L4-all-res}).
This is not surprising, given the fact that \texttt{EfficientSU2} is the least tailored and least expressive ansatz type compared to the others. In specifying the different variants of \texttt{EfficientSU2}, our only choice lies in the selection of the gates in the rotation block, and the entanglement scheme.  In specifying the variants of \texttt{TwoLocal} quantum circuits, not only do we have the same choices as the \texttt{EfficientSU2} case, we also have an additional choice of parameterized  gates for the entanglement block. On the other hand, for the \texttt{EvolvedOperatorAnsatz} quantum circuits, we moved away from the rigid structure of `rotation-entanglement' blocks and had the freedom to use entirely new building blocks made of Pauli string operators, which can be selected to be those forming the Hamiltonian of interest. When compared with the results reported in \cite{prx-22}, which were obtained using the depth-3 \texttt{EfficientSU2} circuits with either $R_Y$ or $R_YR_Z$ rotation blocks and full entanglement pattern, our results are promising in the sense that  while \texttt{EvolvedOperatorAnsatz}  always outperform the results of \cite{prx-22}, \texttt{TwoLocal} ansatzes also do better than the results of \cite{prx-22} in some cases (see Table \ref{L2-all-res} and Table \ref{L4-all-res}).
\item For the case of supersymmetric $SU(2)$ model at Fock space cutoff $\L=2$, we worked only with \\ \texttt{EvolvedOperatorAnsatz} variational quantum circuits and created twelve different ansatzes, three of which are unique and made of building blocks with the largest 15, 20 and 24 operators chosen from the 25 operators making up the $\L=2$ Hamiltonian (see Table \ref{bmn-Hpauli}). The remaining ansatzes are the higher-depth (depth-2, depth-3, depth-4) versions of these first three (see Table \ref{bmn-ev-op}). Using these 12 ansatzes, we performed 24 VQE runs at each coupling using SPSA and COBYLA optimizers (see Fig.\ref{fig_L2_BMN_all_ev_op}), for the same four couplings of 0.2, 0.5, 1.0 and 2.0\footnote{For the $\L=2$ $SU(2)$ supersymmetric model, there were 96 VQE runs in total}. The obtained results consistently show the best variational quantum circuit ansatz as the depth-1 circuit with 20 operators in its building blocks (see Table \ref{L2-BMN-all-res}). Higher-depth circuits actually recorded poorer performances compared to their lower-depth counterparts. When using as benchmarks the results of \cite{prx-22}, which were obtained using deep \texttt{EfficientSU2} circuits (either depth-8 or depth-9) with around 162 or 180 parameters, our best results obtained by using the shallow 20-parameter \texttt{EvolvedOperatorAnsatz} are really competitive, given that for $\l=0.5$ and $\l=2.0$, \texttt{EvolvedOperatorAnsatz} emerged as the best performer, while for $\l=0.2$ and $\l=1.0$, \texttt{EvolvedOperatorAnsatz} obtained very close results to the much deeper \texttt{EfficientSU2} of \cite{prx-22}.
This is again very promising in the sense that by using a tailored architecture without involving either rotation or entanglement building blocks, one can obtain comparably good or better results at a small fraction (around 1/8 or 1/9) of the number of parameters required when using \texttt{EfficientSU2}.
\eni
Overall, the obtained results in this work suggest that given their potential to outperform the well-known and routinely used \texttt{EfficientSU2} in the context of $SU(2)$ matrix model, \texttt{TwoLocal} and \texttt{EvolvedOperatorAnsatz} variational quantum circuits should be considered more often in future quantum simulation studies involving VQE algorithm in high energy physics in general, either alongside or as new alternatives to \texttt{EfficientSU2}. A class of interesting examples of these future studies involves the quantum computing of Schwarzschild-de-Sitter black holes \cite{qc-bh-2202} or the quantum computing of string theory black holes \cite{qc-bh-2207}, all of which employed \texttt{EfficientSU2} circuits as variational ansatzes. Another interesting class of examples involves the benchmarking of VQE algorithm on different types of quantum computing hardware as done in \cite{benchmarking-2305} in which the authors also employed \texttt{EfficientSU2} quantum circuits (referred to as the $R_Y$-CNOT ansatz).
\\\\
Perhaps of more immediate relevance to this work is the possibility of applying \texttt{TwoLocal} and \texttt{EvolvedOperator} to $SU(N)$ matrix models with $N>2$. As previously discussed in Section \ref{vqe-s}, higher $SU(N)$ matrix models are much more computationally intensive than $SU(2)$ model due to the exponentially increasing size of the Hilbert spaces of these models. While it is possible to run VQE experiments involving $SU(3)$ matrix model with generic ansatzes like \texttt{EfficientSU2} and \texttt{TwoLocal} with the circular entanglement pattern which scale linearly \footnote{and \texttt{TwoLocal} variants with the full entanglement pattern which scale quadratically} in the number of qubits, we note that, to run just the simulator, this requires substantial computing resources typically possible only with an access to real quantum hardware or a cloud computing platform.
This is where \texttt{EvolvedOperatorAnsatz} circuits might turn out to be an especially good candidate for a trial wavefunction, since unlike \texttt{TwoLocal} and \texttt{EfficientSU2}, they do not scale in the number of qubits (but depend only on the number of operators used in their construction), potentially making it possible to tackle the problem without involving large computing powers. Furthermore, we note that a more streamlined method of constructing the tailored \texttt{EvolvedOperatorAnsatz} circuits would involve the ADAPT-VQE algorithm that implements the iterative adjustment process to fine tune the operators to be included in the final form of the ansatz. This bypasses the need to manually construct different variants of \texttt{EvolvedOperatorAnsatz} and could be more efficient when dealing with more complex $SU(N)$ matrix models. 
\\\\
 We hope to be able to return to these issues in future works.
 \\\\
\textbf{Acknowledgements}: The author is an unaffiliated and independent researcher. This work is possible thanks to the open-source IBM quantum computing platform \texttt{Qiskit} \cite{ibm-qiskit}. 
\newpage
\appendix
\section{$\Lambda=2$ bosonic $SU(2)$ model: Full results}
\label{sec-L2-full-res}
\subsection{\texttt{EfficientSU2} \& \texttt{TwoLocal}}
\label{sec-L2-full-res-1}
\begin{table}[!ht]
\centering
\begin{tabular}{c l c c c}
\hline\hline
&Ansatz  & Energy (COBYLA) & Energy (SPSA) &\\\hline\hline
&\texttt{effsu2\_Ry\_c} &    3.19141 &3.15703 &\\
&\texttt{effsu2\_Rz\_c} &    3.14980 &3.14980 &\\
& \texttt{effsu2\_RyRz\_c} &    3.15801 &3.16816 &\\
& \texttt{effsu2\_RyY\_c} &    3.15977 &3.16641 &\\
& \texttt{effsu2\_Ry\_f} &    3.15918 &3.15332 &\\
& \texttt{effsu2\_Rz\_f} &    3.14980 &3.14980 &\\
& \texttt{effsu2\_RyRz\_f} &    3.16211 &3.15137 &\\
& \texttt{effsu2\_RyY\_f} &    3.15664 &3.15137 &\\
& \texttt{tl\_Ry\_c} &    3.14863 & \textbf{3.14941} &\\
& \texttt{tl\_Rz\_c} &    3.14980 &3.14980 &\\
& \texttt{tl\_RyRz\_c} &    3.15762 &3.15645 &\\
& \texttt{tl\_RyY\_c} &    3.14863 &3.14941 &\\
& \texttt{tl\_Ry\_f} &    \textbf{3.14844} &3.14980 &\\
& \texttt{tl\_Rz\_f} &    3.14980 &3.14980 &\\
& \texttt{tl\_RyRz\_f} &    3.16992 &3.15605 &\\
& \texttt{tl\_RyY\_f} &   3.14844 &3.14980&\\
\hline
\end{tabular}
\caption{Full results from the VQE experiments $\lf(H^{\L=2}_{\l=0.2}, \texttt{EfficientSU2/TwoLocal}, \text{COBYLA/SPSA}\rr)$. The exact energy is $E^{\L=2}_{\l=0.2} = 3.14808$. The best result from each optimizer is noted in bold.}\label{qve-l2-l02-ef-tl}
\end{table}
\begin{figure}[!ht]
\centering
\includegraphics[width=.7\textwidth]{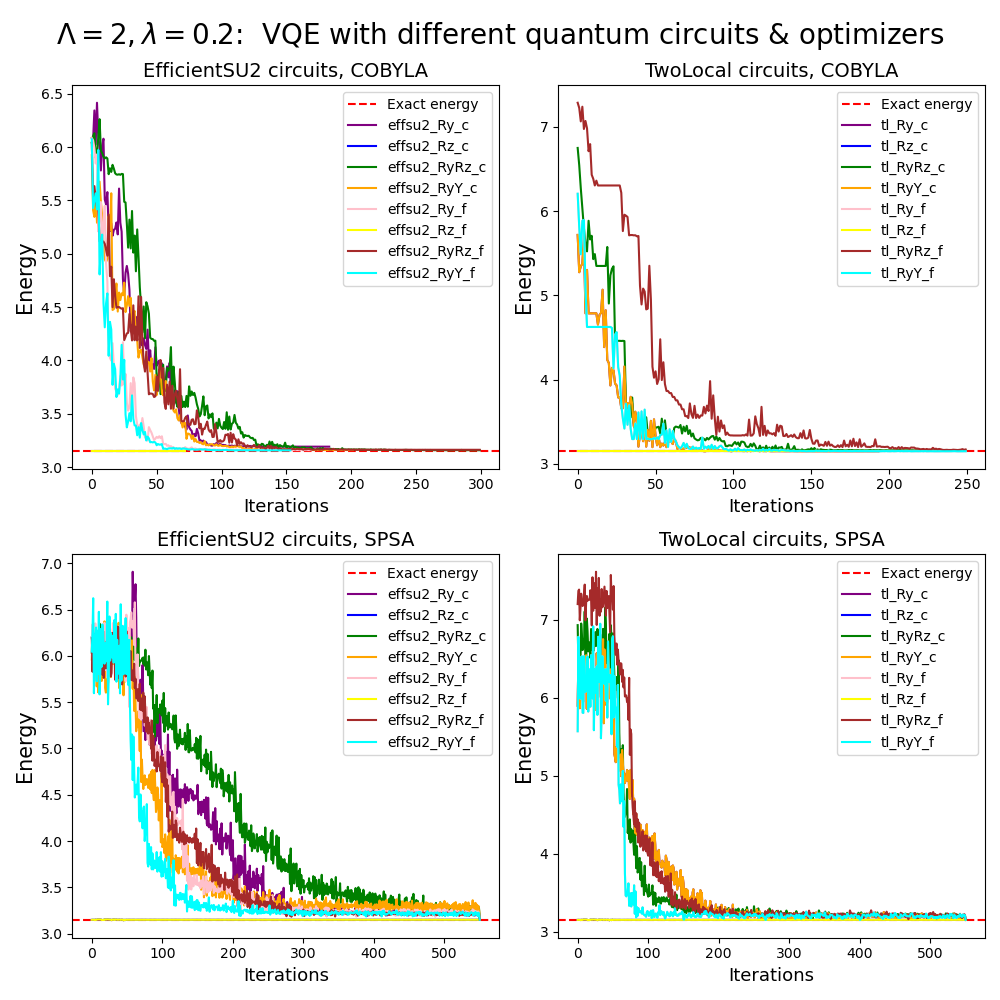}
\caption{Convergence curves of the energy for the VQE experiments involving $H^{\L=2}_{\lambda=0.2}$. Clockwise from top left: ($H^{\L=2}_{\lambda=0.2}$, \texttt{EfficientSU2}, COBYLA), ($H^{\L=2}_{\lambda=0.2}$, \texttt{TwoLocal}, COBYLA), ($H^{\L=2}_{\lambda=0.2}$, \texttt{TwoLocal}, SPSA), ($H^{\L=2}_{\lambda=0.2}$, \texttt{EfficientSU2}, SPSA).}
\label{fig_L2_l02_ef_tl}
\end{figure}
\newpage
\begin{table}[!ht]
\centering
\begin{tabular}{cl c c c}
\hline\hline
&Ansatz  & Energy (COBYLA) & Energy (SPSA) &\\\hline\hline
&\texttt{effsu2\_Ry\_c} &   3.37158 & 3.38623&\\
&\texttt{effsu2\_Rz\_c} &   3.37451 &3.37451&\\
&\texttt{effsu2\_RyRz\_c} &   3.36963 &4.41553&\\
&\texttt{effsu2\_RyY\_c} &   3.39014 & 3.40088&\\
&\texttt{effsu2\_Ry\_f} &   3.37549 & 3.37305&\\
&\texttt{effsu2\_Rz\_f} &   3.37451 & 3.37451&\\
&\texttt{effsu2\_RyRz\_f} &   3.40283 & 3.37646&\\
&\texttt{effsu2\_RyY\_f} &    3.37549 &3.37891&\\
&\texttt{tl\_Ry\_c} &  \textbf{3.36475} & \textbf{3.37207}&\\
&\texttt{tl\_Rz\_c} &   3.37451 & 3.37451&\\
&\texttt{tl\_RyRz\_c} &   3.37012 & 3.37939&\\
&\texttt{tl\_RyY\_c} &   3.36475   & 3.37207&\\
&\texttt{tl\_Ry\_f} &   3.36523 & 3.37646&\\
&\texttt{tl\_Rz\_f} &   3.37451 & 3.37451&\\
&\texttt{tl\_RyRz\_f} &   3.39502 & 3.38379&\\
&\texttt{tl\_RyY\_f} &   3.36523 & 3.37646&\\
\hline
\end{tabular}
\caption{Full results from the VQE experiments $\lf(H^{\L=2}_{\l=0.5}, \texttt{EfficientSU2/TwoLocal}, \text{COBYLA/SPSA}\rr)$. The exact energy is $E^{\L=2}_{\l=0.5} = 3.36254$. The best result from each optimizer is noted in bold.}\label{qve-l2-l05-ef-tl}
\end{table}
\begin{figure}[!ht]
\centering
\includegraphics[width=.7\textwidth]{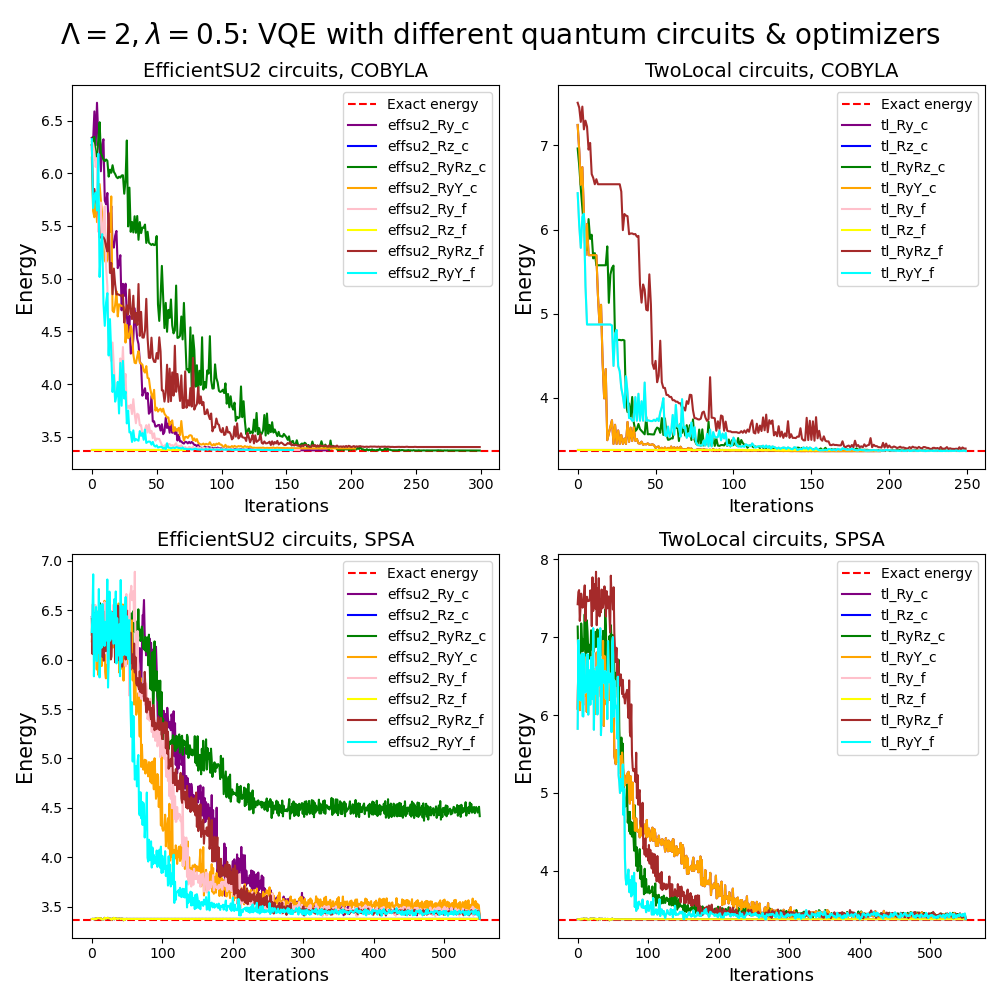}
\caption{Convergence curves of the energy for the VQE experiments involving $H^{\L=2}_{\lambda=0.5}$. Clockwise from top left: ($H^{\L=2}_{\lambda=0.5}$, \texttt{EfficientSU2}, COBYLA), ($H^{\L=2}_{\lambda=0.5}$, \texttt{TwoLocal}, COBYLA), ($H^{\L=2}_{\lambda=0.5}$, \texttt{TwoLocal}, SPSA), ($H^{\L=2}_{\lambda=0.5}$, \texttt{EfficientSU2}, SPSA). }
\label{fig_L2_l05_ef_tl}
\end{figure}
\newpage
\begin{table}[!ht]
\centering
\begin{tabular}{cl c c c}
\hline\hline
&Ansatz  & Energy (COBYLA) & Energy (SPSA) &\\\hline\hline
&\texttt{effsu2\_Ry\_c} & 3.77051 &3.76953 &\\
&\texttt{effsu2\_Rz\_c} &  3.74902 &3.74902&\\
&\texttt{effsu2\_RyRz\_c} &  3.78906 &3.79199&\\
&\texttt{effsu2\_RyY\_c} &  3.79297 &3.77832&\\
&\texttt{effsu2\_Ry\_f} & 3.80469 & \textbf{3.74316}&\\
&\texttt{effsu2\_Rz\_f} & 3.74902 &3.74902&\\
&\texttt{effsu2\_RyRz\_f} & 3.76465 &3.75098&\\
&\texttt{effsu2\_RyY\_f} & 3.75879 &3.74902&\\
&\texttt{tl\_Ry\_c} &  \textbf{3.73730} &3.75098&\\
&\texttt{tl\_Rz\_c} &    3.74902 &3.74902&\\
&\texttt{tl\_RyRz\_c} &   3.74414 &3.75293&\\
&\texttt{tl\_RyY\_c} &    3.73730 &3.75098&\\
&\texttt{tl\_Ry\_f} &   3.74121 &3.75195&\\
&\texttt{tl\_Rz\_f} &    3.74902 &3.74902&\\
&\texttt{tl\_RyRz\_f} &   3.76953 &3.74707&\\
&\texttt{tl\_RyY\_f} &   3.74121 &3.75195&\\
\hline
\end{tabular}
\caption{Full results from the VQE experiments $\lf(H^{\L=2}_{\l=1.0}, \texttt{EfficientSU2/TwoLocal}, \text{COBYLA/SPSA}\rr)$. The exact energy is $E^{\L=2}_{\l=1.0} = 3.69722$. The best result from each optimizer is noted in bold.}\label{qve-l2-l10-ef-tl}
\end{table}
\begin{figure}[!ht]
\centering
\includegraphics[width=.7\textwidth]{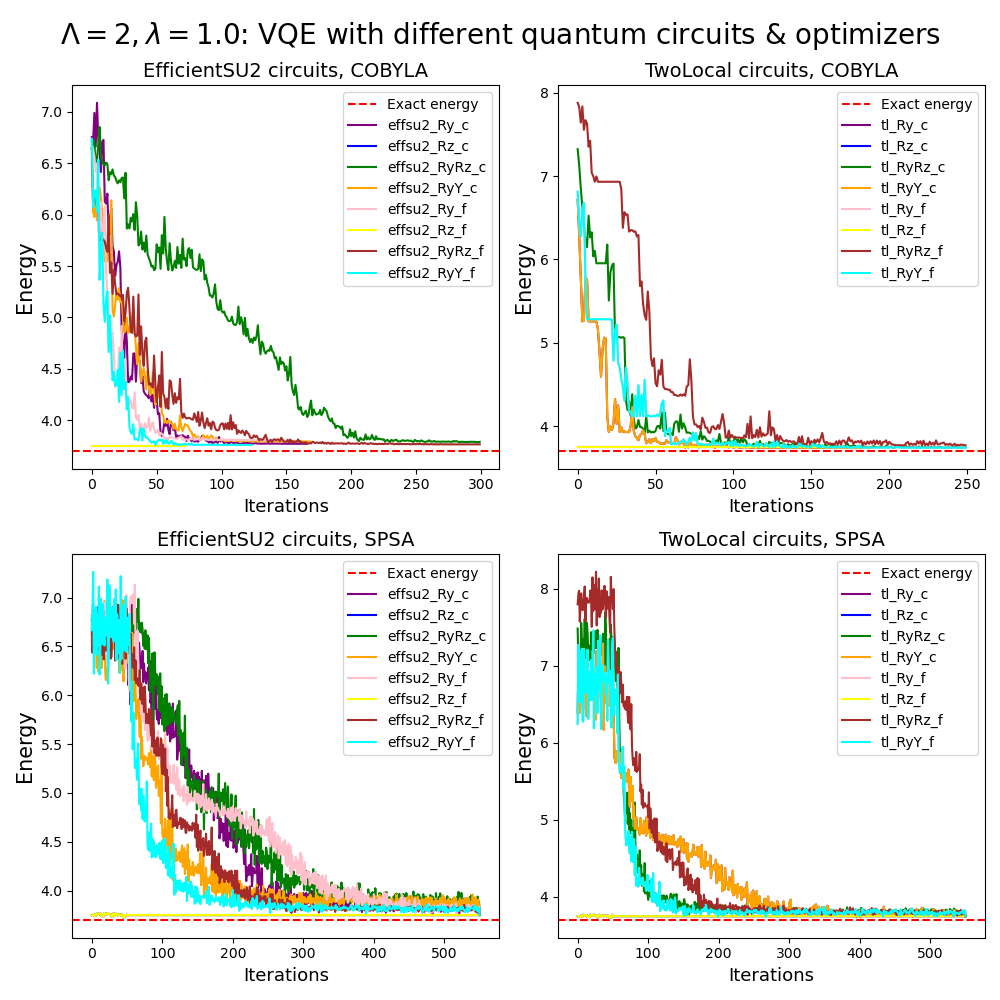}
\caption{Convergence curves of the energy for the VQE experiments involving $H^{\L=2}_{\lambda=1.0}$. Clockwise from top left: ($H^{\L=2}_{\lambda=1.0}$, \texttt{EfficientSU2}, COBYLA), ($H^{\L=2}_{\lambda=1.0}$, \texttt{TwoLocal}, COBYLA), ($H^{\L=2}_{\lambda=1.0}$, \texttt{TwoLocal}, SPSA), ($H^{\L=2}_{\lambda=1.0}$, \texttt{EfficientSU2}, SPSA).  }
\label{fig_L2_l10_ef_tl}
\end{figure}

\newpage
\begin{table}[!ht]
\centering
\begin{tabular}{cl c c c}
\hline\hline
&Ansatz  & Energy (COBYLA) & Energy (SPSA) &\\\hline\hline
&\texttt{effsu2\_Ry\_c} &  5.61816 & 5.54297 &\\
&\texttt{effsu2\_Rz\_c} &  4.49805 & 4.49805 &\\
&\texttt{effsu2\_RyRz\_c} &  4.46973 & 4.49609 &\\
&\texttt{effsu2\_RyY\_c} &  4.88574 & 4.52148&\\ 
&\texttt{effsu2\_Ry\_f} &  4.49121 & 4.51172&\\
&\texttt{effsu2\_Rz\_f} &  4.49805 & 4.49805&\\
&\texttt{effsu2\_RyRz\_f} &  4.45508 & 4.52051&\\
&\texttt{effsu2\_RyY\_f} &  4.54004 & 4.50098&\\
&\texttt{tl\_Ry\_c} &  \textbf{4.41895} & \textbf{4.48535}&\\
&\texttt{tl\_Rz\_c} &   4.49805 & 4.49805&\\
&\texttt{tl\_RyRz\_c} &   4.52441 & 4.50684&\\
&\texttt{tl\_RyY\_c} &   4.41895 & 4.48535&\\
&\texttt{tl\_Ry\_f} &   4.44922 & 4.51562&\\
&\texttt{tl\_Rz\_f} &   4.49805 &4.49805&\\
&\texttt{tl\_RyRz\_f} &   4.48730 &4.49121&\\
&\texttt{tl\_RyY\_f} &   4.44922 & 4.51562&\\
\hline
\end{tabular}
\caption{Full results from the VQE experiments $\lf(H^{\L=2}_{\l=2.0}, \texttt{EfficientSU2/TwoLocal}, \text{COBYLA/SPSA}\rr)$. The exact energy is $E^{\L=2}_{\l=2.0} = 4.26795$. The best result from each optimizer is noted in bold.}\label{qve-l2-l20-ef-tl}
\end{table}
\begin{figure}[!ht]
\centering
\includegraphics[width=.7\textwidth]{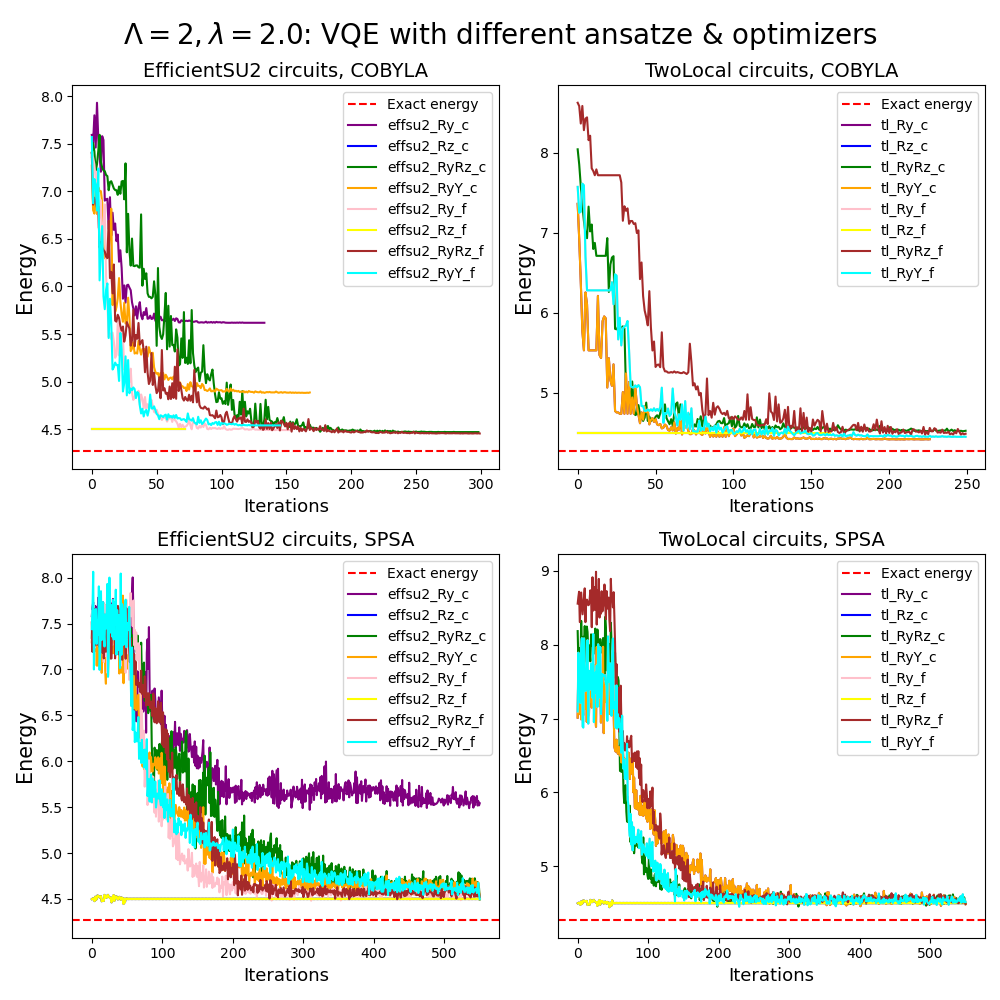}
\caption{Convergence curves of the energy for the VQE experiments involving $H^{\L=2}_{\lambda=2.0}$. Clockwise from top left: ($H^{\L=2}_{\lambda=2.0}$, \texttt{EfficientSU2}, COBYLA), ($H^{\L=2}_{\lambda=2.0}$, \texttt{TwoLocal}, \text{COBYLA}), ($H^{\L=2}_{\lambda=2.0}$, \texttt{TwoLocal}, SPSA), ($H^{\L=2}_{\lambda=2.0}$, \texttt{EfficientSU2}, SPSA).  }
\label{fig_L2_l20_ef_tl}
\end{figure}
\FloatBarrier
\newpage
\subsection{\texttt{EvolvedOperatorAnsatz}}
\label{sec-L2-full-res-2}
\begin{table}[!ht]
\centering
\begin{tabular}{cc c c c}
\hline\hline
&Ansatz  & Energy (COBYLA) & Energy (SPSA) &\\
\hline\hline
&\texttt{ev\_op\_r} &  3.14980 &3.14863&\\
&\texttt{ev\_op\_r3} &   3.15078& \textbf{3.14844}&\\
&\texttt{ev\_op\_H} &  3.15000 &3.14863&\\
&\texttt{ev\_op\_H\_2f} & 3.15000& 3.15488&\\
&\texttt{ev\_op\_H\_3f} & 3.14883 &3.15332&\\
&\texttt{ev\_op\_Hp} & 3.15059 &3.14980&\\
&\texttt{ev\_op\_Hp2} &3.15156 &3.15156&\\
&\texttt{ev\_op\_Hp3}  &3.14902 & 3.15254&\\
&\texttt{ev\_op\_Hp4}  &\textbf{3.14844} &3.15273&\\
\hline
\end{tabular}
\caption{Full results from the VQE experiments $\lf(H^{\L=2}_{\l=0.2}, \texttt{EvolvedOperatorAnsatz}, \text{COBYLA/SPSA}\rr)$. The exact energy is $E^{\L=2}_{\l=0.2} = 3.14808$. The best result from each optimizer is noted in bold.}\label{qve-l2-l02-ev-op}
\end{table}

\begin{table}[!ht]
\centering
\begin{tabular}{cc c c c}
\hline\hline
&Ansatz  & Energy (COBYLA) & Energy (SPSA) &\\
\hline\hline
&\texttt{ev\_op\_r} &  3.37451 & 3.37158&\\
&\texttt{ev\_op\_r3} & 3.37109 & 3.37305&\\
&\texttt{ev\_op\_H} & 3.36572 & 3.37158&\\
&\texttt{ev\_op\_H\_2f} &   \textbf{3.36328} & 3.37695&\\
&\texttt{ev\_op\_H\_3f} &  3.36426 & 3.37354&\\
&\texttt{ev\_op\_Hp} &   3.37451 & 3.37451&\\
&\texttt{ev\_op\_Hp3} &  3.36768 & \textbf{3.36719}&\\
&\texttt{ev\_op\_Hp4} & 3.37695 & 3.37109&\\
\hline
\end{tabular}
\caption{Full results from the VQE experiments $\lf(H^{\L=2}_{\l=0.5}, \texttt{EvolvedOperatorAnsatz}, \text{COBYLA/SPSA}\rr)$. The exact energy is $E^{\L=2}_{\l=0.5} = 3.36254$. The best result from each optimizer is noted in bold.}\label{qve-l2-l05-ev-op}
\end{table}

\newpage
\begin{table}[!ht]
\centering
\begin{tabular}{cc c c c}
\hline\hline
&Ansatz  & Energy (COBYLA) & Energy (SPSA) &\\
\hline\hline
&\texttt{ev\_op\_r} &  3.74902 & 3.74902 &\\
&\texttt{ev\_op\_r3} &    3.74414 & 3.74512&\\
&\texttt{ev\_op\_H} & 3.72656 & 3.72949&\\
&\texttt{ev\_op\_H\_2f} &  3.71484 & 3.73242&\\
&\texttt{ev\_op\_H\_3f} &  3.71387 & 3.72461&\\
&\texttt{ev\_op\_Hp} &   3.74902 & 3.73828&\\
&\texttt{ev\_op\_Hp2} &   3.73242 & \textbf{3.72266}&\\
&\texttt{ev\_op\_Hp3} &   3.71582 & 3.73242&\\
&\texttt{ev\_op\_Hp4} &   \textbf{3.70508} & 3.73926&\\
\hline
\end{tabular}
\caption{Full results from the VQE experiments $\lf(H^{\L=2}_{\l=1.0}, \texttt{EvolvedOperatorAnsatz}, \text{COBYLA/SPSA}\rr)$. The exact energy is $E^{\L=2}_{\l=1.0} = 3.69722$. The best result from each optimizer is noted in bold.}\label{qve-l2-l10-ev-op}
\end{table}
\begin{table}[!ht]
\centering
\begin{tabular}{cc c c c}
\hline\hline
&Ansatz  & Energy (COBYLA) & Energy (SPSA) &\\
\hline\hline
&\texttt{ev\_op\_r} &4.49805 & 4.49805 &\\
&\texttt{ev\_op\_r3}  & 4.48535 & 4.48926 &\\
&\texttt{ev\_op\_H} & 4.29297 & 4.31055 &\\
&\texttt{ev\_op\_H\_2f} & \textbf{4.28906} &\textbf{4.30664}&\\
&\texttt{ev\_op\_H\_3f} & 4.29102 &4.30859&\\
&\texttt{ev\_op\_Hp}  & 4.44141 & 4.44141&\\
&\texttt{ev\_op\_Hp2}  &  4.39453 &4.32227&\\
&\texttt{ev\_op\_Hp3}  & 4.29883 &4.34766&\\
&\texttt{ev\_op\_Hp4} & 4.28906 & 4.33008&\\
\hline
\end{tabular}
\caption{Full results from the VQE experiments $\lf(H^{\L=2}_{\l=2.0}, \texttt{EvolvedOperatorAnsatz}, \text{COBYLA/SPSA}\rr)$. The exact energy is $E^{\L=2}_{\l=2.0} = 4.26795$. The best result from each optimizer is noted in bold.}\label{qve-l2-l20-ev-op}
\end{table}
\newpage
\begin{figure}[!ht]
\centering
\includegraphics[width=.75\textwidth]{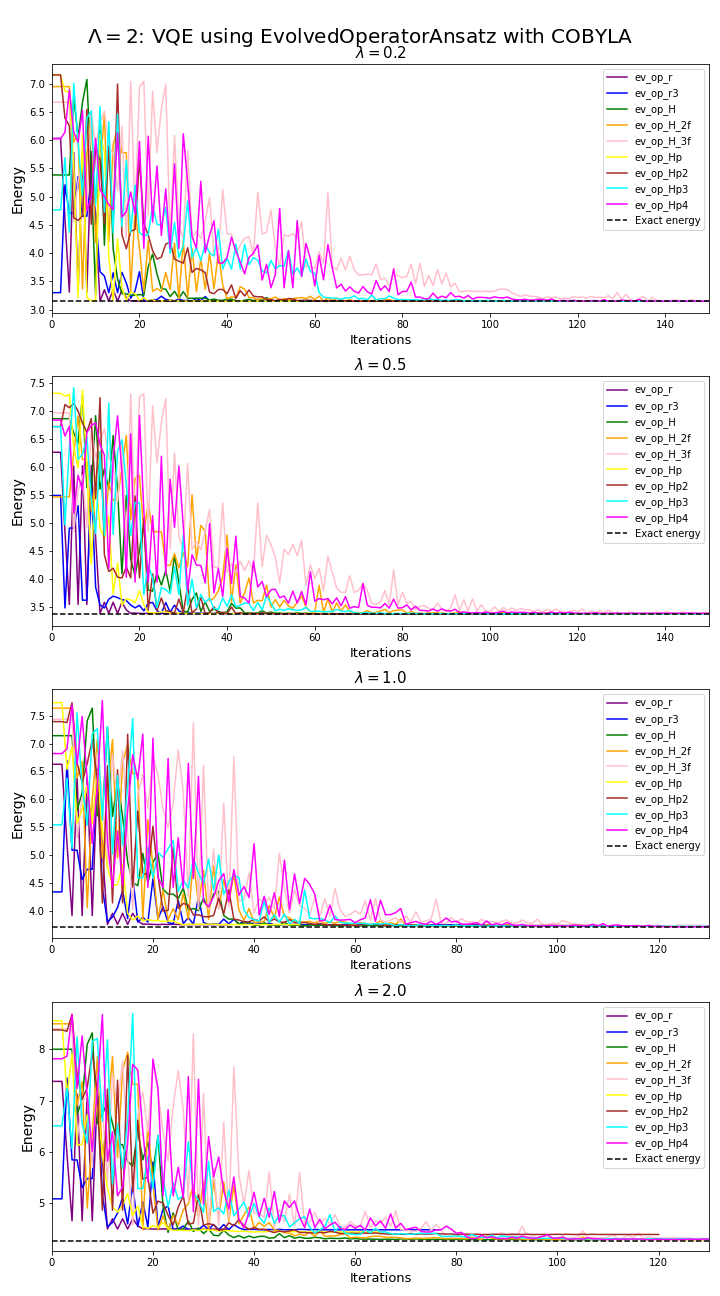}
\caption{Convergence curves from the VQE experiments $\lf(H^{\L=2}_{\l}, \texttt{EvolvedOperatorAnsatz}, \text{COBYLA}\rr)$. From top to bottom: $\l=0.2, 0.5, 1.0, 2.0$.}
\label{fig_L2_ev_op_cobyla}
\end{figure}
\begin{figure}[!ht]
\centering
\includegraphics[width=.75\textwidth]{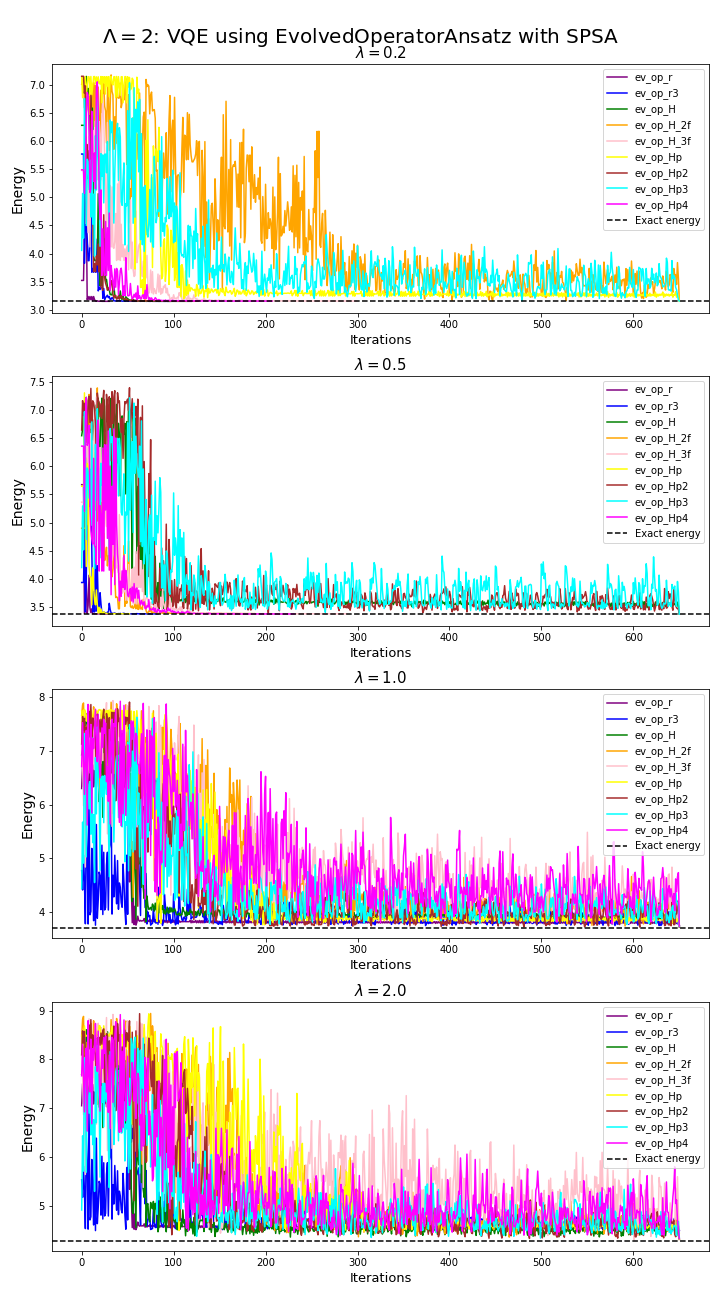}
\caption{Convergence curves from the VQE experiments $\lf(H^{\L=2}_{\l}, \texttt{EvolvedOperatorAnsatz}, \text{SPSA}\rr)$. From top to bottom: $\l=0.2, 0.5, 1.0, 2.0$.}
\label{fig_L2_ev_op_spsa}
\end{figure}
\FloatBarrier
\clearpage

\section{$\Lambda=4$ bosonic $SU(2)$ model: Full results}
\label{sec-L4-full-res}
\subsection{\texttt{EfficientSU2} \& \texttt{TwoLocal}}
\label{sec-L4-full-res-1}
\begin{table}[!ht]
\centering
\begin{tabular}{cl c c c}
\hline\hline
&Ansatz  & Energy (COBYLA) & Energy (SPSA) &\\
\hline\hline
&\texttt{effsu2\_Ry\_c} &  3.34450 & 3.23204 &\\
&\texttt{effsu2\_Rz\_c} &    \textbf{3.17910} &3.17910&\\
&\texttt{effsu2\_RyRz\_c} &    7.35898 & 4.89771&\\
&\texttt{effsu2\_RyY\_c} &    4.93931 & 3.14605&\\
&\texttt{effsu2\_Ry\_f} &    3.37026 & 3.19555&\\
&\texttt{effsu2\_Rz\_f} &    3.17910 & 3.17910&\\
&\texttt{effsu2\_RyRz\_f} &   9.85983 & 3.11959&\\
&\texttt{effsu2\_RyY\_f} &    3.38528 & 4.24528&\\
&\texttt{tl\_Ry\_c} &     3.18228 & 3.18339&\\
&\texttt{tl\_Rz\_c} &     3.17910 &3.17910&\\
&\texttt{tl\_RyRz\_c} &    3.38940 & 3.62318&\\
&\texttt{tl\_RyY\_c} &   3.21248 & \textbf{3.13679}&\\
&\texttt{tl\_Ry\_f} &    3.49465 & 3.16617&\\
&\texttt{tl\_Rz\_f} &    3.17910 &3.17910&\\
&\texttt{tl\_RyRz\_f} &   4.07165 &4.10208&\\
&\texttt{tl\_RyY\_f} &    3.58869 &3.14366&\\
\hline
\end{tabular}
\caption{Full results of the VQE experiments $\lf(H^{\L=4}_{\l=0.2}, \texttt{EfficientSU2}\&\texttt{TwoLocal}, \text{COBYLA/SPSA}\rr)$ involving the  $SU(2)$ bosonic matrix model at Fock cut-off $\Lambda = 4$ and coupling $\lambda=0.2$ using \texttt{EfficientSU2}/\texttt{TwoLocal} quantum circuit with SPSA \& COBYLA optimizers. The exact energy is $E^{\L=4}_{\l=0.2} = 3.13406$. The best result for each type of optimizers is noted in bold.}\label{qve-l4-l02-es-tl}
\end{table}   
\begin{figure}[!ht]
\centering
\includegraphics[width=.7\textwidth]{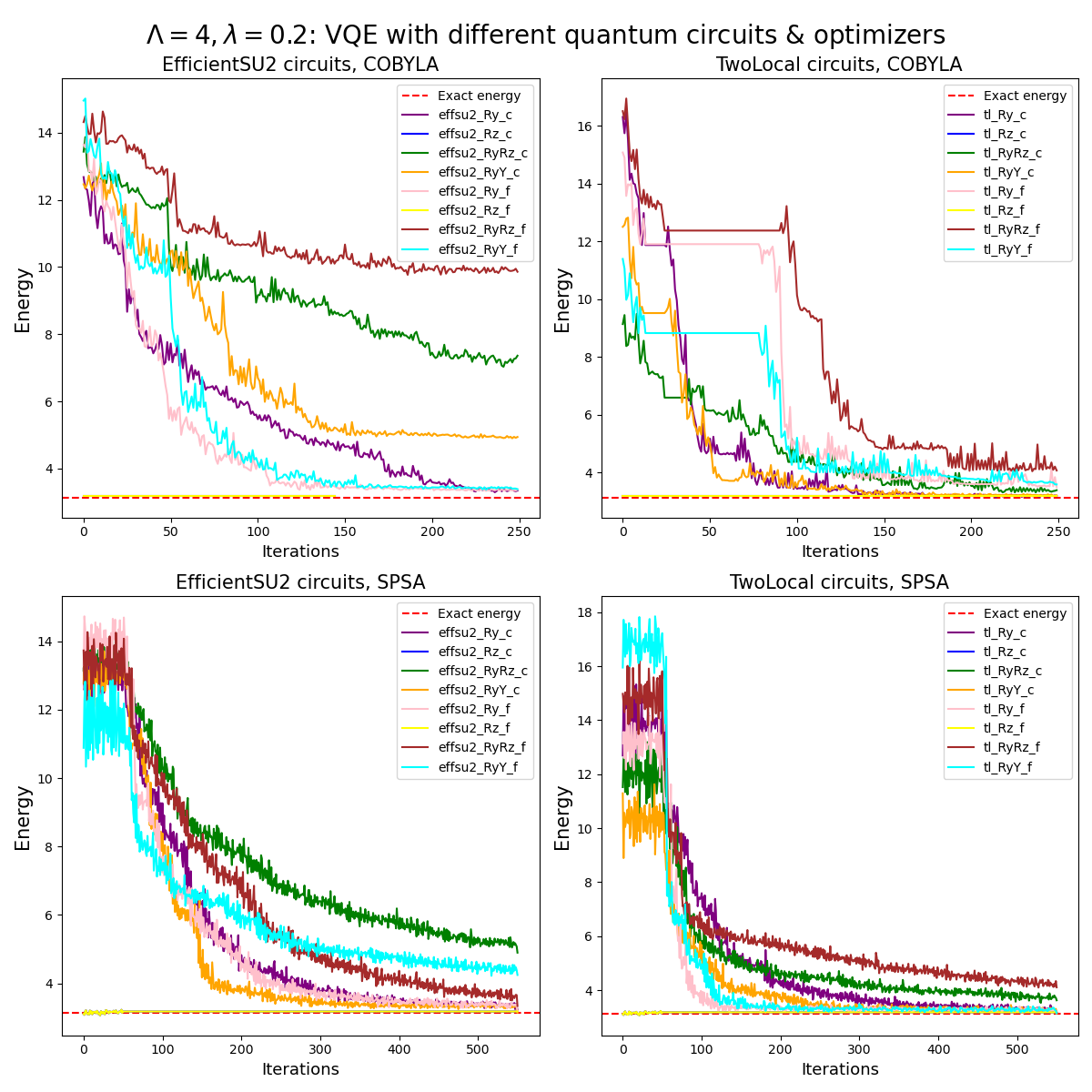}
\caption{Convergence curves of the energy for the VQE experiments involving $H^{\L=4}_{\lambda=0.2}$. Clockwise from top left: ($H^{\L=4}_{\lambda=0.2}$, \texttt{EfficientSU2}, COBYLA), ($H^{\L=4}_{\lambda=0.2}$, \texttt{TwoLocal}, COBYLA), ($H^{\L=4}_{\lambda=0.2}$, \texttt{TwoLocal}, SPSA), ($H^{\L=4}_{\lambda=0.2}$, \texttt{EfficientSU2}, SPSA).  }
\label{fig_L4_l02_ef_tl}
\end{figure}
\begin{table}[!ht]
\centering
\begin{tabular}{cl c c c}
\hline\hline
&Ansatz  & Energy (COBYLA) & Energy (SPSA) &\\
\hline\hline
&\texttt{effsu2\_Ry\_c} &   4.18830 & 3.61873&\\
&\texttt{effsu2\_Rz\_c} &    3.44775 &3.44775&\\
&\texttt{effsu2\_RyRz\_c} &    7.68608 &5.96494&\\
&\texttt{effsu2\_RyY\_c} &    3.71764 &3.66239&\\
&\texttt{effsu2\_Ry\_f} &    3.95226 &3.40536&\\
&\texttt{effsu2\_Rz\_f} &  3.44775 &3.44775&\\
&\texttt{effsu2\_RyRz\_f} &    3.99438 & 4.53325&\\
&\texttt{effsu2\_RyY\_f} &  3.58207 & 3.55380&\\
&\texttt{tl\_Ry\_c} &   3.21974 & 3.49978&\\
&\texttt{tl\_Rz\_c} &    3.44775 & 3.44775&\\
&\texttt{tl\_RyRz\_c} &   3.60629 & 3.39842&\\
&\texttt{tl\_RyY\_c} &   \textbf{3.27478} &3.38926&\\
&\texttt{tl\_Ry\_f} &   3.53397 & 3.32111&\\
&\texttt{tl\_Rz\_f} &  3.44775 & 3.44775&\\
&\texttt{tl\_RyRz\_f} &   4.35256 & 3.50001&\\
&\texttt{tl\_RyY\_f} &   3.47803 & \textbf{3.30641}&\\
\hline
\end{tabular}
\caption{Full results of the VQE experiments $\lf(H^{\L=4}_{\l=0.5}, \texttt{EfficientSU2}\&\texttt{TwoLocal}, COBYLA/SPSA\rr)$ involving the  $SU(2)$ bosonic matrix model at Fock cut-off $\Lambda = 4$ and coupling $\lambda=0.2$ using \texttt{EfficientSU2}/\texttt{TwoLocal} quantum circuit with SPSA \& COBYLA optimizers. The exact energy is $E^{\L=4}_{\l=0.5} = 3.29894$. The best result for each type of optimizers is noted in bold.}
\label{qve-l4-l05-es-tl}
\end{table} 
\begin{figure}[!ht]
\centering
\includegraphics[width=.7\textwidth]{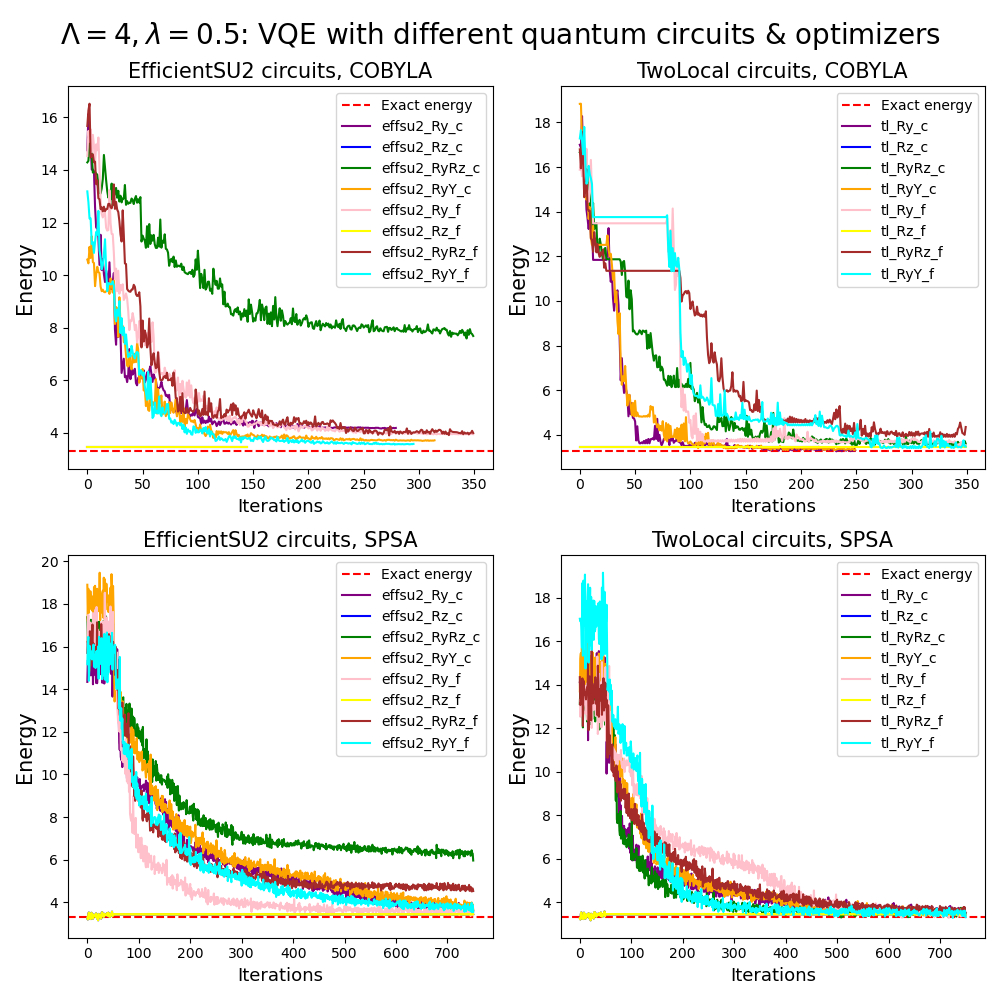}
\caption{Convergence curves of the energy for the VQE experiments involving $H^{\L=4}_{\lambda=0.5}$. Clockwise from top left: ($H^{\L=4}_{\lambda=0.5}$, \texttt{EfficientSU2}, COBYLA), ($H^{\L=4}_{\lambda=0.5}$, \texttt{TwoLocal}, COBYLA), ($H^{\L=4}_{\lambda=0.5}$, \texttt{TwoLocal}, SPSA), ($H^{\L=4}_{\lambda=0.5}$, \texttt{EfficientSU2}, SPSA).  }
\label{fig_L4_l05_ef_tl}
\end{figure}
\begin{table}[!ht]
\centering
\begin{tabular}{cl c c c}
\hline\hline
&Ansatz  & Energy (COBYLA) & Energy (SPSA) &\\
\hline\hline
&\texttt{effsu2\_Ry\_c}&   5.64314 &4.22957&\\
&\texttt{effsu2\_Rz\_c}&   3.89550 &3.89550&\\
&\texttt{effsu2\_RyRz\_c}&  6.94346 & 8.01018&\\
&\texttt{effsu2\_RyY\_c}&   6.65659 & 5.67670&\\
&\texttt{effsu2\_Ry\_f}&   6.30709 &4.19845&\\
&\texttt{effsu2\_Rz\_f}&  3.89550 &3.89550&\\
&\texttt{effsu2\_RyRz\_f}&   6.11065 &3.78857&\\
&\texttt{effsu2\_RyY\_f}&   4.30899 &4.10703&\\
&\texttt{tl\_Ry\_c}&   \textbf{3.53869} &5.26669&\\
&\texttt{tl\_Rz\_c}&   3.89550 &3.89550&\\
&\texttt{tl\_RyRz\_c}&    6.18099 &\textbf{3.55374}&\\
&\texttt{tl\_RyY\_c}&   4.96284 &4.83891&\\
&\texttt{tl\_Ry\_f}&   3.56694 &3.75099&\\
&\texttt{tl\_Rz\_f}&  3.89550 &3.89550&\\
&\texttt{tl\_RyRz\_f}& 4.60628 &3.92772&\\
&\texttt{tl\_RyY\_f}&  3.74414 &4.97329&\\
\hline
\end{tabular}
\caption{Full results of the VQE experiments $\lf(H^{\L=4}_{\l=1.0}, \texttt{EfficientSU2}\&\texttt{TwoLocal}, COBYLA/SPSA\rr)$ involving the  $SU(2)$ bosonic matrix model at Fock cut-off $\Lambda = 4$ and coupling $\lambda=1.0$ using \texttt{EfficientSU2}/\texttt{TwoLocal} quantum circuit with SPSA \& COBYLA optimizers. The exact energy is $E^{\L=4}_{\l=1.0} = 3.52625$. The best result for each type of optimizers is noted in bold.}
\label{qve-l4-l10-es-tl}
\end{table} 
\begin{figure}[!ht]
\centering
\includegraphics[width=.7\textwidth]{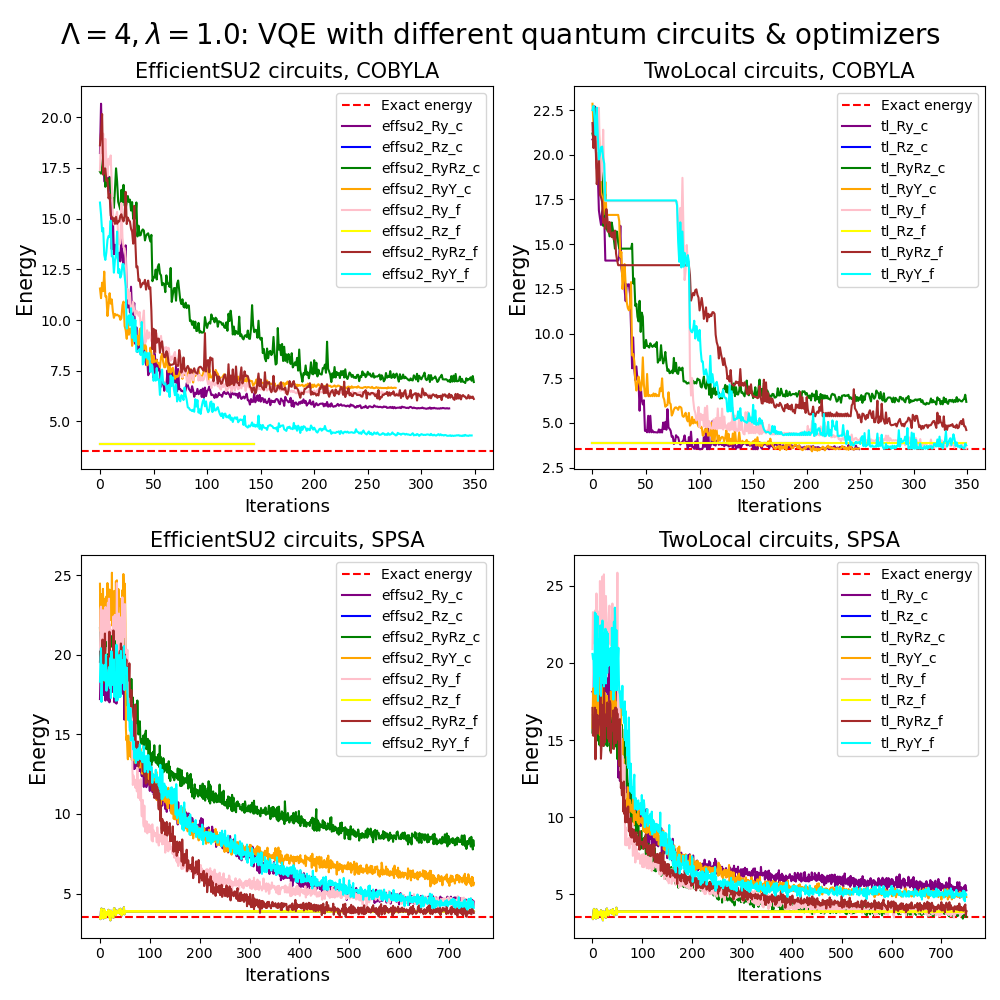}
\caption{Convergence curves of the energy for the VQE experiments involving $H^{\L=4}_{\lambda=1.0}$. Clockwise from top left: ($H^{\L=4}_{\lambda=1.0}$, \texttt{EfficientSU2}, COBYLA), ($H^{\L=4}_{\lambda=1.0}$, \texttt{TwoLocal}, COBYLA), ($H^{\L=4}_{\lambda=1.0}$, \texttt{TwoLocal}, SPSA), ($H^{\L=4}_{\lambda=1.0}$, \texttt{EfficientSU2}, SPSA).  }
\label{fig_L4_l10_ef_tl}
\end{figure}
\begin{table}[!ht]
\centering
\begin{tabular}{cl c c c}
\hline\hline
&Ansatz  & Energy (COBYLA) & Energy (SPSA) &\\
\hline\hline
&\texttt{effsu2\_Ry\_c} &   10.89956  &  6.92786&\\
&\texttt{effsu2\_Rz\_c} &  4.79100 &4.79100&\\
&\texttt{effsu2\_RyRz\_c} &  11.05988  &  4.32874&\\
&\texttt{effsu2\_RyY\_c} &  9.68253 &4.94393&\\
&\texttt{effsu2\_Ry\_f} &   8.01434 & 4.20670&\\
&\texttt{effsu2\_Rz\_f} &   4.79100 & 4.79100&\\
&\texttt{effsu2\_RyRz\_f} &   7.32845 &5.87643&\\
&\texttt{effsu2\_RyY\_f} &   7.79339 &5.54155&\\
&\texttt{tl\_Ry\_c} &   6.39657 &5.45181&\\
&\texttt{tl\_Rz\_c} &    4.79100 &4.79100&\\
&\texttt{tl\_RyRz\_c} &   7.28931 &4.61735&\\
&\texttt{tl\_RyY\_c} &    4.26378 &\textbf{3.94466}&\\
&\texttt{tl\_Ry\_f} &   4.29829 &6.29251&\\
&\texttt{tl\_Rz\_f} &    4.79100& 4.79100&\\
&\texttt{tl\_RyRz\_f} &   7.99626 &5.42990&\\
&\texttt{tl\_RyY\_f} &    \textbf{4.16062} &5.61301&\\
\hline
\end{tabular}
\caption{Full results of the VQE experiments $\lf(H^{\L=4}_{\l=2.0}, \texttt{EfficientSU2}\&\texttt{TwoLocal}, COBYLA/SPSA\rr)$ involving the  $SU(2)$ bosonic matrix model at Fock cut-off $\Lambda = 4$ and coupling $\lambda=2.0$ using \texttt{EfficientSU2}/\texttt{TwoLocal} quantum circuit with SPSA \& COBYLA optimizers. The exact energy is $E^{\L=4}_{\l=2.0} = 3.52625$. The best result for each type of optimizers is noted in bold.}\label{qve-l4-l20-es-tl}
\end{table}
\begin{figure}[!ht]
\centering
\includegraphics[width=.7\textwidth]{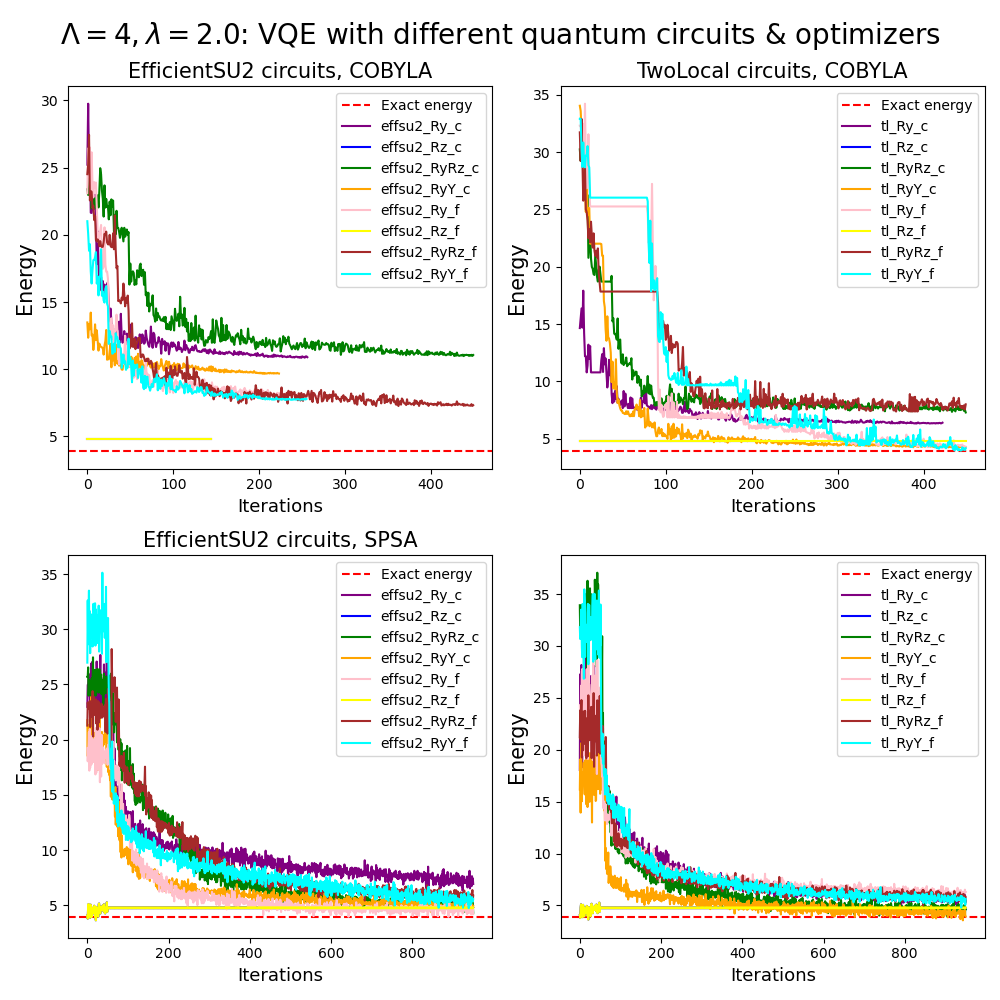}
\caption{Convergence curves of the energy for the VQE experiments involving $H^{\L=4}_{\lambda=2.0}$. Clockwise from top left: ($H^{\L=4}_{\lambda=2.0}$, \texttt{EfficientSU2}, COBYLA), ($H^{\L=4}_{\lambda=2.0}$, \texttt{TwoLocal}, COBYLA), ($H^{\L=4}_{\lambda=2.0}$, \texttt{TwoLocal}, SPSA), ($H^{\L=4}_{\lambda=2.0}$, \texttt{EfficientSU2}, SPSA).  }
\label{fig_L4_l20_ef_tl}
\end{figure}
\FloatBarrier
\clearpage
\subsection{\texttt{EvolvedOperatorAnsatz}}
\label{sec-L4-full-res-2}
\begin{table}[!ht]
\centering
\begin{tabular}{cc c c c}
\hline\hline
&Ansatz  & Energy (COBYLA) & Energy (SPSA) &\\
\hline\hline
&\texttt{ev\_op\_Hp15}  &3.27813 & 3.17320&\\
&\texttt{ev\_op\_Hp20} & 3.18835 & 3.15612&\\
&\texttt{ev\_op\_Hp25} &3.17782 & 3.14337&\\
&\texttt{ev\_op\_Hp30}  &3.21665 & 3.14084&\\
&\texttt{ev\_op\_Hp15\_2f} &  \textbf{3.15952} & 3.15614&\\
&\texttt{ev\_op\_Hp20\_2f} &  3.23598 & 3.19462&\\
&\texttt{ev\_op\_Hp25\_2f} & 3.33135 & \textbf{3.13421}&\\
&\texttt{ev\_op\_Hp30\_2f} &  3.47127 & 3.15462&\\
\hline
\end{tabular}
\caption{Results of the VQE experiments ($H^{\L=4}_{\l=0.2}$, \texttt{EvolvedOperatorAnsatz} from Table \ref{L4_ev_op_quantum circuits}, SPSA \& COBYA). The exact energy is $E= 3.13406$. The best result from each type of optimizers is noted in bold.}
\label{L4_l02_eo}
\end{table}
\begin{table}[!ht]
\centering
\begin{tabular}{cc c c c}
\hline\hline
&Ansatz  & Energy (COBYLA) & Energy (SPSA) &\\
\hline\hline
&\texttt{ev\_op\_Hp15} &   4.03464 &3.41548&\\
&\texttt{ev\_op\_Hp20} &  3.35601  & 3.30692&\\
&\texttt{ev\_op\_Hp25}  &\textbf{3.29968} &3.42603&\\
&\texttt{ev\_op\_Hp30}  &33.51242 & 3.30582&\\
&\texttt{ev\_op\_Hp15\_2f}   &3.30153 &3.34375&\\
&\texttt{ev\_op\_Hp20\_2f}  & 3.54492 & 3.30794&\\
&\texttt{ev\_op\_Hp25\_2f}  & 3.30028 & \textbf{3.29896}&\\
&\texttt{ev\_op\_Hp30\_2f}  & 3.77042 & 3.33945 &\\
\hline
\end{tabular}
\caption{Full results of the VQE experiments ($H^{\L=4}_{\l=0.5}$, \texttt{EvolvedOperatorAnsatz} from Table \ref{L4_ev_op_quantum circuits}, SPSA \& COBYA). The exact energy is $E= 3.29894$. The best result from each type of optimizers is noted in bold.}
\label{L4_l05_eo}
\end{table}

\begin{table}[!ht]
\centering
\begin{tabular}{cc c c c}
\hline\hline
&Ansatz  & Energy (COBYLA) & Energy (SPSA) &\\
\hline\hline
&\texttt{ev\_op\_Hp15} &  3.62082  &3.67193&\\
&\texttt{ev\_op\_Hp20} &  3.84073 & 3.60327&\\
&\texttt{ev\_op\_Hp25}  & \textbf{3.53512} & 3.96745&\\
&\texttt{ev\_op\_Hp30}  &   3.77387 & \textbf{3.54551}&\\
&\texttt{ev\_op\_Hp15\_2f} & 4.35314 & 3.71765&\\
&\texttt{ev\_op\_Hp20\_2f}&  3.75998 & 3.59577&\\
&\texttt{ev\_op\_Hp25\_2f}  &  4.92616 & 3.68810&\\
& \texttt{ev\_op\_Hp30\_2f}  &  4.75916 & 3.63468&\\
\hline
\end{tabular}
\caption{Full results of the VQE experiments ($H^{\L=4}_{\l=1.0}$, \texttt{EvolvedOperatorAnsatz} from Table \ref{L4_ev_op_quantum circuits}, SPSA \& COBYA). The exact energy is $E= 3.52625$. The best result from each type of optimizers is noted in bold.}
\label{L4_l10_eo}
\end{table}
\begin{table}[!ht]
\centering
\begin{tabular}{cc c c c}
\hline\hline
&Ansatz  & Energy (COBYLA) & Energy (SPSA) &\\
\hline\hline
&\texttt{ev\_op\_Hp15}  & 4.86152 & 4.69522 &\\
&\texttt{ev\_op\_Hp20} & \textbf{4.16425} & 4.62906 &\\
&\texttt{ev\_op\_Hp25}  & 6.16766 & 4.37152 &\\
& \texttt{ev\_op\_Hp30}  & 4.88188 & 4.12944 &\\
&\texttt{ev\_op\_Hp15\_2f} &  4.94432 & \textbf{3.93348} &\\
&\texttt{ev\_op\_Hp20\_2f}&4.93465 &4.10240 &\\
&\texttt{ev\_op\_Hp25\_2f}  & 7.39414 &4.51453 &\\
&\texttt{ev\_op\_Hp30\_2f}  &  6.01181& 4.76864 &\\
\hline
\end{tabular}
\caption{Results of the VQE experiments ($H^{\L=4}_{\l=2.0}$, \texttt{EvolvedOperatorAnsatz} from Table \ref{L4_ev_op_quantum circuits}, SPSA \& COBYA). The exact energy is $E= 3.89548$. The best result from each type of optimizers is noted in bold.}\label{L4_l20_eo}
\end{table}

\begin{figure}[!ht]
\centering
\includegraphics[width=.8\textwidth]{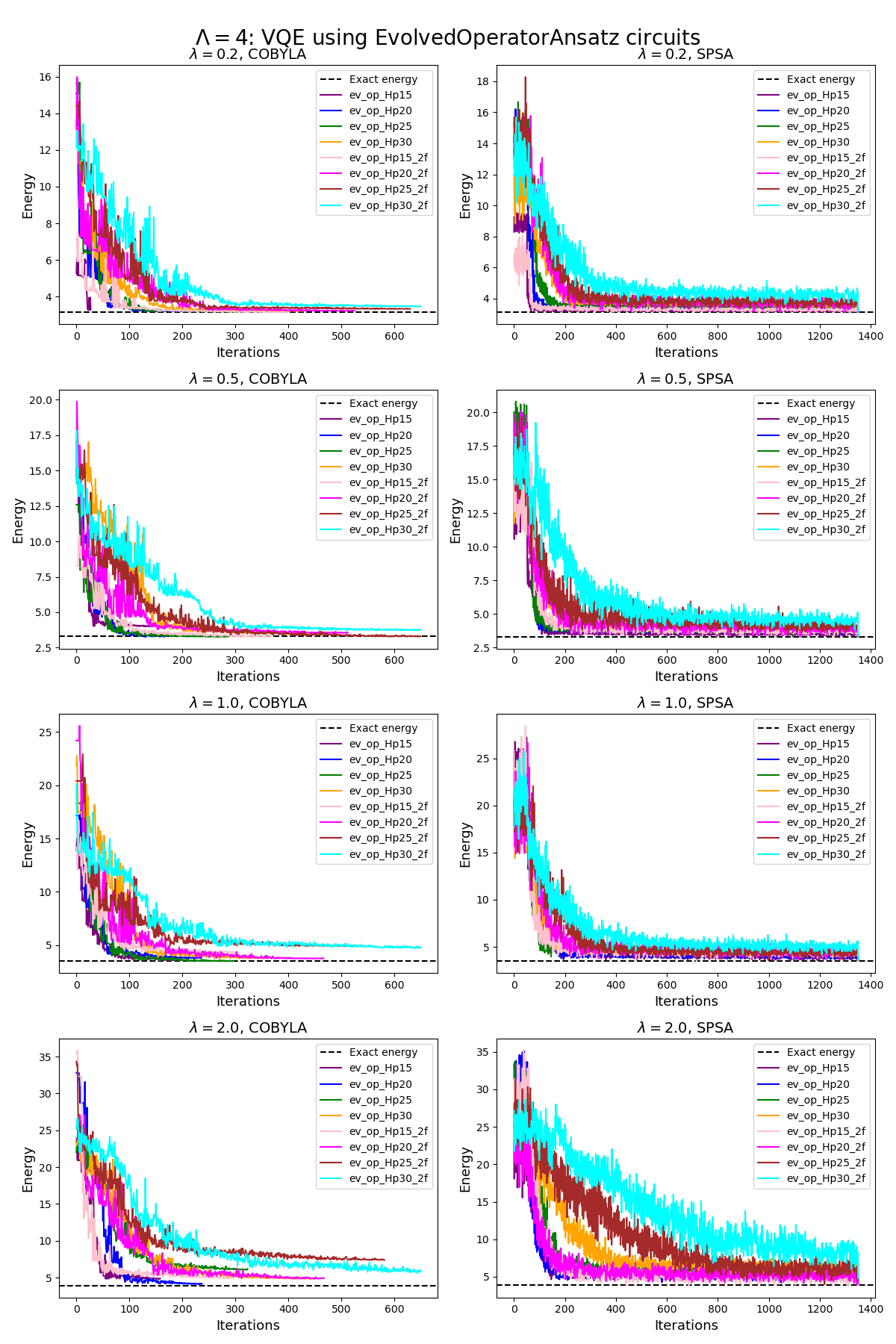}
\caption{Convergence curves of the energy for the VQE experiments ($H^{\L=4}_{\l=0.2, 0.5, 1.0, 2.0}$, \texttt{EvolvedOperatorAnsatz} from Table \ref{L4_ev_op_quantum circuits}, SPSA \& COBYA). First column: ($H^{\L=4}_{\l}$, \texttt{EvolvedOperatorAnsatz}, COBYA). Second column:  ($H^{\L=4}_{\l}$, \texttt{EvolvedOperatorAnsatz}, SPSA).  }
\label{fig_L4_eo_curves}
\end{figure}
\FloatBarrier
\clearpage
\section{$\L=2$ supersymmetric $SU(2)$ model: Full results}
\label{sec-L2-bmn-full-res}
\begin{table}[!ht]
\centering
\begin{tabular}{cc c c c}
\hline\hline
&Ansatz  & Energy (COBYLA) & Energy (SPSA) &\\
\hline\hline
& \texttt{ev\_op\_Hp15} & 2.05509  &   0.09354 &\\
& \texttt{ev\_op\_Hp20}  &0.07807  & \textbf{0.01228}&\\
& \texttt{ev\_op\_H}    &2.14721   &0.11827&\\
& \texttt{ev\_op\_Hp15\_2f} &  0.09291  & 0.09172&\\
& \texttt{ev\_op\_Hp20\_2f}&  0.07201  & 0.05251&\\
&\texttt{ev\_op\_H\_2f} & \textbf{0.03099}  & 1.53414&\\
& \texttt{ev\_op\_Hp15\_3f} &  0.09358  & 0.09763&\\
& \texttt{ev\_op\_Hp20\_3f} & 0.16704  & 0.18184&\\
& \texttt{ev\_op\_H\_3f} &2.20486   &2.20766&\\
& \texttt{ev\_op\_Hp15\_4f}  & 0.10025 &  0.13584&\\
& \texttt{ev\_op\_Hp20\_4f} & 0.57030 &  0.94922&\\
& \texttt{ev\_op\_H\_4f}& 3.28990  & 3.04627&\\
\hline
\end{tabular}
\caption{Full results of the VQE experiments $\lf(H^{(S)\L=2}_{\l=0.2},\texttt{EvolvedOperatorAnsatz}, \text{COBYLA/SPSA}\rr)$. All \texttt{EvolvedOperatorAnsatz} variants are described in Table \ref{bmn-ev-op}. The exact energy is $E=0.003287$. The best result from each optimizer is noted in bold.}\label{bmn_L2_l02_eo}
\end{table}
\begin{table}[!ht]
\centering
\begin{tabular}{cc c c c}
\hline\hline
&Ansatz  & Energy (COBYLA) & Energy (SPSA) &\\
\hline\hline
&\texttt{ev\_op\_Hp15}  &2.14404  & 0.22217 &\\
&\texttt{ev\_op\_Hp20}  &0.24023  & \textbf{0.01953}&\\
&\texttt{ev\_op\_H}  &2.36133      & 0.33105&\\
&\texttt{ev\_op\_Hp15\_2f}  & \textbf{0.19482} &  0.23730&\\
&\texttt{ev\_op\_Hp20\_2f} & 0.22119  & 0.18799&\\
&\texttt{ev\_op\_H\_2f}& 0.39746    & 2.15186&\\
&\texttt{ev\_op\_Hp15\_3f} &   0.20605  & 0.26318&\\
&\texttt{ev\_op\_Hp20\_3f} & 1.22900  & 0.32031&\\
&\texttt{ev\_op\_H\_3f} & 2.87256    &1.84668&\\
&\texttt{ev\_op\_Hp15\_4f} &  0.24854 &   0.26904&\\
&\texttt{ev\_op\_Hp20\_4f}&  1.33350 &  0.65479&\\
&\texttt{ev\_op\_H\_4f}& 3.59375  & 3.48535&\\
\hline
\end{tabular}
\caption{Full results of the VQE experiments $\lf(H^{(S)\L=2}_{\l=0.5},\texttt{EvolvedOperatorAnsatz}, \text{COBYLA/SPSA}\rr)$. All \texttt{EvolvedOperatorAnsatz} variants are described in Table \ref{bmn-ev-op}. The exact energy is $E=0.01690$. The best result from each optimizer is noted in bold.}\label{bmn_L2_l05_eo}
\end{table}

\begin{table}[!ht]
\centering
\begin{tabular}{cc c c c}
\hline\hline
&Ansatz  & Energy (COBYLA) & Energy (SPSA) &\\
\hline\hline
&\texttt{ev\_op\_Hp15}  & 2.34535 &  0.47291&\\
&\texttt{ev\_op\_Hp20}  &0.46279 &  \textbf{0.10229}&\\
&\texttt{ev\_op\_H}   &2.69001 &  0.67807&\\
&\texttt{ev\_op\_Hp15\_2f} &  0.55453  & 0.44484&\\
&\texttt{ev\_op\_Hp20\_2f}&  0.40886 &  0.13280&\\
&\texttt{ev\_op\_H\_2f} & 0.64810  & 1.29156&\\
&\texttt{ev\_op\_Hp15\_3f} &  \textbf{0.39722}  & 0.50091&\\
&\texttt{ev\_op\_Hp20\_3f}&  2.34058  & 0.44396&\\
&\texttt{ev\_op\_H\_3f} & 3.73377  & 2.00362&\\
&\texttt{ev\_op\_Hp15\_4f} &  0.57642  & 0.55918&\\
&\texttt{ev\_op\_Hp20\_4f} & 3.24279  & 0.47511&\\
&\texttt{ev\_op\_H\_4f} &3.34156  & 3.08163&\\
\hline
\end{tabular}
\caption{Full results of the VQE experiments $\lf(H^{(S)\L=2}_{\l=1.0},\texttt{EvolvedOperatorAnsatz}, \text{COBYLA/SPSA}\rr)$. All \texttt{EvolvedOperatorAnsatz} variants are described in Table \ref{bmn-ev-op}. The exact energy is $E=0.04829$. The best result from each optimizer is noted in bold.}\label{bmn_L2_l10_eo}
\end{table}
\begin{table}[!ht]
\centering
\begin{tabular}{cc c c c}
\hline\hline
&Ansatz  & Energy (COBYLA) & Energy (SPSA) &\\
\hline\hline
&\texttt{ev\_op\_Hp15}  &2.87744  & 0.87207&\\
&\texttt{ev\_op\_Hp20}  &0.67041  & \textbf{0.15918}&\\
&\texttt{ev\_op\_H}  &3.51416  & 1.40527&\\
& \texttt{ev\_op\_Hp15\_2f} &   0.92383  & 0.88135&\\
&\texttt{ev\_op\_Hp20\_2f} & 3.04639  & 0.41748&\\
&\texttt{ev\_op\_H\_2f} &3.05371   &1.17822&\\
&\texttt{ev\_op\_Hp15\_3f} &   0.88379  & 0.90771&\\
&\texttt{ev\_op\_Hp20\_3f} &  \textbf{0.62500}  & 0.82568&\\
&\texttt{ev\_op\_H\_3f} &3.80322   &2.17969&\\
&\texttt{ev\_op\_Hp15\_4f} &  0.95947 &   0.93457&\\
&\texttt{ev\_op\_Hp20\_4f} & 2.89307  & 1.25342&\\
&\texttt{ev\_op\_H\_4f} &2.73145  & 2.16602&\\
\hline
\end{tabular}
\caption{Full results of the VQE experiments $\lf(H^{(S)\L=2}_{\l=2.0},\texttt{EvolvedOperatorAnsatz}, \text{COBYLA/SPSA}\rr)$. All \texttt{EvolvedOperatorAnsatz} variants are described in Table \ref{bmn-ev-op}. The exact energy is $E=0.08385$. The best result from each optimizer is noted in bold.}\label{bmn_L2_l20_eo}
\end{table}
\begin{figure}[!ht]
\centering
\includegraphics[width=.75\textwidth]{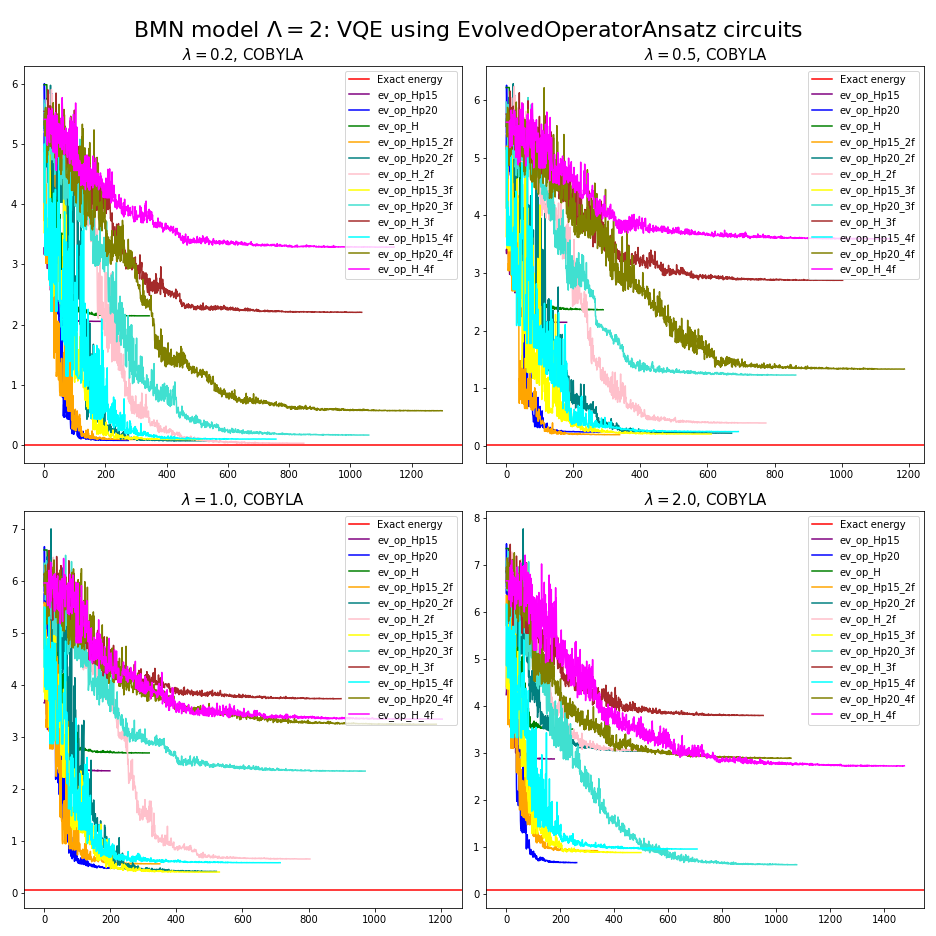}
\caption{VQE experiments $\lf(H^{(S)\L=2}_{\l},\texttt{EvolvedOperatorAnsatz}, \text{COBYLA}\rr)$: Convergence curves of the energy values. Clockwise from top left: $\l=0.2, 0.5, 2.0, 1.0$. In all 4 subfigures, depth-4 circuits perform much worse than their lower depth versions, as is evident from the corresponding convergence curves.  }
\label{fig_L2_bmn_eo_curves_cobyla}
\end{figure}

\begin{figure}[!ht]
\centering
\includegraphics[width=.5\textwidth]{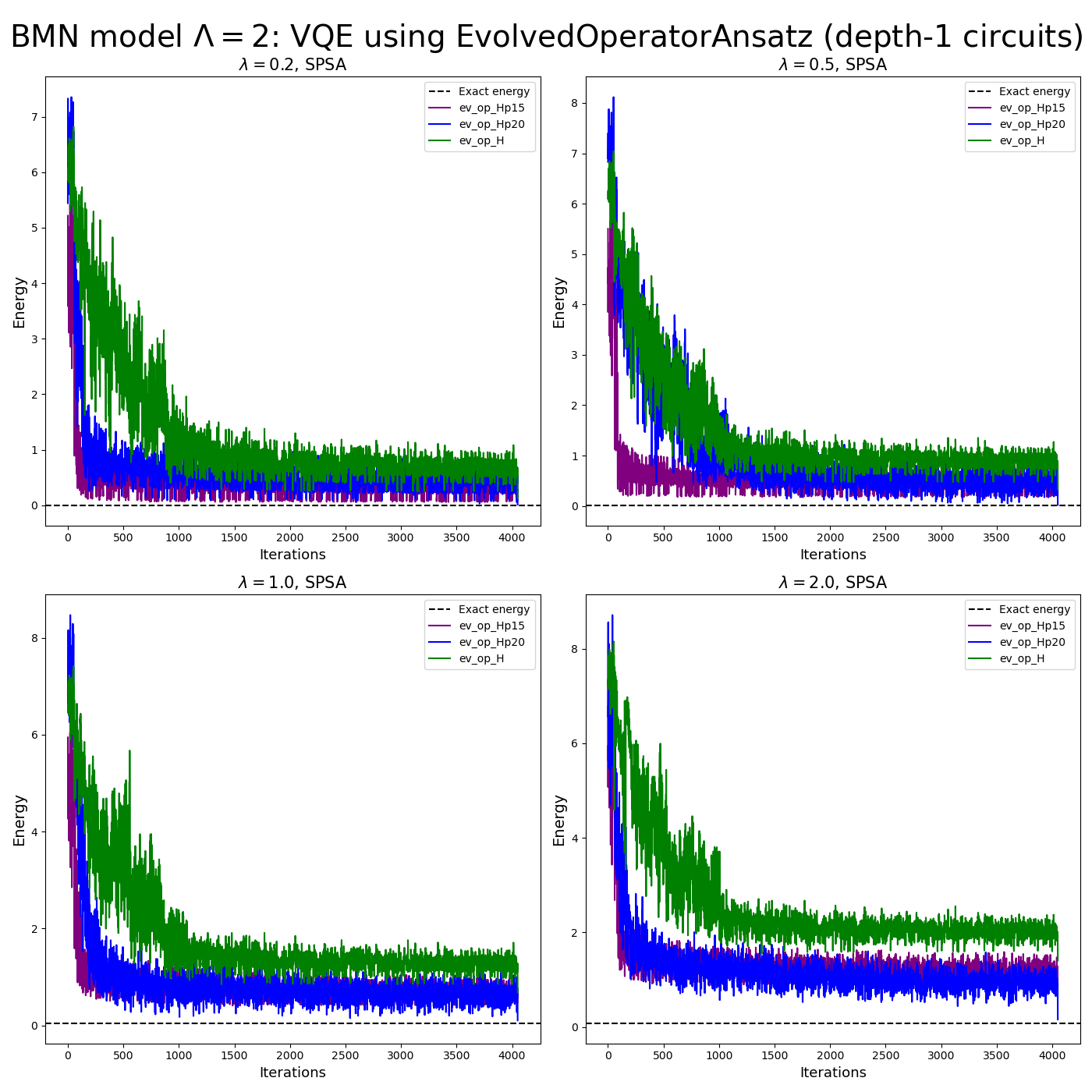}
\caption{VQE experiments $\lf(H^{(S)\L=2}_{\l},\texttt{EvolvedOperatorAnsatz}, \text{SPSA}\rr)$: Convergence curves of the energy values. Only depth-1 circuits from Table \ref{bmn-ev-op} are plotted. Clockwise from top left: $\l=0.2, 0.5, 2.0, 1.0$. In all 4 subfigures, \texttt{ev\_op\_Hp20} is the best performing variant while \texttt{ev\_op\_H} is the worst performing variant, as is evident from how close their corresponding convergence curves are to the exact energy denoted by the horizontal black dashed line.}
\label{fig_L2_bmn_eo_curves_spsa_1f}
\end{figure}

\begin{figure}[!ht]
\centering
\includegraphics[width=.5\textwidth]{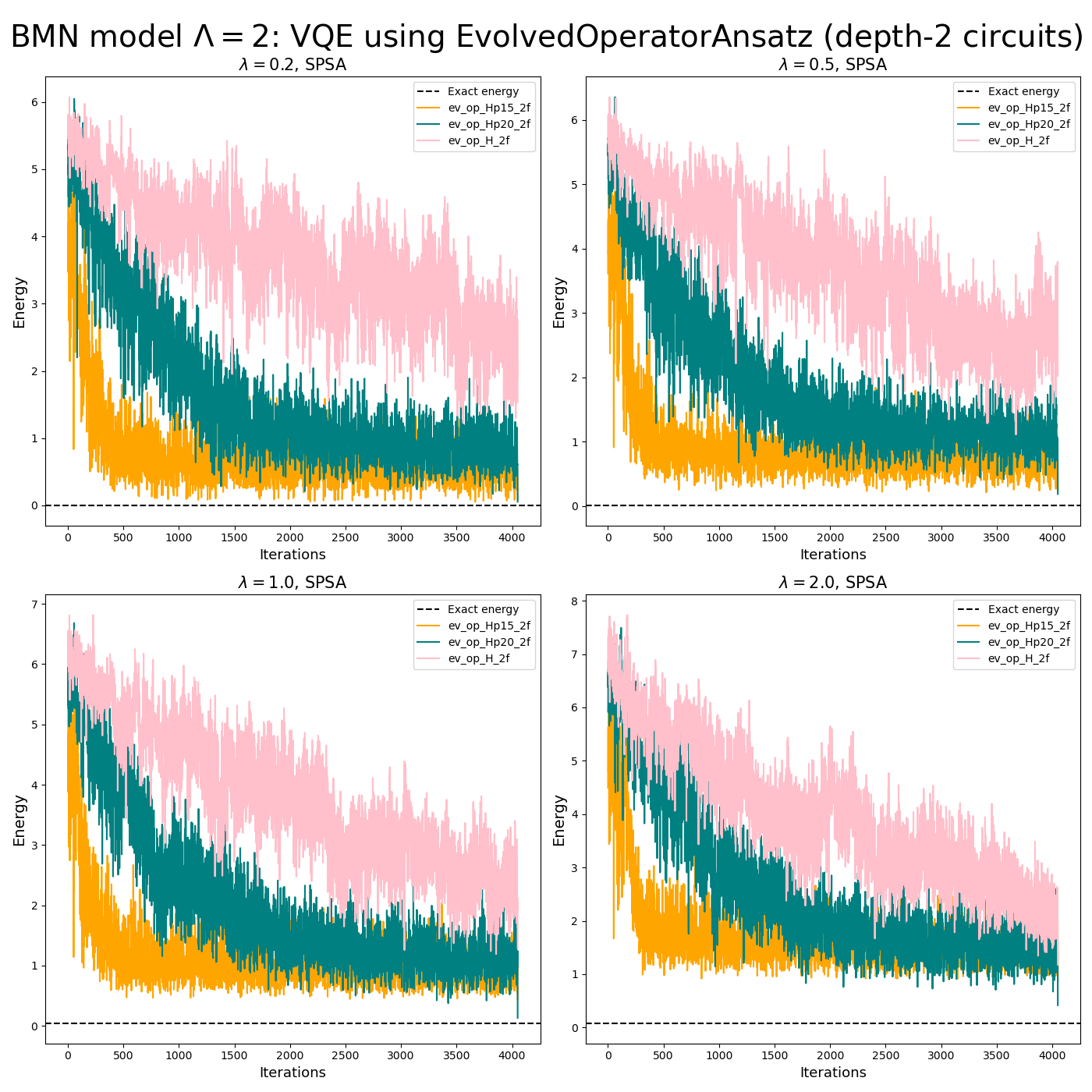}
\caption{VQE experiments $\lf(H^{(S)\L=2}_{\l},\texttt{EvolvedOperatorAnsatz}, \text{SPSA}\rr)$: Convergence curves of the energy values. Only depth-2 circuits from Table \ref{bmn-ev-op} are plotted. Clockwise from top left: $\l=0.2, 0.5, 2.0, 1.0$. In all 4 subfigures, \texttt{ev\_op\_Hp20} is the best performing variant while \texttt{ev\_op\_H} is the worst performing variant, as is evident from how close their corresponding convergence curves are to the exact energy denoted by the horizontal black dashed line.}
\label{fig_L2_bmn_eo_curves_spsa_2f}
\end{figure}

\begin{figure}[!ht]
\centering
\includegraphics[width=.5\textwidth]{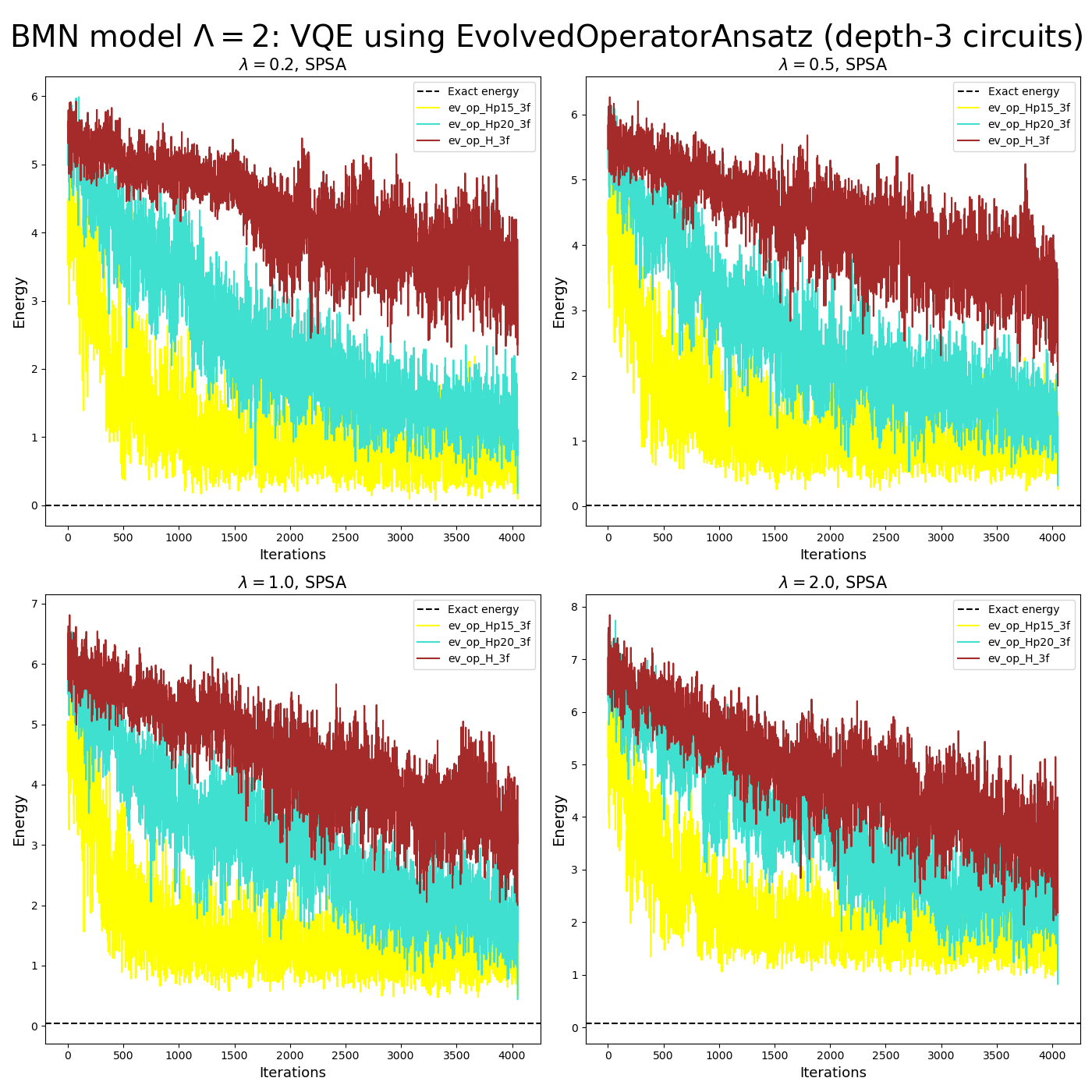}
\caption{VQE experiments $\lf(H^{(S)\L=2}_{\l},\texttt{EvolvedOperatorAnsatz}, \text{SPSA}\rr)$: Convergence curves of the energy values. Only depth-3 circuits from Table \ref{bmn-ev-op} are plotted. Clockwise from top left: $\l=0.2, 0.5, 2.0, 1.0$. In all 4 subfigures, \texttt{ev\_op\_Hp20} is the best performing variant while \texttt{ev\_op\_H} is the worst performing variant, as is evident from how close their corresponding convergence curves are to the exact energy denoted by the horizontal black dashed line.}
\label{fig_L2_bmn_eo_curves_spsa_3f}
\end{figure}

\begin{figure}[!ht]
\centering
\includegraphics[width=.5\textwidth]{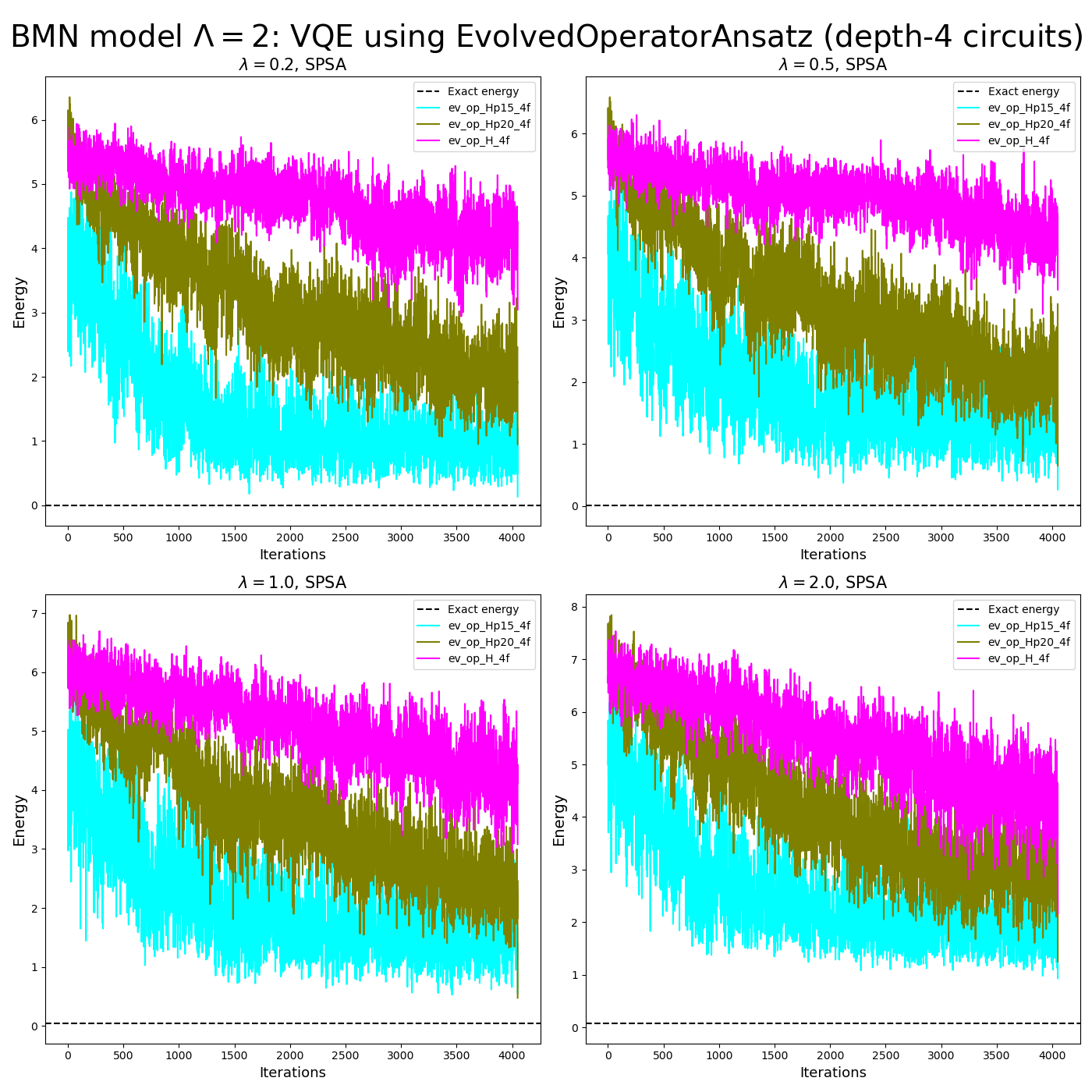}
\caption{VQE experiments $\lf(H^{(S)\L=2}_{\l},\texttt{EvolvedOperatorAnsatz}, \text{SPSA}\rr)$: Convergence curves of the energy values. Only depth-4 circuits from Table \ref{bmn-ev-op} are plotted. Clockwise from top left: $\l=0.2, 0.5, 2.0, 1.0$. In all 4 subfigures, \texttt{ev\_op\_Hp20} is the best performing variant while \texttt{ev\_op\_H} is the worst performing variant, as is evident from how close their corresponding convergence curves are to the exact energy denoted by the horizontal black dashed line.}
\label{fig_L2_bmn_eo_curves_spsa_4f}
\end{figure}
\FloatBarrier
\clearpage
\section{\texttt{TwoLocal} versus \texttt{EfficientSU2}} \label{sec-tl-vs-es2}
\subsection{$\Lambda=2$ bosonic $SU(2)$ model}
\label{sec-L2-tl-vs-effsu2}
\begin{figure}[!ht]
\centering
\includegraphics[width=.8\textwidth]{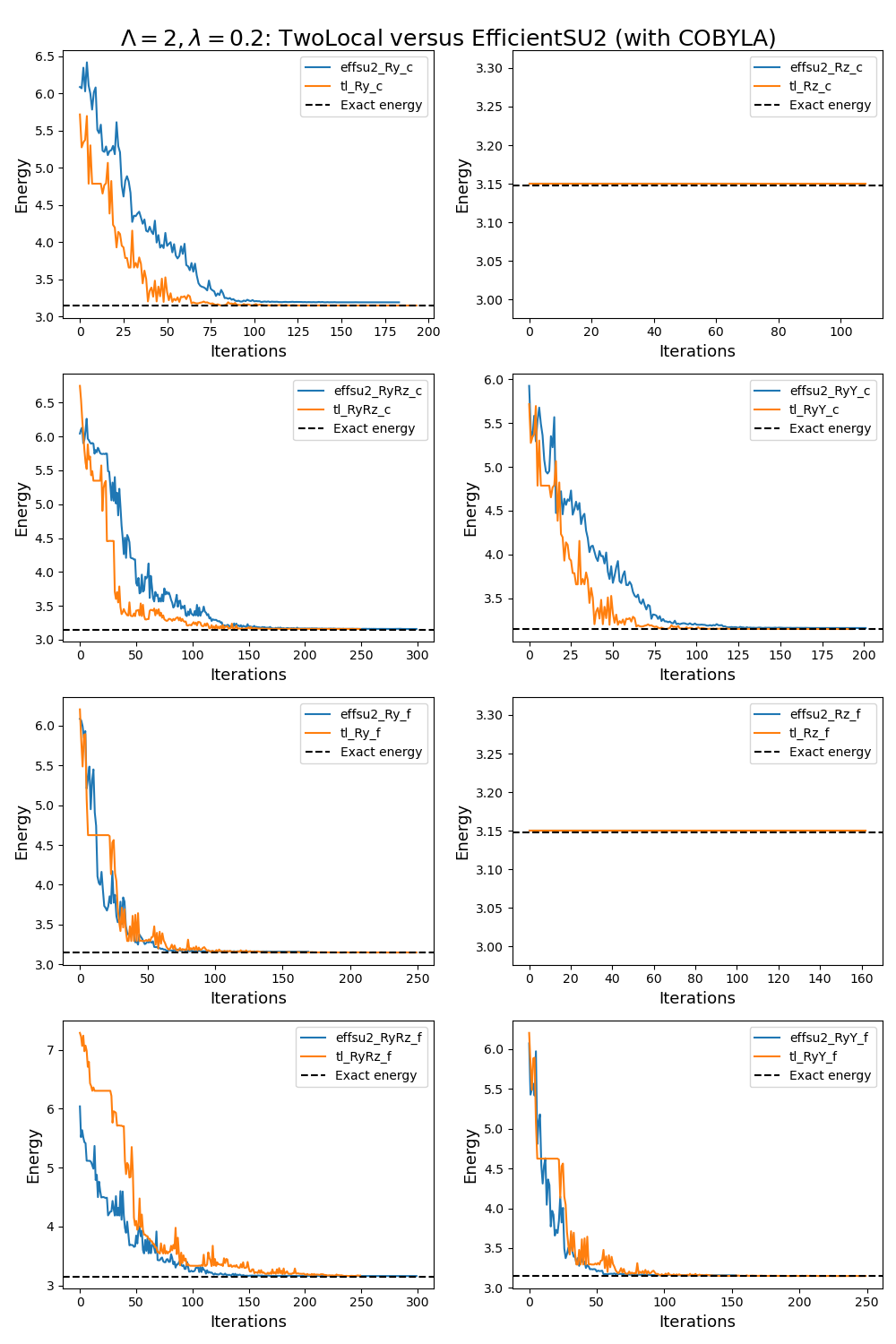}
\caption{VQE experiments $\lf(H^{\L=2}_{\l=0.2}, \texttt{EfficientSU2}\& \texttt{TwoLocal}, \text{COBYLA}\rr)$: Comparison of the performances of \texttt{TwoLocal} circuits and \texttt{EfficientSU2}, variant by variant using COBYLA optimizer. All 8 variants of \texttt{TwoLocal} outperform or are on par with the corresponding 8 variants of \texttt{EfficientSU2}, as is evident from the orange line representing the \texttt{TwoLocal} variant converges at a lower/the same value than/as the blue line representing the \texttt{EfficientSU2} variant. Both \texttt{TwoLocal} \& \texttt{EfficientSU2} variants involving $R_Z$ rotation block fail to be optimized with COBYLA as their convergence curves are just straight lines (first row \& third row, right subfigure). }
\label{fig_L2_l02_tl_vs_es_cobyla}
\end{figure}
\begin{figure}[!ht]
\centering
\includegraphics[width=.8\textwidth]{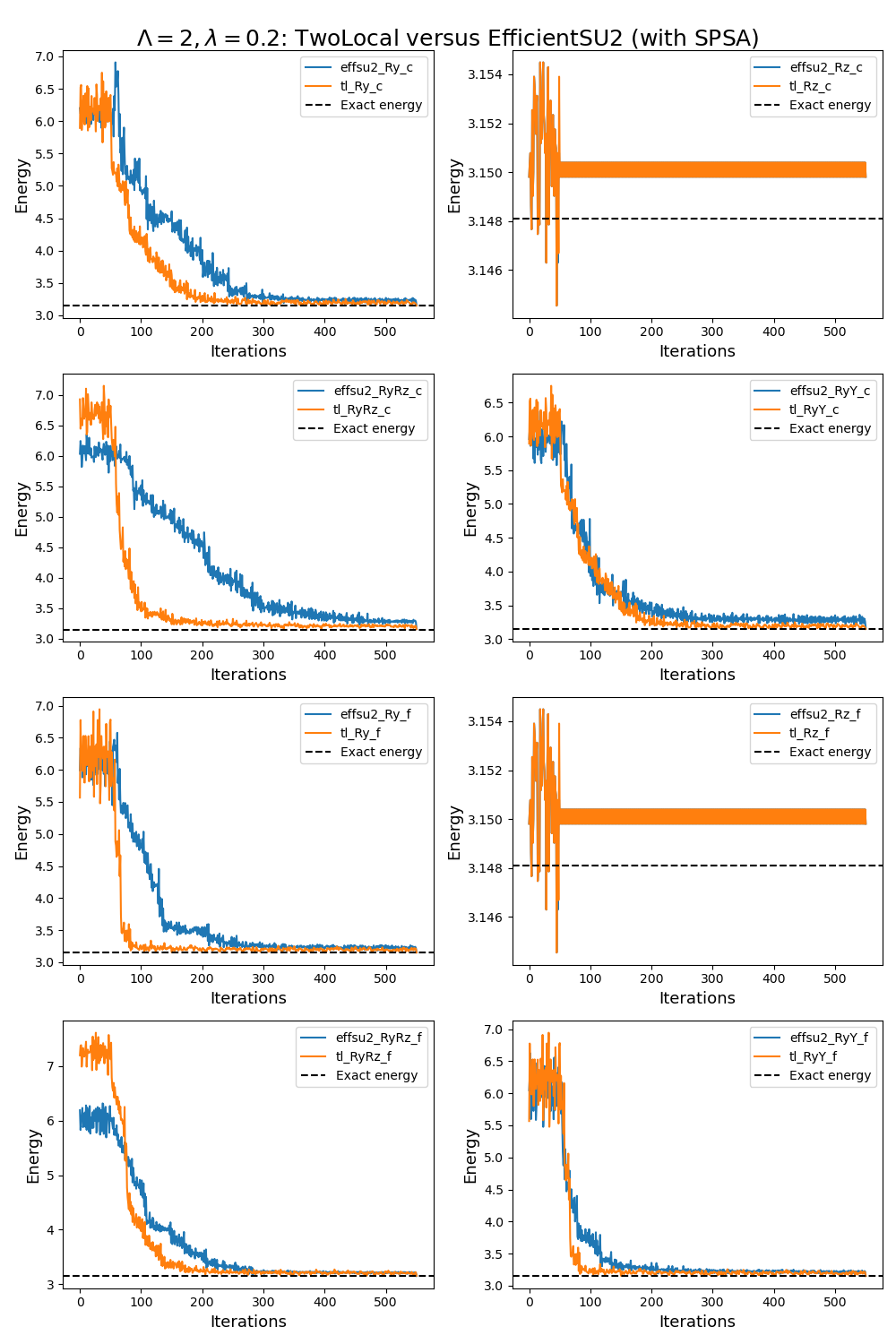}
\caption{VQE experiments $\lf(H^{\L=2}_{\l=0.2}, \texttt{EfficientSU2} \& \texttt{TwoLocal}, \text{SPSA}\rr)$: Comparison of the performances of \texttt{TwoLocal} circuits and \texttt{EfficientSU2}, variant by variant using SPSA optimizer. All 8 variants of \texttt{TwoLocal} outperform or are on par with the corresponding 8 variants of \texttt{EfficientSU2}, as is evident from the orange line representing the \texttt{TwoLocal} variant converges at a lower/the same value than/as the blue line representing the \texttt{EfficientSU2} variant. Both \texttt{TwoLocal} \& \texttt{EfficientSU2} variants involving $R_Z$ rotation block fail to be optimized with SPSA as their convergence curves are practically just straight lines coinciding with each other (first row \& third row, right subfigure). }
\label{fig_L2_l02_tl_vs_es_spsa}
\end{figure}
\begin{figure}[!ht]
\centering
\includegraphics[width=.8\textwidth]{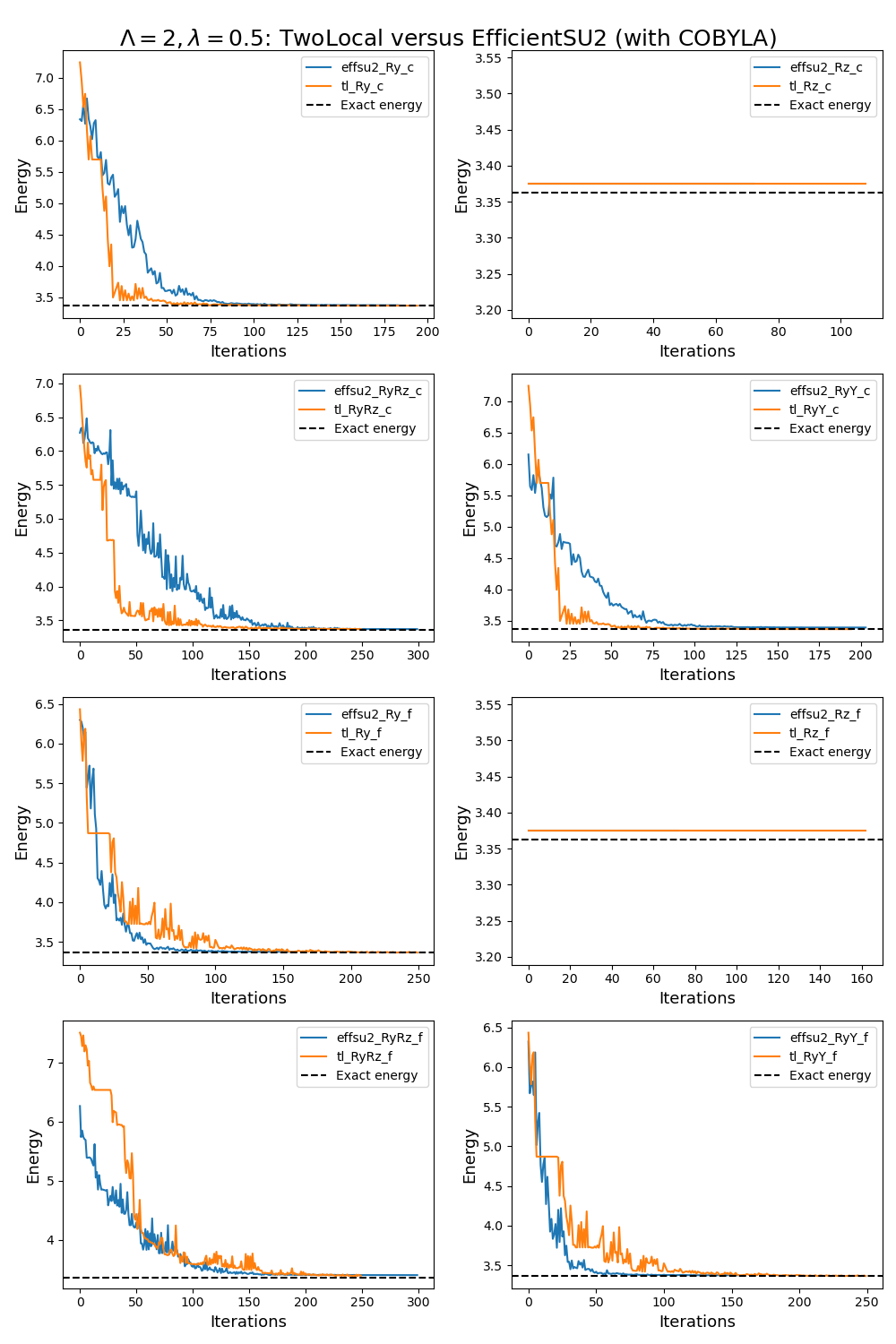}
\caption{VQE experiments $\lf(H^{\L=2}_{\l=0.5}, \texttt{EfficientSU2} \& \texttt{TwoLocal}, \text{COBYLA}\rr)$: Comparison of the performances of \texttt{TwoLocal} circuits and \texttt{EfficientSU2}, variant by variant using COBYLA optimizer. All 8 variants of \texttt{TwoLocal} outperform or are on par with the corresponding 8 variants of \texttt{EfficientSU2}, as is evident from the orange line representing the \texttt{TwoLocal} variant converges at a lower/the same value than/as the blue line representing the \texttt{EfficientSU2} variant. Both \texttt{TwoLocal} \& \texttt{EfficientSU2} variants involving $R_Z$ rotation block fail to be optimized with COBYLA as their convergence curves are just straight lines (first row \& third row, right subfigure).}
\label{fig_L2_l05_tl_vs_es_cobyla}
\end{figure}
\begin{figure}[!ht]
\centering
\includegraphics[width=.8\textwidth]{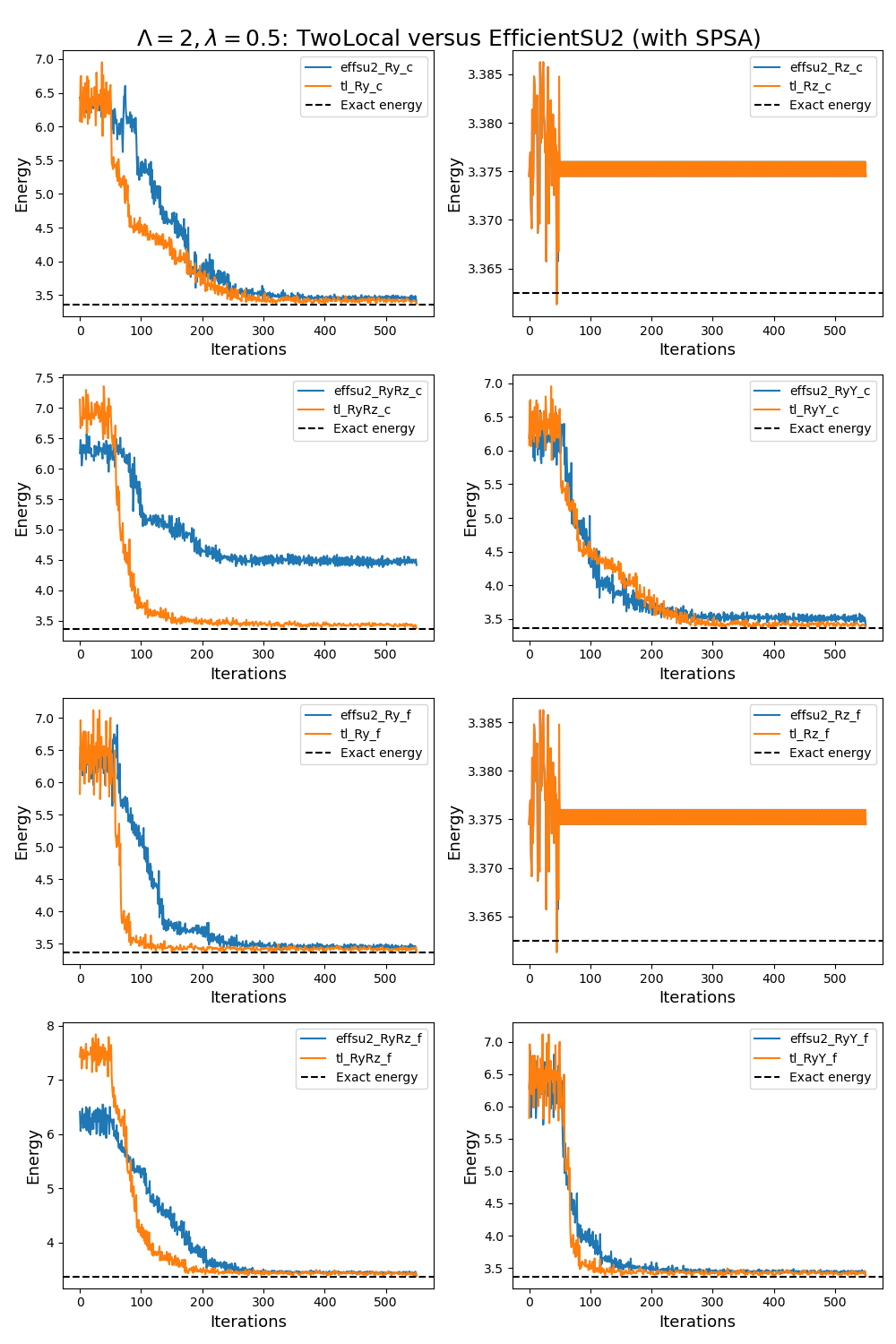}
\caption{VQE experiments $\lf(H^{\L=2}_{\l=0.5}, \texttt{EfficientSU2} \& \texttt{TwoLocal}, \text{SPSA}\rr)$: Comparison of the performances of \texttt{TwoLocal} circuits and \texttt{EfficientSU2}, variant by variant using SPSA optimizer. All 8 variants of \texttt{TwoLocal} outperform or are on par with the corresponding 8 variants of \texttt{EfficientSU2}, as is evident from the orange line representing the \texttt{TwoLocal} variant converges at a lower/the same value than/as the blue line representing the \texttt{EfficientSU2} variant. Both \texttt{TwoLocal} \& \texttt{EfficientSU2} variants involving $R_Z$ rotation block fail to be optimized with SPSA as their convergence curves are practically just straight lines coinciding with each other (first row \& third row, right subfigure).}
\label{fig_L2_l05_tl_vs_es_spsa}
\end{figure}
\begin{figure}[!ht]
\centering
\includegraphics[width=.8\textwidth]{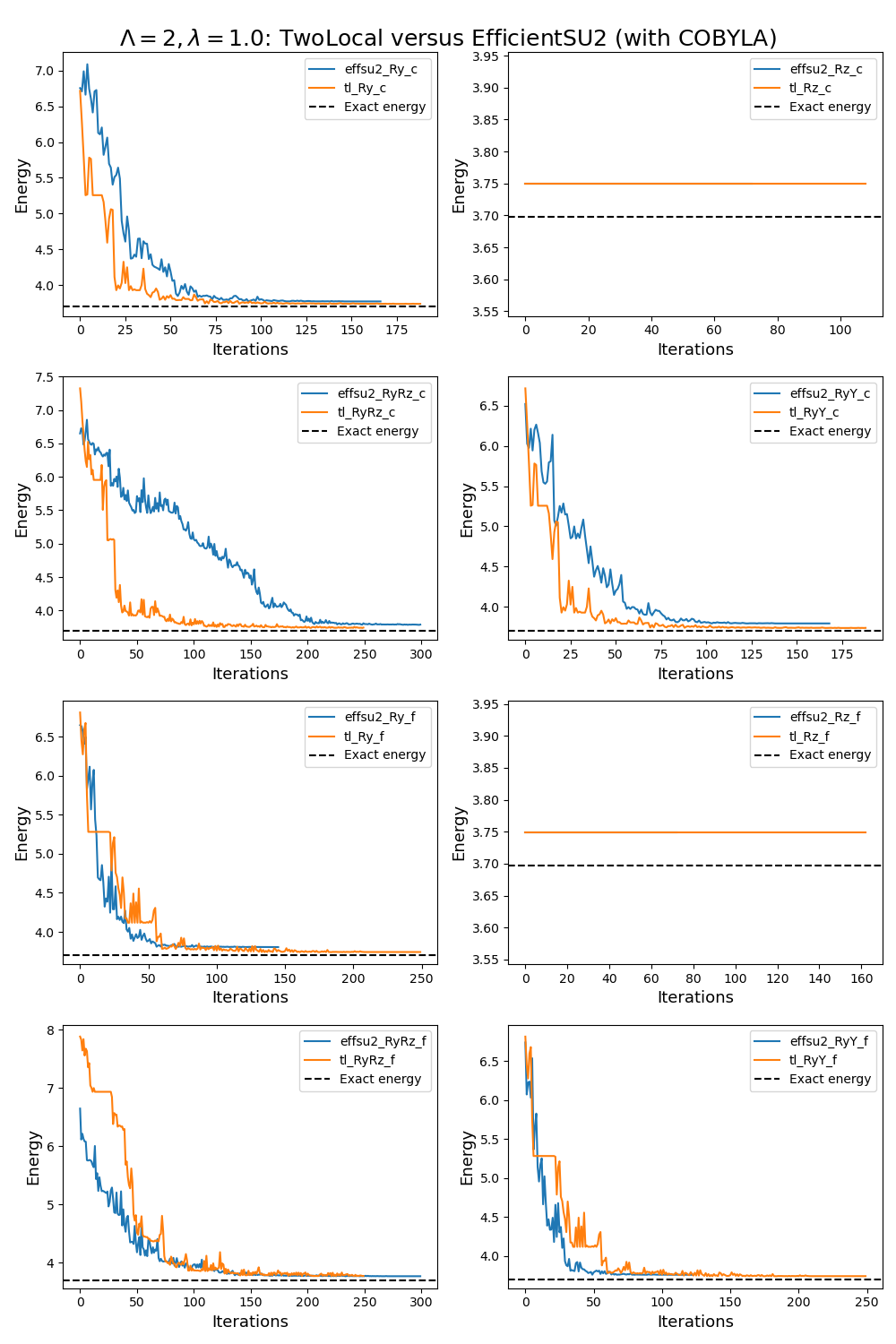}
\caption{VQE experiments $\lf(H^{\L=2}_{\l=1.0}, \texttt{EfficientSU2} \& \texttt{TwoLocal}, \text{COBYLA}\rr)$: Comparison of the performances of \texttt{TwoLocal} circuits and \texttt{EfficientSU2}, variant by variant using COBYLA optimizer. All 8 variants of \texttt{TwoLocal} outperform or are on par with the corresponding 8 variants of \texttt{EfficientSU2}, as is evident from the orange line representing the \texttt{TwoLocal} variant converges at a lower/the same value than/as the blue line representing the \texttt{EfficientSU2} variant. Both \texttt{TwoLocal} \& \texttt{EfficientSU2} variants involving $R_Z$ rotation block fail to be optimized with COBYLA as their convergence curves are just straight lines (first row \& third row, right subfigure).}
\label{fig_L2_l10_tl_vs_es_cobyla}
\end{figure}
\begin{figure}[!ht]
\centering
\includegraphics[width=.8\textwidth]{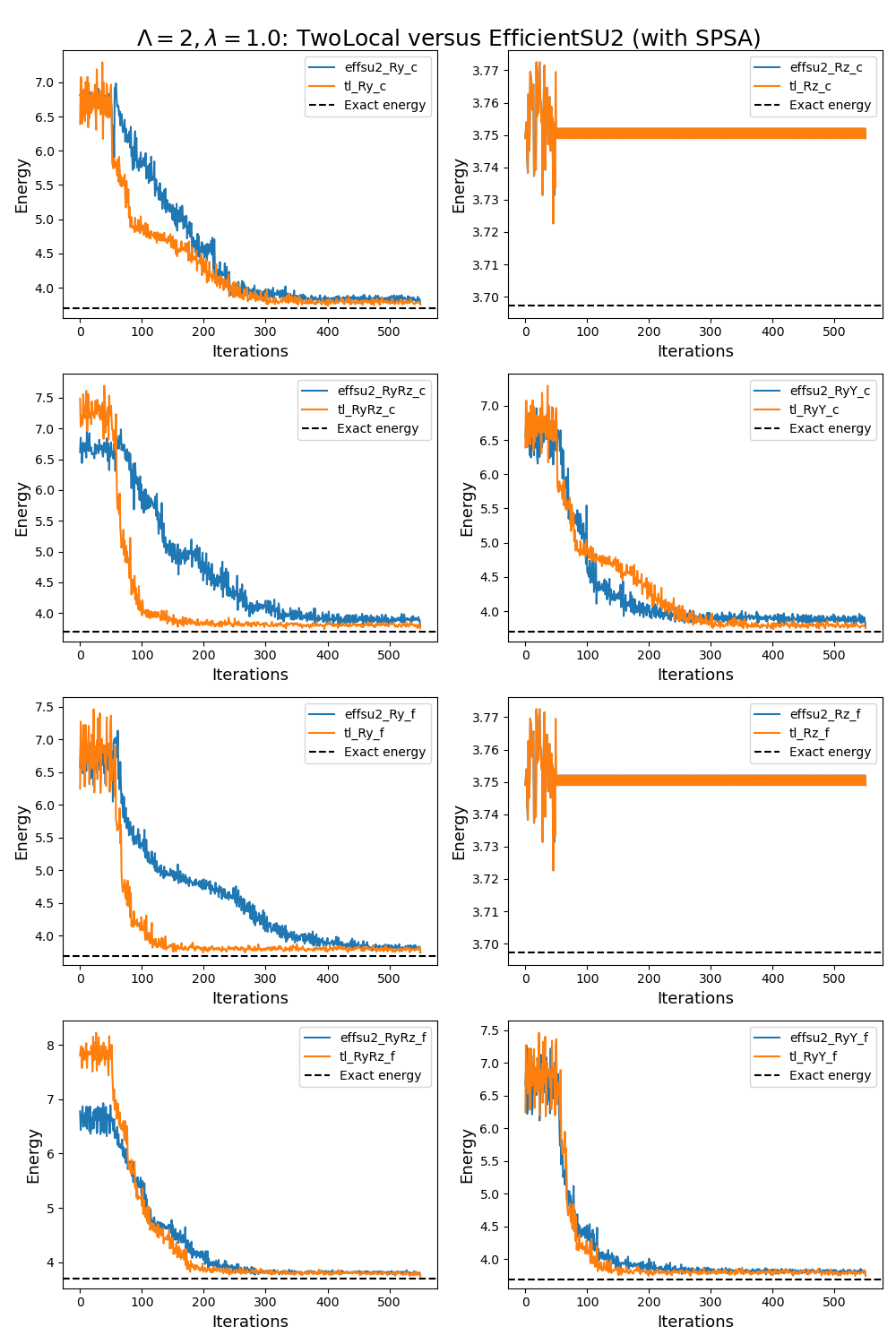}
\caption{VQE experiments $\lf(H^{\L=2}_{\l=1.0}, \texttt{EfficientSU2} \& \texttt{TwoLocal}, \text{SPSA}\rr)$: Comparison of the performances of \texttt{TwoLocal} circuits and \texttt{EfficientSU2}, variant by variant using SPSA optimizer. All 8 variants of \texttt{TwoLocal} outperform or are on par with the corresponding 8 variants of \texttt{EfficientSU2}, as is evident from the orange line representing the \texttt{TwoLocal} variant converges at a lower/the same value than/as the blue line representing the \texttt{EfficientSU2} variant. Both \texttt{TwoLocal} \& \texttt{EfficientSU2} variants involving $R_Z$ rotation block fail to be optimized with SPSA as their convergence curves are practically just straight lines coinciding with each other (first row \& third row, right subfigure).}
\label{fig_L2_l10_tl_vs_es_spsa}
\end{figure}
\begin{figure}[!ht]
\centering
\includegraphics[width=.8\textwidth]{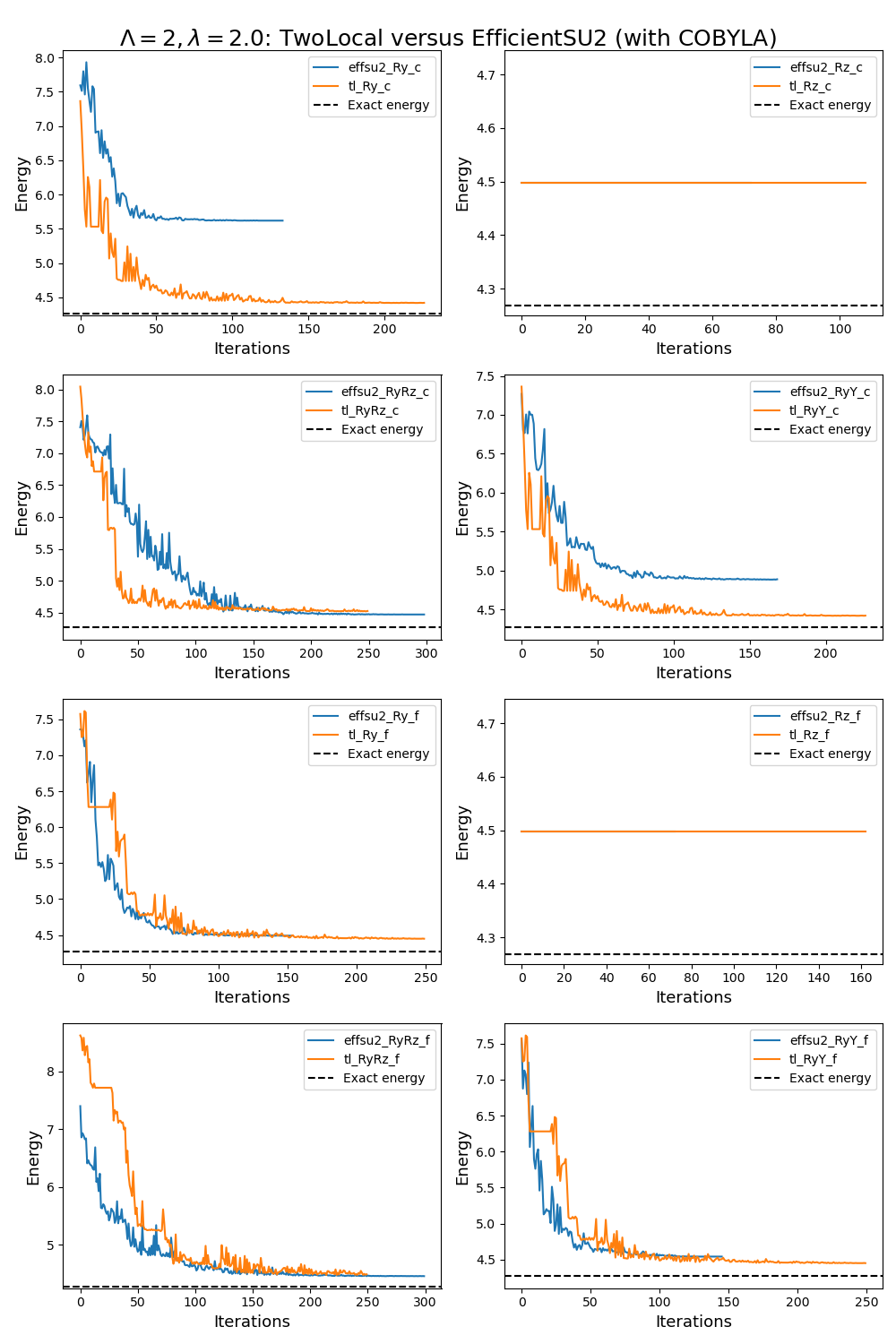}
\caption{VQE experiments $\lf(H^{\L=2}_{\l=2.0}, \texttt{EfficientSU2} \& \texttt{TwoLocal}, \text{COBYLA}\rr)$: Comparison of the performances of \texttt{TwoLocal} circuits and \texttt{EfficientSU2}, variant by variant using COBYLA optimizer. All 8 variants of \texttt{TwoLocal} outperform or are on par with the corresponding 8 variants of \texttt{EfficientSU2}, as is evident from the orange line representing the \texttt{TwoLocal} variant converges at a lower/the same value than/as the blue line representing the \texttt{EfficientSU2} variant. Both \texttt{TwoLocal} \& \texttt{EfficientSU2} variants involving $R_Z$ rotation block fail to be optimized with COBYLA as their convergence curves are just straight lines (first row \& third row, right subfigure).}
\label{fig_L2_l20_tl_vs_es_cobyla}
\end{figure}
\begin{figure}[!ht]
\centering
\includegraphics[width=.8\textwidth]{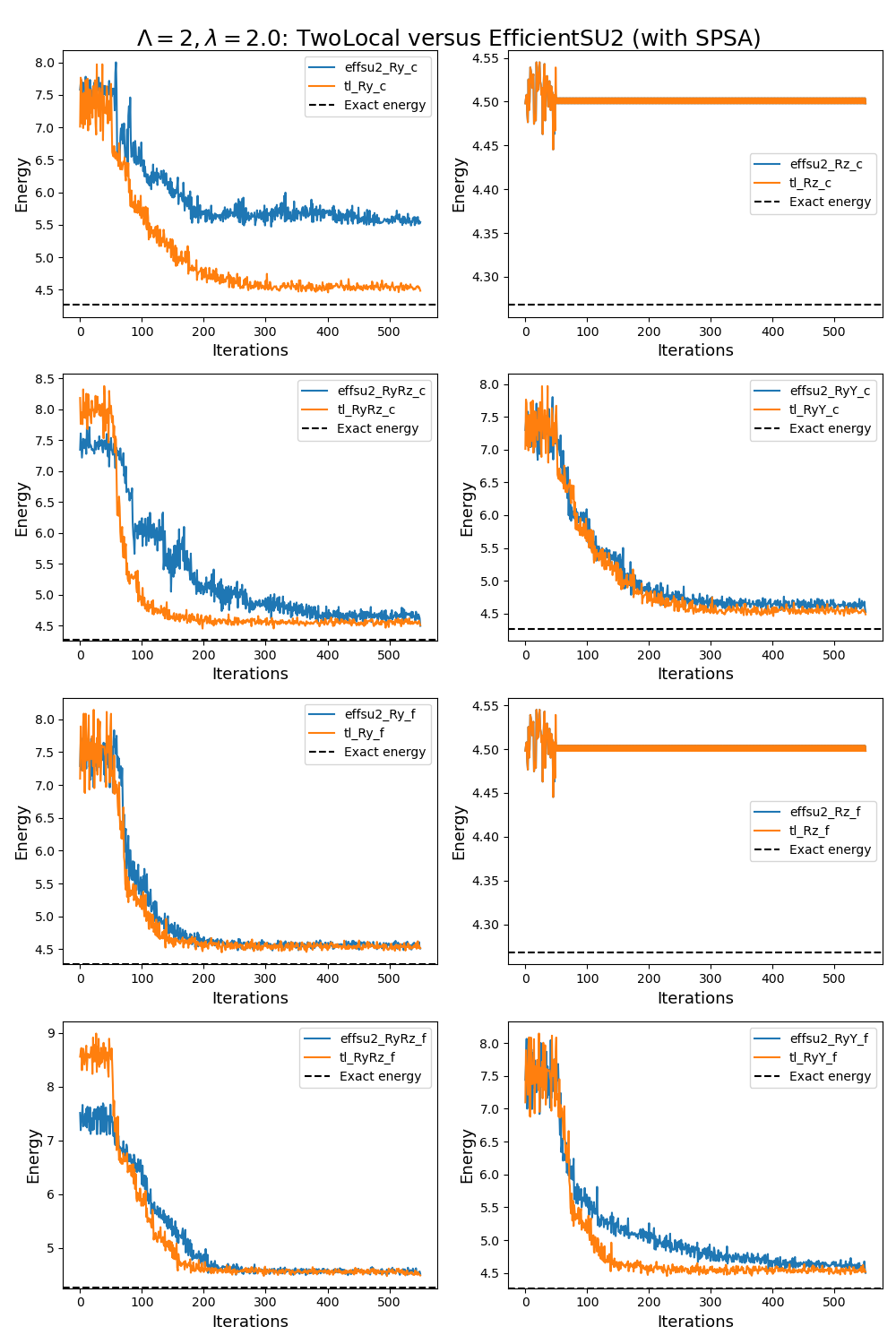}
\caption{CVQE experiments $\lf(H^{\L=2}_{\l=2.0}, \texttt{EfficientSU2} \& \texttt{TwoLocal}, \text{SPSA}\rr)$: Comparison of the performances of \texttt{TwoLocal} circuits and \texttt{EfficientSU2}, variant by variant using SPSA optimizer. All 8 variants of \texttt{TwoLocal} outperform or are on par with the corresponding 8 variants of \texttt{EfficientSU2}, as is evident from the orange line representing the \texttt{TwoLocal} variant converges at a lower/the same value than/as the blue line representing the \texttt{EfficientSU2} variant. Both \texttt{TwoLocal} \& \texttt{EfficientSU2} variants involving $R_Z$ rotation block fail to be optimized with SPSA as their convergence curves are practically just straight lines coinciding with each other (first row \& third row, right subfigure).}
\label{fig_L2_l20_tl_vs_es_spsa}
\end{figure}
\FloatBarrier
\clearpage
\subsection{$\Lambda=4$ bosonic $SU(2)$ model}
\label{sec-L4-tl-vs-effsu2}
\begin{figure}[!ht]
\centering
\includegraphics[width=.8\textwidth]{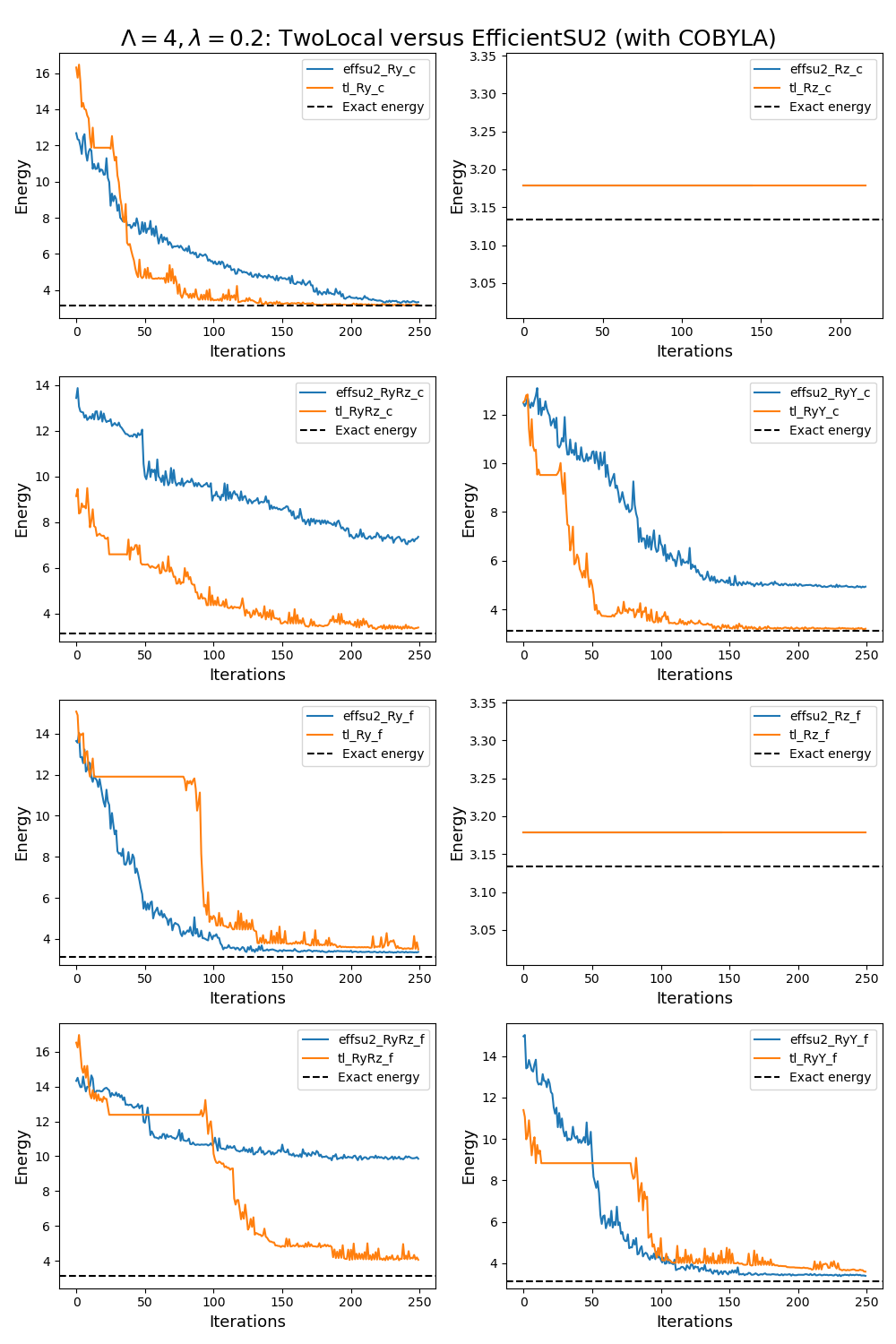}
\caption{VQE experiments $\lf(H^{\L=4}_{\l=0.2}, \texttt{EfficientSU2} \& \texttt{TwoLocal}, \text{COBYLA}\rr)$: Comparison of the performances of \texttt{TwoLocal} circuits and \texttt{EfficientSU2}, variant by variant using COBYLA optimizer. Apart from \texttt{tl\_Ry\_f} ($3^\text{rd}$ row, left subfigure) and \texttt{tl\_RyY\_f} ($4^\text{th}$ row, right subfigure), the remaining 6 variants of \texttt{TwoLocal} outperform the corresponding 8 variants of \texttt{EfficientSU2}, as is evident from the orange line representing the \texttt{TwoLocal} variant converges at a lower value than the blue line representing the \texttt{EfficientSU2} variant. Both \texttt{TwoLocal} \& \texttt{EfficientSU2} variants involving $R_Z$ rotation block fail to be optimized with COBYLA as their convergence curves are just straight lines coinciding with each other ($1^\text{st}$ row \& $3^\text{rd}$ row, right subfigure).}
\label{fig_L4_l02_tl_vs_es_cobyla}
\end{figure}
\begin{figure}[!ht]
\centering
\includegraphics[width=.8\textwidth]{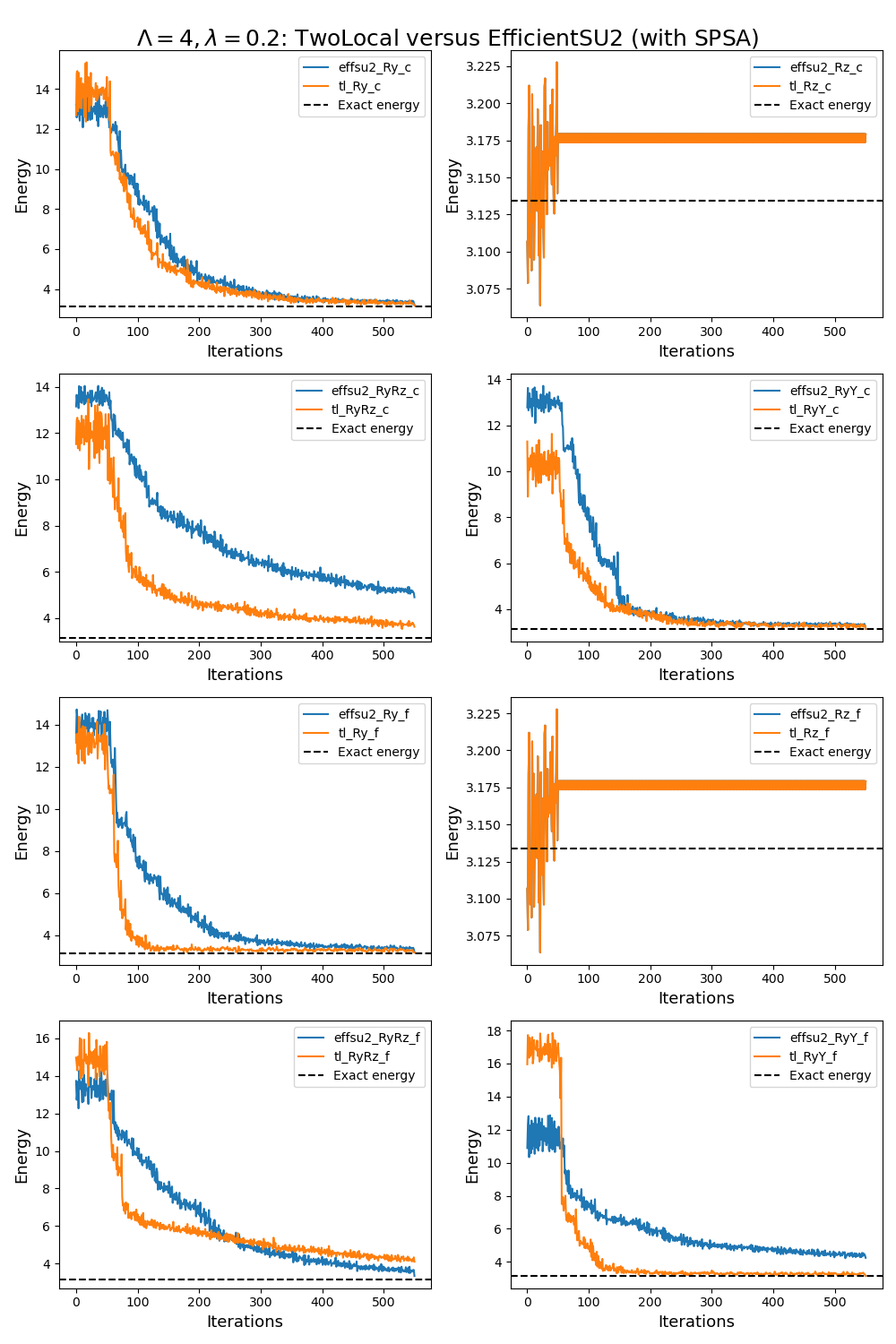}
\caption{VQE experiments $\lf(H^{\L=4}_{\l=0.2}, \texttt{EfficientSU2} \& \texttt{TwoLocal}, \text{SPSA}\rr)$: Comparison of the performances of \texttt{TwoLocal} circuits and \texttt{EfficientSU2}, variant by variant using COBYLA optimizer. Apart from \texttt{tl\_RyRz\_f}  ($4^\text{th}$ row, left subfigure), the remaining 7 variants of \texttt{TwoLocal} outperform the corresponding 8 variants of \texttt{EfficientSU2}, as is evident from the orange line representing the \texttt{TwoLocal} variant converges at a lower value than the blue line representing the \texttt{EfficientSU2} variant. Both \texttt{TwoLocal} \& \texttt{EfficientSU2} variants involving $R_Z$ rotation block fail to be optimized with COBYLA as their convergence curves are just straight lines coinciding with each other ($1^\text{st}$ row \& $3^\text{rd}$ row, right subfigure).}
\label{fig_L4_l02_tl_vs_es_spsa}
\end{figure}
\begin{figure}[!ht]
\centering
\includegraphics[width=.8\textwidth]{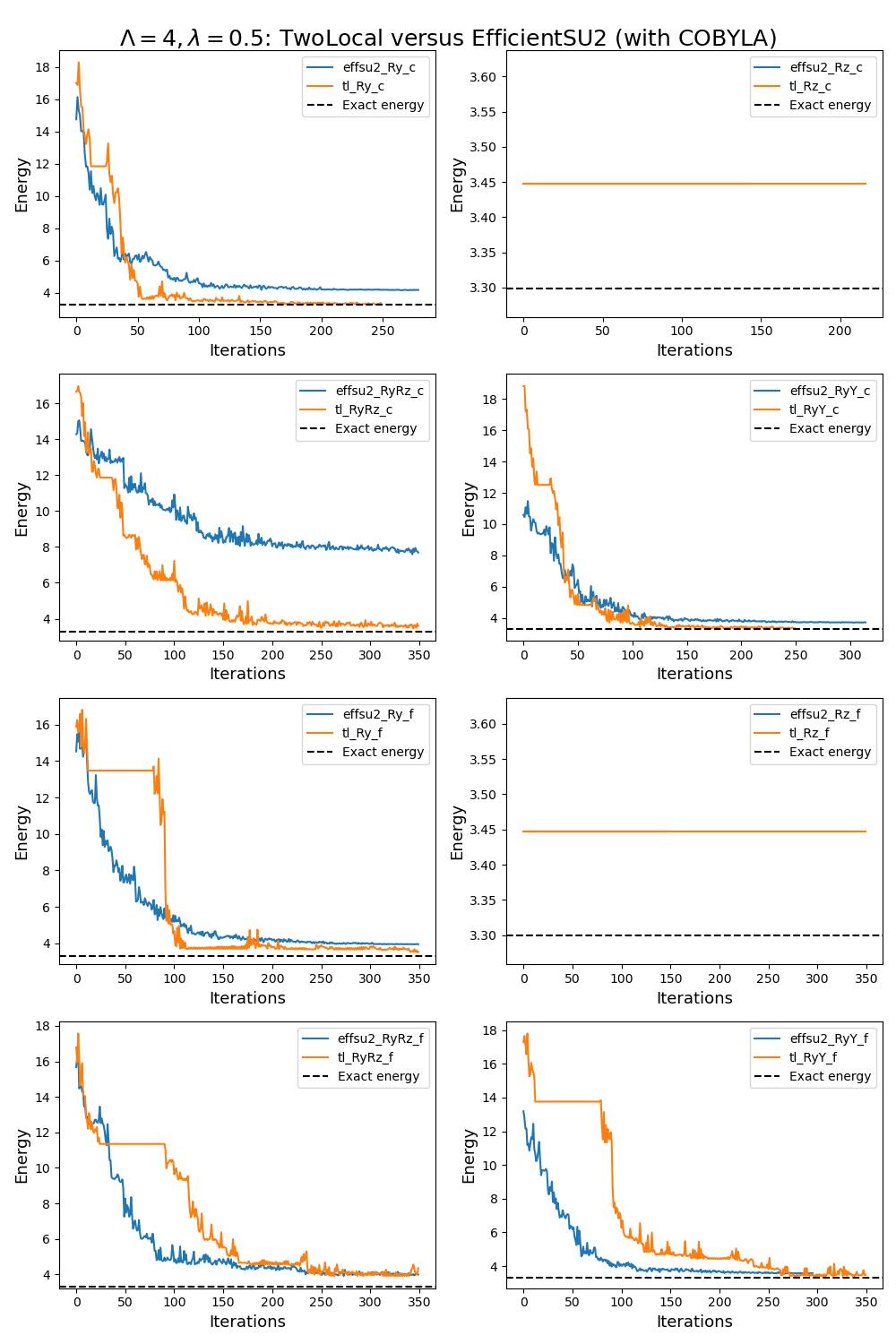}
\caption{VQE experiments $\lf(H^{\L=4}_{\l=0.5}, \texttt{EfficientSU2} \& \texttt{TwoLocal}, \text{COBYLA}\rr)$: Comparison of the performances of \texttt{TwoLocal} circuits and \texttt{EfficientSU2}, variant by variant using COBYLA optimizer. All 8 variants of \texttt{TwoLocal} outperform or are on par with the corresponding 8 variants of \texttt{EfficientSU2}, as is evident from the orange line representing the \texttt{TwoLocal} variant converges at a lower/the same value than/as the blue line representing the \texttt{EfficientSU2} variant. Both \texttt{TwoLocal} \& \texttt{EfficientSU2} variants involving $R_Z$ rotation block fail to be optimized with COBYLA as their convergence curves are just straight lines coinciding with each other ($1^\text{st}$ row \& $3^\text{rd}$ row, right subfigure).}
\label{fig_L4_l05_tl_vs_es_cobyla}
\end{figure}
\begin{figure}[!ht]
\centering
\includegraphics[width=.8\textwidth]{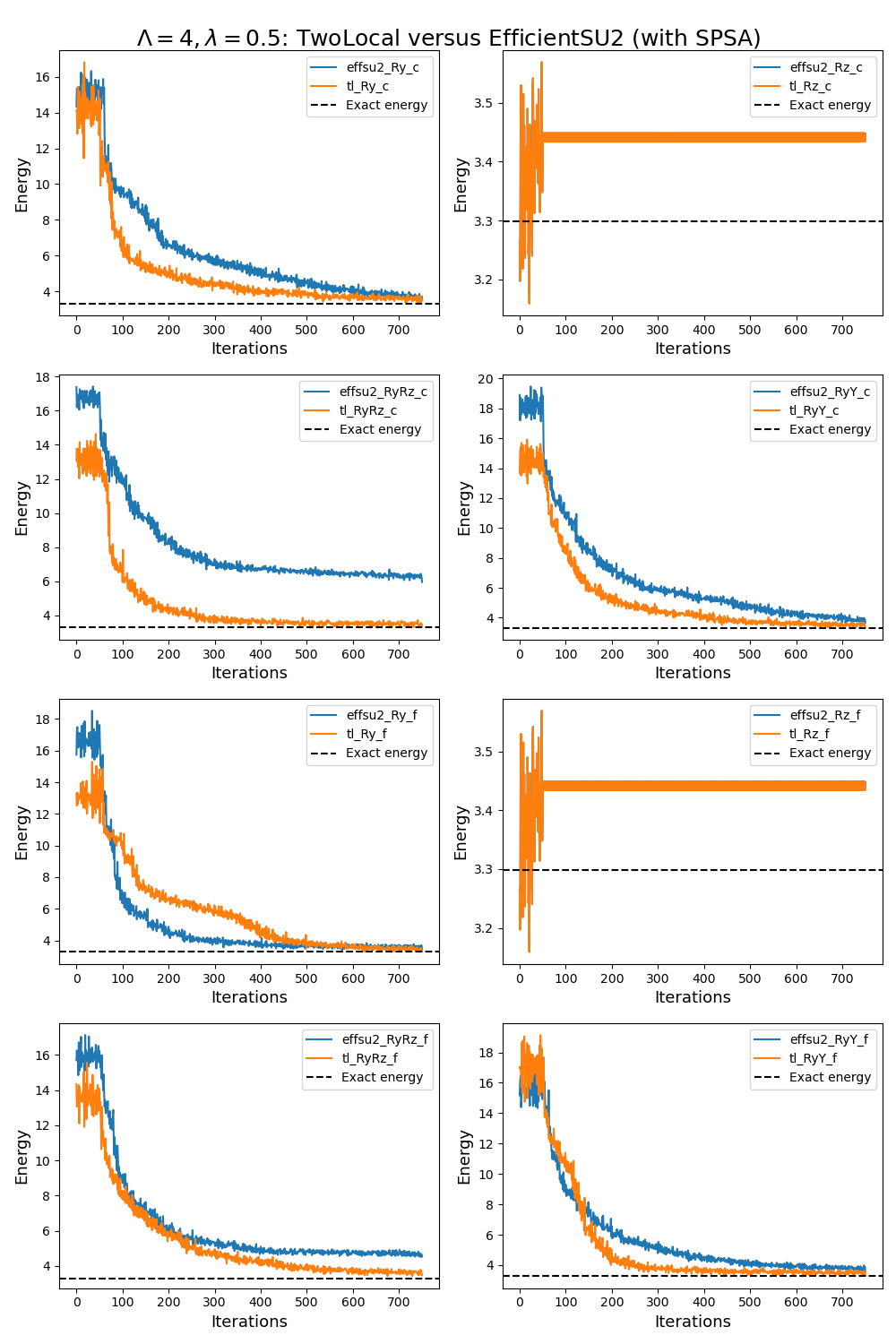}
\caption{VQE experiments $\lf(H^{\L=4}_{\l=0.5}, \texttt{EfficientSU2} \& \texttt{TwoLocal}, \text{SPSA}\rr)$: Comparison of the performances of \texttt{TwoLocal} circuits and \texttt{EfficientSU2}, variant by variant using SPSA optimizer. All 8 variants of \texttt{TwoLocal} outperform or are on par with the corresponding 8 variants of \texttt{EfficientSU2}, as is evident from the orange line representing the \texttt{TwoLocal} variant converges at a lower/the same value than/as the blue line representing the \texttt{EfficientSU2} variant. Both \texttt{TwoLocal} \& \texttt{EfficientSU2} variants involving $R_Z$ rotation block fail to be optimized with SPSA as their convergence curves are just straight lines coinciding with each other ($1^\text{st}$ row \& $3^\text{rd}$ row, right subfigure).}
\label{fig_L4_l05_tl_vs_es_spsa}
\end{figure}
\begin{figure}[!ht]
\centering
\includegraphics[width=.8\textwidth]{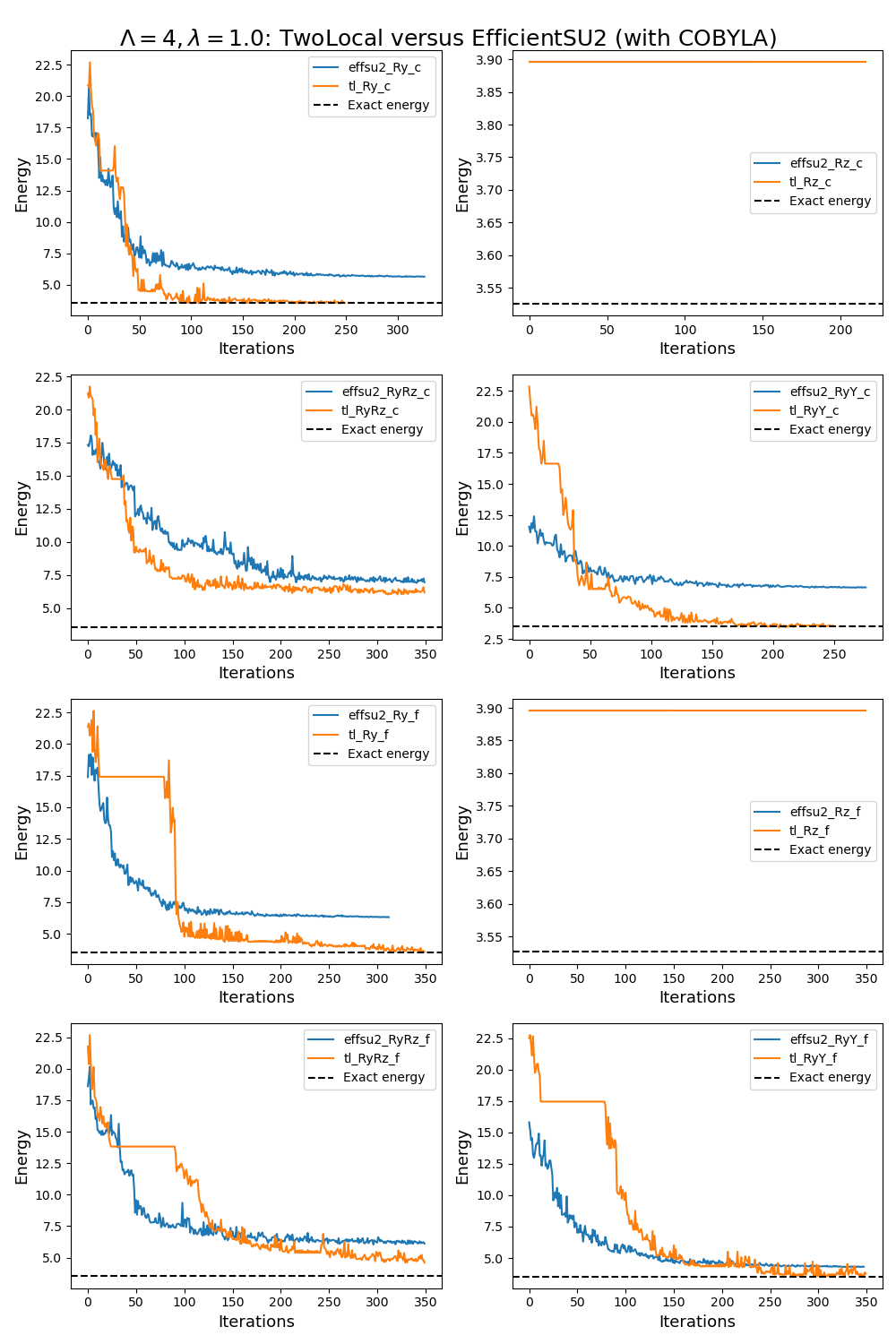}
\caption{VQE experiments $\lf(H^{\L=4}_{\l=1.0}, \texttt{EfficientSU2} \& \texttt{TwoLocal}, \text{COBYLA}\rr)$: Comparison of the performances of \texttt{TwoLocal} circuits and \texttt{EfficientSU2}, variant by variant using COBYLA optimizer. All 8 variants of \texttt{TwoLocal} outperform the corresponding 8 variants of \texttt{EfficientSU2}, as is evident from the orange line representing the \texttt{TwoLocal} variant converges at a lower value than the blue line representing the \texttt{EfficientSU2} variant. Both \texttt{TwoLocal} \& \texttt{EfficientSU2} variants involving $R_Z$ rotation block fail to be optimized with COBYLA as their convergence curves are just straight lines coinciding with each other ($1^\text{st}$ row \& $3^\text{rd}$ row, right subfigure).}
\label{fig_L4_l10_tl_vs_es_cobyla}
\end{figure}
\begin{figure}[!ht]
\centering
\includegraphics[width=.8\textwidth]{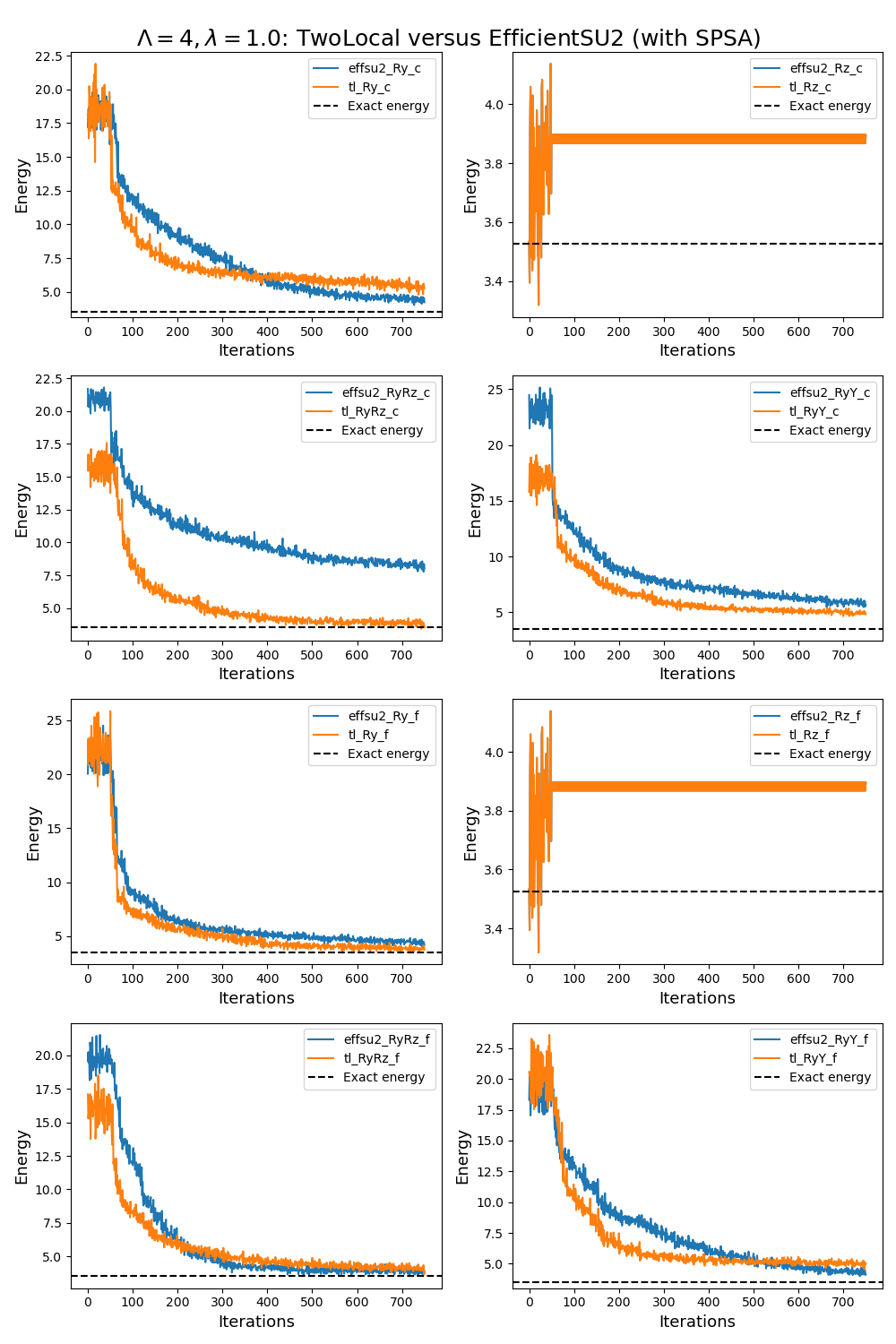}
\caption{VQE experiments $\lf(H^{\L=4}_{\l=1.0}, \texttt{EfficientSU2} \& \texttt{TwoLocal}, \text{SPSA}\rr)$: Comparison of the performances of \texttt{TwoLocal} circuits and \texttt{EfficientSU2}, variant by variant using SPSA optimizer. Apart from \texttt{tl\_Ry\_c} (first row, left subfigure) and \texttt{tl\_RyY\_f} ($4^\text{th}$ row, right subfigure), the remaining 6 variants of \texttt{TwoLocal} outperform or are on par with the corresponding 8 variants of \texttt{EfficientSU2}, as is evident from the orange line representing the \texttt{TwoLocal} variant converges at a lower/the same value than/as the blue line representing the \texttt{EfficientSU2} variant. Both \texttt{TwoLocal} \& \texttt{EfficientSU2} variants involving $R_Z$ rotation block fail to be optimized with SPSA as their convergence curves are just straight lines coinciding with each other ($1^\text{st}$ row \& $3^\text{rd}$ row, right subfigure).}
\label{fig_L4_l10_tl_vs_es_spsa}
\end{figure}
\begin{figure}[!ht]
\centering
\includegraphics[width=.8\textwidth]{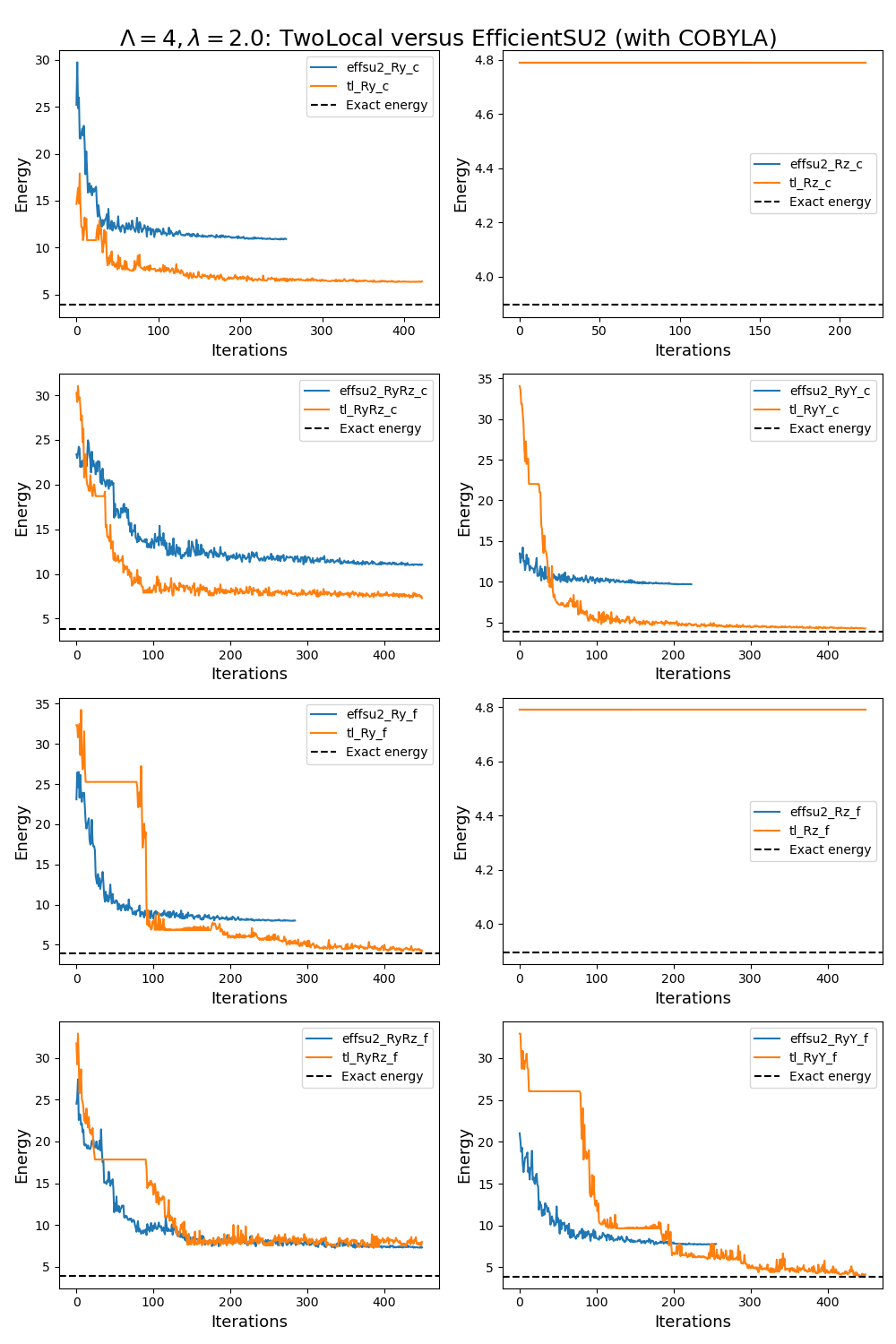}
\caption{VQE experiments $\lf(H^{\L=4}_{\l=2.0}, \texttt{EfficientSU2} \& \texttt{TwoLocal}, \text{COBYLA}\rr)$: Comparison of the performances of \texttt{TwoLocal} circuits and \texttt{EfficientSU2}, variant by variant using COBYLA optimizer. All 8 variants of \texttt{TwoLocal} outperform or are on par with the corresponding 8 variants of \texttt{EfficientSU2}, as is evident from the orange line representing the \texttt{TwoLocal} variant converges at a lower/the same value than/as the blue line representing the \texttt{EfficientSU2} variant. Both \texttt{TwoLocal} \& \texttt{EfficientSU2} variants involving $R_Z$ rotation block fail to be optimized with COBYLA as their convergence curves are just straight lines coinciding with each other ($1^\text{st}$ row \& $3^\text{rd}$ row, right subfigure).}
\label{fig_L4_l20_tl_vs_es_cobyla}
\end{figure}
\begin{figure}[!ht]
\centering
\includegraphics[width=.8\textwidth]{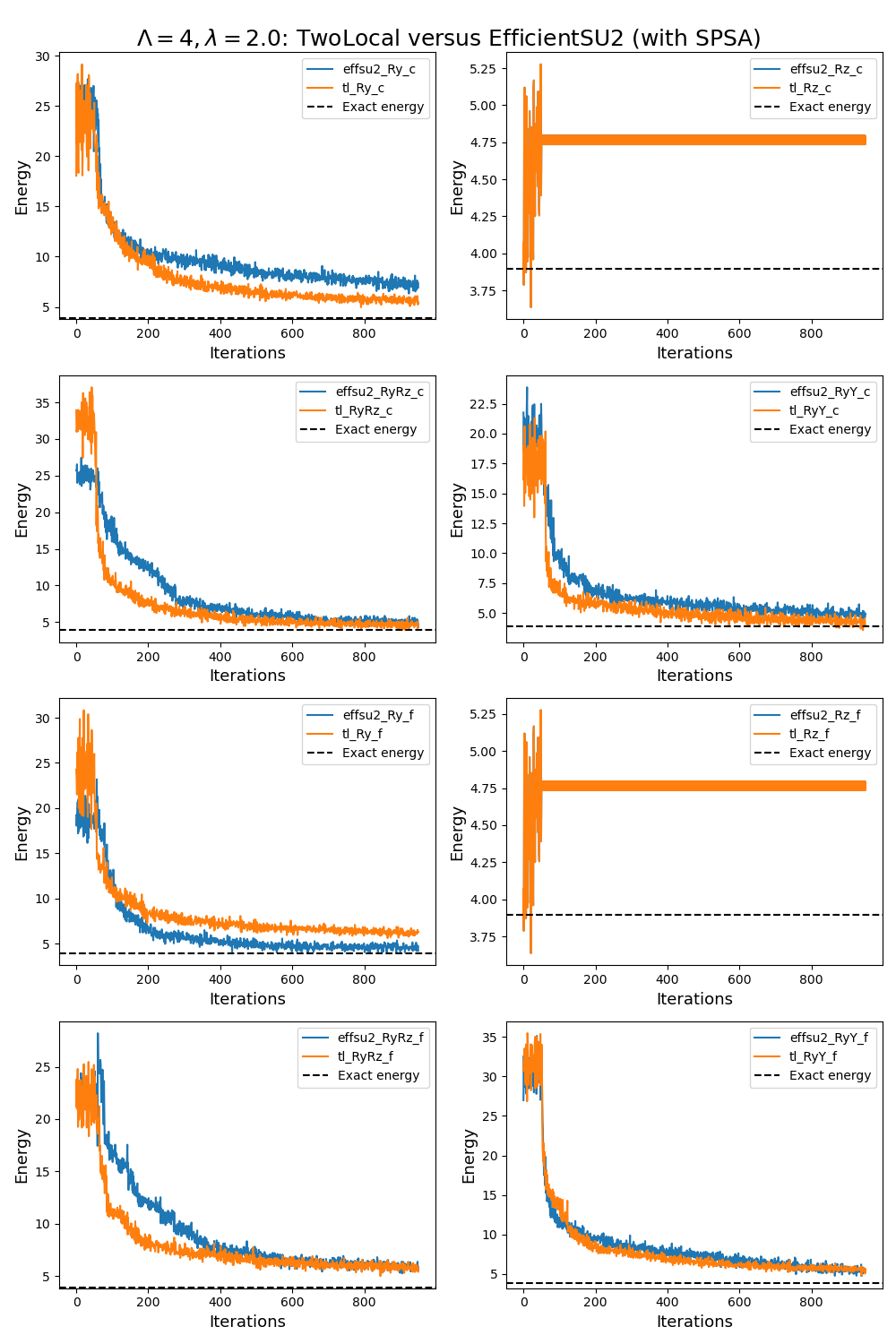}
\caption{VQE experiments $\lf(H^{\L=4}_{\l=2.0}, \texttt{EfficientSU2} \& \texttt{TwoLocal}, \text{SPSA}\rr)$: Comparison of the performances of \texttt{TwoLocal} circuits and \texttt{EfficientSU2}, variant by variant using SPSA optimizer. Apart from  \texttt{tl\_Ry\_f} ($3^\text{rd}$ row, left subfigure), the remaining 7 variants of \texttt{TwoLocal} outperform or are on par with the corresponding 8 variants of \texttt{EfficientSU2}, as is evident from the orange line representing the \texttt{TwoLocal} variant converges at a lower/the same value than/as the blue line representing the \texttt{EfficientSU2} variant. Both \texttt{TwoLocal} \& \texttt{EfficientSU2} variants involving $R_Z$ rotation block fail to be optimized with SPSA as their convergence curves are just straight lines coinciding with each other ($1^\text{st}$ row \& $3^\text{rd}$ row, right subfigure).}
\label{fig_L4_l20_tl_vs_es_spsa}
\end{figure}
\FloatBarrier
\clearpage


\end{document}